\pdfoutput=1

\documentclass[12pt]{article}
\usepackage{amssymb,amsmath,amsthm,epsfig,a4,verbatim,pstricks,ifthen,setspace} 
\usepackage{url,ifpdf}
\usepackage[ruled]{algorithm2e}

\definecolor{mygreen}{rgb}{0.05,0.6,0.05}

\newcommand{\R}{{\mathbb R}}

\newcommand{\dx}{\; {\rm d}x}
\newcommand{\ds}{\; {\rm d}s}
\renewcommand{\vec}{}
\newcommand{\unitn}{\vec{\rm n}}
\newcommand{\nabs}{\nabla_{\!s}}
\def\conduct{\mathcal{K}}

\newcommand{\dd}[1]{\frac{\rm d}{{\rm d}#1}}
\newcommand{\ddt}{\dd{t}}

\def\epsilon{\varepsilon} 
\newcommand{\uD}{u_D}
\newcommand{\cPsi}{c_\Psi}
\newcommand{\Varrho}{{\rm P}}

\newcommand{\PFEM}{PFEM}
\newcommand{\PF}{PF$^{\rm obs}$-FEM}
\newcommand{\PFi}{PF$_{\rm (i)}^{\rm obs}$-FEM}
\newcommand{\PFii}{PF$_{\rm (ii)}^{\rm obs}$-FEM}
\newcommand{\PFiii}{PF$_{\rm (iii)}^{\rm obs}$-FEM}

\newcommand{\cPFiii}{$\widehat{\rm PF}\rule{0pt}{0pt}^{\rm obs}_{\rm (iii)}$-FEM}
\newcommand{\PFq}{PF$^{\rm qua}$-FEM}
\newcommand{\PFqi}{PF$_{\rm (i)}^{\rm qua}$-FEM}

\newcommand{\PFqiii}{PF$_{\rm (iii)}^{\rm qua}$-FEM}

\newcommand{\cPFqiv}{$\widehat{\rm PF}\rule{0pt}{0pt}_{\rm (iv)}^{\rm qua}$-FEM}

\newcommand{\errorRr}[2]{\|r_{#1}^h - r_{#2}\|_{L^\infty}}

\newcommand{\errorWu}{\|w^h - u\|_{L^\infty}}
\newcommand{\LerrorWu}{\|w^h - u\|_{L^2}}

\newcommand{\errorUu}{\|u^h - u\|_{L^\infty}}
\newcommand{\LerrorUu}{\|u^h - u\|_{L^2}}
\newenvironment{AMS}%
{{\upshape\bfseries AMS subject classifications. }\ignorespaces}{}
\newenvironment{keywords}{{\upshape\bfseries Key words. }\ignorespaces}{}

\def\vL{L\kern-0.08cm\char39}

\begin{document}
\title{
Phase Field Models versus \\ Parametric Front Tracking Methods: \\
Are they accurate and computationally efficient?
}

\author{John W. Barrett\footnotemark[2] \and 
        Harald Garcke\footnotemark[3]\ \and 
        Robert N\"urnberg\footnotemark[2]}

\renewcommand{\thefootnote}{\fnsymbol{footnote}}
\footnotetext[2]{Department of Mathematics, 
Imperial College London, London, SW7 2AZ, UK}
\footnotetext[3]{Fakult{\"a}t f{\"u}r Mathematik, Universit{\"a}t Regensburg, 
93040 Regensburg, Germany}

\date{}

\maketitle

\begin{abstract}
We critically compare the practicality and accuracy of numerical approximations
of phase field models and sharp interface models of solidification. Particular
emphasis is put on Stefan problems, and their quasi-static variants, with 
applications to crystal growth.
New approaches with a high mesh quality for the parametric
approximations of the resulting free boundary problems and new stable
discretizations of the anisotropic phase field system are taken into
account in a comparison involving benchmark problems based on exact
solutions of the free boundary problem. 
\end{abstract} 



\renewcommand{\thefootnote}{\arabic{footnote}}

\section{Introduction} \label{sec:0}

The solidification of a liquid or the melting of a solid lead to 
complex free boundary problems involving many different physical
effects. For example, latent heat is set free at the interface and a
competition between surface energy and diffusion leads to instabilities
like the Mullins--Sekerka instability. The resulting model is a Stefan
problem with boundary conditions taking surface energy effects and
kinetic effects at the interface into account, see
e.g.\ \cite{Gurtin93, Davis01}. Crystals forming in an undercooled melt
lead to very complex patterns and, in particular, dendritic growth can
be observed since the growth is typically diffusion limited, see
\cite{BenJacob93}.

The numerical simulation of time-dependent Stefan problems, or
variants of it, is a formidable task since the evolving free boundary
has to be computed. 
Hence, direct front tracking type numerical methods need to 
adequately capture the geometry of the interface and to evolve the interface
approximation, often with a coupling to other physical fields. This
coupling, in particular, represents a significant initial hurdle towards
obtaining practical implementations, and thus numerical simulations for the
problem at hand. Examples of the 
implementation of such direct methods can be found in e.g.\ 
\cite{KesslerKL84,Strain89,BanschS92,Almgren93,Schmidt93,RoosenT94,Schmidt96,%
JuricT96,Schmidt98,dendritic}.

A further drawback of direct front tracking methods has been the
inability of most direct methods to deal with so-called mesh effects, or to
prevent them altogether. When a discrete approximation of an interface,
for example a polygonal curve in the plane, evolves in time, then in general it
is possible for the approximation to deteriorate or to break down. Examples of
such pathologies are self-crossings and vertex coalescence. While for simple
isotropic problems in the plane these issues can be dealt with, for example by
frequent remeshings or by using clever formulations that only allow
equidistributed approximations, see e.g.\ \cite{KesslerKL84,Strain89}, until
very recently there has been no remedy for fully anisotropic problems in two
and three space dimensions.

However, building on their work for isotropic problems in 
\cite{triplej,triplejMC,gflows3d}, the present authors recently introduced 
stable parametric finite element schemes for the direct approximation of
anisotropic geometric evolution equations in \cite{triplejANI,ani3d}, 
for which good mesh properties can be guaranteed. In particular, even for the
simulation of interface evolutions in the presence of strong anisotropies, no
remeshing or redistribution of vertices is needed in practice. These schemes,
in which only the interface without a coupling to bulk quantities is modelled,
have been extended to approximations of the Stefan problem with
fully anisotropic Gibbs--Thomson law and kinetic undercooling in
\cite{dendritic}. The novel method from \cite{dendritic} can be shown to be
stable and to have good mesh properties. We remark that these
approaches extend earlier ideas from \cite{Dziuk91,Schmidt93,Schmidt96}. 
Here we recall the pioneering work of Schmidt \cite{Schmidt93,Schmidt96},
where the full Stefan problem in three dimensions was solved within a 
sharp interface framework for the first time.

Phase field methods are an alternative approach to solve
solidification phenomena in the framework of continuum modelling. In
phase field approaches a new non-conserved order parameter $\varphi$
is introduced, which in the two phases is close to two different prescribed
values and which smoothly changes its value across a small diffuse
interfacial region. 
A parabolic partial differential equation for $\varphi$ is then
coupled to an energy balance, which results in a diffusion equation for
the temperature taking latent heat effects into account. We refer to
\cite{Langer86,CollinsL85,Caginalp86,PenroseF90,WangSWMCM93} and
to the five review articles
\cite{BoettingerWBK02,Chen02,McFadden02,SingerLoginovaS08,Steinbach09} 
for further details. 
In particular, it can be shown that solutions to the phase field
equations converge to classical sharp interface problems, see e.g.\ 
\cite{Caginalp89,AlikakosBC94,Soner95,CaginalpC98} and the references
therein. 

The popularity of phase field methods, often also called diffuse interface
methods, can be
explained by two characteristic features that they share with the level set
method, which is another sharp interface computational tool. 
Firstly, phase field methods
naturally allow for changes of topology. And secondly, computing simulations
for phase field models only requires the solution of partial differential
equations on standard Cartesian domains. 
The fact that these can usually be implemented and solved in a
relatively straightforward
way makes the phase field method particularly appealing.

It is the aim of this article to investigate the practical merits of
phase field models compared to the recently introduced sharp interface 
algorithm for the approximation of Stefan problems from \cite{dendritic};
see also
\cite{crystal,jcg}. Of particular interest will be the relative accuracy of the
two methods, in situations where a true solution to the sharp interface problem
is known. In a phase field simulation the observed error is made up of
contributions from
\begin{itemize}
\item the asymptotic error,
\item the spatial and temporal discretization errors,
\item rounding errors and solver tolerances.
\end{itemize}
Here the asymptotic error is induced by the choice of interfacial parameter
$\epsilon>0$. In general one can formally show that the asymptotic error
converges to zero as $\epsilon \to 0$, see e.g.\ \cite{Caginalp86}. 
For certain phase field models and under certain conditions it
can be rigorously shown that the asymptotic error vanishes as $\epsilon\to0$,
see e.g.\ \cite{CaginalpC98}. 
In computations for sharp interface approximations, on the other hand, the
observed error is made up of the latter two contributions only, i.e.\ of 
discretization and rounding errors. A disadvantage of phase field models is
that the resulting PDEs become stiff for decreasing $\epsilon$, leading
to a requirement for very fine spatial and temporal discretizations. Hence it
becomes computationally challenging to produce very accurate phase field
simulations. In any case, the available computational resources will often set
a limit on the smallest interfacial parameter $\epsilon$ that one can compute
for.
Hence another aspect that needs to be taken into account in an objective
comparison between phase field simulations and sharp interface approximations
is the overall CPU time that is needed to obtain the results. While it can
often be formally shown that phase field computations can attain an arbitrarily
high accuracy, the existing limitations on computer hardware often mean that
in practice very fine computations cannot be performed. In addition,
as discussed in \cite{KarmaR96}, the early computational approaches
were limited as they could only be used in the presence of interfacial kinetics
in the Gibbs--Thomson law.

Historically these limitations of the phase field model have long been known,
and as a result a different underlying interpretation of the 
model, the so-called ``thin
interface limit'', has been introduced and analyzed 
by Karma and Rappel \cite{KarmaR96,KarmaR98}.
Their approach 
made it possible to do
efficient computations with a smaller capillary length to interface
thickness ratio, and to study the physically relevant case of small or
zero kinetic coefficients. Later the findings of Karma and Rappel were
reinterpreted as second order convergence with respect to the
interfacial thickness, see \cite{Almgren99,GarckeS06,ChenCE06}. 

The first successful phase field computations of dendritic growth were
performed by Kobayashi \cite{Kobayashi93}, and his computations demonstrated
the importance of anisotropy for dendritic growth. 
Since then many successful improvements with respect to
numerical simulations have appeared in the literature. We refer to
\cite{ElliottG96preprint,WheelerMS93,KarmaR96,ProvatasGD98,Nestler05}
and the references therein.

Finally, we would like to mention work on the numerical analysis of
phase field and sharp interface approaches. 
Numerical analysis of discretizations of phase field models can be found in 
e.g.\
\cite{CaginalpL87,FixL88,Lin88,ChenH94,Yue96,FengP04a,BartelsM10}.
Numerical analysis of discretizations of sharp interface models can be found in
\cite{Veeser99,dendritic,crystal}.  
We also remark that level set methods are another possible way to
solve Stefan problems and related free boundary problems. We refer to
\cite{OsherF03,Sethian} for more details on how the level
set method can be used to solve free boundary problems. 

The remainder of the paper is organized as follows.  In
Section~\ref{sec:2} we state the sharp interface formulation of the
two phase Stefan problem with kinetic undercooling and an anisotropic
Gibbs--Thomson law. In Section~\ref{sec:3} we state the corresponding
phase field model and recall the finite element algorithms from
\cite{vch}. In Section~\ref{sec:4} we numerically compare the sharp
interface method from \cite{dendritic} with the phase field algorithms
from Section~\ref{sec:3} for some isotropic benchmark problems with
known true solutions. Computations for a phase field model with a
correction term, for which a second order convergence property can
formally be shown, are presented in Section~\ref{sec:43}.
Finally, we compare the sharp interface and phase field methods
for a variety of anisotropic model problems in Section~\ref{sec:5}.

\setcounter{equation}{0} 
\section{Sharp interface problem} \label{sec:2}

In this paper we concentrate on interfacial problems in materials
science in which one driving force is due to capillary effects. In
applications the interface often separates a solid and a liquid phase,
say, or a solid phase and a gas phase. Let $\Gamma(t) \subset \R^d$,
$d=2,3$, denote this sharp interface.
Then the surface energy of $\Gamma(t)$ is defined as 
\begin{equation} \label{eq:EGamma}
\int_{\Gamma(t)} \gamma(\unitn) \ds \,,
\end{equation} 
where $\unitn$ denotes the unit normal of $\Gamma(t)$, and where 
the anisotropic density function $\gamma : \R^d \to \R_{\geq 0}$ with 
$\gamma\in C^2(\R^d \setminus \{ \vec 0 \}) \cap C(\R^d)$ is assumed to be 
absolutely homogeneous of degree one, i.e.\
\begin{equation*}
\gamma(\lambda\,\vec{p}) = |\lambda|\gamma(\vec{p}) \quad \forall \
\vec{p}\in \R^d,\ \forall\ \lambda \in\R \quad \Rightarrow
\quad \gamma'(\vec{p})\,.\,\vec{p} = \gamma(\vec{p})
\quad \forall\ \vec{p}\in \R^d\setminus\{\vec0\}, 
\end{equation*}
with $\gamma'$ denoting the gradient of $\gamma$. For all the considerations in
this paper we assume that $\gamma$ is of the class of anisotropies first
introduced by the authors in \cite{triplejANI,ani3d}; see also
\cite{dendritic,vch}.

Relevant for our considerations is the first variation, $-\kappa_\gamma$, of
(\ref{eq:EGamma}), which can be computed as
\begin{equation*}
\kappa_\gamma := - \nabs\,.\, \gamma'(\unitn)\,;
\end{equation*}
where $\nabs .$ is the tangential divergence of $\Gamma$, see e.g.\
\cite{CahnH74,ani3d,dendritic}.
Note that $\kappa_\gamma$ reduces to $\kappa$, 
the sum of the principal curvatures of $\Gamma$, 
in the isotropic case, i.e.\ when $\gamma$ satisfies
\begin{equation} \label{eq:iso}
\gamma(\vec p) = |\vec p| \qquad \forall\ \vec p \in \R^d\,.
\end{equation}

\subsection{Stefan problem}

Then the full Stefan problem we want to consider in this paper is given as
follows, where $\Omega\subset\mathbb{R}^d$ is a given fixed domain with
boundary $\partial\Omega$ and outer normal $\vec\nu$.
Find $u : \Omega \times [0,T] \to \R$ 
and the interface $(\Gamma(t))_{t\in[0,T]}$ such that
for all $t\in (0,T]$ the following conditions hold:
\begin{subequations}
\begin{alignat}{2}
\vartheta\,u_t - \conduct_-\,\Delta u & = 0 
\qquad \mbox{in } \Omega_-(t)\,, \qquad
\vartheta\,u_t - \conduct_+\,\Delta u = 0 
\qquad &&\mbox{in } \Omega_+(t)\,, \label{eq:1a} \\
\left[ \conduct\,\frac{\partial u}{\partial \unitn} \right]_{\Gamma(t)}
&=-\lambda\,{\cal V} \qquad &&\mbox{on } \Gamma(t)\,, 
\label{eq:1b} \\ 
\frac{\rho\,{\cal V}}{\beta(\unitn)} &= \alpha\,\kappa_\gamma - 
a\,u \qquad &&\mbox{on } \Gamma(t)\,, 
\label{eq:1c} \\ 
\frac{\partial u}{\partial \vec\nu} &= 0 
\qquad \mbox{on } \partial_N\Omega, \qquad
u = \uD \qquad &&\mbox{on }
\partial_D \Omega \,, \label{eq:1d} \\
\Gamma(0) & = \Gamma_0 \,, \qquad\qquad 
\vartheta\,u(\cdot,0) = \vartheta\,u_0 \quad && \mbox{in } 
\Omega \,. \label{eq:1e} 
\end{alignat}
\end{subequations}
In the above $u$ denotes the deviation from the melting temperature $T_M$,
i.e.\ $T_M$ is the melting temperature for a planar interface.
In addition, $\Omega_-(t)$ is the solid region, with boundary
$\Gamma(t) = \partial \Omega_-(t)$, so that the liquid region is given 
by $\Omega_+(t):=\Omega\setminus\overline{\Omega_-(t)}$. 
Moreover, here and throughout this paper, for
a quantity $v$ defined on $\Omega$, we use the shorthand notations
$v_- := v\!\mid_{\Omega_-}$ and $v_+ := v\!\mid_{\Omega_+}$.
The parameters $\vartheta\geq0$, $\lambda>0$, $\rho\geq0$, $\alpha>0$, $a>0$ 
are assumed to be constant, while $\conduct>0$ is
assumed to be constant in each phase. 
The mobility coefficient $\beta:\R^d \to \R_{\geq0}$ is assumed to satisfy
$\beta(\vec p) > 0$ for all $\vec p \not= \vec 0$ and to be positively
homogeneous of degree one. We note that in the isotropic case (\ref{eq:iso}) it
is often also assumed that
\begin{equation} \label{eq:betaiso}
\beta(p) = |p| \qquad \forall\ p \in \R^d \quad\Rightarrow\quad
\beta(\unitn) = 1\,.
\end{equation}
In addition 
$[\conduct\,\frac{\partial u}{\partial \unitn}]_{\Gamma(t)}(\vec{z}) := 
(\conduct_+\,\frac{\partial u_+}{\partial \unitn} - 
\conduct_-\,\frac{\partial u_-}{\partial \unitn})
(\vec{z})$ for all $\vec{z}\in\Gamma(t)$, and
$\mathcal{V}$ is the velocity of $\Gamma(t)$ in the direction of its normal
$\unitn$, which from now on we assume is pointing into $\Omega_+(t)$. 
Finally, $\partial\Omega =
\overline{\partial_N\Omega}\cup\overline{\partial_D\Omega}$ with
$\partial_N\Omega\cap\partial_D\Omega = \emptyset$, 
$\uD \in \R_{\leq 0}$ is the applied supercooling at the boundary,
and $\Gamma_0 \subset \overline\Omega$ and
$u_0 : \Omega \to \R$ are given initial data. Here we use the convention that
$\uD = 0$ if $\partial\Omega = \partial_N\Omega$.

The model (\ref{eq:1a}--e) can be derived for example
within the theory of rational thermodynamics and we refer to
\cite{Gurtin88a} for details. We remark that a derivation from thermodynamics
would lead to the identity $a = \frac{\lambda}{T_M}$.
We note that
(\ref{eq:1b}) is the well-known Stefan condition, while (\ref{eq:1c}) is the
Gibbs--Thomson condition, with kinetic undercooling if $\rho>0$. The
case $\vartheta>0$, $\rho>0$, $\alpha >0$ leads to the Stefan
problem with the Gibbs--Thomson law and kinetic undercooling.
In some models in the literature, see e.g.\ \cite{Luckhaus90}, the
kinetic undercooling is set to zero, i.e.\ $\rho=0$. Setting
$\vartheta=\rho=0$ but keeping $\alpha>0$ leads to the
Mullins--Sekerka problem with the Gibbs--Thomson law, see
\cite{MullinsS63}. 

We recall from \cite{dendritic} that 
for a solution $u$ and $\Gamma$ to
(\ref{eq:1a}--e) it can be shown that the following equality holds
\begin{equation}
 \ddt\mathcal{F}(\Gamma, u) = - (\conduct\,\nabla\,u, \nabla\,u) 
-\frac{\lambda\,\rho}{a}\, \int_{\Gamma(t)} \frac{{\cal V}^2}{\beta(\unitn)} \ds
\leq  0 \,, \label{eq:testD}
\end{equation}
where
\begin{equation} \label{eq:F}
\mathcal{F}(\Gamma, u) :=
\frac\vartheta2\,|u-\uD|^2_0 + 
\frac{\lambda\,\alpha}{a}\, \int_{\Gamma(t)} \gamma(\unitn)\ds
-\lambda\,\uD\,|\Omega_+(t)|
\end{equation}
and where $(\cdot,\cdot)$ denotes the $L^2$--inner product
over $\Omega$, with the corresponding norm given by $|\cdot|_0$, and where
$|\Omega_+(t)| := \int_{\Omega_+(t)} 1 \dx$.

\subsection{Parametric method \PFEM}

Traditional front tracking methods for sharp interface problems had a
major drawback, in that the meshes used to describe the interface seriously
deteriorated during the evolution. In addition, introducing 
mesh smoothing during the evolution
is difficult, see e.g.\ the discussion in \cite{Schmidt96}. For 
interfaces in the plane 
it is possible to formulate a non-trivial method such that
mesh points are nearly equally distributed during the evolution, see
\cite{HouLS94,MikulaS01}. The
present authors introduced a novel parametric finite element method for 
problems involving curves and surfaces evolving in time, which has a simple
variational structure and which leads to good mesh properties, see
\cite{triplej,triplejMC,gflows3d}. In fact, 
for curves a semi-discrete variant leads to equally
distributed mesh points in the isotropic case, while in the general anisotropic
setting equidistribution with respect to some anisotropic weight function is
obtained, see \cite{triplejANI}. 
For surfaces the resulting meshes have also
good properties, which in the isotropic case can be explained by using ideas 
from conformal geometry. In particular, no remeshing is needed during the
evolution, even in the general anisotropic situation. 
An example triangulation obtained during the simulation of dendritic
growth in three space dimensions can be seen in Figure~\ref{fig:mesh}, below.
In addition, as the mesh for the parameterization of 
the interface is decoupled from the bulk mesh, no 
deformation of the bulk mesh is required 
in order to contain the interface at predefined locations on it. 

The novel 
and stable parametric finite element approximation of (\ref{eq:1a}--e)
in the case $\conduct_+ = \conduct_- > 0$ has been introduced by the
present authors in \cite{dendritic}, and this scheme has been extended
to the more general case $\conduct_\pm \geq 0$ in
\cite{crystal}. Throughout this paper we will refer to these variants
as \PFEM. The algorithm \PFEM\ features the discretization parameters
$h_\Gamma$, $h_f$, $h_c$ and $\tau$. Here $h_\Gamma$ refers to the
fineness of the triangulated approximation of $\Gamma(t)$, for which
isoparametric piecewise linear finite elements are employed. In particular,
a simple mesh refinement strategy allows for the natural growth of the
interface, i.e.\ elements of the triangulated approximation of $\Gamma(t)$ are
refined when they become too large.
Moreover, the temperature in the bulk is approximated with standard
continuous piecewise linear finite elements, and $h_f$ and $h_c$ refer to bulk
mesh parameters for fine regions close to the interface and coarser
regions far away from it. For all the computations presented in this
paper we fix $h_c = 8\,h_f$ and, unless stated otherwise, 
we let $h_f \approx h_\Gamma$. Finally,
$\tau$ denotes a uniform time step size.  The linear discrete systems
of equations are solved with preconditioned conjugate gradient solvers
of suitable Schur complement formulations.  We refer to
\cite{dendritic,crystal} for more details. As indicated earlier, no
remeshing of the discrete interface is necessary for the scheme \PFEM,
and all the numerical results presented in this paper for this scheme
are performed without any redistribution of mesh points.

\setcounter{equation}{0} 
\section{Phase field model} \label{sec:3}

We now state the phase field model that we are going to consider in this paper. 
To this end, for $\vec p \in \R^d$, let
\begin{equation*}
A(\vec p) = \tfrac12\,|\gamma(\vec p)|^2 
\quad\Rightarrow\quad
A'(\vec p) = 
\begin{cases}
\gamma(\vec p)\,\gamma'(\vec p) & \vec p \not= 0\,, \\
\vec 0 & \vec p = \vec 0\,,
\end{cases} 
\end{equation*}
and define
\begin{equation} \label{eq:mu}
\mu(\vec p) = \begin{cases}
\dfrac{\gamma(\vec p)}{\beta(\vec p)} & \vec p \not = \vec 0\,, \\
\bar\mu & \vec p = \vec 0\,,
\end{cases}
\end{equation}
where $\bar\mu\in\R_{>0}$ is a constant satisfying
$\min_{\vec{p} \not= \vec 0} \frac{\gamma(\vec p)}{\beta(\vec p)}
\leq \bar\mu \leq 
\max_{\vec{p} \not= \vec 0} \frac{\gamma(\vec p)}{\beta(\vec p)}$.

Moreover, let $\varphi:\Omega \times (0,T) \to \R$
be the phase field variable, so that the sets
$\{ x \in \Omega : \pm \varphi(x,t) > 0\}$ are approximations to
$\Omega_\pm(t)$, with the zero level set of $\varphi(\cdot,t)$ approximating
the interface $\Gamma(t)$.
On introducing the small interfacial parameter $\epsilon>0$, it can be shown
that
$$
\frac1\cPsi\,\mathcal{E}_\epsilon(\varphi) \approx 
\int_{\Gamma} \gamma(\unitn) \ds\,,
$$
for $\epsilon$ sufficiently small, where
\begin{equation} \label{eq:Eg}
{\cal E}_\epsilon(\varphi) := 
\int_\Omega \tfrac\epsilon2\,|\gamma(\nabla\, \varphi)|^2 +
\epsilon^{-1}\,\Psi(\varphi) \dx
\quad \text{with} \quad
\cPsi := \int_{-1}^1 \sqrt{2\,\Psi(s)}\;{\rm d}s\,.
\end{equation}
Here $\Psi : \R \to [0,\infty]$ is a double well potential, which 
we assume to be symmetric and to have its global minima at $\pm1$.
The canonical example is
\begin{equation} \label{eq:quartic}
\Psi(s) := \tfrac14\,(s^2 - 1)^2
\qquad\Rightarrow\qquad
\Psi'(s) = s^3-s
\quad \text{and}\quad
\cPsi = \tfrac13\,{2^{\frac32}}\,.
\end{equation}
Another possibility is to choose
\begin{equation} \label{eq:obstacle}
\Psi(s):= \begin{cases}
\textstyle \frac 12 \left(1-s^2\right)  & |s|\leq 1\,,\\
\infty & |s|> 1\,,
\end{cases} 
\qquad\Rightarrow\qquad
\cPsi = \tfrac\pi2\,;
\end{equation}
see e.g.\ \cite{BloweyE92,BloweyE94,ElliottG96preprint,Elliott97}. 
Clearly the obstacle potential (\ref{eq:obstacle}),
which in contrast to the smooth quartic potential (\ref{eq:quartic}) 
forces $\varphi$ to stay within the interval $[-1,1]$,
is not differentiable at $\pm1$. 
Hence, whenever we write $\Psi'(s)$ in the case (\ref{eq:obstacle}) in this
paper, we mean that the
expression holds only for $|s|<1$, and that in general a variational inequality
needs to be employed.

Our phase field model for (\ref{eq:1a}--e) is then given by the coupled system
\begin{subequations}
\begin{alignat}{2} \label{eq:heata} 
\vartheta\,w_t + \lambda\,\varrho(\varphi)\,\varphi_t -
\nabla \,.\, (b(\varphi)\,\nabla\, w ) & = 0 \qquad && \mbox{in } 
\Omega_T:=\Omega\times(0,T)\,, \\
w & = \uD 
&& \mbox{on}\;\;\partial_D \Omega\times(0,T)\,, \label{eq:heatb} \\
b(\varphi)\,\frac{\partial w}{\partial \vec\nu} & = 0 \qquad
&& \mbox{on}\;\;\partial_N \Omega\times(0,T)\,, \label{eq:heatc} \\
\vartheta\,w(\cdot,0) & = \vartheta\,w_0 \quad && \mbox{in } 
\Omega \,, \label{eq:heatd} 
\end{alignat}
\end{subequations}
with
\begin{subequations}
\begin{alignat}{2}
\cPsi\,\frac{a}\alpha\,\varrho(\varphi)\,w & = 
\epsilon\,\frac\rho\alpha\,\mu(\nabla\,\varphi)\,\varphi_t
-\epsilon\,\nabla \,.\, A'(\nabla\, \varphi) +\epsilon^{-1}\,\Psi'(\varphi) 
\qquad && \mbox{in} \;\;\Omega_T\,, \label{eq:ACa} \\ 
\frac{\partial \varphi}{\partial \vec\nu} & = 0 \qquad
&& \mbox{on}\;\;\partial \Omega\times(0,T)\,, \label{eq:ACb} \\
\qquad \varphi(\cdot,0) & = \varphi_0 && \mbox{in} \;\;\Omega\,, \label{eq:ACc}
\end{alignat}
\end{subequations}
where  
\begin{equation*} 
b(s) = \tfrac12\,(1+s)\,\conduct_+ + \tfrac12\,(1-s)\,\conduct_-\,,
\end{equation*}
and where the function $\varrho \in C^1(\R)$ is such that
\begin{equation} \label{eq:Varrho}
\varrho(s) \geq 0\quad\forall\ s \in[-1,1]\,,\quad
\int^1_{-1} \varrho(y) \; {\rm d}y = 1 \quad \text{and} \quad
\Varrho(s) := \int^s_{-1} \varrho(y) \; {\rm d}y\,.
\end{equation}
We note that $\Varrho$, which is a monotonically increasing function over the 
interval $[-1,1]$ with $\Varrho(-1) = 0$ and $\Varrho(1) = 1$, 
is often called the interpolation
function. In this paper, we follow the convention 
from \cite{ElliottG96preprint}, where
$\varrho = \Varrho'$ is called the shape function. 
Possible choices of $\varrho$ that will be considered in this paper are
\begin{equation} \label{eq:varrho}
\text{(i)}\ 
\varrho(s) = \tfrac12\,,\quad
\text{(ii)}\ 
\varrho(s) = \tfrac12\,(1 - s) \,,\quad
\text{(iii)}\ 
\varrho(s) = \tfrac{15}{16}\,(s^2 - 1)^2\,,\quad
\text{(iv)}\ 
\varrho(s) = \tfrac34\,(1 - s^2) \,.
\end{equation}
More details on interpolation functions $\Varrho$, respectively
shape functions $\varrho$, can be found in e.g.\
\cite{WangSWMCM93,GarckeS06,CaginalpCE08}. 
In particular, if one also assumes that $\varrho$ is symmetric, i.e.\
\begin{subequations}
\begin{equation} \label{eq:varrhosym}
\varrho(s) = \varrho(-s) \qquad \forall\ s \in \R\,,
\end{equation}
and that
\begin{equation} \label{eq:varrhoroot}
\varrho(1) = \varrho(-1) = 0\,,
\end{equation}
\end{subequations}
then a faster rate of convergence of the phase field model to the sharp 
interface limit, as $\epsilon\to0$, can be shown in the isotropic 
case (\ref{eq:iso}),
(\ref{eq:betaiso}) on replacing $\rho$ in (\ref{eq:ACa}) with the first order
correction
\begin{equation} \label{eq:corrho}
\widehat\rho := \rho + \epsilon\,\rho_1\,,
\end{equation}
where $\rho_1$ is defined in (\ref{eq:rho1}) in Section~\ref{sec:43}, below.
The condition (\ref{eq:varrhoroot}) is one motivation for the latter two
choices in (\ref{eq:varrho}), with the choice (\ref{eq:varrho})(iv) also
satisfying the stronger condition (\ref{eq:varrhoPsi}), below, for the quartic
potential (\ref{eq:quartic}).  
An error analysis for a fully discrete approximation of the phase field model 
(\ref{eq:heata}--d), (\ref{eq:ACa}--c) with the quartic potential
(\ref{eq:quartic}) and the shape function (\ref{eq:varrho})(i) 
in the isotropic case (\ref{eq:iso}), (\ref{eq:betaiso}) 
with $\partial_N\Omega=\partial\Omega$ and $\conduct_+=\conduct_- > 0$
has been performed in \cite{FengP04a}. These authors also show convergence of
the phase field discretizations to the underlying sharp interface problem as
$\epsilon,\, h,\,\tau \to 0$,
where $h$ and $\tau$ denote the discretization parameters in space and time,
respectively. 
However, to our knowledge, no convergence rates are known for the convergence
of discretizations of the phase field model to the sharp interface problem
(\ref{eq:1a}--e).
Here we recall that for the much simpler situation of planar curvature flow,
as the sharp interface limit of the isotropic Allen--Cahn equation, such
convergence rates have been obtained in \cite{NochettoV97}. 
In particular, it can be shown that the
zero level sets of discretizations of
\begin{equation*} 
\epsilon\,\varphi_t = \epsilon\,\Delta\,\varphi - \epsilon^{-1}\,\Psi'(\varphi)
\end{equation*}
for the obstacle potential (\ref{eq:obstacle}) 
converge with $\mathcal{O}(\epsilon)$ to the sharp interface limit moving by
$\mathcal{V} = \kappa$ if 
\begin{equation} \label{eq:NV97a}
\tau = \mathcal{O}(h^2) = \mathcal{O}(\epsilon^4)\,.
\end{equation}
While no such result is known for the full phase field model
(\ref{eq:heata}--d), (\ref{eq:ACa}--c) even in the isotropic setting,
it is natural to expect constraints of the form (\ref{eq:NV97a}) in order to
observe $\mathcal{O}(\epsilon)$ in practice.

We remark that the phase field analogue of the sharp interface energy identity
(\ref{eq:testD}) is given by the formal energy bound
\begin{equation}\label{eq:pf2Lyap}
 \ddt\mathcal{F}_\epsilon(\varphi, w) 
= - (b(\varphi)\,\nabla\,w, \nabla w)
- \epsilon\,
\frac{\lambda\,\rho}{a}\,\frac1\cPsi
\left(\mu(\nabla\,\varphi), (\varphi_t)^2\right)  \leq 0
\end{equation}
for the phase field model (\ref{eq:heata}--d), (\ref{eq:ACa}--c), where
\begin{equation} \label{eq:Fg}
\mathcal{F}_\epsilon(\varphi, w) := 
\frac\vartheta2\,|w - \uD|_0^2 +
\frac{\lambda\,\alpha}{a}\,\frac1\cPsi\,\mathcal{E}_\epsilon(\varphi) - 
\lambda\,\uD\,\int_\Omega \Varrho(\varphi) \dx\,.
\end{equation}
Phase field models that satisfy such an inequality, in analogy to the sharp
interface energy identity (\ref{eq:testD}), are often called thermodynamically
consistent, see \cite{PenroseF90,WangSWMCM93,GarckeS06}.

\subsection{Phase field methods \PF\ and \PFq} \label{sec:31}

Unconditionally stable, fully practical finite element approximations of 
(\ref{eq:heata}--d), \linebreak \mbox{(\ref{eq:ACa}--c)} 
with either (\ref{eq:quartic}) or (\ref{eq:obstacle}) 
have been introduced by the authors in \cite{vch}.
Here stable means that they satisfy a discrete analogue of the formal energy
bound (\ref{eq:pf2Lyap}). 
Throughout this paper we will refer to the approximations
from \cite{vch} for (\ref{eq:quartic}) and (\ref{eq:obstacle}) 
as \PFq\ and \PF, 
respectively, 
where the inclusion of a subscript refers to
the choice of shape function in (\ref{eq:varrho}), e.g.\ \PFi.
We recall from \cite{vch} that 
a side effect of the interpolation function $\Varrho$ in (\ref{eq:Fg})
is that the function
\begin{equation*} 
G(s) = \alpha\,(a\,\cPsi\,\epsilon)^{-1}\,\Psi(s) - \uD\,\Varrho(s)
\end{equation*}
need no longer have local minima at $s = \pm1$
if $\uD \not=0$. This can result, for example, 
in undesired, artificial boundary layers for strong supercoolings, i.e.\ when
$-\uD$ is large.
For the smooth potential $\Psi$ from (\ref{eq:quartic}), 
sufficient conditions for $s=\pm1$ to be local minimum points of $G(s)$ are 
\begin{equation} \label{eq:quarticcond}
\varrho(\pm1)=\varrho'(\pm1)=0\,,
\end{equation}
which is evidently satisfied by (\ref{eq:varrho})(iii). 
In fact, in applications phase field models for solidification almost
exclusively use 
this shape function; see e.g.\ \cite{BoettingerWBK02,Chen02,McFadden02}. 
For the obstacle potential (\ref{eq:obstacle}) the situation is similar,
although there is more flexibility in the possible choices of $\varrho$. 
In particular, here a
sufficient condition for $G(s)$ to have local minima at $s=\pm1$ is given by
\begin{equation} \label{eq:obstcond}
\alpha\,(a\,\cPsi\,\epsilon)^{-1} \pm \uD\,\varrho(\pm1) \geq 0\,.
\end{equation}
On recalling that $\uD \leq 0$ we see that for (\ref{eq:obstcond}) to hold it
is sufficient to require that $\varrho(1) = 0$, which is evidently
satisfied by (\ref{eq:varrho})(ii), (\ref{eq:varrho})(iii) and 
(\ref{eq:varrho})(iv).
A major advantage of (\ref{eq:varrho})(ii)
over (\ref{eq:varrho})(iii) and (\ref{eq:varrho})(iv) 
is that for (\ref{eq:varrho})(ii) it is possible to
derive almost linear finite element approximations that are 
unconditionally stable. The corresponding unconditionally stable schemes for
the nonlinear shape functions (\ref{eq:varrho})(iii) and (\ref{eq:varrho})(iv),
on the other hand, turn out to be highly nonlinear. 
See \cite{vch} for more details.

We remark that even when (\ref{eq:quarticcond}) and (\ref{eq:obstcond}) are
satisfied for the potentials (\ref{eq:quartic}) and (\ref{eq:obstacle}),
respectively, it is possible that mushy interfacial regions are observed in
practice for approximations of the phase field model (\ref{eq:heata}--d),
(\ref{eq:ACa}--c); see e.g.\ Figure~\ref{fig:Stefanii_16pi}, below.
That is particularly the case in situations where the
instability of the moving free boundary is strong, i.e.\ when
$-\uD\,a\,\alpha^{-1}$ is large, recall (\ref{eq:F}) and see e.g.\ 
\cite{MullinsS63}.
Then $\epsilon$ needs to be chosen small, recall (\ref{eq:Fg}), 
so that the phase field variable $\varphi$ admits well-defined interfacial 
regions that approximate the sharp interface $\Gamma(t)$. This gives rise to a
formal constraint of the form
\begin{equation} \label{eq:epscond}
\epsilon \leq C\,\alpha\,(-\uD\,a)^{-1} \qquad \mbox{if }\ \uD < 0\,,
\end{equation}
for the choice of the interfacial parameter $\epsilon$ in terms of the physical
parameters for the sharp interface problem (\ref{eq:1a}--e), 
irrespective of the choice of $\varrho$. The reason for this is that
in the estimate (\ref{eq:pf2Lyap}) the double
well term $\epsilon^{-1}\,\int_\Omega \Psi(\varphi) \dx$
in $\mathcal{E}_\epsilon(\varphi)$ is for large $\epsilon$ not
strong enough to bound the unstable term involving $\Varrho(\varphi)$,
which encourages the growth of the diffuse interface.

The two algorithms \PF\ and \PFq,
which use continuous piecewise linear finite elements in space,
feature the
discretization parameters $h_f$, $h_c$ and $\tau$. Here $h_f$ and $h_c$ 
are mesh parameters for fine triangulations inside the diffuse interfacial 
region and coarser triangulations far away from it. Meaningful phase field
simulations need to resolve the interfacial regions, whose width is of the 
order $\epsilon$, and so a constraint of the form
\begin{equation} \label{eq:hf}
h_f \leq C\,\epsilon
\end{equation}
needs to be enforced. For (\ref{eq:obstacle}) the asymptotic interface width is
$\pi\,\epsilon$ in the isotropic case (\ref{eq:iso}),
and in this paper we always choose 
$h_f \leq \frac{\pi\,\epsilon}8$ with $h_c = 8\,h_f \leq \pi\,\epsilon$.
Unless otherwise stated we let $h_f = \frac{\pi\,\epsilon}8$.
Finally, $\tau$ denotes a uniform time step size. Here we recall that the
schemes \PF\ and \PFq\ employ a semi-implicit discretization in time, 
which utilizes convex/concave splittings of the nonlinearity arising from the
potential $\Psi$ and from the interpolation function $\Varrho$. Such a
splitting for $\Psi$ was first proposed in \cite{ElliottS93}, see also
\cite{BarrettBG99}, and the idea generalizes naturally to $\Varrho$;
see Section~\ref{sec:32}, below, for details.
This means that
for the shape function choices (\ref{eq:varrho})(i) and (\ref{eq:varrho})(ii)
almost linear schemes are obtained, while
the choices (\ref{eq:varrho})(iii) and (\ref{eq:varrho})(iv)
give rise to more nonlinear finite element approximations;
see \cite{vch} for details. 
The discrete systems of linear equations and variational inequalities 
arising from the schemes \PF\ are 
solved with the Uzawa-multigrid solver from \cite{voids3d}, while the systems
of nonlinear equations arising from \PFq\ are 
solved with a Newton method. We refer to \cite{vch} for more details.

For completeness we briefly describe the choice of the initial profile 
$\varphi_0$
in (\ref{eq:ACc}) in our numerical computations. Given the initial interface
$\Gamma_0$ from (\ref{eq:1e}), we let $d_0 : \Omega \to \R$ denote the signed
distance function of $\Gamma_0$. Then,
on recalling the asymptotic phase field profiles from e.g.\ \cite{Elliott97}, 
we define
\begin{subequations}
\begin{equation} \label{eq:phi0}
 \varphi_0(x) = 
 \Phi(\epsilon^{-1}\,d_0(x))\,,\quad \text{where}\quad
 \Phi(s) :=
\begin{cases} 
 -1 &  s \leq - \textstyle\frac{\pi}2\,, \\
 \sin(s) & |s| < \frac{\pi}2\,, \\
 1 & s \geq \textstyle\frac{\pi}2\,, \\
\end{cases}
\end{equation}
for the obstacle potential (\ref{eq:obstacle}),
while for the smooth quartic potential (\ref{eq:quartic}) we use
\begin{equation} \label{eq:sphi0}
 \varphi_0(x) =
 \Phi(\epsilon^{-1}\,d_0(x))\,,\quad \text{where}\quad
 \Phi(s) := \tanh (2^{-\frac12}\,s) \,.
\end{equation}
\end{subequations}
For simplicity we use the profiles (\ref{eq:phi0},b) also in the anisotropic
setting, where it would be more appropriate to replace $d_0$ with a suitably
defined anisotropic distance function $d_{\gamma}$, see \cite{DeckelnickDE05}
for details. Finally, if $\vartheta > 0$, we fix $w_0 = u_0$.

\subsection{Possible time discretizations} \label{sec:32}
In Section~\ref{sec:4} we will investigate the accuracy and the 
efficiency of several discretizations of the phase field model 
(\ref{eq:heata}--d), (\ref{eq:ACa}--c) in the isotropic case (\ref{eq:iso}),
(\ref{eq:betaiso}). In addition to the schemes \PF\ and \PFq\ from \cite{vch},
which use a semi-implicit discretization in time, we will also look at a more 
implicit discretization and at a fully explicit discretization. For later
reference, we now state the three different time discretizations, and for
simplicity we do so on ignoring spatial discretization. 
A strong formulation of
the time discretization from \cite{vch} is given as
follows. Let $\Psi = \Psi^+ + \Psi^-$, with $\Psi^+$ being convex on $\R$
and $\Psi^-$ being concave, and let $\Varrho = \Varrho^+ + \Varrho^-$ be a
similar splitting that is convex/concave on a suitable superset of $[-1,1]$,
where we recall that $\varphi$ need not remain in $[-1,1]$ for the quartic
potential (\ref{eq:quartic}). 
We also define $\varrho^\pm := (\Varrho^\pm)'$. 
For the schemes \PF\ and \PFq\ we set $\Psi^-(s) = -\frac12\,s^2$ and
\begin{equation*} 
\text{(i)}\ 
\varrho^+(s) = 0\,,\qquad
\text{(ii)}\ 
\varrho^+(s) = 0 \,,\qquad
\text{(iii)}\ 
\varrho^+(s) = \tfrac32\,s\,,\qquad
\text{(iv)}\ 
\varrho^+(s) = 2\,s 
\end{equation*}
for the choices of $\varrho$ in (\ref{eq:varrho}). The semi-implicit time 
discretization employed by the schemes \PF\ and \PFq\ can then be formulated
as:
\begin{subequations}
\begin{align}
& 
\vartheta\,(w_n - w_{n-1}) + \lambda\,(\varrho^+(\varphi_n) +
\varrho^-(\varphi_{n-1}))\,(\varphi_n - \varphi_{n-1}) -
\tau\,\nabla \,.\, (b(\varphi_{n-1})\,\nabla\, w_n ) = 0 \,,\label{eq:semi1}\\
& \cPsi\,\frac{a}\alpha\,(\varrho^+(\varphi_n) + \varrho^-(\varphi_{n-1}))
\,w_n = 
\epsilon\,\frac\rho\alpha\,\frac{\varphi_n- \varphi_{n-1}}{\tau}
-\epsilon\,\Delta\, \varphi_n + \epsilon^{-1}\,((\Psi^+)'(\varphi_n) +
(\Psi^-)'(\varphi_{n-1})) \,. \label{eq:semi2} 
\end{align}
\end{subequations}
With this time discretization existence of a unique solution 
$(\varphi^h_n, w^h_n)$ to the fully
discrete scheme can be shown for arbitrary time step sizes $\tau$ if
$\varrho^+=0$, where $w^h_n$ may not be unique if $\vartheta=0$ and
$\partial_N\Omega = \partial\Omega$ in very rare circumstances.
Moreover, any solution to the semi-implicit schemes \PF\ and \PFq\ 
is stable; see \cite{vch} for details.
The semi-implicit time discretization in (\ref{eq:semi1},b) can be 
modified to an implicit time discretization by replacing (\ref{eq:semi2}) with
\begin{equation} \label{eq:impl}
\cPsi\,\frac{a}\alpha\,(\varrho^+(\varphi_n) + \varrho^-(\varphi_{n-1}))
\,w_n = 
\epsilon\,\frac\rho\alpha\,\frac{\varphi_n- \varphi_{n-1}}{\tau}
-\epsilon\,\Delta\, \varphi_n + \epsilon^{-1}\,\Psi'(\varphi_n) \,,
\end{equation}
which then gives rise to a time step size constraint of the form
\begin{equation} \label{eq:tsc}
\tau < \rho\,\alpha^{-1}\,\epsilon^2 \quad\text{if}\ \rho > 0\,,
\end{equation}
in order to ensure the existence of a unique, stable solution
in the case $\varrho^+=0$ and $\rho > 0$.
In the situation $\vartheta = \rho =0$ and $\varrho(s) = \frac12$, with
$\partial_N\Omega = \partial\Omega$, which
has been treated by the present authors in \cite{eck}, 
a stronger time step constraint
of the form $\tau = \mathcal{O}(\epsilon^3)$ arises for the implicit
discretization (\ref{eq:impl}); see also \cite{BloweyE92,BlankBG11}. 
Here we recall that it is often observed 
that implicit time discretizations of Allen--Cahn and Cahn--Hilliard
type equations yield a better accuracy in time
compared to semi-implicit time discretizations as in (\ref{eq:semi1},b);
see e.g.\ \cite{BlankBG11,BlankGSS12,GraserKS11preprint}.
We will present several computations for a variant of \PFq\ with the implicit
time discretization (\ref{eq:semi1}), (\ref{eq:impl}) in Sections~\ref{sec:4}
and \ref{sec:5}.

Finally, fully explicit approximations, as advanced in e.g.\ 
\cite{NochettoV96,GarckeNS99,Nestler05}, can also be considered. Here we
replace (\ref{eq:semi1},b) by
\begin{subequations}
\begin{align}
& \vartheta\,(w_n - w_{n-1}) + \lambda\,\varrho(\varphi_{n-1})\,
(\varphi_n - \varphi_{n-1}) -
\tau\,\nabla \,.\, (b(\varphi_{n-1})\,\nabla\, w_{n-1} ) = 0 \,,
\label{eq:expl1}\\
& \cPsi\,\frac{a}\alpha\,\varrho(\varphi_{n-1})
\,w_{n-1} = \epsilon\,\frac\rho\alpha\,\frac{\varphi_n- \varphi_{n-1}}{\tau}
-\epsilon\,\Delta\, \varphi_{n-1} + \epsilon^{-1}\,
((\Psi^+)'(\varphi_n) + (\Psi^-)'
(\varphi_{n-1})) \,. \label{eq:expl2} 
\end{align}
\end{subequations}
If $\vartheta > 0$ and $\rho > 0$, then the above fully explicit time 
discretization is well-defined, and in this
case stability of the fully discrete scheme can be shown if
\begin{equation} \label{eq:tauh}
\tau = \mathcal{O}(h^2)\,.
\end{equation}
In particular, in the case of the obstacle potential (\ref{eq:obstacle}), if 
$\tau \leq \tfrac12\,\vartheta\,\epsilon\,\rho\,(\lambda^2\,\alpha)^{-1}\,
\varrho_{\max}^{-2}$,
then the solutions to the fully discrete variant of (\ref{eq:expl1},b) are
stable if 
\begin{equation} \label{eq:tauh2}
\tau\,h^{-2} \leq C_\star\,\min \{ \tfrac12\,\vartheta\,\conduct_{\max}^{-1} ,
\rho\,\alpha^{-1} \}\,,
\end{equation}
where $\varrho_{\max} := \max_{s \in [-1,1]} |\varrho(s)|$, 
$\conduct_{\max} := \max \{\conduct_+, \conduct_-\}$ and where $C_\star$ is a
constant only depending on the spatial mesh.
The advantage of (\ref{eq:expl1},b) over (\ref{eq:semi1},b) is that the
discretized systems of equations now decouple in space, which leads to huge
efficiency gains when the computations are performed in parallel on a large 
cluster. 
However, in practice this advantage is often negated because (\ref{eq:tauh}),
together with e.g.\ (\ref{eq:hf}), enforces that very small time steps need to
be taken.
In Section~\ref{sec:42} and in Section~\ref{sec:5}
we will present computations for a variant of \PF\ with
an explicit time discretization as in (\ref{eq:expl1},b).

\setcounter{equation}{0} 
\section{Quantitative comparison for isotropic problems} \label{sec:4}
\begin{figure}
\center
\ifpdf
\includegraphics[angle=-0,width=0.3\textwidth]{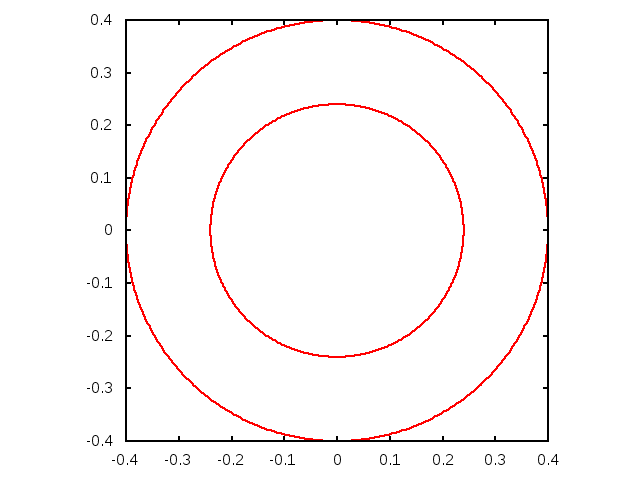}
\includegraphics[angle=-0,width=0.3\textwidth]{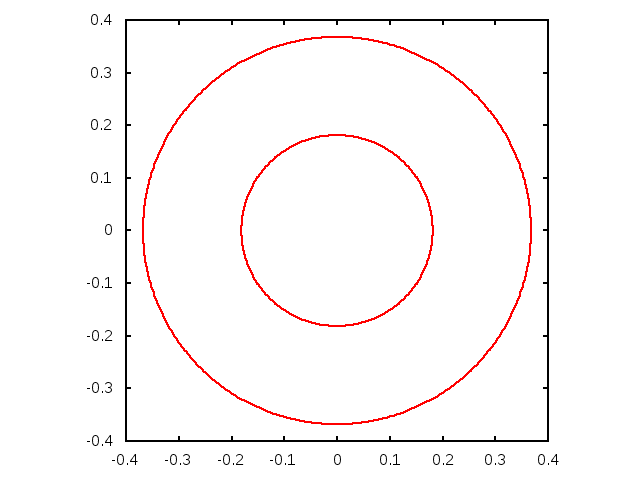}
\includegraphics[angle=-0,width=0.3\textwidth]{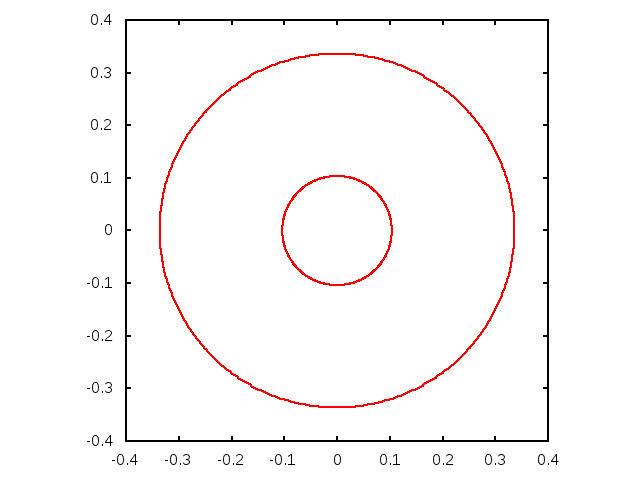}
\fi
\caption{
The true solution from (\ref{eq:ODEa}) at times
$t=0,\,10^{-3},\,2\times10^{-3}$.
}
\label{fig:MStrue}
\end{figure}%

A standard validation used for phase field models in the literature is the
comparison of tip velocities between the computed phase field discretizations
and real world measurements from the laboratory, see e.g.\ 
\cite{KarmaR96,ProvatasGD98}. Here the physical parameters in the phase field
model have to be chosen appropriately, so that they correspond to the physical
properties of the material in question. However, often the exact values of
these parameters are unknown or they are themselves based on measurements.
Here we propose a much simpler quantitative validation, which makes use of
known radially symmetric solutions to the underlying sharp interface problem
in the isotropic case.
We would argue that such a simple comparison should be part of the validation
of every phase field method to be proposed in the literature. It gives an
indication of the accuracy of the overall method and it helps to fine-tune the
discretization parameters that should be used for the anisotropic physical
applications.

In particular, in this section we consider the following isotropic variant
of (\ref{eq:1a}--e). Find $u : \Omega \times [0,T] \to \R$ 
and the interface $(\Gamma(t))_{t\in[0,T]}$ such that
for all $t\in (0,T]$ it holds that:
\begin{subequations}
\begin{alignat}{2}
\vartheta\,u_t - \Delta u & = f
\qquad && \mbox{in } \Omega \setminus \Gamma(t)\,, \label{eq:i1a} \\
\left[ \frac{\partial u}{\partial \unitn} \right]_{\Gamma(t)}
&=-\mathcal{V} \qquad &&\mbox{on } \Gamma(t), 
\label{eq:i1b} \\ 
\rho\,\mathcal{V} &= \alpha\,\kappa - u 
\qquad &&\mbox{on } \Gamma(t), 
\label{eq:i1c} \\ 
\frac{\partial u}{\partial \vec\nu} &= 0 
\qquad \mbox{on } \partial_N\Omega, \qquad
u = \uD \qquad &&\mbox{on }
\partial_D \Omega \,, \label{eq:i1d} \\
\Gamma(0) & = \Gamma_0 \,, \qquad\qquad 
\vartheta\,u(\cdot,0) = \vartheta\,u_0 \quad && \mbox{in } 
\Omega \,. \label{eq:i1e} 
\end{alignat}
\end{subequations}
Here $f: [0,T] \to \R$ in (\ref{eq:i1a}) is a given spatially homogeneous 
forcing term. In the phase field approximation (\ref{eq:heata}--d),
(\ref{eq:ACa}--c) this forcing appears analogously as a right hand side term
$f$ in (\ref{eq:heata}).  
Note that for $f=0$ the above system (\ref{eq:i1a}--e) corresponds to
(\ref{eq:1a}--e) with (\ref{eq:iso}), (\ref{eq:betaiso}) and
$\conduct_\pm=a=\lambda=1$.

\subsection{Mullins--Sekerka problem} \label{sec:41}
\begin{figure}
\center
\ifpdf
\includegraphics[angle=-90,width=0.45\textwidth]{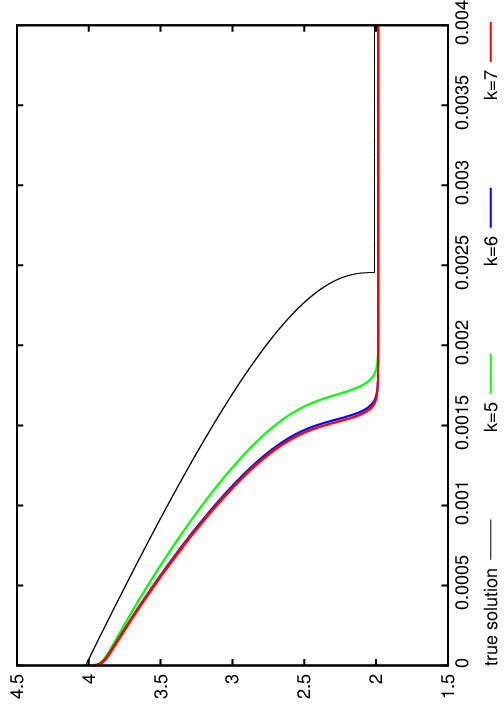}
\includegraphics[angle=-90,width=0.45\textwidth]{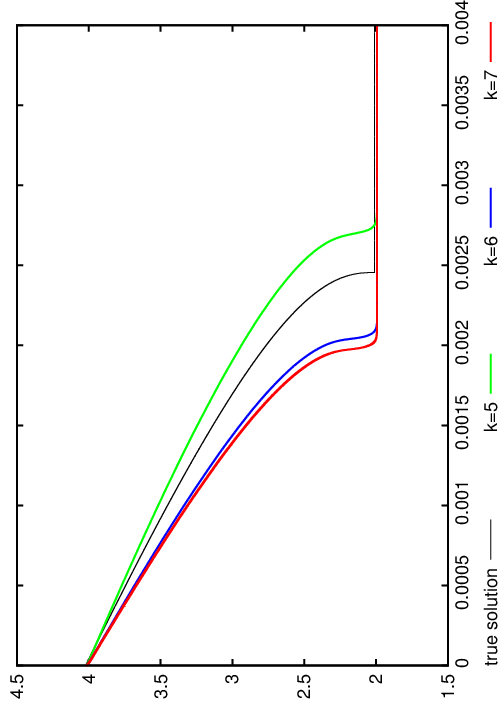}
\includegraphics[angle=-90,width=0.45\textwidth]{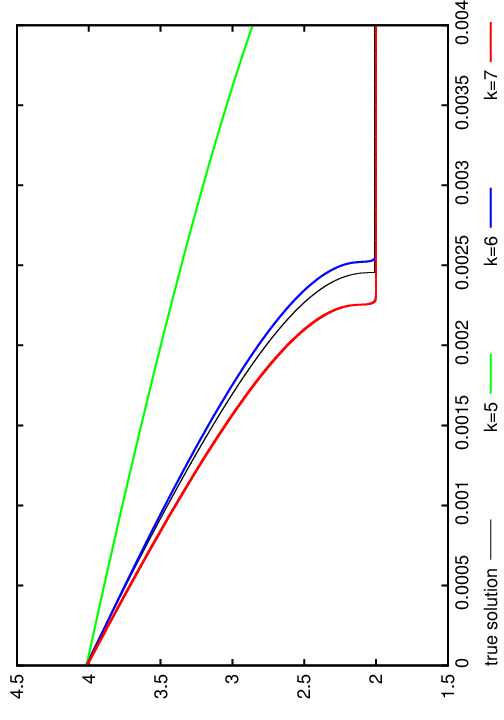}
\includegraphics[angle=-90,width=0.45\textwidth]{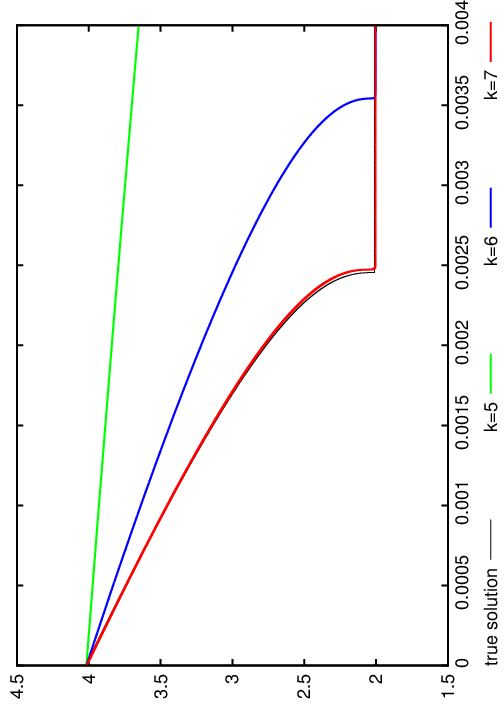}
\fi
\caption{(\PFqi, 
$\epsilon^{-1} = 8\,\pi,\,16\,\pi,\,32\,\pi,\,64\,\pi$)
Comparison of the energies $\mathcal{F}$ and $\mathcal{F}_\epsilon^h$ for
the benchmark problem~\ref{bm:MS} with $T=4\times10^{-3}$. 
The uniform time step sizes are chosen as $\tau=10^{-k}$, $k=5\to7$.
}
\label{fig:tuneqi}
\end{figure}%
\begin{figure}
\center
\ifpdf
\includegraphics[angle=-90,width=0.45\textwidth]{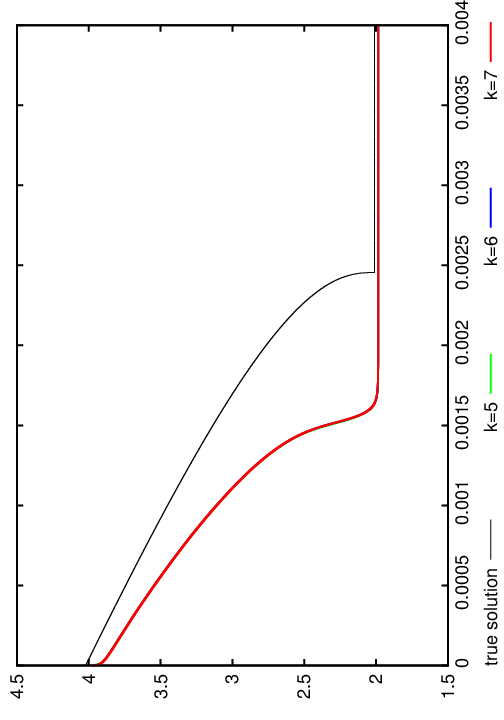}
\includegraphics[angle=-90,width=0.45\textwidth]{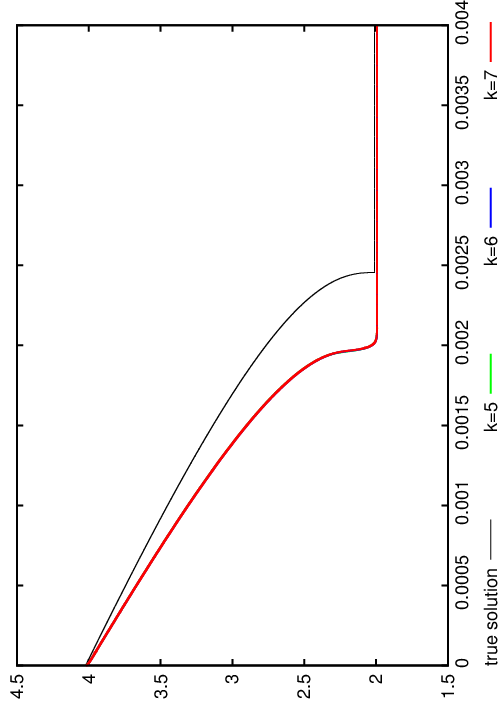}
\includegraphics[angle=-90,width=0.45\textwidth]{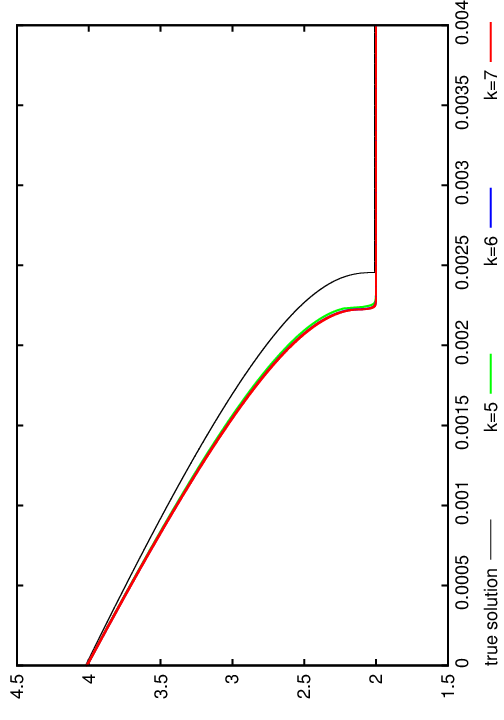}
\includegraphics[angle=-90,width=0.45\textwidth]{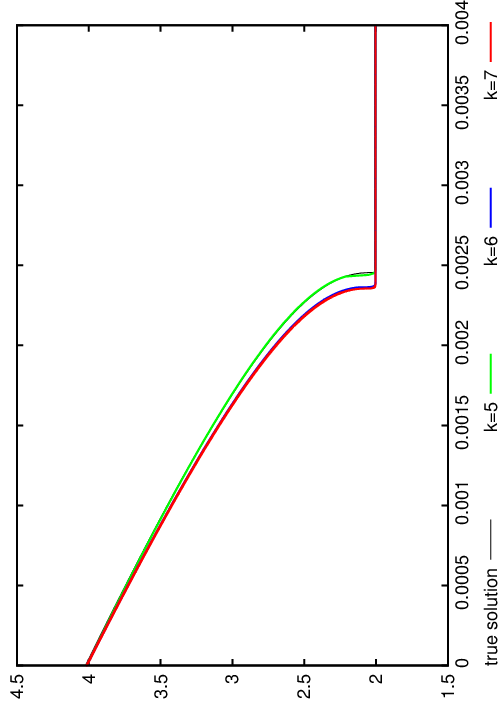}
\fi
\caption{(\PFqi\ with the implicit time discretization from (\ref{eq:impl}), 
$\epsilon^{-1} = 8\,\pi,\,16\,\pi,\,32\,\pi,\,64\,\pi$)
Comparison of the energies $\mathcal{F}$ and $\mathcal{F}_\epsilon^h$ for
the benchmark problem~\ref{bm:MS} with $T=4\times10^{-3}$. 
The uniform time step sizes are chosen as $\tau=10^{-k}$, $k=5\to7$.
}
\label{fig:tuneqi_impl}
\end{figure}%

We first consider the quasi-static case $\vartheta=\rho=0$.
To this end we take the known solution of an annular region $\Omega_-(t)$, for
which the inner boundary shrinks to a point so that $\Omega_-(t)$ becomes a 
disk for sufficiently large $t$.
Here we take $\alpha = 1$ and let 
$\partial_N\Omega = \partial\Omega$.
In addition, $\Gamma(0)=\Gamma_0$ consists of two concentric
circles/spheres. 
It is then not difficult to show that the two radii $r_1 < r_2$
satisfy the following system of nonlinear ODEs:
In the case $d=2$ we have
\begin{subequations}
\begin{equation}
[r_1]_t = - \frac1{r_1}\frac{\frac1{r_1} + \frac1{r_2}}{\ln\frac{r_2}{r_1}} 
\quad\mbox{ and }\quad
[r_2]_t = - \frac1{r_2}\frac{\frac1{r_1} + \frac1{r_2}}{\ln\frac{r_2}{r_1}} = 
\frac{r_1}{r_2}\,[r_1]_t \qquad \forall\ t
\in[0,T_0)\,,
\label{eq:ODEa}
\end{equation}
while for $d=3$ it holds that
\begin{equation}
[r_1]_t = - \frac2{r_1^2}\frac{r_1 + r_2}{r_2-r_1} \quad\mbox{ and }\quad
[r_2]_t = - \frac2{r_2^2}\frac{r_1 + r_2}{r_2-r_1} = \frac{r_1^2}{r_2^2}\,
[r_1]_t \qquad \forall\ t
\in[0,T_0)\,,
\label{eq:ODEb}
\end{equation}
\end{subequations}
where $T_0$ is the extinction time of the smaller sphere, 
i.e.\ $\lim_{t\to T_0} r_1(t) = 0$, 
see e.g.\ \cite{BatesCD95,Stoth96}. 
Note that the corresponding solution $u$ satisfying (\ref{eq:1a}--e) is given
by the radially symmetric function
\begin{equation}
u(\vec{x},t) = \begin{cases}
 -\frac{d-1}{r_2(t)} & |\vec{x}| \geq r_2(t)\,, \\
\begin{cases}
\frac1{r_1(t)} - \ln\frac{|\vec{x}|}{r_1(t)}\,
\dfrac{\frac1{r_1(t)} + \frac1{r_2(t)}}{\ln\frac{r_2(t)}{r_1(t)}} & d = 2 \\
-\frac4{r_2(t) - r_1(t)} + 
\frac2{|\vec{x}|}\,\frac{{r_1(t)} + {r_2(t)}}{r_2(t)-r_1(t)} & d = 3
\end{cases}
& r_1(t) \leq |\vec{x}| \leq r_2(t)\,, \\
 \frac{d-1}{r_1(t)} & |\vec{x}| \leq r_1(t)\,.
\end{cases} 
\label{eq:ODE_u}
\end{equation}
As (\ref{eq:ODEa},b) does not appear to be analytically solvable, it needs to 
be integrated numerically to compute the solution $(r_1,r_2)(t)$, 
for $t\in[0,T]$, where $T<T_0$.
Possible strategies to integrate (\ref{eq:ODEa},b) to a high accuracy 
are described in \cite{dendritic}. 
We visualize the evolution of $\Gamma(t)$ over the time
interval $[0, 2\times10^{-3}]$ in Figure~\ref{fig:MStrue}. The above true
solution forms the basis of our first benchmark problem:

\begin{algorithm}[H]
\caption{2d Mullins--Sekerka with $\vartheta = \rho = 0$.}
\label{bm:MS}
True solution (\ref{eq:ODEa}), (\ref{eq:ODE_u}) to (\ref{eq:i1a}--e) with
$\vartheta = \rho = 0$ and $\alpha=1$. \\
Initial data $(r_1,r_2)(0) = (0.24, 0.4)$. \\
Domain $\Omega = (-\frac12, \frac12)^2$ with 
$\partial_N\Omega = \partial\Omega$. \\
Time interval $[0,T]$ with $T=10^{-3}$ so that 
$(r_1,r_2)(T) \approx (0.18, 0.37)$.
\end{algorithm}

\begin{figure}
\center
\ifpdf
\includegraphics[angle=-90,width=0.55\textwidth]{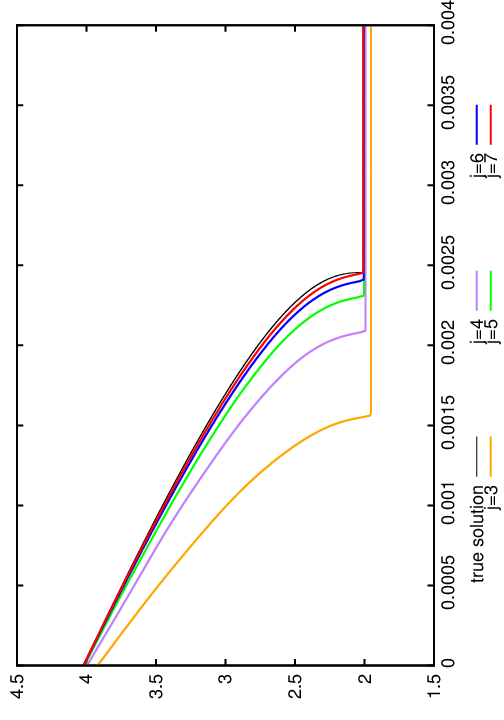}
\fi
\caption{(\PFEM
)
Comparison of the energies $\mathcal{F}$ and $\mathcal{F}^h$ for
the benchmark problem~\ref{bm:MS} with $T=4\times10^{-3}$. 
The uniform time step size is chosen as $\tau=10^{-5}$, 
while the spatial discretizations are proportional to $2^{-j}$, $j=3\to7$.
}
\label{fig:tunep}
\end{figure}

In Figure~\ref{fig:tuneqi}
we compare the energy $\mathcal{F}$ of the true sharp
interface solution, recall (\ref{eq:F}), 
to the corresponding energies $\mathcal{F}_\epsilon^h \approx 
\mathcal{F}_\epsilon$, 
recall (\ref{eq:Fg}), of the finite element approximations
from the algorithm \PFqi\ on the time interval $[0,4 \times 10^{-3}]$.
Here we recall that for the given data and
the given evolution it holds that
\begin{equation*} 
\mathcal{F}(\Gamma, u) = 
|\Gamma(t)| := \int_{\Gamma(t)} 1 \ds = 2\,\pi\,(r_1(t) + r_2(t)) \,.
\end{equation*}
Very similar energy plots can be obtained for the other variants of
\PFq\ and those of \PF.
We note that for decreasing
$\epsilon$, the time step size $\tau$ needs to be chosen smaller and smaller in
order to capture the correct time scaling of the evolution. 
We compare this computation for \PFqi, which uses a semi-implicit
discretization in time, now with one computation for the implicit time
discretization as in (\ref{eq:impl}). Here we recall that better 
accuracy for such discretizations has been reported in \cite{BlankBG11}
for the isotropic Cahn--Hilliard equation, i.e.\ 
(\ref{eq:heata}), (\ref{eq:ACa}) for (\ref{eq:iso})  
with (\ref{eq:obstacle}), (\ref{eq:varrho})(i) and with $\vartheta = \rho = 0$.
In fact, here we observe a similar behaviour. See Figure~\ref{fig:tuneqi_impl}, 
where even for very large choices of $\tau$ the time evolution of the phase 
field model seems to be captured accurately.
Finally, we show some discrete energies $\mathcal{F}^h\approx\mathcal{F}$ 
from the parametric scheme \PFEM\
in Figure~\ref{fig:tunep}. Here we choose rather crude discretization
parameters, since otherwise the
discrete energies $\mathcal{F}^h$ would lie virtually 
on top of the true energy $\mathcal{F}$.

In Table~\ref{tab:tuneiL2} we present the errors in $r_1$ and in $u$ for the
scheme \PFi\ for a selection of interface parameters $\epsilon$ and for a range
of discretization parameters $h_f$ and $\tau$ for the 
benchmark problem~\ref{bm:MS}.
The displayed error quantities are defined as
$\LerrorWu := (\tau\,\sum_{0 \leq n \leq T / \tau} 
\|w^h(\cdot,n\,\tau) - u(\cdot,n\,\tau)\|_{L^2(\Omega)}^2)^\frac12$
and 
$\errorRr{x_1}{1} := \max_{0 \leq n \leq T / \tau} 
|r_{x_1}^h(n\,\tau) - r_1(n\,\tau)|$,
where
$r_{x_1}^h(t) := \inf \{s \geq 0 : \varphi^h(s\,{\rm e}_1, t) = 0 \}$,
with ${\rm e}_1 = (1,0)^T$ being the first unit vector in $\R^2$, denotes the
phase field approximation of the inner radius. We also show the number of
degrees of freedom (DOFs) for the calculation of the discrete solution for the 
final time step at time $t=T$.
The presented overall CPU times are for a single-thread computation on an
Intel i7-860 (2.8 GHz) processor with 8 GB of main memory.
\begin{table}
\center 
\begin{tabular}{c|c|c|c|c|c|r}
 $\epsilon^{-1}$ & $2^\frac12 /h_{f}$ & $\tau$ & 
 $\errorRr{x_1}{1}$ &  $\LerrorWu$ & DOFs$(T)$ & CPU time \\ 
 \hline 
 $8\,\pi$ 
 &64  & $10^{-5}$ & 1.3004e-02 & 1.8448e-03 & 4522 & 5 secs\\
 &64  & $10^{-6}$ & 1.3154e-02 & 9.9918e-04 & 4402 & 39 secs\\
 &64  & $10^{-7}$ & 1.3175e-02 & 3.2861e-04 & 4354 & 6:54 mins\\
 &128 & $10^{-7}$ & 1.4283e-02 & 3.5051e-04 & 15626 & 37:22 mins\\ 
 \hline
 $16\,\pi$ 
 &128 & $10^{-5}$ & 1.4139e-02 & 2.1678e-03 & 10378 & 15 secs \\ 
 &128 & $10^{-6}$ & 2.6476e-03 & 2.5879e-04 & 10130 & 1:58 mins\\ 
 &128 & $10^{-7}$ & 4.9843e-03 & 1.0553e-04 & 10106 & 17:06 mins\\ 
 &256 & $10^{-7}$ & 5.5154e-03 & 1.1024e-04 & 34682 & 1:23 hours\\ 
 \hline
 $32\,\pi$ 
 &256 & $10^{-5}$ & 3.4984e-02 & 5.6999e-03 & 23362 & 50 secs\\ 
 &256 & $10^{-6}$ & 6.2022e-03 & 3.4521e-04 & 21650 & 6:28 mins\\ 
 &256 & $10^{-7}$ & 8.8958e-04 & 3.7115e-05 & 21082 & 50:03 mins\\ 
 &512 & $10^{-7}$ & 1.8543e-03 & 3.3740e-05 & 74322 & 5:04 hours \\
 \hline
 $64\,\pi$ 
 &512 & $10^{-5}$ & 5.0345e-02 & 8.2731e-03 & 52602 & 1:54 mins\\ 
 &512 & $10^{-6}$ & 2.2346e-02 & 1.2118e-03 & 49370 & 17:19 mins\\ 
 &512 & $10^{-7}$ & 4.7000e-03 & 1.1020e-04 & 46922 & 2:14 hours\\ 
 &1024& $10^{-7}$ & 2.1723e-03 & 3.5938e-05 & 166498 & 15:50 hours\\ 
 \hline
 $128\,\pi$ 
 &1024& $10^{-5}$ & 5.6324e-02 & 9.1597e-03 & 122314 & 7:25 mins \\
 &1024& $10^{-6}$ & 4.2515e-02 & 2.2390e-03 & 118818 & 1:01 hours\\
 &1024& $10^{-7}$ & 1.5232e-02 & 2.8138e-04 & 112794 & 9:50 hours \\
 &2048& $10^{-7}$ & 1.0919e-02 & 1.8112e-04 & 404962 & 68:23 hours \\
\end{tabular}
\caption{Benchmark problem~\ref{bm:MS} for \PFi.}
\label{tab:tuneiL2}
\end{table}%
For the benchmark problem~\ref{bm:MS} the
remaining variants of \PF\ and \PFq\ exhibit very similar errors to the ones in 
Table~\ref{tab:tuneiL2}, and so we do not present them here. 
In later computations we will also employ the stronger norm
$\errorWu := \max_{0 \leq n \leq T / \tau} 
\|w^h(\cdot,n\,\tau) - u(\cdot,n\,\tau)\|_{L^\infty(\Omega)}$ for the
temperature error.
However, for the experiments in Table~\ref{tab:tuneiL2} no convergence can 
be observed in the $L^\infty(\Omega_T)$-error for the true temperature 
(\ref{eq:ODE_u}) for the phase field
approximations. In fact, for the computations in Table~\ref{tab:tuneiL2} the
errors $\errorWu$ are 
in the interval $[1,16]$, where we note that the
true solution (\ref{eq:ODE_u}) itself remains in the range $[-2.8,5.6]$ over
the computed time interval. 
It is for this reason that we report the weaker error norms $\LerrorWu$ in
Table~\ref{tab:tuneiL2}.

A repeat of the computation in Table~\ref{tab:tuneiL2}, but now for an implicit
time discretization of \PFqi\ can be seen in
Table~\ref{tab:tuneqi_impl}. 
One clearly observes that, for fixed $\epsilon$,
the errors $\errorRr{x_1}{1}$ and $\LerrorWu$ soon appear to be almost
independent of the time step size $\tau$. This indicates that the implicit time
discretization from (\ref{eq:impl}) manages to eliminate the temporal
discretization error relatively quicker than the semi-implicit discretization
from (\ref{eq:semi1},b). 
Moreover, for small $\epsilon$ and fixed $\tau$, the error in the 
approximation of the sharp interface problem (\ref{eq:i1a}--e) is in general
significantly smaller for the implicit time discretization.
We remark that the converged errors in Table~\ref{tab:tuneqi_impl} 
appear to indicate a convergence of $\mathcal{O}(\epsilon)$ in the
error $\errorRr{x_1}{1}$, with a similar convergence rate for the error
$\LerrorWu$, if discretization errors are neglected.

\begin{table}
\center 
\begin{tabular}{c|c|c|c|c|c|r}
 $\epsilon^{-1}$ & $2^\frac12 /h_{f}$ & $\tau$ & 
 $\errorRr{x_1}{1}$ &  $\LerrorWu$ & DOFs$(T)$ & CPU time \\ 
 \hline 
 $8\,\pi$ 
 &64  & $10^{-5}$ & 3.5385e-02 & 7.8725e-03 & 5018 & 10 secs\\
 &64  & $10^{-6}$ & 3.5228e-02 & 2.4832e-03 & 5018 & 1:06 mins\\
 &64  & $10^{-7}$ & 3.5211e-02 & 7.8429e-04 & 5018 & 12:30 mins\\
 &128 & $10^{-7}$ & 3.6235e-02 & 7.9552e-04 & 18658 & 50:07 mins \\
 \hline
 $16\,\pi$ 
 &128 & $10^{-5}$ & 1.4453e-02 & 2.2343e-03 & 11514 & 22 secs\\
 &128 & $10^{-6}$ & 1.4667e-02 & 7.2292e-04 & 11530 & 2:20 mins\\ 
 &128 & $10^{-7}$ & 1.4675e-02 & 2.2894e-04 & 11514 & 27:29 mins\\ 
 &256 & $10^{-7}$ & 1.5018e-02 & 2.3527e-04 & 38858 & 1:31 hours\\ 
 \hline
 $32\,\pi$ 
 &256 & $10^{-5}$ & 5.6651e-03 & 8.2274e-04 & 24802 & 58 secs \\ 
 &256 & $10^{-6}$ & 6.4112e-03 & 2.9604e-04 & 24290 & 7:23 mins\\ 
 &256 & $10^{-7}$ & 6.4838e-03 & 9.6155e-05 & 24322 & 1:06 hours\\ 
 &512 & $10^{-7}$ & 6.6777e-03 & 1.0156e-04 & 83690 & 5:41 hours\\
 \hline
 $64\,\pi$ 
 &512 & $10^{-5}$ & 3.1282e-04 & 1.3540e-04 & 54298 & 2:20 mins \\ 
 &512 & $10^{-6}$ & 2.5224e-03 & 1.1540e-04 & 53466 & 17:31 mins\\ 
 &512 & $10^{-7}$ & 2.8135e-03 & 4.4061e-05 & 53114 & 3:27 hours\\ 
 &1024& $10^{-7}$ & 3.0098e-03 & 4.2577e-05 & 185410 & 20:03 hours\\
 \hline
 $128\,\pi$ 
 &1024& $10^{-5}$ & 8.3850e-03 & 1.4662e-03 & 128866 & 15:56 mins \\
 &1024& $10^{-6}$ & 3.3622e-04 & 2.2218e-05 & 123770 & 1:17 hours\\
 &1024& $10^{-7}$ & 8.3440e-04 & 1.5965e-05 & 123402 & 10:46 hours\\
 &2048& $10^{-7}$ & 1.1768e-03 & 1.6903e-05 & 437330 & 116:48 hours \\
\end{tabular}
\caption{Benchmark problem~\ref{bm:MS} for \PFqi\ with the implicit time
discretization from (\ref{eq:impl}).}
\label{tab:tuneqi_impl}
\end{table}%
Finally we note that the numbers in Tables~\ref{tab:tuneiL2} and
\ref{tab:tuneqi_impl} indicate that refining the mesh discretization
parameters $h_f$, and hence $h_c$, 
in general does not reduce the error. Hence choosing
$h_c = 8\,h_f = \pi\,\epsilon$ appears to be sufficient for classical phase
field model computations, and we will restrict ourselves to this choice from
now on in this paper.

We compare these convergence experiments with the corresponding errors for 
the sharp interface
algorithm \PFEM\ in Table~\ref{tab:tunep}. For these sets of experiments we
always choose $h_\Gamma \approx h_f$.
Here the error quantities 
$\errorUu$ and $\LerrorUu$ are defined as $\errorWu$ and $\LerrorWu$ as before,
but with $w^h$ replaced by $u^h$. In addition, we let
$\errorRr{1}{1} := \max_{0 \leq n \leq T / \tau} 
\max_{p \in \Gamma_1^h(n\,\tau)} | |p| - r_1(n\,\tau)|$, 
with $\Gamma_1^h(t)$ denoting the parametric approximation of the inner circle
of the true solution $\Gamma(t)$. Note that the norm in the definition of 
$\errorRr{1}{1}$ employed here is much stronger than the phase field equivalent 
$\errorRr{x_1}{1}$ introduced earlier, where the difference between the true 
interface position 
$r_1(t)$ and the phase field approximation is measured in the $x_1$-coordinate
direction only.
All of the error quantities shown in Table~\ref{tab:tunep} appear to be
converging with order at least $\mathcal{O}(h)$ if the time discretization
errors are neglected.
\begin{table}
\center
\begin{tabular}{c|c|c|c|c|c|r}
 $2^\frac12/h_{f}$ & $\tau$ & $\errorRr11$ & $\errorUu$ & $\LerrorUu$ & 
 DOFs$(T)$ & CPU time \\ 
 \hline 
    8 & $10^{-4}$ & 3.6782e-02 & 2.8260e-00 & 3.5916e-02 & 113 & 0 secs \\
    8 & $10^{-5}$ & 4.2433e-02 & 4.8862e-00 & 4.1925e-02 & 113 & 0 secs \\
 \hline
   16 & $10^{-4}$ & 1.1028e-02 & 9.5419e-01 & 1.2123e-02 & 285 & 0 secs \\
   16 & $10^{-5}$ & 1.3394e-02 & 1.7413e-00 & 1.5531e-02 & 285 & 0 secs \\
 \hline
   32 & $10^{-4}$ & 3.6288e-03 & 4.3183e-01 & 4.8601e-03 & 693 & 0 secs \\
   32 & $10^{-5}$ & 5.7298e-03 & 6.3045e-01 & 7.1794e-03 & 657 & 1 secs \\
 \hline
   64 & $10^{-4}$ & 7.7318e-04 & 2.9301e-01 & 2.8734e-03 & 1585 & 0 secs \\
   64 & $10^{-5}$ & 2.4803e-03 & 2.9384e-01 & 3.2140e-03 & 1473 & 1 secs \\
 \hline
  128 & $10^{-4}$ & 5.8266e-04 & 3.1247e-01 & 3.2813e-03 & 3553 & 0 secs \\
  128 & $10^{-5}$ & 1.1406e-03 & 1.1283e-01 & 1.4618e-03 & 3213 & 3 secs \\
  128 & $10^{-6}$ & 1.3262e-03 & 1.5152e-01 & 1.7016e-03 & 3173 & 32 secs \\
 \hline
  256 & $10^{-4}$ & 1.1384e-03 & 3.4935e-01 & 3.6955e-03 & 8289 & 1 secs \\
  256 & $10^{-5}$ & 4.7583e-04 & 5.9495e-02 & 6.7789e-04 & 6945 & 7 secs \\
  256 & $10^{-6}$ & 6.4981e-04 & 7.6398e-02 & 8.4454e-04 & 6777 & 1:11 mins \\
 \hline
  512 & $10^{-4}$ & 1.4098e-03 & 3.6302e-01 & 3.8770e-03 & 20381 & 3 secs \\
  512 & $10^{-5}$ & 1.3003e-04 & 3.4588e-02 & 3.4619e-04 & 16109 & 19 secs \\
  512 & $10^{-6}$ & 2.9886e-04 & 3.4579e-02 & 3.9724e-04 & 15649 & 3:02 mins \\
 \hline
 1024 & $10^{-4}$ & 1.4724e-03 & 3.6426e-01 & 3.9180e-03 & 43133 & 6 secs \\
 1024 & $10^{-5}$ & 3.8716e-05 & 3.1143e-02 & 3.2082e-04 & 41757 & 57 secs \\
 1024 & $10^{-6}$ & 1.3093e-04 & 1.4765e-02 & 1.8090e-04 & 39493 & 9:31 mins \\
 \hline
 2048 & $10^{-4}$ & 1.5121e-03 & 3.6991e-01 & 3.9402e-03 & 93385 & 18 secs \\
 2048 & $10^{-5}$ & 1.1372e-04 & 3.7037e-02 & 3.6346e-04 & 120429 & 3:35 mins\\
 2048 & $10^{-6}$ & 5.3291e-05 & 6.9988e-03 & 7.7964e-05 & 112285 & 32:22 mins\\
\end{tabular}
\caption{Benchmark problem~\ref{bm:MS} for \PFEM.}
\label{tab:tunep}
\end{table}%

The numbers in Tables~\ref{tab:tuneiL2}--\ref{tab:tunep} convey a very clear
message. Firstly, we recall that the experiments in Tables~\ref{tab:tuneiL2}
and \ref{tab:tuneqi_impl} do not converge in the norm $\errorWu$, whereas 
$\errorUu$ in Table~\ref{tab:tunep} does appear to converge with
$\mathcal{O}(h)$.
Secondly, we can see that even with computations that take almost $5$ days, the
phase field schemes \PFi\ and \PFqi\ cannot reduce the error in the radius to
below $3\times 10^{-4}$.
Yet, better accuracies for the radius can be achieved with the
sharp interface approximation \PFEM, running on the same computing hardware, in
less than a minute. 
Hence, for this measurement, the simulations with \PFEM\ are at least
$7\,000$ times faster than the computations with \PF\ and \PFq.
The main reason behind this very slow convergence appears to be that the
biggest contribution to the observed error comes from the interfacial parameter
$\epsilon$. Hence in order to obtain reasonable errors, 
$\epsilon$ needs to be taken very
small, which on recalling (\ref{eq:hf}) 
implies that the
discretization parameters need to be chosen very small as well;
recall also (\ref{eq:NV97a}). Unfortunately,
phase field computations thus soon reach the limit of what is computable on
today's computer hardware.
As an aside we note that when comparing CPU times between e.g.\ \PFi\ and
\PFEM\ in terms of degrees of freedom, then it is crucial to take into account
the value of $\tau$, as this will be indirectly proportional to the number of
algebraic systems that need to be solved during the whole simulation. In fact,
for the same number of degrees of freedom and the same value of $\tau$, 
the CPU times between the two different algorithms are similar.

\begin{figure}
\center
\ifpdf
\includegraphics[angle=-0,width=0.55\textwidth]{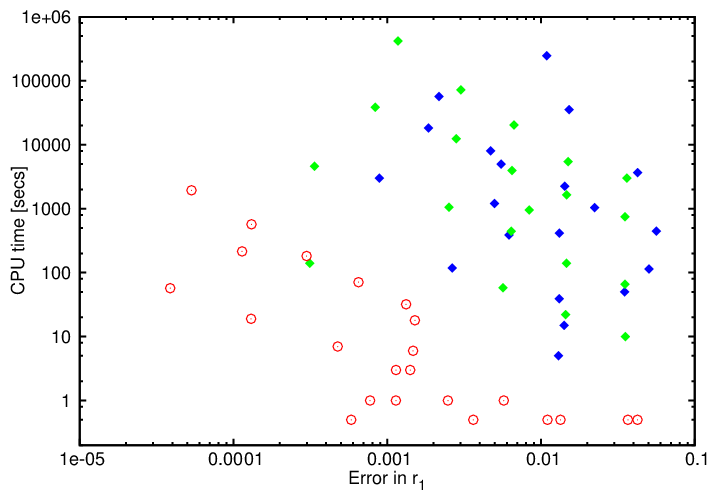}
\fi
\caption{(Benchmark problem~\ref{bm:MS})
Log-log scatter plot of $\errorRr{x_1}1$ and $\errorRr11$
against the CPU time for
the entries in Table~\ref{tab:tuneiL2} (\PFi, blue rhombi), 
Table~\ref{tab:tuneqi_impl} (\PFqi, green rhombi) and 
Table~\ref{tab:tunep} (\PFEM, red circles).
}
\label{fig:scatter123}
\end{figure}

To better visualize the relative performances of \PFi\ in
Table~\ref{tab:tuneiL2}, \PFqi\ in Table~\ref{tab:tuneqi_impl} and \PFEM\ in
Table~\ref{tab:tunep}, we present a plot of the errors in the radius $r_1$
against the necessary CPU time for all the entries in the three tables in
Figure~\ref{fig:scatter123}. This plot underlines the superiority of the sharp
interface algorithm \PFEM\ over the phase field methods.

For the 3d solution (\ref{eq:ODEb}), (\ref{eq:ODE_u}) it is virtually
impossible to perform a meaningful convergence experiment for the phase field
approximations \PF\ and \PFq. The reason is that for the same values of
$\epsilon$ and for comparable discretization parameters as in e.g.\
Table~\ref{tab:tuneiL2}, the simulations in three space dimensions 
do not finish in a reasonable
amount of time. For an example of a convergence test for the solution
(\ref{eq:ODEb}), (\ref{eq:ODE_u}) for the scheme \PFEM\ we refer to
\cite[Table~6]{dendritic}.

\subsection{Stefan problem} \label{sec:42}
In this section we consider the full Stefan problem (\ref{eq:i1a}--e) with
$\vartheta = 1$ and $\rho \geq 0$.
Here we adapt the
expanding circle/sphere solution introduced in \cite[p.\ 303--304]{Schmidt96},
where the radius of the circle/sphere is given by $r(t)$ with
\begin{subequations}
\begin{equation} \label{eq:TSr}
r(t) = (r^2(0) + t)^\frac12 \,. 
\end{equation}
In particular, we let 
\begin{equation*} 
z(t) = -\frac{\alpha\,(d-1) + \frac12\,\rho}{r(t)}, \qquad
v(s) = - \frac{e^{\frac14}}2\,\int_1^s \frac{e^{-\frac14\,y^2}}{y^{d-1}}\;{\rm
d}y\,. 
\end{equation*}
Then it is easy to see that 
\begin{equation}
u(\vec{x},t) = \begin{cases}
z(t) & |\vec{x}| \leq r(t)\,, \\
z(t) + v\bigl(\frac{|\vec{x}|}{r(t)}\bigr) & |\vec{x}| > r(t) \,,
\end{cases}
\label{eq:TrueSchmidtu}
\end{equation}
\end{subequations}
is the solution to (\ref{eq:i1a}--e) with
\begin{equation} \label{eq:f}
f(t) = \ddt z(t) = \frac{\alpha\,(d-1) + \frac12\,\rho}{2\,r^3(t)}\,,
\end{equation}
and with $\uD$ in (\ref{eq:i1d}) replaced by
$u\!\mid_{\partial_D\Omega}$ on $\partial_D\Omega = \partial\Omega$.

For the Stefan problem with kinetic undercooling we propose the following
benchmark problem, where on recalling e.g.\ (\ref{eq:epscond})
we note that increasing the parameter $\ell \in \mathbb{N}$ leads
to the benchmark problem becoming computationally more and more challenging.

\renewcommand{\thealgocf}{\arabic{algocf}$^{(\ell)}$}
\begin{algorithm}[H]
\caption{2d Stefan problem with $\vartheta = 1$ and $\rho > 0$.}
\label{bm:Stefan1}
True solution (\ref{eq:TSr},b) to 
(\ref{eq:i1a}--e) with (\ref{eq:f}) and with $\vartheta = 1$ and
$\alpha = \rho = 10^{-\ell}$. \\
Initial data $r(0) = \frac14$ and $u_0 = u(\cdot,0)$ from 
(\ref{eq:TrueSchmidtu}). \\
Domain $\Omega = (-1, 1)^2$ with 
$\partial_D\Omega = \partial\Omega$ and $\uD = u\!\mid_{\partial_D\Omega}$
from (\ref{eq:TrueSchmidtu}). \\
Time interval $[0,T]$ with $T=\frac12$, so that $r(T) = \frac34$.
\end{algorithm}

\begin{figure}
\center
\ifpdf
\includegraphics[angle=-90,width=0.45\textwidth]{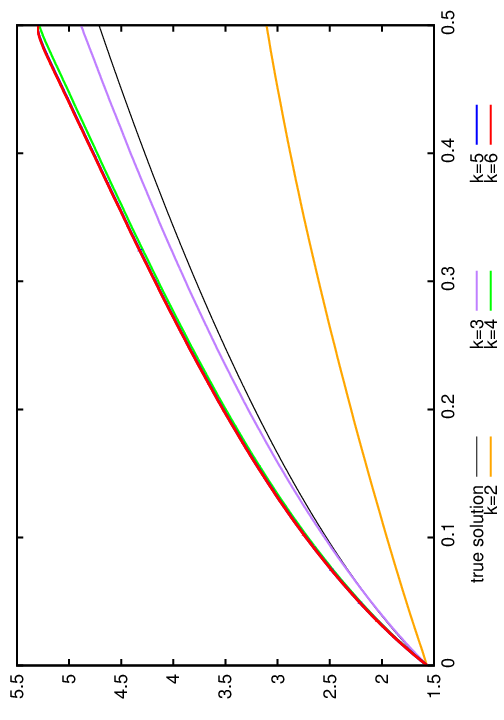}
\includegraphics[angle=-90,width=0.45\textwidth]{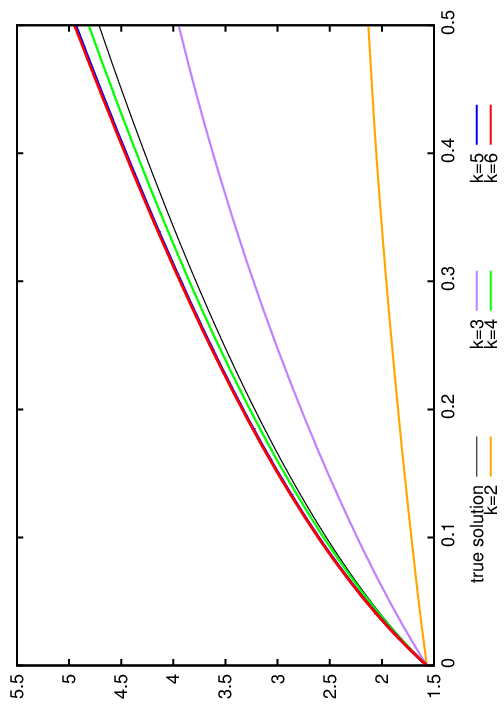}
\includegraphics[angle=-90,width=0.45\textwidth]{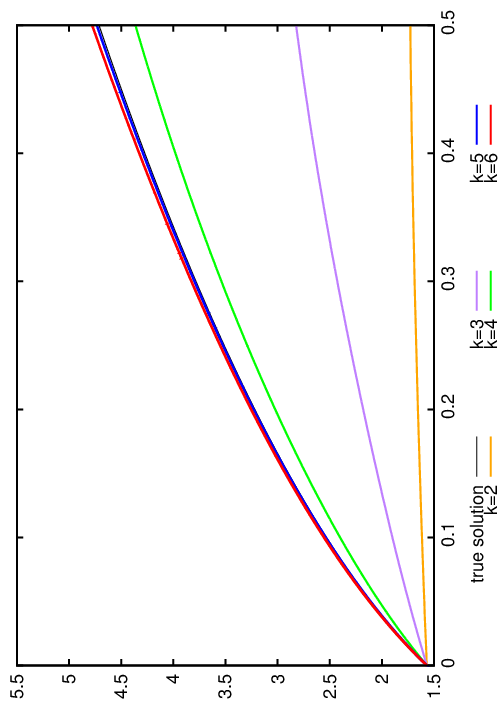}
\includegraphics[angle=-90,width=0.45\textwidth]{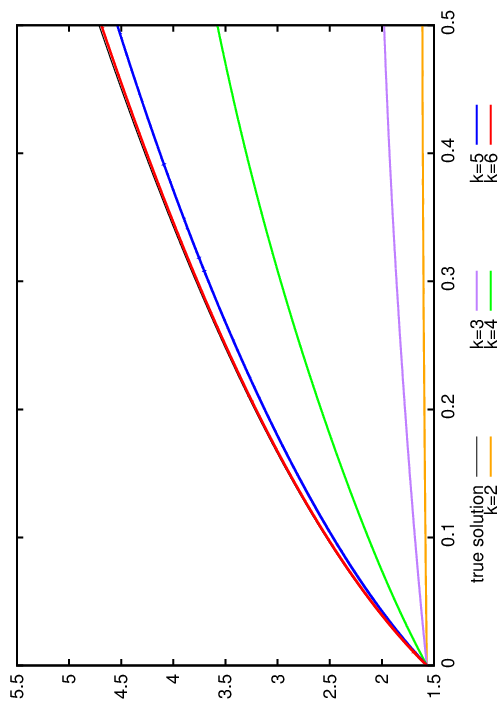}
\fi
\caption{(\PFqiii, 
$\epsilon^{-1} = 4\,\pi,\,8\,\pi,\,16\,\pi,\,32\,\pi$)
Comparison of $|\Gamma(t)|$ and $\cPsi^{-1}\,\mathcal{E}_\epsilon^h$ for
the benchmark problem~\ref{bm:Stefan1} with $\ell=1$.
The uniform time step sizes are chosen as $\tau=10^{-k}$, $k=2\to6$.
}
\label{fig:TSqiii}
\end{figure}

\begin{figure}
\center
\ifpdf
\includegraphics[angle=-90,width=0.45\textwidth]{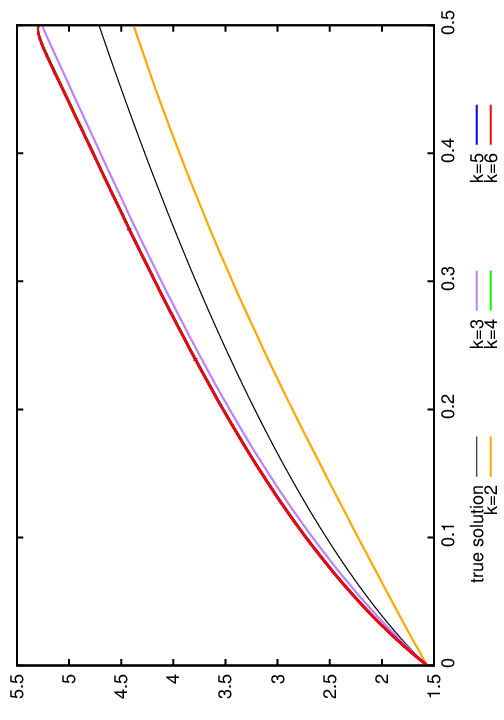}
\includegraphics[angle=-90,width=0.45\textwidth]{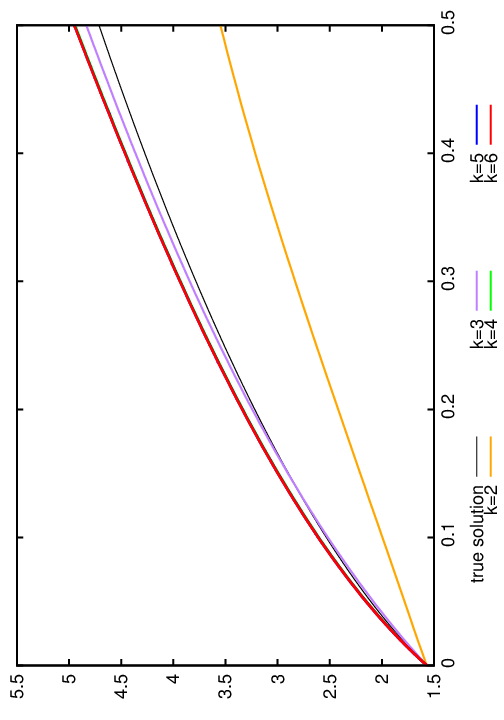}
\includegraphics[angle=-90,width=0.45\textwidth]{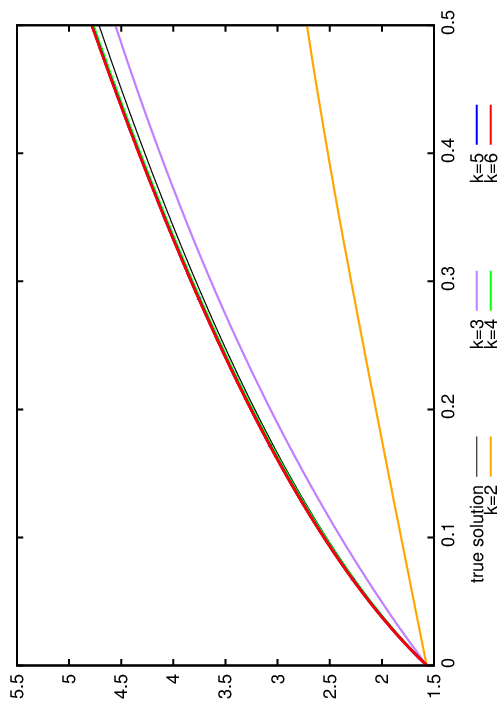}
\includegraphics[angle=-90,width=0.45\textwidth]{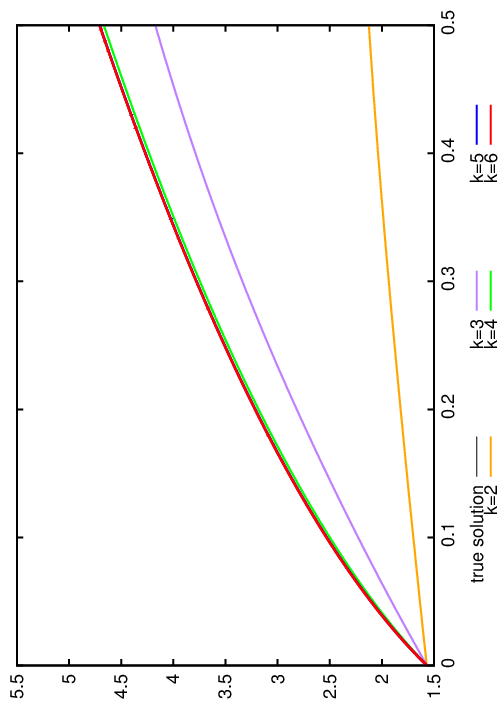}
\fi
\caption{(\PFqiii\ with the implicit time discretization from (\ref{eq:impl}), 
$\epsilon^{-1} = 4\,\pi,\,8\,\pi,\,16\,\pi,\,32\,\pi$)
Comparison of $|\Gamma(t)|$ and $\cPsi^{-1}\,\mathcal{E}_\epsilon^h$ for
the benchmark problem~\ref{bm:Stefan1} with $\ell=1$. 
The uniform time step sizes are chosen as $\tau=10^{-k}$, $k=2\to6$.
}
\label{fig:TSqiii_impl}
\end{figure}

For the benchmark problem~\ref{bm:Stefan1} with $\ell = 1$,
all of the phase field schemes were able 
to compute the necessary evolutions reasonably well. 
As an example we show the results for the scheme \PFqiii\ in 
Table~\ref{tab:TSqiii1}.
Here the definition of $\errorRr{x_1}{}$ is the same as $\errorRr{x_1}{1}$ with
$r_1$ replaced by $r$.
In order to be able to assess the absolute temperature errors $\errorWu$, 
we note that for this benchmark problem the true solution 
(\ref{eq:TrueSchmidtu}) remains in the range $[-0.95,-0.15]$.
The same computation now for the implicit time discretization from
(\ref{eq:impl}) is shown in Table~\ref{tab:TSqiii1_impl}.
In order to visualize the different behaviour of the two different time
discretizations, we plot the scaled phase field approximations 
$\mathcal{E}^h_\epsilon \approx \mathcal{E}_\epsilon$, recall (\ref{eq:Eg}),
together with $|\Gamma(t)| = 2\,\pi\,r(t)$ in Figures~\ref{fig:TSqiii}
and \ref{fig:TSqiii_impl}.
Similarly to Section~\ref{sec:41} it can be seen that the implicit time
discretization eliminates the time discretization error in the approximation of
the phase field equations sooner than the semi-implicit discretization.
Moreover, for the smallest value of $\epsilon$ the absolute errors 
$\errorRr{x_1}{}$
appear to be significantly smaller for the implicit scheme.
We remark that the converged errors in Table~\ref{tab:TSqiii1_impl} 
appear to indicate a convergence of $\mathcal{O}(\epsilon)$ in the
error $\errorRr{x_1}{}$, with a similar convergence rate for the error
$\LerrorWu$, if discretization errors are neglected.
\begin{table}
\center
\begin{tabular}{c|c|c|c|c|c|r}
 $\epsilon^{-1}$ & $2^\frac12 /h_{f}$ & $\tau$ & $\errorRr{x_1}{}$ & $\errorWu$
 & DOFs$(T)$ & CPU time \\ 
 \hline
 $4\,\pi$ 
 &32  & $10^{-2}$ & 2.5442e-01 & 2.1420e-01 & 2722 & 2 secs \\ 
 &32  & $10^{-3}$ & 3.0555e-02 & 3.6722e-02 & 4082 & 37 secs \\ 
 &32  & $10^{-4}$ & 1.2407e-01 & 9.6043e-02 & 4378 & 4:53 mins \\ 
 &32  & $10^{-5}$ & 1.5339e-01 & 1.3470e-01 & 4346 & 47:32 mins \\ 
 &32  & $10^{-6}$ & 1.5785e-01 & 1.3983e-01 & 4346 & 13:38 hours\\
 \hline
 $8\,\pi$ 
 &64  & $10^{-2}$ & 4.1068e-01 & 2.6415e-01 & 4082 & 4 secs\\ 
 &64  & $10^{-3}$ & 1.2126e-01 & 1.1494e-01 & 7074 & 3:06 mins\\
 &64  & $10^{-4}$ & 1.6550e-02 & 1.5965e-02 & 8418 & 32:17 mins\\
 &64  & $10^{-5}$ & 3.6656e-02 & 2.7052e-02 & 8610 & 3:04 hours\\
 &64  & $10^{-6}$ & 3.8850e-02 & 2.8693e-02 & 8530 & 22:05 hours\\
 \hline
 $16\,\pi$ 
 &128 & $10^{-2}$ & 4.7480e-01 & 2.8186e-01 & 7786 & 9 secs\\ 
 &128 & $10^{-3}$ & 3.0057e-01 & 2.1518e-01 & 11386 & 1:44 mins\\ 
 &128 & $10^{-4}$ & 5.5585e-02 & 5.2878e-02 & 16530 & 35:19 mins\\ 
 &128 & $10^{-5}$ & 4.0354e-03 & 8.8896e-03 & 17594 & 4:42 hours\\ 
 &128 & $10^{-6}$ & 1.1098e-02 & 9.9938e-03 & 18058 & 37:23 hours\\
 \hline
 $32\,\pi$ 
 &256 & $10^{-2}$ & 4.9344e-01 & 2.8715e-01 & 18970 & 23 secs \\ 
 &256 & $10^{-3}$ & 4.3490e-01 & 2.6236e-01 & 21346 & 6:31 mins\\ 
 &256 & $10^{-4}$ & 1.8070e-01 & 1.4732e-01 & 31882 & 1:22 hours\\ 
 &256 & $10^{-5}$ & 2.7810e-02 & 2.6160e-02 & 38186 & 13:55 hours\\
 &256 & $10^{-6}$ & 3.2260e-03 & 6.0793e-03 & 39178 & 93:40 hours \\
\end{tabular}
\caption{Benchmark problem~\ref{bm:Stefan1} with $\ell=1$ for \PFqiii.}
\label{tab:TSqiii1}
\end{table}%
\begin{table}
\center
\begin{tabular}{c|c|c|c|c|c|r}
 $\epsilon^{-1}$ & $2^\frac12 /h_{f}$ & $\tau$ & $\errorRr{x_1}{}$ & $\errorWu$
 & DOFs$(T)$ & CPU time \\ 
 \hline
 $4\,\pi$ 
 &32  & $10^{-2}$ & 5.6463e-02 & 1.4990e-01 & 3634 & 5 secs \\
 &32  & $10^{-3}$ & 1.0972e-01 & 7.4356e-02 & 4378 & 42 secs \\
 &32  & $10^{-4}$ & 1.5077e-01 & 1.3089e-01 & 4394 & 5:29 mins \\
 &32  & $10^{-5}$ & 1.5750e-01 & 1.3943e-01 & 4346 & 47:32 mins \\ 
 &32  & $10^{-6}$ & 1.5834e-01 & 1.4032e-01 & 4346 & 8:54 hours \\
 \hline
 $8\,\pi$ 
 &64  & $10^{-2}$ & 1.8482e-01 & 2.1201e-01 & 6394 & 8 secs \\
 &64  & $10^{-3}$ & 2.0140e-02 & 3.5234e-02 & 8498 & 1:20 mins \\
 &64  & $10^{-4}$ & 3.7523e-02 & 2.7306e-02 & 8578 & 14:27 mins \\
 &64  & $10^{-5}$ & 3.8793e-02 & 2.8669e-02 & 8530 & 2:02 hours \\
 &64  & $10^{-6}$ & 3.9062e-02 & 2.8841e-02 & 8530 & 15:51 hours\\
 \hline
 $16\,\pi$ 
 &128 & $10^{-2}$ & 3.1725e-01 & 2.5152e-01 & 11178 & 18 secs \\
 &128 & $10^{-3}$ & 2.4690e-02 & 6.9721e-02 & 16954 & 4:05 mins \\
 &128 & $10^{-4}$ & 9.0828e-03 & 1.1014e-02 & 17698 & 26:51 mins \\
 &128 & $10^{-5}$ & 1.1510e-02 & 1.0304e-02 & 17898 & 5:02 hours \\
 &128 & $10^{-6}$ & 1.1842e-02 & 1.0627e-02 & 17826 & 38:07 hours\\
 \hline
 $32\,\pi$ 
 &256 & $10^{-2}$ & 4.1167e-01 & 2.7209e-01 & 22506 & 54 secs \\
 &256 & $10^{-3}$ & 8.5869e-02 & 1.2716e-01 & 36050 & 11:59 mins \\
 &256 & $10^{-4}$ & 7.0679e-03 & 1.8667e-02 & 38938 & 1:32 hours \\
 &256 & $10^{-5}$ & 8.7275e-04 & 5.8758e-03 & 38906 & 10:39 hours \\
 &256 & $10^{-6}$ & 3.5699e-04 & 5.0530e-03 & 38922 & 91:29 hours \\
\end{tabular}
\caption{Benchmark problem~\ref{bm:Stefan1} with $\ell=1$ for \PFqiii\
with the implicit time discretization from (\ref{eq:impl}).}
\label{tab:TSqiii1_impl}
\end{table}%

The results for the same benchmark problem for the scheme \PFii\ with the
explicit time discretization from (\ref{eq:expl1},b) are shown in
Table~\ref{tab:TSii1_expl}. Here the reported CPU times are given as guidelines
only, because we did not employ any parallelization. Recall that the discrete
systems of equations decouple in space, and so a significant speedup of the
computations can be expected if they are run in parallel on a large
cluster.
Firstly, we see that the obtained results appear to
confirm the stability constraint (\ref{eq:tauh2}), i.e.\ 
$\tau \leq \frac12\,C_\star\,h^2$. Secondly, it can be observed that once the
explicit method is stable, there is hardly any variation in the numerical 
results when decreasing $\tau$ further. And finally, it is clear from
Table~\ref{tab:TSii1_expl} that there is no convergence in the reported error
quantities, i.e.\ the phase field simulations do not converge to the sharp
interface problem (\ref{eq:i1a}--e) as $\epsilon,\,h_f,\,\tau \to 0$. We
conjecture that this phenomenon is due to the sensitivity of the explicit
method to the employed mesh adaptation strategy, recall Section~\ref{sec:31}.
This is confirmed by repeating the simulations for the explicit scheme on
uniform grids, see Table~\ref{tab:TSii1_expl_unif}. Now the errors appear to be
converging, and the absolute errors agree with the corresponding converged 
errors from the semi-implicit and implicit variants of \PFii, on which we do 
not report here. 
\begin{table}
\center
\begin{tabular}{c|c|c|c|c|c|r}
 $\epsilon^{-1}$ & $2^\frac12 /h_{f}$ & $\tau$ & $\errorRr{x_1}{}$ & $\errorWu$
 & DOFs$(T)$ & CPU time \\ 
 \hline
 $4\,\pi$ 
 &32  & $10^{-4}$ & 4.2499e-02 & 8.9335e-02 & 3114 & 1:02 mins\\ 
 &32  & $10^{-5}$ & 4.2868e-02 & 8.9236e-02 & 3114 & 10:04 mins\\ 
 &32  & $10^{-6}$ & 4.2906e-02 & 8.9228e-02 & 3114 & 1:39 hours\\ 
 &32  & $10^{-7}$ & 4.2909e-02 & 8.9227e-02 & 3114 & 16:29 hours\\ 
 \hline
 $8\,\pi$ 
 &64  & $10^{-4}$ & --- & --- & --- &  unstable \\
 &64  & $10^{-5}$ & 2.6258e-02 & 5.2713e-02 & 6850 & 21:47 mins\\ 
 &64  & $10^{-6}$ & 2.6311e-02 & 5.2678e-02 & 6850 & 3:19 hours\\ 
 &64  & $10^{-7}$ & 2.6317e-02 & 5.2676e-02 & 6850 & 36:27 hours\\ 
 \hline
 $16\,\pi$ 
 &128 & $10^{-4}$ & --- & --- & --- & unstable \\
 &128 & $10^{-5}$ & 2.4416e-02 & 3.4926e-02 & 14706 & 50:02 mins\\ 
 &128 & $10^{-6}$ & 2.4453e-02 & 3.4777e-02 & 14706 & 8:32 hours\\ 
 &128 & $10^{-7}$ & 2.4458e-02 & 3.4767e-02 & 14706 & 84:39 hours\\ 
 \hline
 $32\,\pi$ 
 &256 & $10^{-4}$ & --- & --- & --- & unstable \\
 &256 & $10^{-5}$ & --- & --- & --- & unstable \\
 &256 & $10^{-6}$ & 3.6152e-02 & 3.5620e-02 & 33074 & 22:05 hours\\ 
 &256 & $10^{-7}$ & 3.6150e-02 & 3.5595e-02 & 33138 & 220:11 hours \\ 
\end{tabular}
\caption{Benchmark problem~\ref{bm:Stefan1} with $\ell=1$ for \PFii\
with the explicit time discretization from (\ref{eq:expl1},b).}
\label{tab:TSii1_expl}
\end{table}%
\begin{table}
\center
\begin{tabular}{c|c|c|c|c|c|r}
 $\epsilon^{-1}$ & $2^\frac12 /h$ & $\tau$ & $\errorRr{x_1}{}$ & $\errorWu$
 & DOFs$(T)$ & CPU time \\ 
 \hline
 $4\,\pi$ 
 &32  & $10^{-5}$ & 4.7109e-02 & 8.7382e-02 & 8450 & 33:11 mins\\ 
 &32  & $10^{-6}$ & 4.7154e-02 & 8.7376e-02 & 8450 & 5:32 hours\\ 
 \hline
 $8\,\pi$ 
 &64  & $10^{-5}$ & 2.6105e-02 & 5.0006e-02 & 33282 & 3:24 hours\\ 
 &64  & $10^{-6}$ & 2.6185e-02 & 5.0006e-02 & 33282 & 32:59 hours \\ 
 \hline
 $16\,\pi$ 
 &128 & $10^{-5}$ & 1.7399e-02 & 3.0422e-02 & 132098 & 14:26 hours\\ 
 &128 & $10^{-6}$ & 1.7520e-02 & 3.0420e-02 & 132098 & 139:29 hours \\ 
\end{tabular}
\caption{Benchmark problem~\ref{bm:Stefan1} with $\ell=1$ for \PFii\
with the explicit time discretization from (\ref{eq:expl1},b) and with a
uniform spatial mesh.}
\label{tab:TSii1_expl_unif}
\end{table}%

We compare the above phase field errors with the corresponding errors for 
the sharp interface
algorithm \PFEM. Here $\errorRr{}{}$ is defined as $\errorRr11$, but with
$\Gamma_1^h$ replaced by $\Gamma^h$ and with $r_1$ replaced by $r$.
The errors for the benchmark problem~\ref{bm:Stefan1} with $\ell=1$ are 
reported in Table~\ref{tab:TS1p}.
\begin{table}
\center
\begin{tabular}{c|c|c|c|c|r}
 $2^\frac12/h_{f}$ & $\tau$ & $\errorRr{}{}$ & $\errorUu$ & 
 DOFs$(T)$ & CPU time \\ 
 \hline 
   32 & $10^{-2}$ & 1.1463e-02 & 6.1188e-02 & 1193 & 0 secs\\ 
   32 & $10^{-3}$ & 2.6832e-02 & 1.9387e-02 & 1181 & 3 secs\\ 
   32 & $10^{-4}$ & 2.9267e-02 & 2.1448e-02 & 1193 & 23 secs\\ 
 \hline
   64 & $10^{-2}$ & 1.3678e-02 & 6.3622e-02 & 2585 & 0 secs\\ 
   64 & $10^{-3}$ & 1.1105e-02 & 8.4213e-03 & 2441 & 5 secs\\ 
   64 & $10^{-4}$ & 1.3215e-02 & 1.2807e-02 & 2457 & 52 secs\\ 
 \hline
  128 & $10^{-2}$ & 1.5197e-02 & 6.4894e-02 & 6153 & 1 secs\\ 
  128 & $10^{-3}$ & 3.6275e-03 & 6.6864e-03 & 5513 & 13 secs\\ 
  128 & $10^{-4}$ & 5.5642e-03 & 4.9862e-03 & 5401 & 2:06 mins\\ 
 \hline
  256 & $10^{-2}$ & 1.7421e-02 & 6.4862e-02 & 15497 & 5 secs\\ 
  256 & $10^{-3}$ & 1.0911e-03 & 6.9236e-03 & 13157 & 39 secs\\ 
  256 & $10^{-4}$ & 2.2576e-03 & 1.9430e-03 & 12941 & 6:04 mins\\ 
 \hline
  512 & $10^{-2}$ & 1.7903e-02 & 6.4964e-02 & 37517 & 12 secs\\ 
  512 & $10^{-3}$ & 1.4352e-03 & 7.2994e-03 & 35257 & 2:12 mins\\ 
  512 & $10^{-4}$ & 8.9361e-04 & 8.3770e-04 & 34157 & 20:47 mins\\ 
 \hline
 1024 & $10^{-2}$ & 1.8248e-02 & 6.5448e-02 & 89437 & 45 secs\\ 
 1024 & $10^{-3}$ & 1.6084e-03 & 7.5361e-03 & 105857 & 8:56 mins\\ 
 1024 & $10^{-4}$ & 3.0915e-04 & 6.5204e-04 & 101593 & 1:23 hours\\ 
\end{tabular}
\caption{Benchmark problem~\ref{bm:Stefan1} with $\ell=1$ for \PFEM.}
\label{tab:TS1p}
\end{table}%
Comparing the results in Tables~\ref{tab:TSqiii1}--\ref{tab:TS1p} reveals once
again that the sharp interface approximations from \PFEM\ are more accurate
than the corresponding computations from the phase field schemes
\PF\ and \PFq, and they can be obtained in a fraction of the CPU time.
For example, we see from the Tables~\ref{tab:TSqiii1}, \ref{tab:TSqiii1_impl} 
and \ref{tab:TS1p} that in order to reduce the error in both 
$\Gamma$ and $u$ below $10^{-2}$ requires 13 seconds 
with \PFEM, but it takes around $5$ and $11$ hours, depending on the time
discretization, with \PFqiii. In other words, in this measure 
the parametric front tracking method \PFEM\ is between $1\,300$ and
$2\,900$ times faster than the phase field methods.
A visualization of the numerical results in 
Tables~\ref{tab:TSqiii1}--\ref{tab:TS1p} can be seen in 
Figure~\ref{fig:scatter8all}, below.

If we increase the parameter $\ell$ in the 
benchmark problem~\ref{bm:Stefan1} to $\ell = 3$, 
which on recalling (\ref{eq:epscond}) means that the problem is now
computationally more challenging, then for moderate values of 
$\epsilon$ all of the phase field schemes exhibit mushy regions in which the 
phase field approximations $\varphi^h$ in modulus 
take on values significantly smaller than unity. 
This leads to excessive CPU times, since the adaptive mesh 
strategy uses fine meshes in the interfacial regions. In addition, often the
mushy interfacial region quickly reaches the external boundary
$\partial\Omega$, which creates additional interfaces, so that the phase field
solutions $\varphi^h$ no longer approximate the radially symmetric sharp
interface solution $\Gamma(t)$. 
Hence in what follows we present convergence experiments for the 
benchmark problem~\ref{bm:Stefan1} with $\ell = 3$ 
only for $\epsilon^{-1} \geq 16\,\pi$.
See Table~\ref{tab:TSii} for the results for the scheme \PFii,
where we note that for this benchmark problem the true solution 
(\ref{eq:TrueSchmidtu}) remains in the range $[-0.35,0]$.
Similar results
can be obtained for the other variants of the schemes \PF\ and \PFq\ that
satisfy $\varrho(1)=0$, but we omit them here for brevity.
In particular, the scheme \PFii\ with the implicit time discretization from
(\ref{eq:impl}) yields almost identical results to the ones in
Table~\ref{tab:TSii} for the step sizes $\tau=10^{-k}$, $k=5\to6$.
\begin{table}
\center
\begin{tabular}{c|c|c|c|c|c|r}
 $\epsilon^{-1}$ & $2^\frac12 /h_{f}$ & $\tau$ & $\errorRr{x_1}{}$ & $\errorWu$
 & DOFs$(T)$ & CPU time \\ 
 \hline
 $16\,\pi$ 
 &128 & $10^{-2}$ & 3.7529e-02 & 1.3113e-01 & 15098 & 12 secs \\ 
 &128 & $10^{-3}$ & 1.9000e-02 & 1.7671e-02 & 15970 & 2:02 mins\\ 
 &128 & $10^{-4}$ & 1.2338e-02 & 1.5085e-02 & 15746 & 16:20 mins\\ 
 &128 & $10^{-5}$ & 1.1885e-02 & 1.4764e-02 & 15642 & 2:03 hours\\ 
 &128 & $10^{-6}$ & 1.1836e-02 & 1.4718e-02 & 15626 & 18:57 hours\\
 \hline
 $32\,\pi$ 
 &256 & $10^{-2}$ & 2.2627e-01 & 2.1834e-01 & 26306 & 31 secs\\ 
 &256 & $10^{-3}$ & 6.3959e-03 & 3.5178e-02 & 34642 & 5:31 mins\\ 
 &256 & $10^{-4}$ & 2.5004e-03 & 7.2381e-03 & 34906 & 45:39 mins\\ 
 &256 & $10^{-5}$ & 2.6881e-03 & 7.6427e-03 & 34770 & 6:08 hours\\
 &256 & $10^{-6}$ & 2.7040e-03 & 7.6802e-03 & 34802 & 51:18 hours\\
 \hline
 $64\,\pi$ 
 &512 & $10^{-2}$ & 3.8556e-01 & 2.5703e-01 & 58226 & 2:04 mins\\ 
 &512 & $10^{-3}$ & 7.9248e-02 & 1.0625e-01 & 80330 & 22:12 mins\\ 
 &512 & $10^{-4}$ & 1.0521e-02 & 1.3930e-02 & 85218 & 2:48 hours\\ 
 &512 & $10^{-5}$ & 4.1001e-03 & 5.2299e-03 & 85714 & 27:46 hours\\
 &512 & $10^{-6}$ & 3.3949e-03 & 6.4424e-03 & 85842 &193:11 hours\\
\end{tabular}
\caption{Benchmark problem~\ref{bm:Stefan1} with $\ell=3$ for \PFii.}
\label{tab:TSii}
\end{table}%
We remark that for the numbers in Table~\ref{tab:TSii}
it is somewhat speculative to infer convergence rates in terms of
$\epsilon$,
since the errors $\errorRr{x_1}{}$ and $\errorWu$ for the smallest value of 
$\epsilon$ do not appear to have converged yet in terms of $h_f$ and $\tau$.

We again compare these numbers with the corresponding errors for the sharp 
interface algorithm \PFEM, see Table~\ref{tab:TSp}. 
\begin{table}
\center
\begin{tabular}{c|c|c|c|c|r}
 $2^\frac12/h_{f}$ & $\tau$ & $\errorRr{}{}$ & $\errorUu$ & 
 DOFs$(T)$ & CPU time \\ 
 \hline 
   32 & $10^{-2}$ & 5.9808e-02 & 3.4189e-02 & 1213 & 1 secs \\
   32 & $10^{-3}$ & 4.2215e-02 & 2.9898e-02 & 1169 & 3 secs \\
   32 & $10^{-4}$ & 4.1433e-02 & 2.9813e-02 & 1217 & 27 secs \\
 \hline
   64 & $10^{-2}$ & 2.7229e-02 & 2.9493e-02 & 2705 & 1 secs \\
   64 & $10^{-3}$ & 2.5199e-02 & 1.7501e-02 & 2505 & 7 secs \\
   64 & $10^{-4}$ & 2.6297e-02 & 1.8341e-02 & 2497 & 1:06 mins \\
 \hline
  128 & $10^{-2}$ & 1.1307e-02 & 3.0940e-02 & 6133 & 3 secs \\
  128 & $10^{-3}$ & 1.0799e-02 & 7.7496e-03 & 5529 & 21 secs \\
  128 & $10^{-4}$ & 1.1358e-02 & 8.4107e-03 & 5489 & 2:49 mins \\
 \hline
  256 & $10^{-2}$ & 5.9618e-03 & 3.1694e-02 & 15781 & 8 secs \\
  256 & $10^{-3}$ & 4.6057e-03 & 3.1840e-03 & 13165 & 1:08 mins \\
  256 & $10^{-4}$ & 4.9513e-03 & 3.7938e-03 & 12977 & 10:08 mins \\
 \hline
  512 & $10^{-2}$ & 8.1252e-03 & 3.2015e-02 & 37069 & 21 secs \\
  512 & $10^{-3}$ & 1.9929e-03 & 3.5308e-03 & 35205 & 3:44 mins \\
  512 & $10^{-4}$ & 2.1894e-03 & 1.7126e-03 & 34149 & 33:04 mins \\
 \hline
 1024 & $10^{-2}$ & 4.6688e-03 & 3.2259e-02 & 89301 & 1:09 mins \\
 1024 & $10^{-3}$ & 8.9176e-04 & 3.7858e-03 & 105993 & 14:07 mins \\
 1024 & $10^{-4}$ & 1.0202e-03 & 7.8635e-04 & 101517 & 2:04 hours \\
\end{tabular}
\caption{Benchmark problem~\ref{bm:Stefan1} for $\ell=3$ for \PFEM.}
\label{tab:TSp}
\end{table}%
The results in Tables~\ref{tab:TSii} and \ref{tab:TSp} 
confirm once more that the sharp interface approximations from \PFEM\ are 
more accurate.
For example, in order to reduce both the error in $\Gamma$ and in $u$ to below
$10^{-2}$ requires about a minute with \PFEM, but $46$ minutes with \PFii.
But crucially, it is clear from the numbers in 
Tables~\ref{tab:TSqiii1}, \ref{tab:TSqiii1_impl} and \ref{tab:TSii} 
that only by decreasing $\epsilon$ further can the observed errors 
for the phase field methods be reduced. 
This in turn will lead to enforced reductions in $h_f$ and $\tau$, recall 
e.g.\ (\ref{eq:NV97a}), (\ref{eq:hf}), (\ref{eq:tsc}) and (\ref{eq:tauh}). 
Overall this makes it impossible to
perform these computations in practice. 
On the other hand, 
the presented errors in Tables~\ref{tab:TS1p} and \ref{tab:TSp}
for the scheme \PFEM\ indicate a convergence in the error $\errorRr{}{}$
of order at least $\mathcal{O}(h)$, 
if time discretization effects are neglected. 
Apart from the run for the finest value of $h_f$ in Table~\ref{tab:TS1p}, 
where the time discretization error does not seem to have been eliminated yet, 
the same can be said about the temperature error $\errorUu$. 

\begin{figure}
\center
\ifpdf
\includegraphics[angle=-0,width=0.55\textwidth]{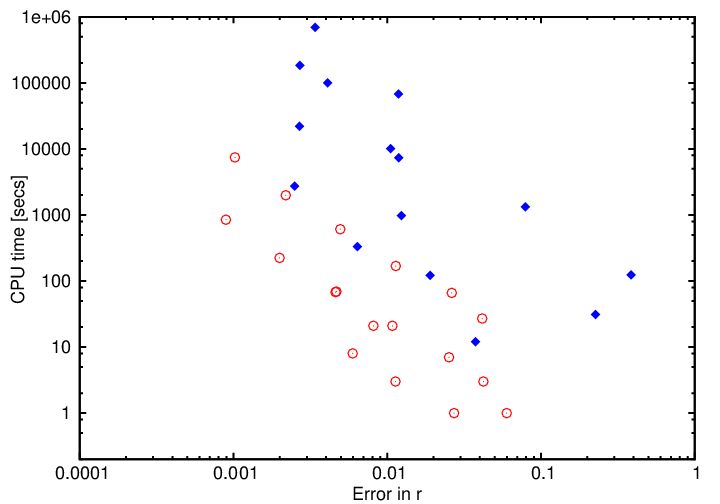}
\fi
\caption{(Benchmark problem~\ref{bm:Stefan1} with $\ell=3$)
Log-log scatter plot of $\errorRr{x_1}{}$ and $\errorRr{}{}$
against the CPU time for
the entries in Table~\ref{tab:TSii} (\PFii, blue rhombi)
and Table~\ref{tab:TSp} (\PFEM, red circles).
}
\label{fig:scatter910}
\end{figure}

In order to visualize the relative performances of \PFii\ in
Table~\ref{tab:TSii} and \PFEM\ in
Table~\ref{tab:TSp}, we present a plot of the errors in the radius $r$
against the necessary CPU time for all the entries in the two tables in
Figure~\ref{fig:scatter910}. 

Similarly to Section~\ref{sec:41}, it is not possible to perform a meaningful
convergence test for the solution (\ref{eq:TSr},b) in
the case $d=3$ for the phase field approximations \PF\ and \PFq. As an example
for such a convergence experiment for the parametric scheme \PFEM\ we now
consider the three-dimensional analogue of the 
benchmark problem~\ref{bm:Stefan1}.

\begin{algorithm}[H]
\caption{3d Stefan problem with $\vartheta = 1$ and $\rho > 0$.}
\label{bm:Stefan3d}
Same as benchmark problem~\ref{bm:Stefan1}, but on the domain
$\Omega = (-1, 1)^3$.
\end{algorithm}

The corresponding errors are shown in Table~\ref{tab:TS3d_coarser}, 
where we let $h_\Gamma \approx 6\,h_f$. Similarly to the two dimensional
benchmark problem, the two errors $\errorRr{}{}$ and $\errorUu$ appear to
converge with order at least $\mathcal{O}(h)$, 
if time discretization effects are neglected. 
\begin{table}
\center
\begin{tabular}{c|c|c|c|c|r}
 $2^\frac12/h_{f}$ & $\tau$ & $\errorRr{}{}$ & $\errorUu$ & 
 DOFs$(T)$ & CPU time \\ 
 \hline 
   32 & $10^{-2}$ & 1.0373e-01 & 4.0392e-02 & 38661 & 3:06 mins\\ 
   32 & $10^{-3}$ & 1.2020e-01 & 6.0675e-02 & 38067 & 23:07 mins\\ 
   32 & $10^{-4}$ & 3.4579e-01 & 1.0935e-01 & 39826 & 3:06 hours \\ 
 \hline
   64 & $10^{-2}$ & 2.4115e-02 & 3.2858e-02 & 162161 & 25:31 mins\\ 
   64 & $10^{-3}$ & 3.0456e-02 & 1.9335e-02 & 147787 & 2:33 hours\\ 
   64 & $10^{-4}$ & 4.7083e-02 & 2.7765e-02 & 148036 & 17:10 hours\\ 
 \hline
  128 & $10^{-2}$ & 1.0819e-02 & 3.1504e-02 & 740635 & 2:08 hours\\
  128 & $10^{-3}$ & 1.1358e-02 & 7.7776e-03 & 603059 & 12:43 hours\\ 
  128 & $10^{-4}$ & 1.6654e-02 & 1.0916e-02 & 593187 & 65:34 hours\\
 \hline
  256 & $10^{-2}$ & 8.2692e-03 & 3.1338e-02 & 3480836 & 9:19 hours\\ 
  256 & $10^{-3}$ & 5.6216e-03 & 3.7231e-03 & 2577960 & 47:19 hours\\
  256 & $10^{-4}$ & 6.4224e-03 & 4.7100e-03 & 2490788 & 308:35 hours\\
\end{tabular}
\caption{Benchmark problem~\ref{bm:Stefan3d} with $\ell=3$ for \PFEM.}
\label{tab:TS3d_coarser}
\end{table}%

Our final benchmark problem is the Stefan problem without interfacial kinetics
in the Gibbs--Thomson law. Here we recall from e.g.\ \cite{KarmaR96} that
often standard, classical phase field methods are not able to deal with 
this case in practice.

\begin{algorithm}[H]
\caption{2d Stefan problem with $\vartheta = 1$ and $\rho = 0$.}
\label{bm:Stefan3}
Same as benchmark problem~\ref{bm:Stefan1}, but with $\alpha = 10^{-\ell}$ and
$\rho = 0$.
\end{algorithm}

\begin{figure}
\center
\ifpdf
\includegraphics[angle=-0,width=0.55\textwidth]{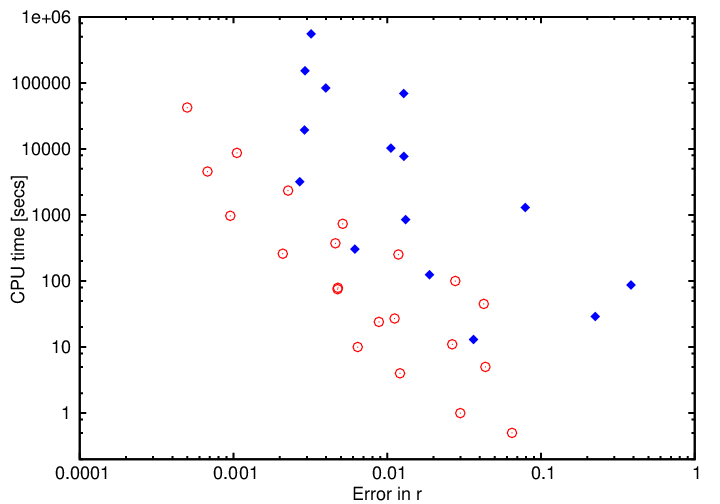}
\fi
\caption{(Benchmark problem~\ref{bm:Stefan3} with $\ell=3$)
Log-log scatter plot of $\errorRr{x_1}{}$ and $\errorRr{}{}$
against the CPU time for
the entries in Table~\ref{tab:TSii_rho0} (\PFii, blue rhombi)
and Table~\ref{tab:TSp_rho0} (\PFEM, red circles).
}
\label{fig:scatter1213}
\end{figure}

We stress that the approach from \cite{vch} for the phase field system has  
no problems in dealing with the case without interfacial kinetics, i.e.\ 
if $\rho=0$. For example, a computation for the scheme \PFii\ can be found in
Table~\ref{tab:TSii_rho0}. We compare these results with the corresponding
computation for the sharp interface algorithm \PFEM\ in 
Table~\ref{tab:TSp_rho0}, where once again it appears that
the error quantities converge with $\mathcal{O}(h)$ if the time 
discretization errors are neglected. 
\begin{table}
\center
\begin{tabular}{c|c|c|c|c|c|r}
 $\epsilon^{-1}$ & $2^\frac12 /h_{f}$ & $\tau$ & $\errorRr{x_1}{}$ & $\errorWu$
 & DOFs$(T)$ & CPU time \\ 
 \hline
 $16\,\pi$ 
 &128 & $10^{-2}$ & 3.6341e-02 & 1.2977e-01 & 15370 & 13 secs \\
 &128 & $10^{-3}$ & 1.8831e-02 & 1.8600e-02 & 16226 & 2:04 mins \\
 &128 & $10^{-4}$ & 1.3159e-02 & 1.5797e-02 & 15818 & 14:12 mins \\
 &128 & $10^{-5}$ & 1.2840e-02 & 1.5560e-02 & 15890 & 2:09 hours \\
 &128 & $10^{-6}$ & 1.2810e-02 & 1.5510e-02 & 15890 & 19:13 hours \\
 \hline
 $32\,\pi$ 
 &256 & $10^{-2}$ & 2.2563e-01 & 2.1820e-01 & 26362 & 29 secs \\ 
 &256 & $10^{-3}$ & 6.1576e-03 & 3.4619e-02 & 34634 & 5:04 mins \\
 &256 & $10^{-4}$ & 2.6943e-03 & 7.5241e-03 & 34938 & 53:05 mins \\
 &256 & $10^{-5}$ & 2.8987e-03 & 7.9607e-03 & 34754 & 5:23 hours \\
 &256 & $10^{-6}$ & 2.9172e-03 & 1.1125e-02 & 34818 & 42:28 hours \\
 \hline
 $64\,\pi$ 
 &512 & $10^{-2}$ & 3.8566e-01 & 2.5703e-01 & 58146 & 1:27 mins \\
 &512 & $10^{-3}$ & 7.9018e-02 & 1.0605e-01 & 80010 & 21:39 mins \\
 &512 & $10^{-4}$ & 1.0560e-02 & 1.3817e-02 & 85266 & 2:51 hours \\
 &512 & $10^{-5}$ & 3.9822e-03 & 5.3896e-03 & 86002 & 23:17 hours \\
 &512 & $10^{-6}$ & 3.1951e-03 & 1.0395e-02 & 86042 & 154:02 hours \\
\end{tabular}
\caption{Benchmark problem~\ref{bm:Stefan3} with $\ell=3$ for \PFii.}
\label{tab:TSii_rho0}
\end{table}%
\begin{table}
\center
\begin{tabular}{c|c|c|c|c|r}
 $2^\frac12/h_{f}$ & $\tau$ & $\errorRr{}{}$ & $\errorUu$ & 
 DOFs$(T)$ & CPU time \\ 
 \hline 
   32 & $10^{-2}$ & 6.4835e-02 & 3.5949e-02 & 1225 & 0 secs\\ 
   32 & $10^{-3}$ & 4.3505e-02 & 3.0646e-02 & 1169 & 5 secs\\ 
   32 & $10^{-4}$ & 4.2440e-02 & 3.0431e-02 & 1217 & 45 secs\\ 
 \hline
   64 & $10^{-2}$ & 2.9889e-02 & 2.9534e-02 & 2613 & 1 secs\\ 
   64 & $10^{-3}$ & 2.6450e-02 & 1.8257e-02 & 2521 & 11 secs\\ 
   64 & $10^{-4}$ & 2.7699e-02 & 1.9103e-02 & 2497 & 1:40 mins\\ 
 \hline
  128 & $10^{-2}$ & 1.2103e-02 & 3.1028e-02 & 6109 & 4 secs\\ 
  128 & $10^{-3}$ & 1.1158e-02 & 8.0117e-03 & 5529 & 27 secs\\ 
  128 & $10^{-4}$ & 1.1813e-02 & 8.7200e-03 & 5489 & 4:12 mins\\ 
 \hline
  256 & $10^{-2}$ & 6.4202e-03 & 3.1800e-02 & 15805 & 10 secs \\ 
  256 & $10^{-3}$ & 4.7724e-03 & 3.3036e-03 & 13221 & 1:19 mins\\ 
  256 & $10^{-4}$ & 5.1346e-03 & 3.9258e-03 & 13001 & 12:16 mins\\ 
 \hline
  512 & $10^{-2}$ & 8.8182e-03 & 3.2129e-02 & 37161 & 24 secs\\ 
  512 & $10^{-3}$ & 2.0921e-03 & 3.5487e-03 & 35117 & 4:19 mins\\ 
  512 & $10^{-4}$ & 2.2635e-03 & 1.7705e-03 & 34165 & 38:59 mins\\ 
 \hline
 1024 & $10^{-2}$ & 4.7558e-03 & 3.2373e-02 & 89269 & 1:15 min\\ 
 1024 & $10^{-3}$ & 9.5286e-04 & 3.8105e-03 & 105997 & 16:11 mins\\ 
 1024 & $10^{-4}$ & 1.0525e-03 & 8.1355e-04 & 101509 & 2:25 hours\\ 
 \hline
 2048 & $10^{-2}$ & 4.6014e-03 & 3.2830e-02 & 295201 & 6:13 min\\ 
 2048 & $10^{-3}$ & 6.7747e-04 & 3.8557e-03 & 352153 & 1:16 hours\\ 
 2048 & $10^{-4}$ & 4.9986e-04 & 3.6634e-04 & 334785 & 11:48 hours\\  
\end{tabular}
\caption{Benchmark problem~\ref{bm:Stefan3} for $\ell=3$ for \PFEM.}
\label{tab:TSp_rho0}
\end{table}%
We visualize the relative performances of \PFii\ in
Table~\ref{tab:TSii_rho0} and \PFEM\ in
Table~\ref{tab:TSp_rho0} in Figure~\ref{fig:scatter1213}. 
As before, the performance of \PFEM\
is vastly superior to the corresponding phase field computations.

\subsection{Second order accurate isotropic phase field model} \label{sec:43}

In this subsection we recall a variant of the phase field model 
(\ref{eq:heata}--d), (\ref{eq:ACa}--c) which in the isotropic setting 
(\ref{eq:iso}), (\ref{eq:betaiso}) yields a second order convergence in
$\epsilon$ to the sharp interface limit; see e.g.\ 
\cite{KarmaR96,Almgren99,GarckeS06,ChenCE06,CaginalpCE08} for details.
In particular, it needs to be assumed that the shape function $\varrho$, recall
(\ref{eq:Varrho}), satisfies (\ref{eq:varrhosym},b).
Clearly, of our examples in (\ref{eq:varrho}) only the choices (iii) and (iv)
satisfy this.

{From} now on we assume that (\ref{eq:iso}) and (\ref{eq:betaiso}) 
hold, and that $\conduct^+ = \conduct^- > 0$. 
Then in place of (\ref{eq:heata}) and (\ref{eq:ACa}) we consider
\begin{subequations}
\begin{align} \label{eq:EGSa} 
\vartheta\,w_t + \lambda\,\widetilde\varrho(\varphi)\,\varphi_t & =
\conduct\,\Delta\, w \,,\\
\cPsi\,a\,\alpha^{-1}\,\varrho(\varphi)\,w & = 
\epsilon\,\alpha^{-1}\,(\rho + \rho_1\,\epsilon)\,\varphi_t
-\epsilon\,\Delta \,\varphi +\epsilon^{-1}\,\Psi'(\varphi)\,, \label{eq:EGSb} 
\end{align}
\end{subequations}
in $\Omega_T$, where $\widetilde\varrho:\R \to \R$ is a second shape
function that satisfies
\begin{equation*} 
\widetilde\varrho(s) \geq 0\quad\forall\ s \in[-1,1]\,,\quad
\widetilde\varrho(s) = \widetilde\varrho(-s) \quad \forall\ s \in \R\,,\quad
\int^1_{-1} \widetilde\varrho(y) \; {\rm d}y = 1 \,.
\end{equation*}
Here $\rho_1$ in (\ref{eq:EGSb}) is a correction term that is given by 
\begin{equation} \label{eq:rho1}
\rho_1 = K\,\frac{\lambda\,a}{\conduct}\,,
\end{equation}
where, see e.g.\ \cite[Eq.\ (35)]{GarckeS06},
\begin{equation} \label{eq:K}
K := \int_\R [1 - \Varrho(\Phi(s))]\,\widetilde\Varrho(\Phi(s)) \; {\rm d}s
\qquad \text{with} \qquad
\widetilde\Varrho(s) := \int^s_{-1} \widetilde\varrho(y) \; {\rm d}y\,.
\end{equation} 
In (\ref{eq:K}) the function $\Phi : \R \to \R$ denotes the unique solution to
\begin{equation*} 
\Phi''(s) - \Psi'(\Phi(s)) = 0 \quad \forall\ s \in \R\,, \qquad
\lim_{s\to\pm\infty}\Phi(s)=\pm1\,,\qquad \int_\R s\,\Phi'(s)\;{\rm d}s = 0\,.
\end{equation*}
For the above choices of $\varrho$, $\widetilde{\varrho}$ and $\rho_1$,
Almgren \cite{Almgren99} 
formally showed second order convergence in the sense that the
approximation of the Gibbs--Thomson law is of $\mathcal{O}(\epsilon^2)$,
whereas in \cite{GarckeS06} it is formally established that
\begin{itemize}
\item the zero level set of the phase field function $\varphi$ approximates the
  interface $\Gamma$ to $\mathcal{O}(\epsilon^2)$,
\item the temperature $w$ in the phase field system approximates the
  temperature $u$ in the sharp interface problem to $\mathcal{O}(\epsilon^2)$.
\end{itemize}
In \cite{ChenCE06}, for the special case $\widetilde\varrho(s) = \frac12$,
the above second order approximation results are shown rigorously. In
particular, on letting $\conduct = a = 1$, and on recalling that in their
notation $G(s) = \cPsi\,\Varrho(s)$, it holds that the expression in
\cite[Eq.\ (1.6)]{ChenCE06} for the correction term $\rho_1$ is given by
\begin{align*}
\rho_1 & = \tfrac12\,\lambda\,
\frac{\int_\R [G(1) - G(\Phi(s))](1+\Phi(s)) \; {\rm d}s}
{\int_\R [\Phi'(s)]^2 \; {\rm d}s}
=
\tfrac12\,\lambda\,\cPsi\,
\frac{\int_\R [1 - \Varrho(\Phi(s))](1+\Phi(s)) \; {\rm d}s}
{\int_\R [\Phi'(s)]^2 \; {\rm d}s} \nonumber\\ &
= \tfrac12\,\lambda\,\int_\R [1 - \Varrho(\Phi(s))](1+\Phi(s)) \; {\rm d}s = 
\lambda\,K\,, 
\end{align*}
and so agrees exactly with (\ref{eq:rho1}). In addition, on assuming the
stronger condition
\begin{equation} \label{eq:varrhoPsi}
\varrho(s) = \tfrac1{\cPsi}\,\sqrt{2\,\Psi(s)}
\end{equation}
in place of (\ref{eq:varrhosym},b), the authors in \cite{ChenCE06} 
also show rigorously that the full phase field converges to second order. 
More precisely, in this case the first order correction to the phase field 
function $\varphi$ is zero. Of course, 
the specific choice $\widetilde{\varrho}(s)=\tfrac{1}{2}$ for the 
interpolation function in the equation for the temperature means that the
overall phase field system considered in \cite{ChenCE06}
is not thermodynamically consistent.
We refer to 
\cite{Almgren99,ChenCE06,GarckeS06} for the precise statements of these 
results.

{From} now on we consider (\ref{eq:EGSa},b) in the case that 
$\widetilde\varrho = \varrho$, which means that we return to
(\ref{eq:heata}), (\ref{eq:ACa}) in the isotropic case (\ref{eq:iso}), 
(\ref{eq:betaiso}) with $\conduct_+ = \conduct_-$. In particular, 
the phase field model (\ref{eq:EGSa},b) is now thermodynamically
consistent, i.e.\ it satisfies (\ref{eq:pf2Lyap}) with (\ref{eq:iso}), 
(\ref{eq:betaiso}) and $\conduct_+ = \conduct_-$.
Since now $\widetilde\varrho = \varrho$, it follows from (\ref{eq:K}) that
\begin{equation*} 
K = \int_\R [1 - \Varrho(\Phi(s))]\,\Varrho(\Phi(s)) \; {\rm d}s\,.
\end{equation*} 
In the case of the obstacle potential (\ref{eq:obstacle}), so that 
$\Phi(s) = \sin(s)$ for $|s|\leq \frac\pi2$ as in (\ref{eq:phi0}), we get 
that 
\begin{equation*} 
K = \int_{-\frac\pi2}^{\frac\pi2} [1 - \Varrho(\sin(s))]\,\Varrho(\sin(s)) 
\; {\rm d}s\,.
\end{equation*}
In particular, for the choice (\ref{eq:varrho})(iii), when
$\Varrho(s) = \frac1{16}\,(3\,s^5-10\,s^3+15\,s+8)$, we have that
\begin{equation} \label{eq:Kiii}
K = \tfrac{4817}{65536}\,\pi = 2^{-16}\,4817\,\pi \approx 0.231\,.
\end{equation}
We refer to Table~\ref{tab:TSciii1} for computations for the benchmark
problem~\ref{bm:Stefan1} with $\ell=1$ for the phase field model 
(\ref{eq:EGSa},b) with the correction term (\ref{eq:rho1}) and 
(\ref{eq:Kiii}).
Here we denote by \cPFiii\ the scheme \PFiii\ but 
for the phase field model (\ref{eq:heata}--d), (\ref{eq:ACa}--c) with $\rho$
in (\ref{eq:ACa}) replaced by (\ref{eq:corrho}).

The numbers in Table~\ref{tab:TSciii1} show that eight points across the
interface are not enough to see $\mathcal{O}(\epsilon^2)$ convergence in the
errors in practice. Here we recall that a similar conclusion can be drawn from
the results reported in \cite[Table~1]{CaginalpCE08}, where a one-dimensional
reformulation of a radially symmetric problem in $\R^3$ is considered. With
$16$ points across the interface the error $\errorRr{x_1}{}$ in 
Table~\ref{tab:TSciii1} appears to converge quadratically in $\epsilon$, while
only linear convergence can be seen in the error $\errorWu$.
We also note that the errors $\errorRr{x_1}{}$ 
shown in Table~\ref{tab:TSciii1} are significantly
smaller than the corresponding errors in Tables~\ref{tab:TSqiii1}
and \ref{tab:TSqiii1_impl} for the classical phase field model, i.e.\ 
for $\rho_1=0$, whereas 
the improvements in the temperature error $\errorWu$ are less pronounced.
\begin{table}
\center
\begin{tabular}{c|c|c|c|c|c|r}
 $\epsilon^{-1}$ & $2^\frac12 /h_{f}$ & $\tau$ & $\errorRr{x_1}{}$ & $\errorWu$
 & DOFs$(T)$ & CPU time \\ 
 \hline
 $4\,\pi$ 
 &32  & $10^{-4}$ & 4.3481e-03 & 2.7780e-02 & 3394 & 5:12 mins\\
 &32  & $10^{-5}$ & 1.9588e-03 & 2.4734e-02 & 3442 & 39:41 mins\\
 &32  & $10^{-6}$ & 1.9743e-03 & 2.4425e-02 & 3474 & 5:40 hours\\
 &64  & $10^{-4}$ & 1.5762e-03 & 2.5180e-02 & 11922 & 24:35 mins\\
 &64  & $10^{-5}$ & 3.9787e-03 & 2.1853e-02 & 11906 & 2:41 hours\\
 &64  & $10^{-6}$ & 4.5109e-03 & 2.1533e-02 & 11970 & 29:05 hours\\
 \hline
 $\sqrt{32}\,\pi$ 
 &64  & $10^{-4}$ & 9.3961e-03 & 2.3676e-02 & 8946 & 19:06 mins\\ 
 &64  & $10^{-5}$ & 9.2783e-04 & 1.7200e-02 & 9074 & 2:25 hours\\
 &64  & $10^{-6}$ & 1.1183e-03 & 1.6723e-02 & 9186 & 19:37 hours\\
 &128 & $10^{-4}$ & 8.5330e-03 & 2.2021e-02 & 32226 & 1:46 hours\\ 
 &128 & $10^{-5}$ & 9.0354e-04 & 1.6249e-02 & 32730 & 13:02 hours\\
 &128 & $10^{-6}$ & 1.7709e-03 & 1.5796e-02 & 32586 & 104:40 hours\\
 \hline
 $8\,\pi$ 
 &64  & $10^{-4}$ & 2.3635e-02 & 2.9991e-02 & 7074 & 15:49 mins\\
 &64  & $10^{-5}$ & 6.8162e-03 & 1.6316e-02 & 7042 & 1:46 hours\\
 &64  & $10^{-6}$ & 5.0485e-03 & 1.5149e-02 & 7066 & 14:08 hours\\
 &128 & $10^{-4}$ & 1.9116e-02 & 2.5065e-02 & 24266 & 1:12 hours\\
 &128 & $10^{-5}$ & 1.9668e-03 & 1.2975e-02 & 24746 & 9:47 hours\\
 &128 & $10^{-6}$ & 4.8430e-04 & 1.2136e-02 & 24874 & 65:32 hours\\
\end{tabular}
\caption{Benchmark problem~\ref{bm:Stefan1} with $\ell=1$ for \protect\cPFiii.}
\label{tab:TSciii1}
\end{table}%

In the case of the quartic potential (\ref{eq:quartic}), so that 
$\Phi(s) = \tanh(2^{-\frac12}\,s)$ as in (\ref{eq:sphi0}), we get that
\begin{equation*} 
K = \sqrt{2}\,\int_\R [1 - \Varrho(\tanh(s))]\,\Varrho(\tanh(s)) \; {\rm d}s\,.
\end{equation*}
In particular, for the choice (\ref{eq:varrho})(iii)
we have that
\begin{equation*} 
K = \tfrac{209\,\sqrt{2}}{840} \approx 0.352\,,
\end{equation*}
while for the choice (\ref{eq:varrho})(iv), on noting that
$\Varrho(s) = \frac14\,(2 + 3\,s - s^3)$, it holds that
\begin{equation} \label{eq:sKquartic}
K = \tfrac{19\,\sqrt{2}}{60} \approx 0.448\,.
\end{equation}
We refer to Table~\ref{tab:TScqiv} for computations for the benchmark
problem~\ref{bm:Stefan1} with $\ell=1$ for the phase field model 
(\ref{eq:EGSa},b) with the correction term (\ref{eq:rho1}) and 
(\ref{eq:sKquartic}).
Here we denote by \cPFqiv\ the scheme \PFq\ 
for the phase field model (\ref{eq:heata}--d), (\ref{eq:ACa}--c) with $\rho$
in (\ref{eq:ACa}) replaced by (\ref{eq:corrho}), and with $\varrho$ given by
(\ref{eq:varrho})(iv). A computation for the scheme  \cPFqiv\ with the implicit
time discretization (\ref{eq:impl}) yielded very similar error numbers, and so
we omit these results here.
\begin{table}
\center
\begin{tabular}{c|c|c|c|c|c|r}
 $\epsilon^{-1}$ & $2^\frac12 /h_{f}$ & $\tau$ & $\errorRr{x_1}{}$ & $\errorWu$
 & DOFs$(T)$ & CPU time \\ 
 \hline
 $4\,\pi$ 
 &32  & $10^{-4}$ & 1.1945e-02 & 4.1523e-02 & 4066 & 7:12 mins\\ 
 &32  & $10^{-5}$ & 1.7315e-02 & 3.9842e-02 & 4090 & 52:56 mins\\ 
 &32  & $10^{-6}$ & 1.7860e-02 & 3.9682e-02 & 4082 & 8:24 hours\\ 
 &64  & $10^{-4}$ & 1.1659e-02 & 4.1790e-02 & 14090 & 29:48 mins\\ 
 &64  & $10^{-5}$ & 1.7071e-02 & 3.9883e-02 & 14130 & 4:18 hours\\
 &64  & $10^{-6}$ & 1.7615e-02 & 3.9686e-02 & 14210 & 43:32 hours\\
 \hline
 $\sqrt{32}\,\pi$ 
 &64  & $10^{-4}$ & 5.1351e-03 & 3.2940e-02 & 10514 & 18:46 mins\\ 
 &64  & $10^{-5}$ & 4.0145e-03 & 2.9244e-02 & 10578 & 2:41 hours\\
 &64  & $10^{-6}$ & 5.0498e-03 & 2.8980e-02 & 10578 & 25:57 hours\\
 &128 & $10^{-4}$ & 5.1726e-03 & 3.2286e-02 & 37194 & 1:09 hours\\ 
 &128 & $10^{-5}$ & 3.7103e-03 & 2.9080e-02 & 37690 & 10:37 hours\\
 &128 & $10^{-6}$ & 4.7916e-03 & 2.8833e-02 & 37626 &109:39 hours\\
 \hline
 $8\,\pi$ 
 &64  & $10^{-4}$ & 1.9976e-02 & 3.2507e-02 & 8082 & 12:14 mins\\
 &64  & $10^{-5}$ & 3.4097e-03 & 2.2891e-02 & 8242 & 2:12 hours\\
 &64  & $10^{-6}$ & 1.4844e-03 & 2.2330e-02 & 8306 & 16:44 hours\\
 &128 & $10^{-4}$ & 1.9474e-02 & 3.0950e-02 & 27826 & 1:15 hours\\
 &128 & $10^{-5}$ & 2.9987e-03 & 2.2277e-02 & 28298 & 7:39 hours\\
 &128 & $10^{-6}$ & 1.0518e-03 & 2.1768e-02 & 28234 & 65:58 hours\\
\end{tabular}
\caption{Benchmark problem~\ref{bm:Stefan1} with $\ell=1$ for \protect\cPFqiv.}
\label{tab:TScqiv}
\end{table}%
What can be clearly seen from Table~\ref{tab:TScqiv} is that once again we have
convergence of order at least $\mathcal{O}(\epsilon^2)$ in $\errorRr{x_1}{}$.

\begin{figure}
\center
\ifpdf
\includegraphics[angle=-0,width=0.55\textwidth]{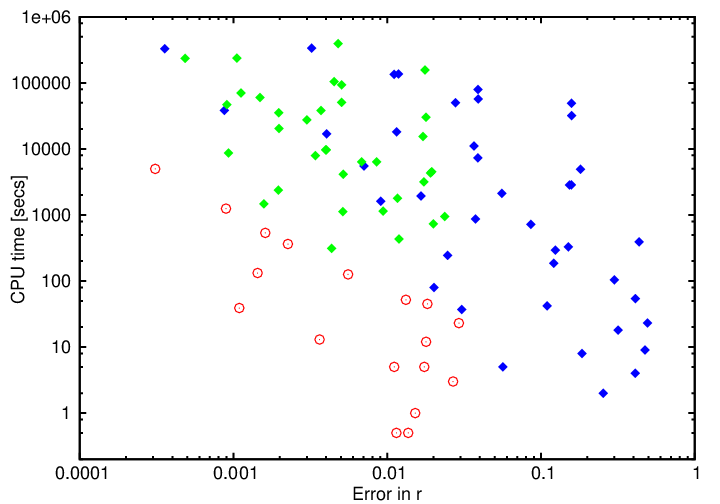}
\fi
\caption{(Benchmark problem~\ref{bm:Stefan1} with $\ell=1$)
Log-log scatter plot of $\errorRr{x_1}{}$ and $\errorRr{}{}$
against the CPU time for
the entries in Tables~\ref{tab:TSqiii1} and \ref{tab:TSqiii1_impl} 
(\PFqiii, blue rhombi), Tables~\ref{tab:TSciii1} and \ref{tab:TScqiv}
(\protect\cPFiii\ and \protect\cPFqiv, green rhombi)
and Table~\ref{tab:TS1p} (\PFEM, red circles).
}
\label{fig:scatter8all}
\end{figure}

In order to visualize the performances of all the considered methods for the
benchmark problem~\ref{bm:Stefan1} with $\ell=1$, i.e.\ including the
computations in Tables~\ref{tab:TSciii1} and \ref{tab:TScqiv} for the second
order accurate phase field model (\ref{eq:EGSa},b), 
we present a log-log plot of the errors in the radius $r$
against the necessary CPU time for all the entries in the appropriate tables in
Figure~\ref{fig:scatter8all}. 
The plot appears to confirm that computations for
the phase field model (\ref{eq:EGSa},b) are on average more efficient than
computations for the standard phase field model, i.e.\ (\ref{eq:EGSa},b) with
$\rho_1=0$. However, the finer meshes needed for computations for 
(\ref{eq:EGSa},b), recall Table~\ref{tab:TSciii1}, mean that due to CPU time
constraints we cannot choose $\epsilon$ as small as in the standard phase field
computations.
Finally, the plot in Figure~\ref{fig:scatter8all}
once again underlines the superiority 
of the sharp interface algorithm \PFEM\ over all the phase field methods.

\setcounter{equation}{0} 
\section{Numerical experiments for anisotropic problems} \label{sec:5}

In this section we present numerical simulations for the anisotropic Stefan
problem \mbox{(\ref{eq:1a}--e)}. Here we always let 
$\partial_D\Omega=\partial\Omega$ and $\beta = \gamma$, 
where we recall (\ref{eq:mu}).
Moreover, we always choose $\lambda = a = 1$ and,
unless otherwise stated, we let $\conduct_\pm = 1$.
In order to appreciate the computational challenges involved with the
different experiments, we recall from (\ref{eq:epscond}), (\ref{eq:hf}) and
(\ref{eq:NV97a}) that for accurate phase field calculations the following 
implications arise:
\begin{equation*}
-\uD\,\alpha^{-1} \ \text{large} \quad \Rightarrow \quad
\epsilon \ \text{small} \quad \Rightarrow \quad
h_f\,,\tau \ \text{small}\,.
\end{equation*}
Moreover, we note that for the fully anisotropic situation a formally second 
order accurate phase field model similarly to Section~\ref{sec:43} involves a
parameter $\rho_1(\nabla\,\varphi)$ in place of (\ref{eq:rho1}) that depends on
$\beta$, $\gamma$ and on $\nabla\,\varphi$, see \cite{KarmaR96,KarmaR98}. 
We stress that these approaches are not well analyzed so far, e.g.\ in the
spirit of \cite{GarckeS06,ChenCE06}. In particular, to our knowledge there are
no formal or rigorous results in the literature on the second order convergence
of phase field models for the fully anisotropic Gibbs--Thomson law.
Moreover, our numerical results 
in the isotropic case showed that the second order models do not give
a large gain in computational efficiency. 
That is why all of 
our phase field computations in this section are for the standard
phase field model (\ref{eq:heata}--d), (\ref{eq:ACa}--c).

For the first simulations that we present we choose as anisotropy the
regularized $l^1$-norm
\begin{equation*}
 \text{\sc ani$_1$:} \qquad
\gamma(\vec{p}) = \sum_{j=1}^2\, \left[ \delta^2\,
|\vec{p}|^2+ p_j^2\,(1-\delta^2) \right]^{\frac12} \,,\quad
\text{with}\quad \delta = 0.3\,. 
\end{equation*}
The radius of the initially circular seed $\Gamma_0$ is chosen as $0.1$, while
we set $\vartheta=0$, $\alpha=0.03$ and $\rho = 0.01$. The supercooling
at the boundary $\partial_D\Omega=\partial\Omega$ 
of $\Omega=(-8,8)^2$ is set to $\uD = -2$.

\begin{figure}
\center
\ifpdf
\includegraphics[angle=-0,width=0.19\textwidth]{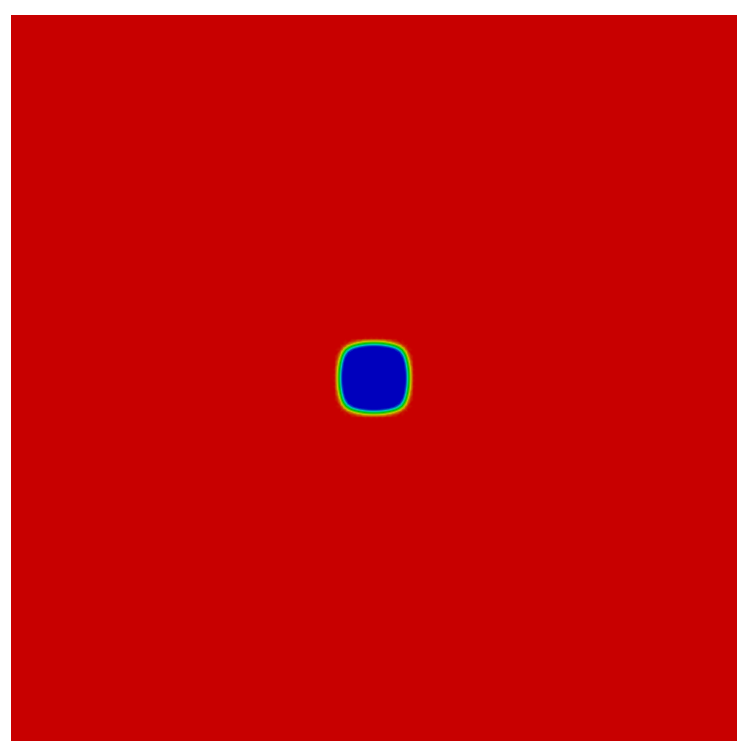}
\includegraphics[angle=-0,width=0.19\textwidth]{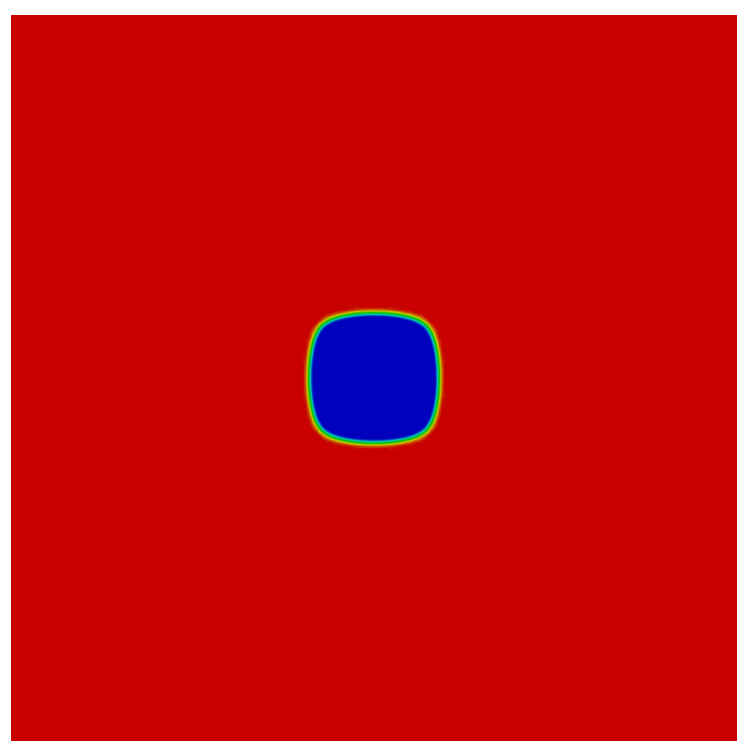}
\includegraphics[angle=-0,width=0.19\textwidth]{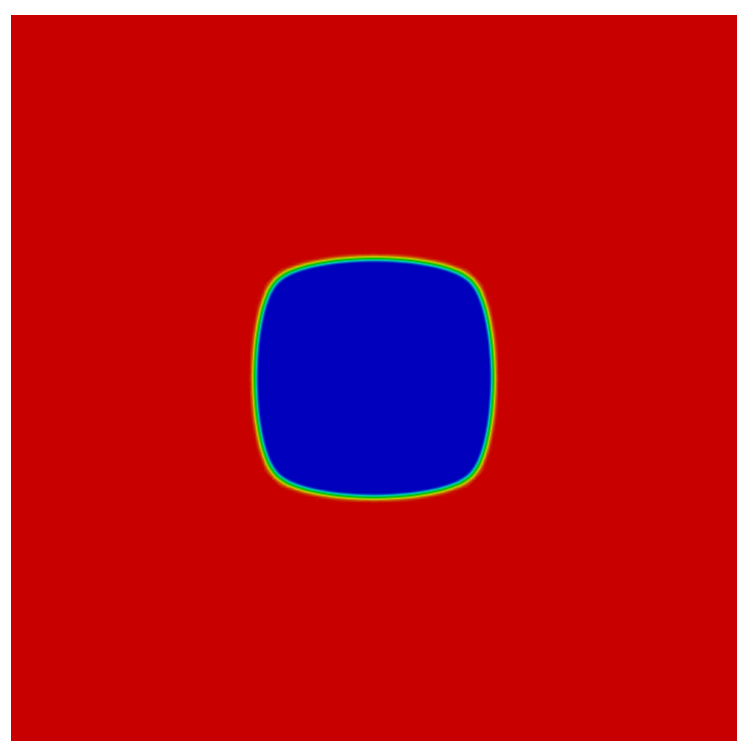}
\includegraphics[angle=-0,width=0.19\textwidth]{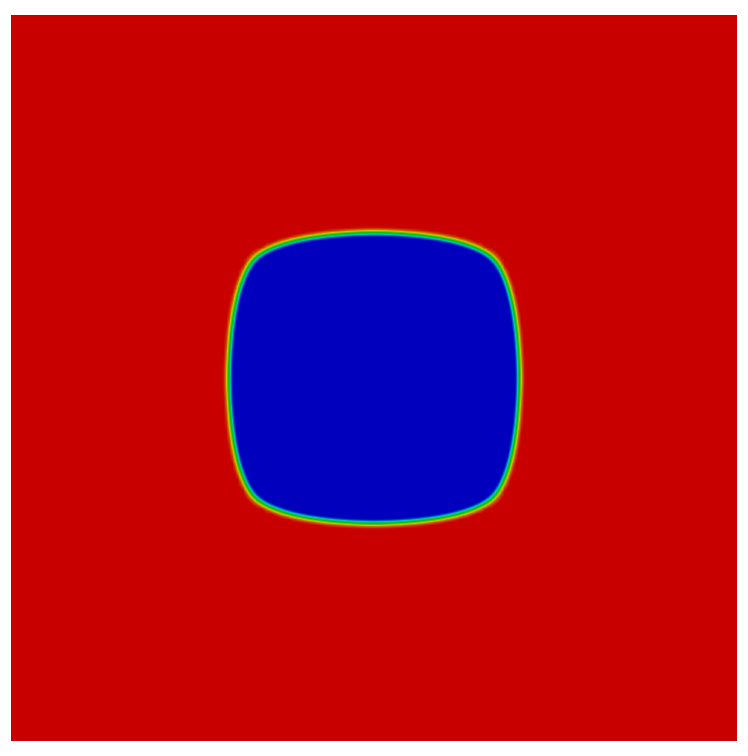}
\includegraphics[angle=-0,width=0.19\textwidth]{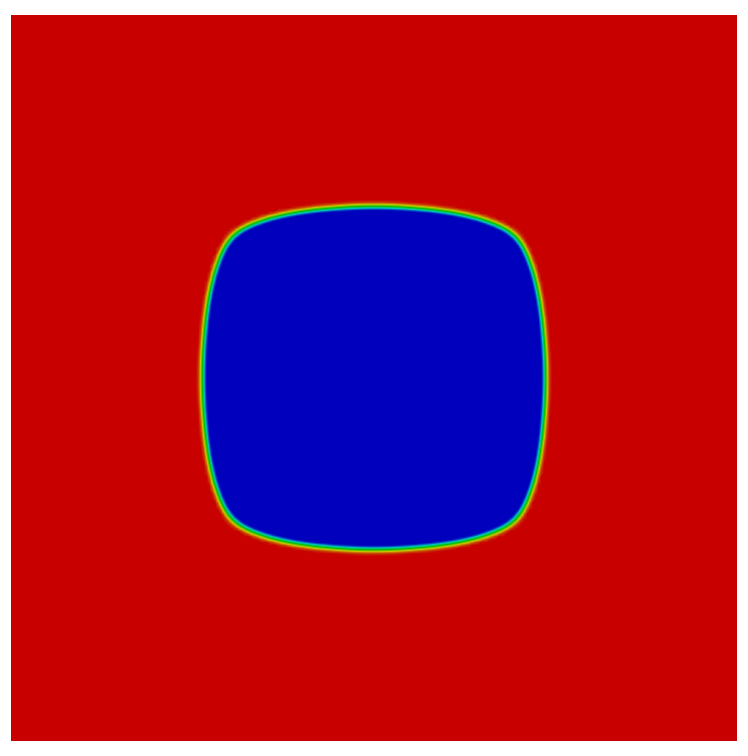}
\includegraphics[angle=-0,width=0.19\textwidth]{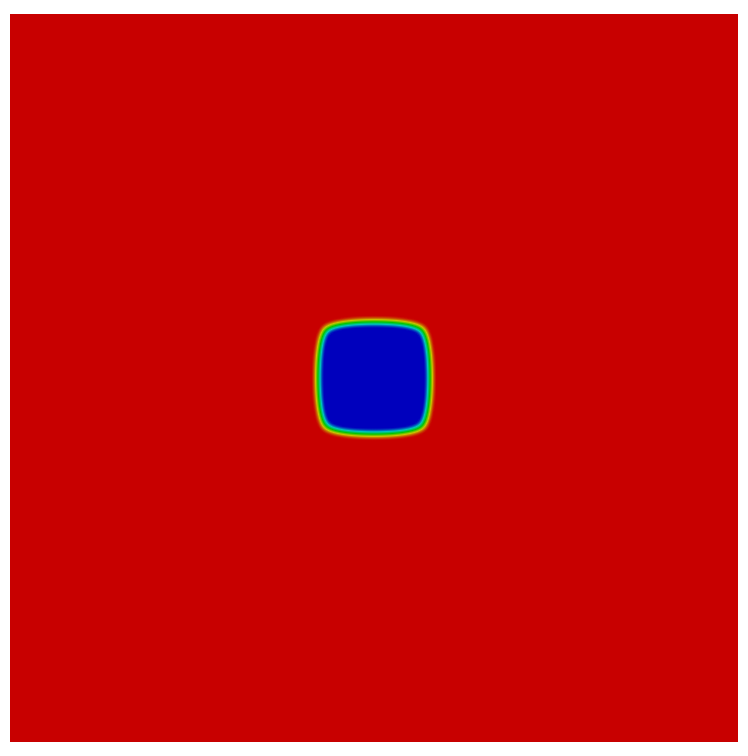}
\includegraphics[angle=-0,width=0.19\textwidth]{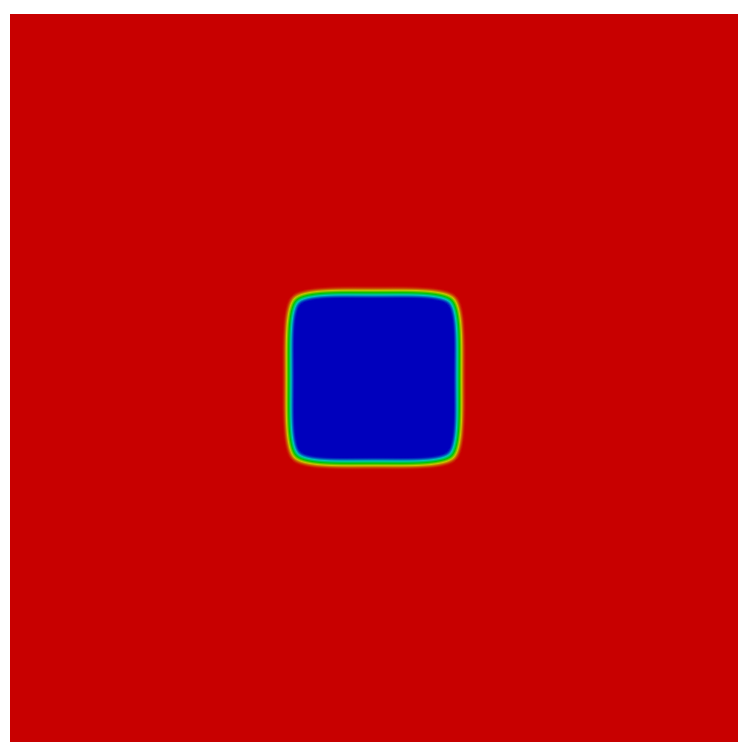}
\includegraphics[angle=-0,width=0.19\textwidth]{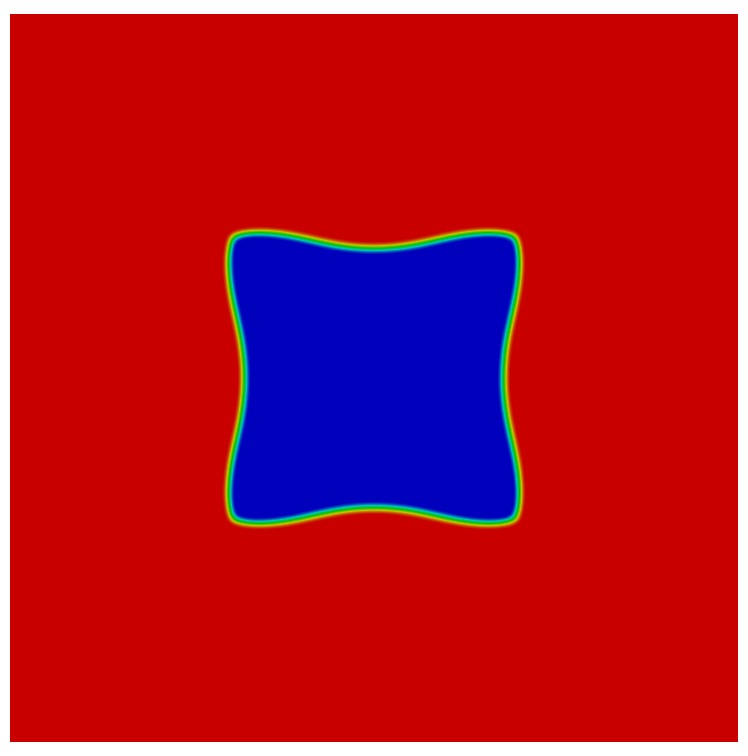}
\includegraphics[angle=-0,width=0.19\textwidth]{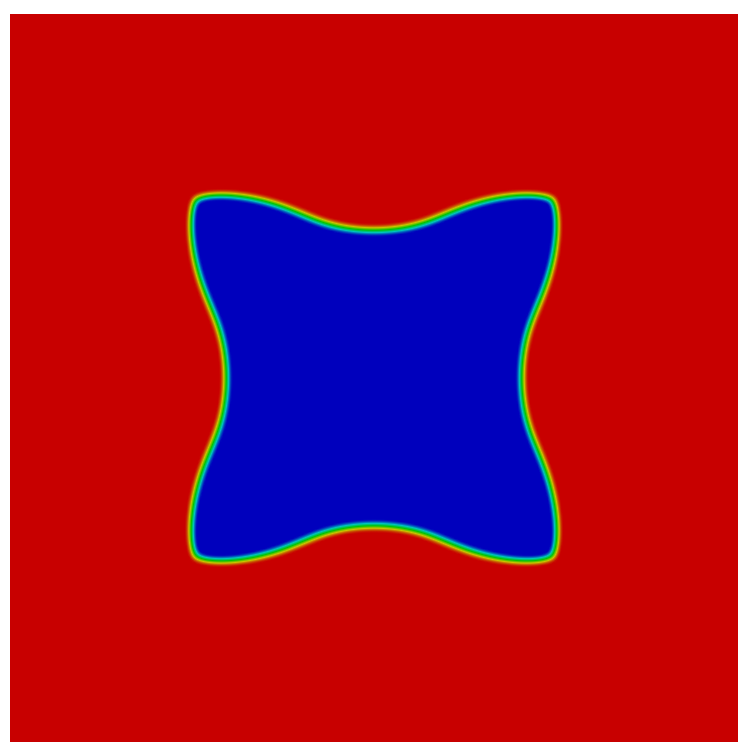}
\includegraphics[angle=-0,width=0.19\textwidth]{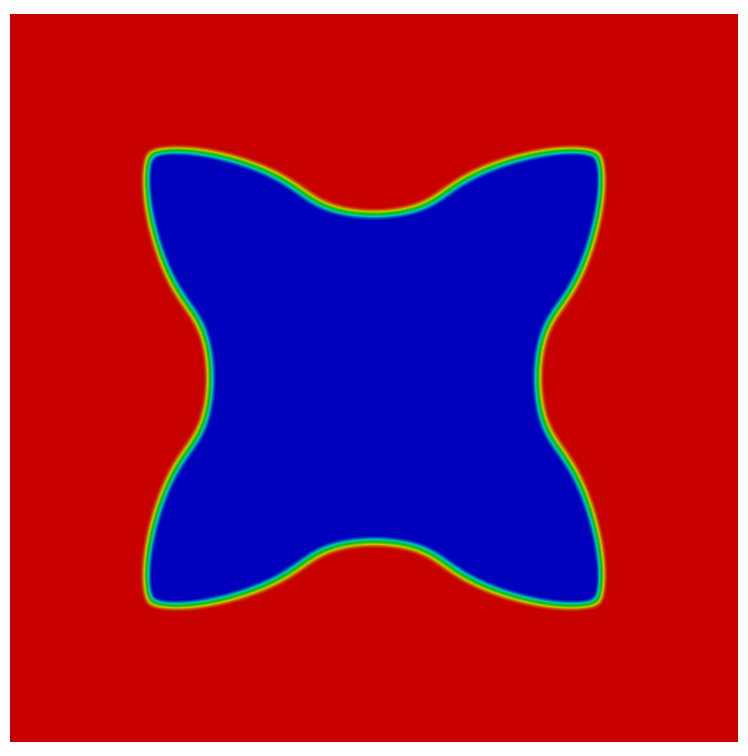}
\includegraphics[angle=-0,width=0.19\textwidth]{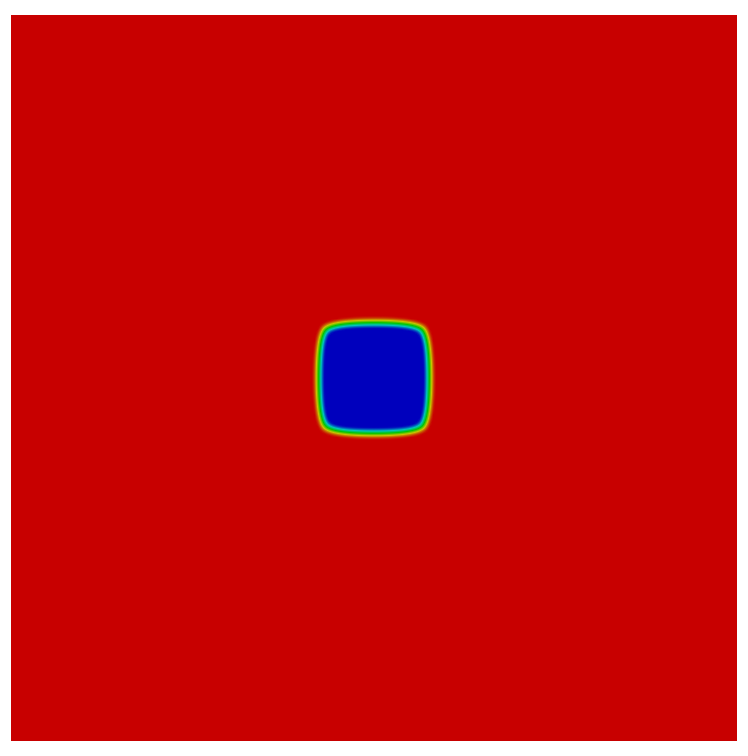}
\includegraphics[angle=-0,width=0.19\textwidth]{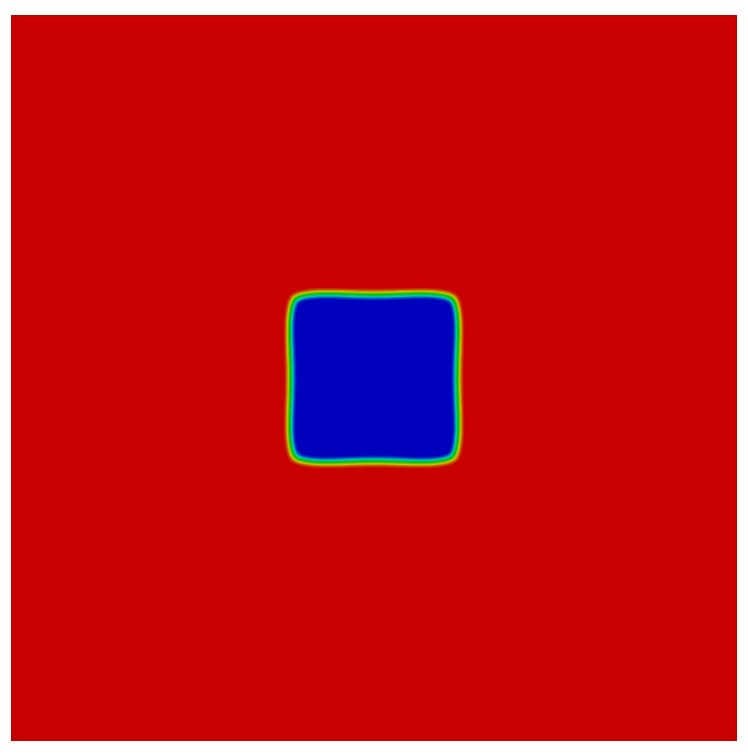}
\includegraphics[angle=-0,width=0.19\textwidth]{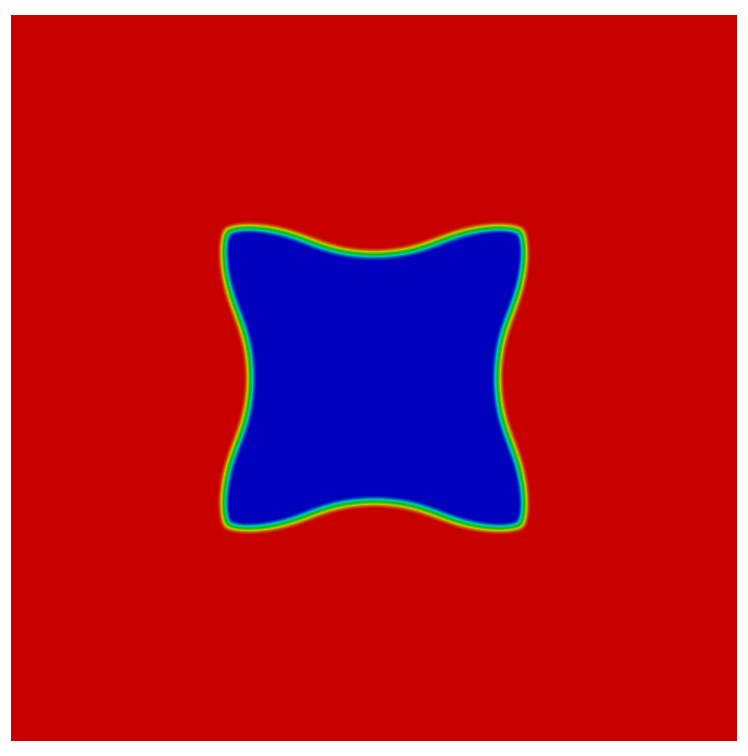}
\includegraphics[angle=-0,width=0.19\textwidth]{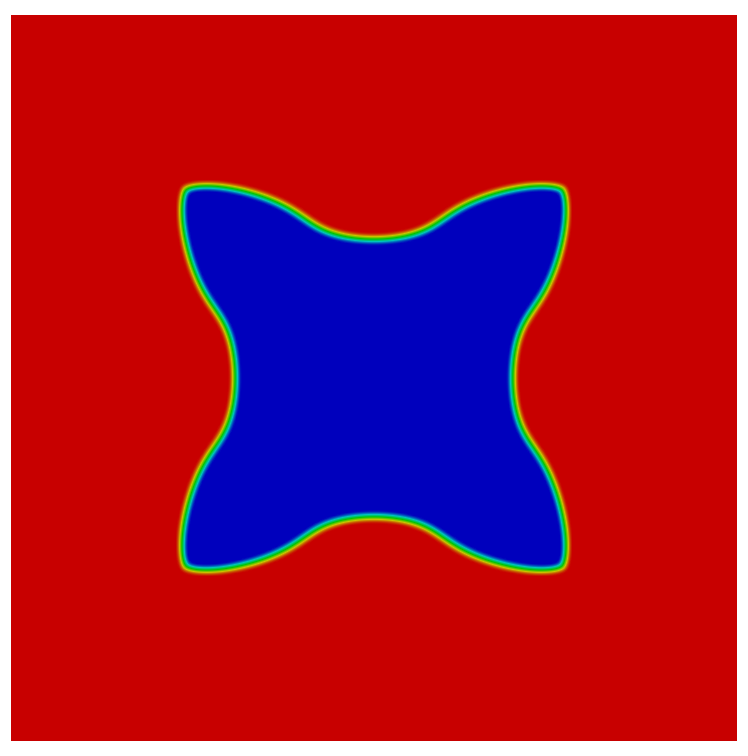}
\includegraphics[angle=-0,width=0.19\textwidth]{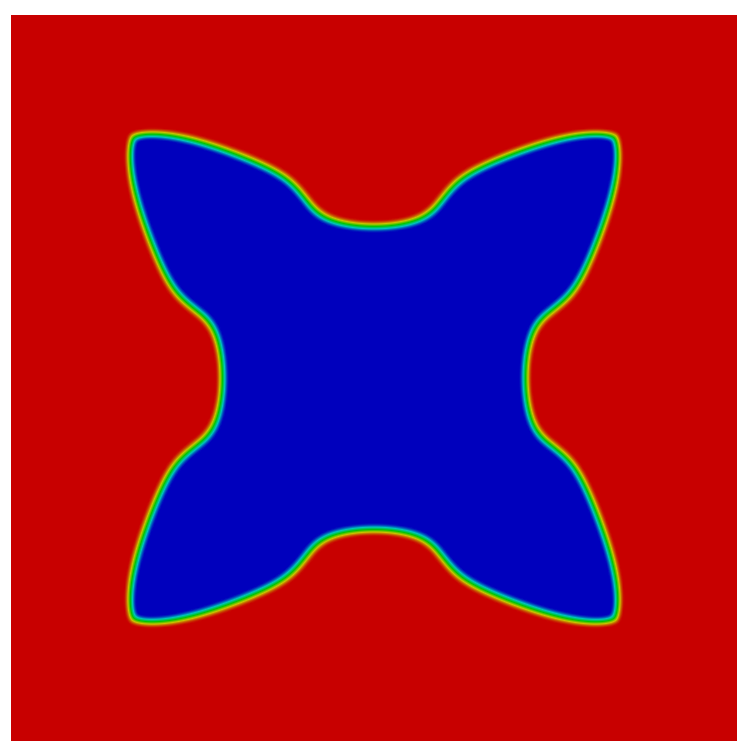}
\fi
\caption{(\PFii, $\epsilon^{-1} = 4\,\pi$, {\sc ani$_1$}, 
$\vartheta=0$, $\alpha=0.03$, $\rho = 0.01$, $\uD = -2$, 
$\Omega=(-8,8)^2$)
Snapshots of the solution at times $t=1,\,2,\,4,\,5,\,6$. From top to bottom
$\tau = 10^{-k}$, $k = 1 \to 3$.
[These computations took $34$ seconds, $7$ minutes and $57$ minutes, 
respectively.]
}
\label{fig:BSrhoii_4pi}
\end{figure}

Three numerical simulations for the scheme \PFii\ with the interfacial
parameter $\epsilon^{-1} = 4\,\pi$ can be seen in Figure~\ref{fig:BSrhoii_4pi}.
It can be seen that varying the time discretization parameter $\tau$ from
$10^{-1}$ to $10^{-3}$ has a significant impact on the observed numerical
results. However, the observed changes for the smallest value of $\tau$ are 
small, which indicates that the simulation appears to be converging. Very
similar results can be obtained for \PFqiii\ with the implicit time
discretization from (\ref{eq:impl}), and so we omit them here.

\begin{figure}
\center
\ifpdf
\includegraphics[angle=-0,width=0.19\textwidth]{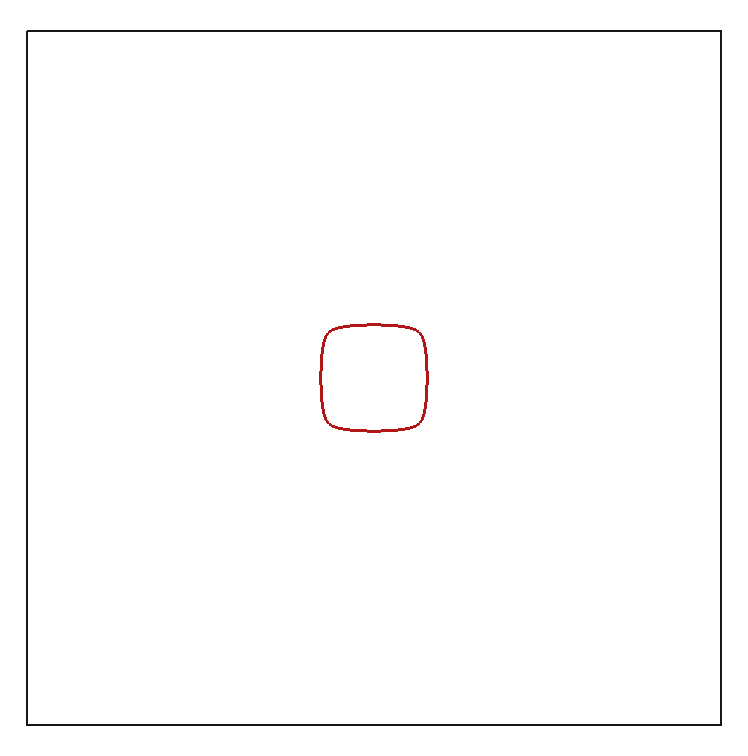}
\includegraphics[angle=-0,width=0.19\textwidth]{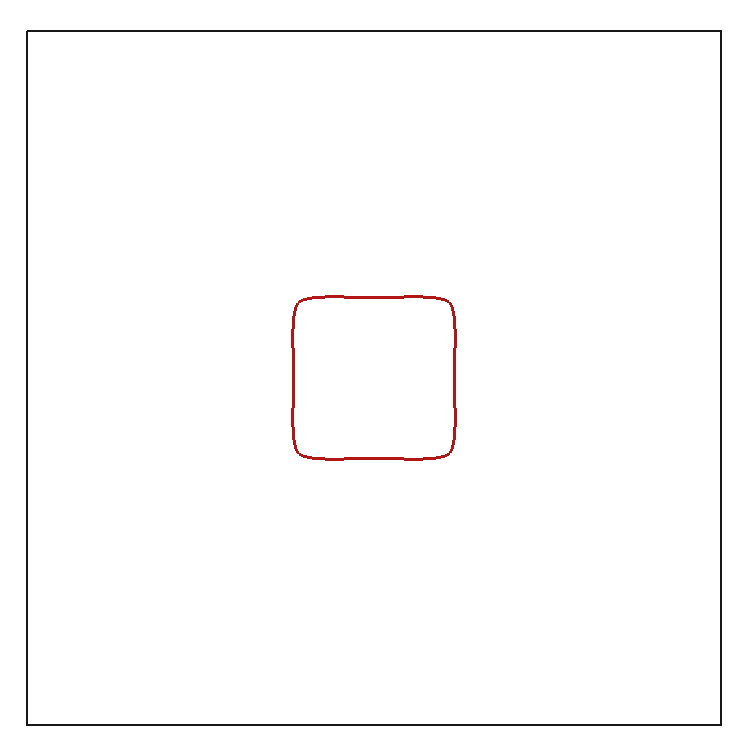}
\includegraphics[angle=-0,width=0.19\textwidth]{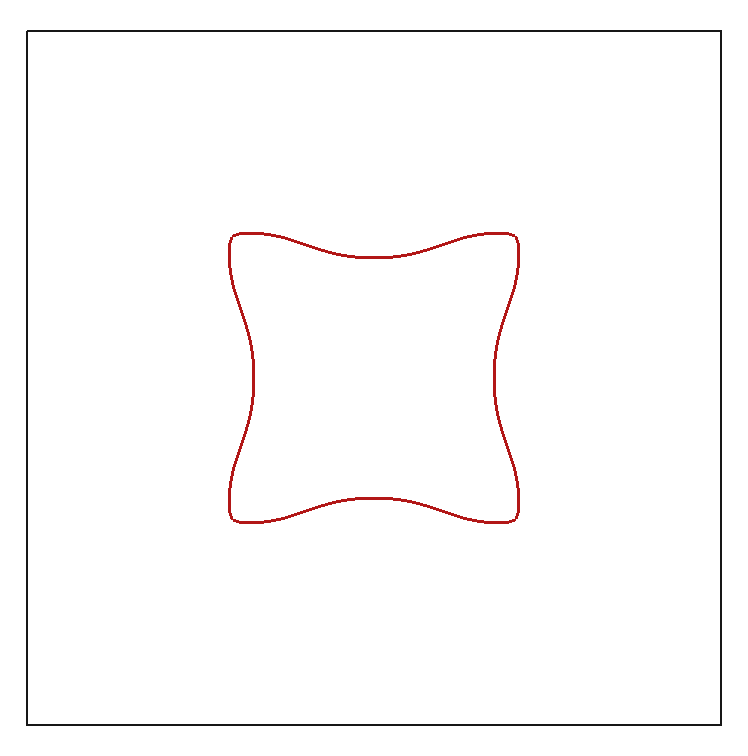}
\includegraphics[angle=-0,width=0.19\textwidth]{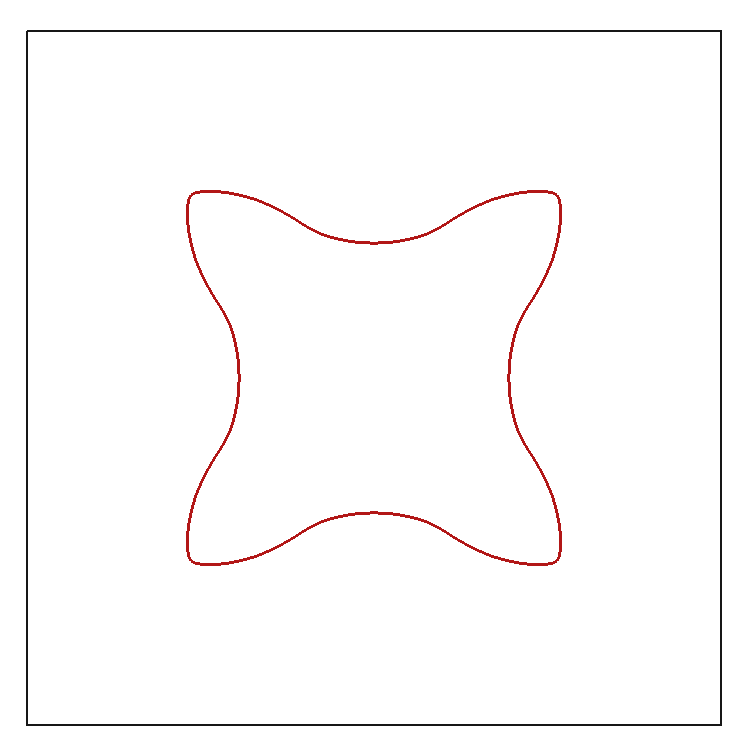}
\includegraphics[angle=-0,width=0.19\textwidth]{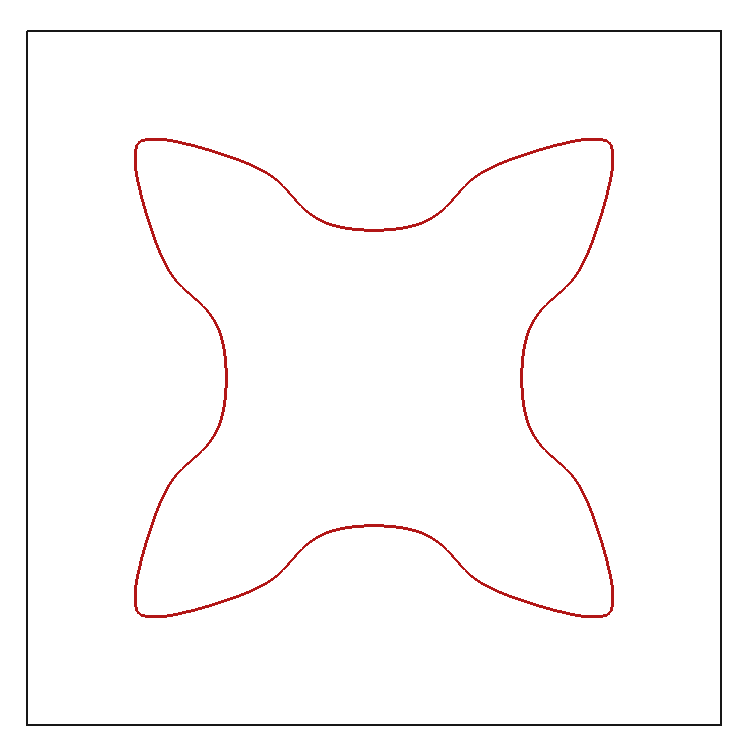}
\includegraphics[angle=-0,width=0.19\textwidth]{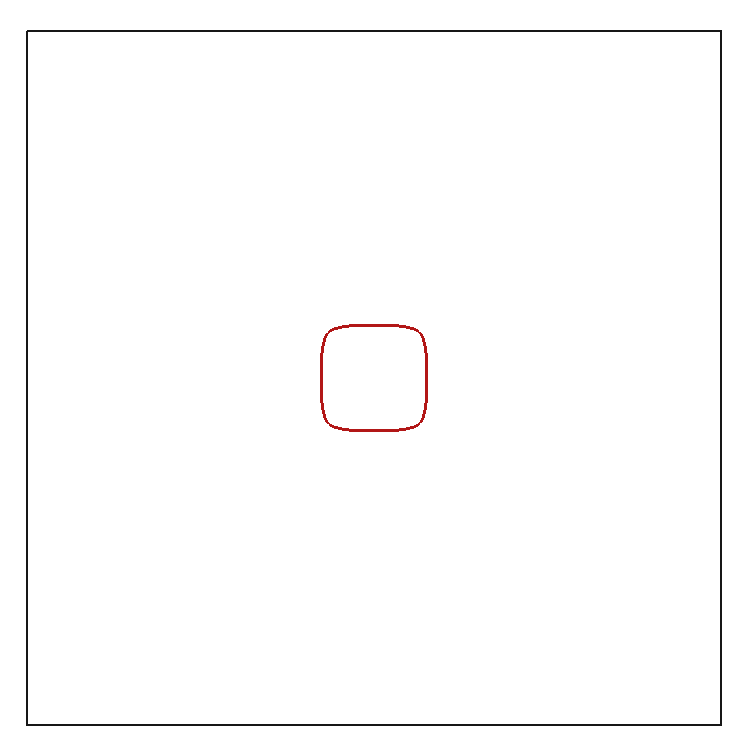}
\includegraphics[angle=-0,width=0.19\textwidth]{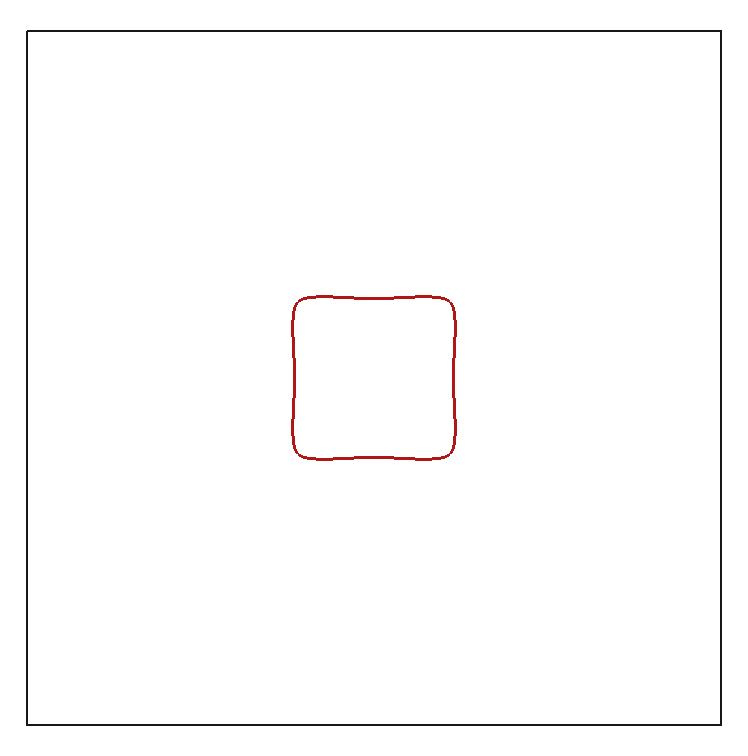}
\includegraphics[angle=-0,width=0.19\textwidth]{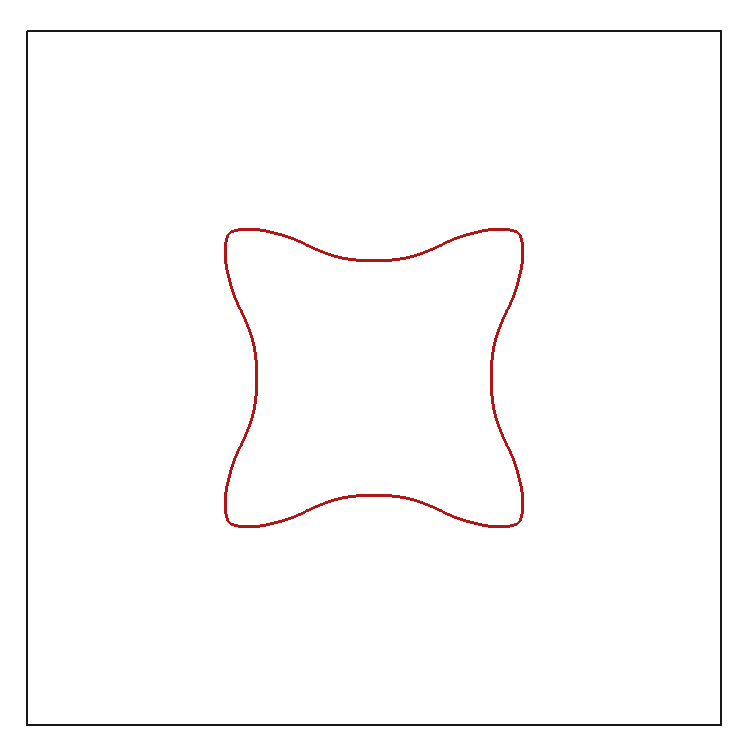}
\includegraphics[angle=-0,width=0.19\textwidth]{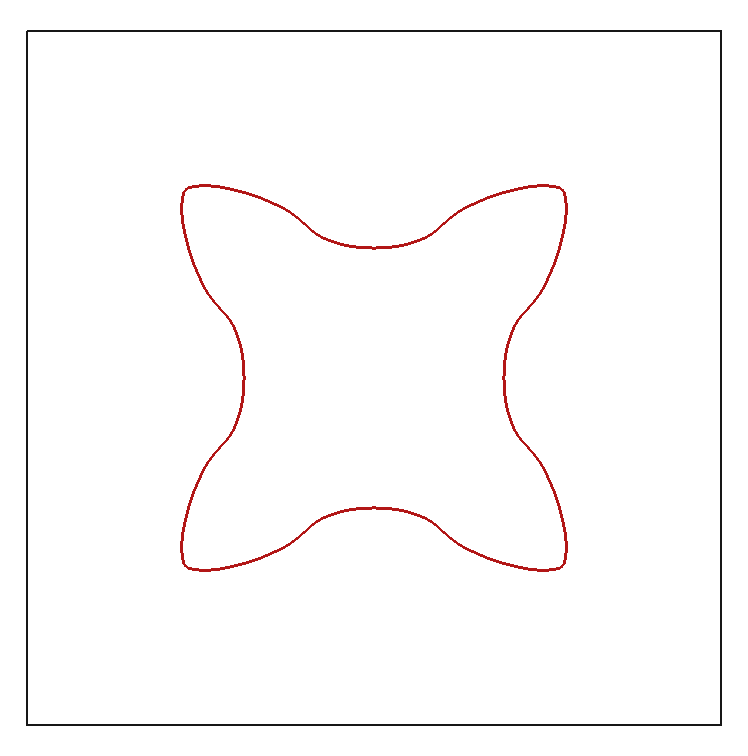}
\includegraphics[angle=-0,width=0.19\textwidth]{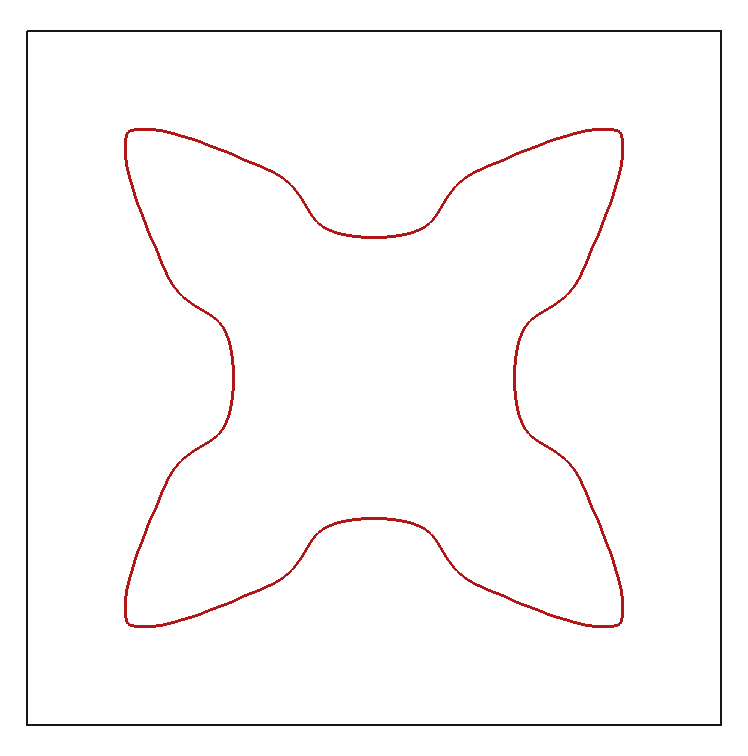}
\includegraphics[angle=-0,width=0.19\textwidth]{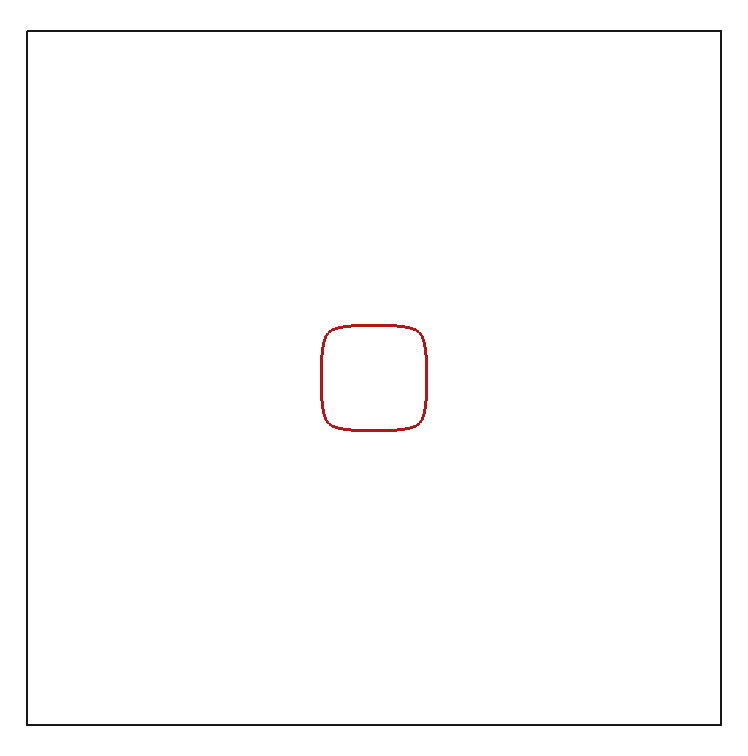}
\includegraphics[angle=-0,width=0.19\textwidth]{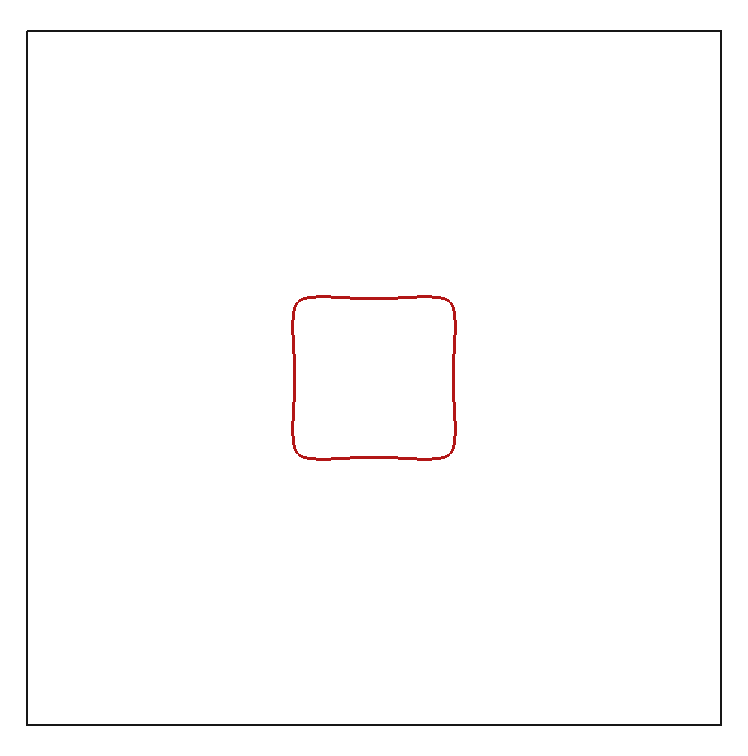}
\includegraphics[angle=-0,width=0.19\textwidth]{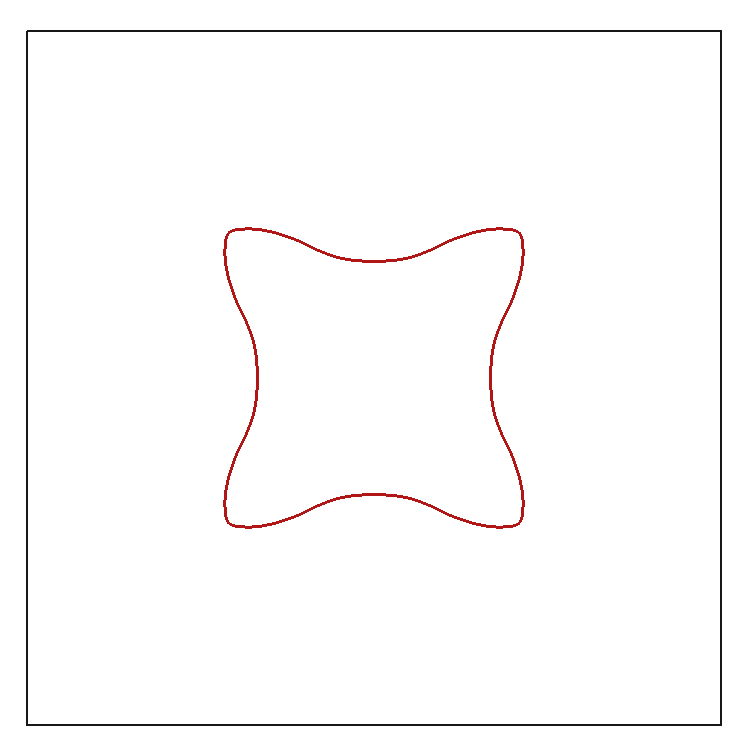}
\includegraphics[angle=-0,width=0.19\textwidth]{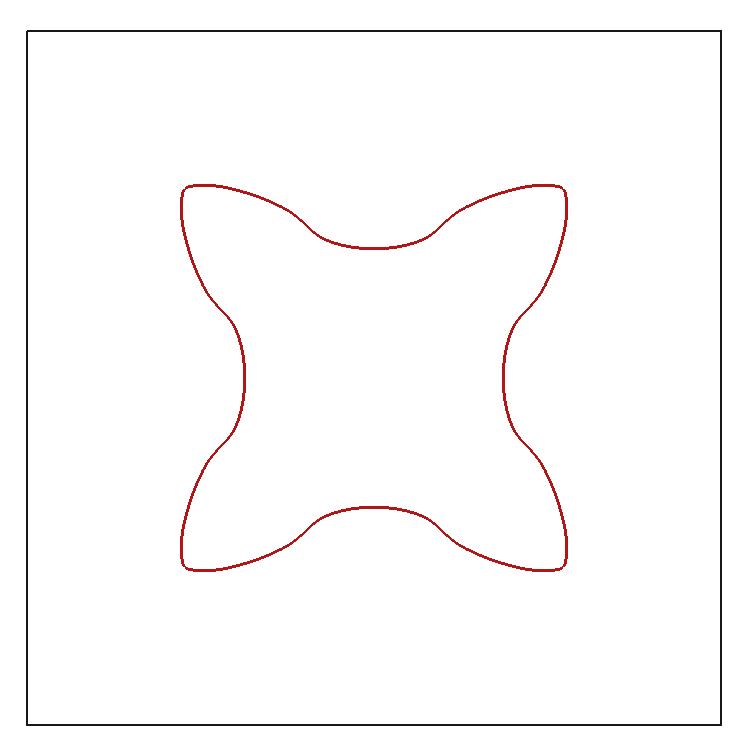}
\includegraphics[angle=-0,width=0.19\textwidth]{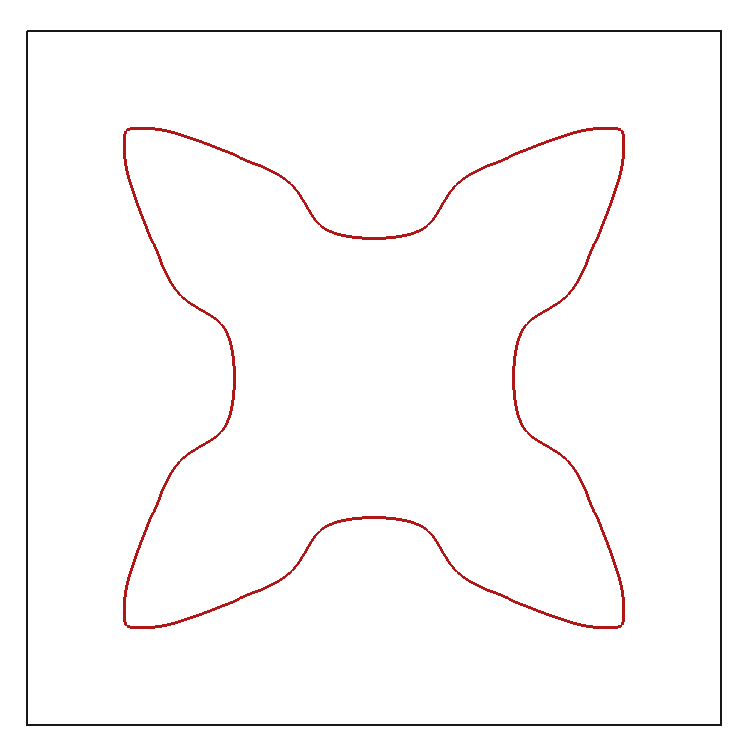}
\fi
\caption{(\PFEM, {\sc ani$_1$}, 
$\vartheta=0$, $\alpha=0.03$, $\rho = 0.01$, $\uD = -2$, 
$\Omega=(-8,8)^2$)
Snapshots of the solution at times $t=1,\,2,\,4,\,5,\,6$. From top to bottom
$\tau = 10^{-k}$, $k = 1 \to 3$.
[These computations took $1$, $6$ and $73$ seconds, 
respectively.]
}
\label{fig:BSrho}
\end{figure}

We compare the above numerical experiments for the phase field method with 
three simulations for the sharp interface approximation
\PFEM\ in Figure~\ref{fig:BSrho}, where we fix the spatial discretization
parameters as $h_\Gamma \approx \tfrac12\,h_f = \frac{\sqrt{2}}{4}$.
Here we observe that even for a very crude
time discretization, the evolution is captured remarkably well, and there is
very little variation in the numerical results from \PFEM\ when $\tau$ is
decreased. Also note that it takes (less than) a second of CPU time with \PFEM\
in order to get a good idea about the evolution of the growing crystal, while 
the phase field methods \PF\ and \PFq\ take at least $400$ times as long.

For the next set of numerical experiments we use the hexagonal anisotropy
\begin{equation} \label{eq:hexgamma2d}
 \text{\sc ani$_2$:} \qquad
\gamma(\vec{p}) =
\sum_{\ell = 1}^3
l(R(\tfrac{\pi}{12} + \ell\,\tfrac{\pi}3)\,\vec{p})\,,
\quad\text{where}\quad
l(\vec{p}) = \left[p_1^2 + 10^{-4}\,p_2^2 \right]^{\frac12}\,,
\end{equation}
and where $R(\theta)=
\left(\!\!\!\scriptsize
\begin{array}{rr} \cos\theta & \sin\theta \\
-\sin\theta & \cos\theta \end{array}\!\! \right)$ 
denotes a clockwise rotation through the angle $\theta$.
Moreover, we use the parameters $\vartheta=1$, $\alpha=5\times10^{-4}$,
$\rho = 0.01$ and $\uD = -\frac12$ on the boundary 
$\partial_D\Omega=\partial\Omega$ of $\Omega=(-2,2)^2$.
The radius of the initially circular seed is again chosen as $R_0=0.1$, and we
set
\begin{equation}
u_0(\vec{x}) = \begin{cases} 0 & |\vec{x}| \leq R_0\,, \\
\dfrac{u_D}{1 - e^{R_0 - H}}\left(1 - e^{R_0 - |\vec{x}|}\right) 
& R_0 < |\vec{x}| < H\,, \\
u_D & |\vec{x}| \geq H \,,
\end{cases}
\label{eq:u0exp}
\end{equation}
with $H:=2$.

\begin{figure}
\center
\ifpdf
\includegraphics[angle=-0,width=0.19\textwidth]{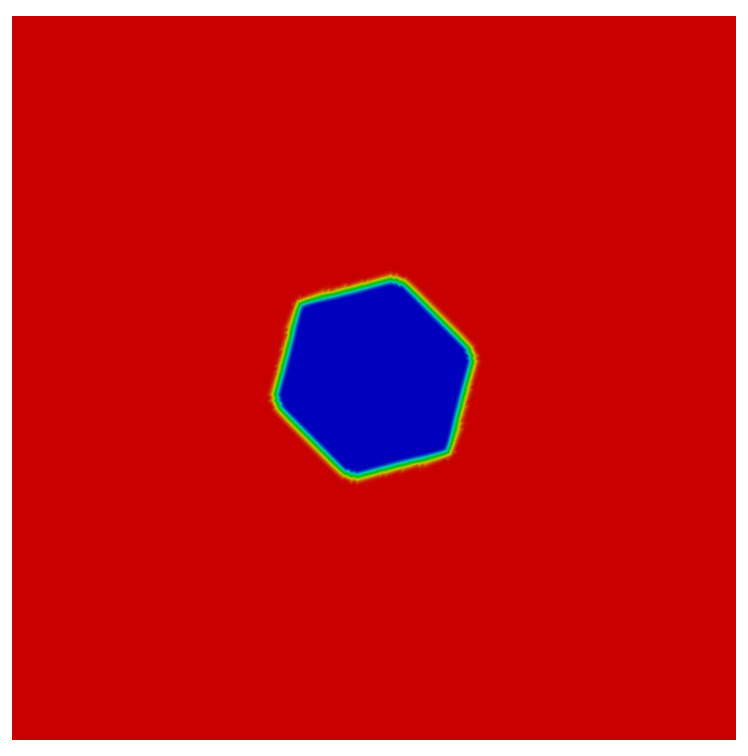}
\includegraphics[angle=-0,width=0.19\textwidth]{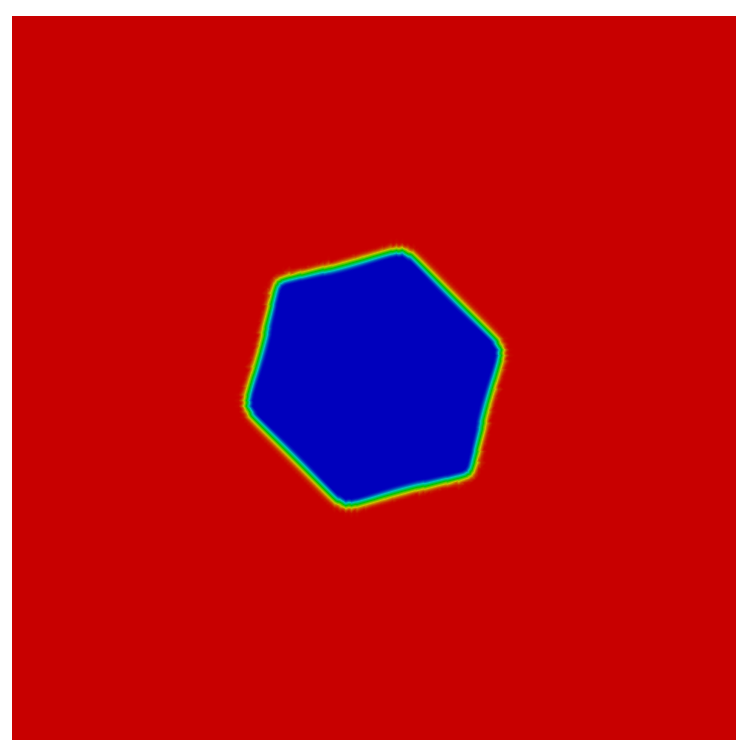}
\includegraphics[angle=-0,width=0.19\textwidth]{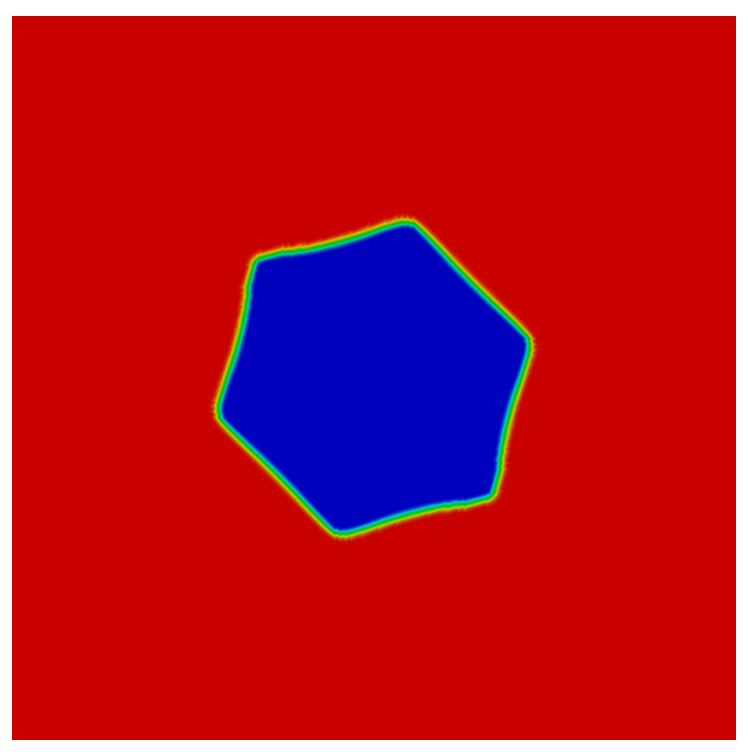}
\includegraphics[angle=-0,width=0.19\textwidth]{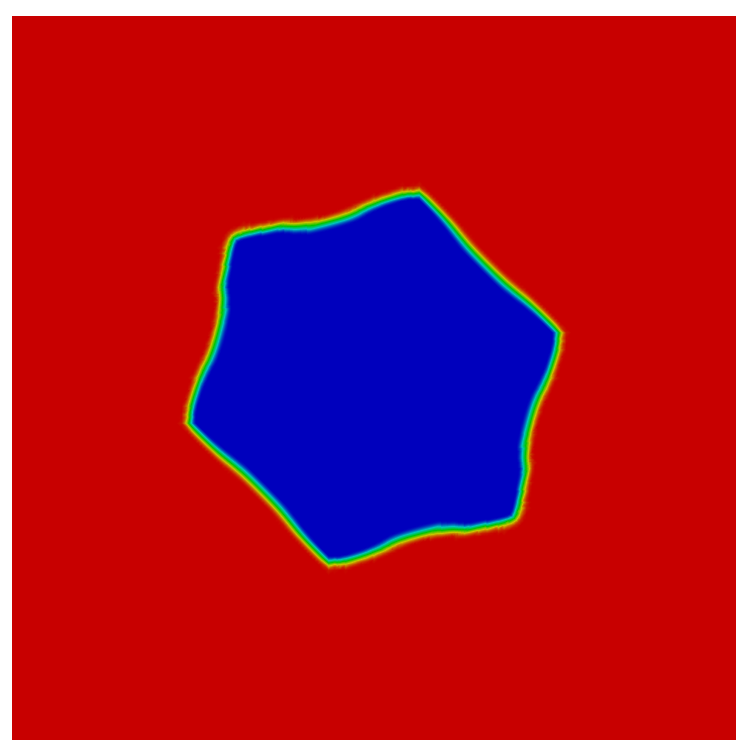}
\includegraphics[angle=-0,width=0.19\textwidth]{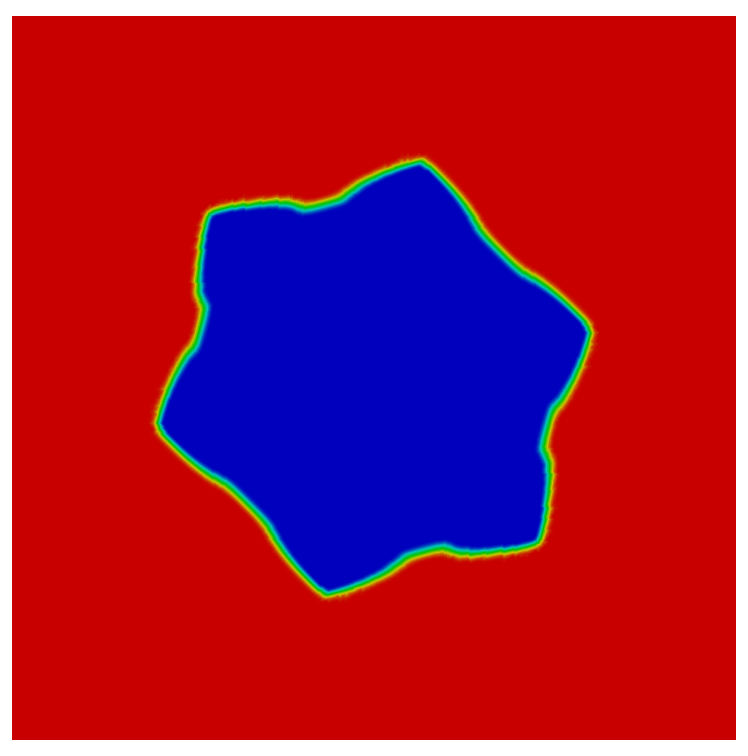}
\includegraphics[angle=-0,width=0.19\textwidth]{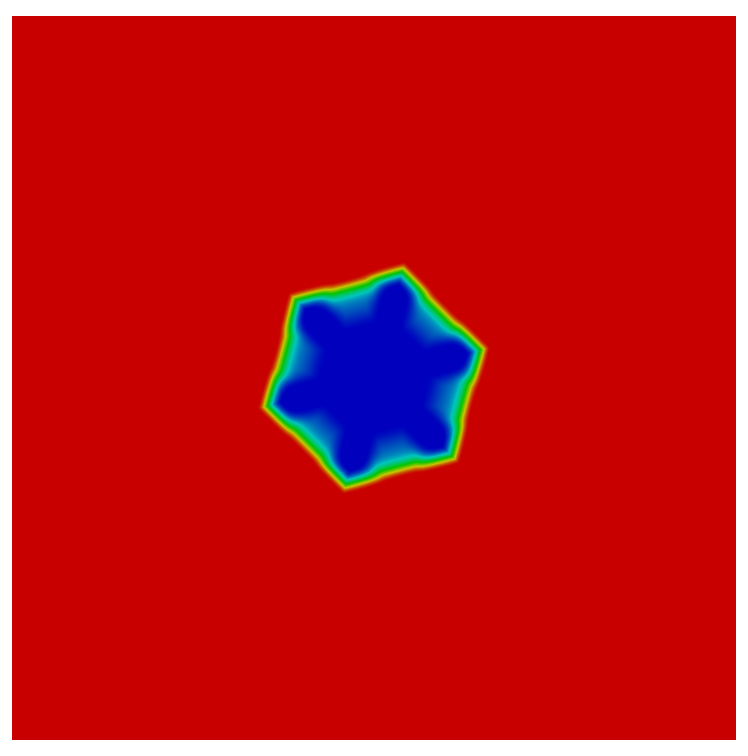}
\includegraphics[angle=-0,width=0.19\textwidth]{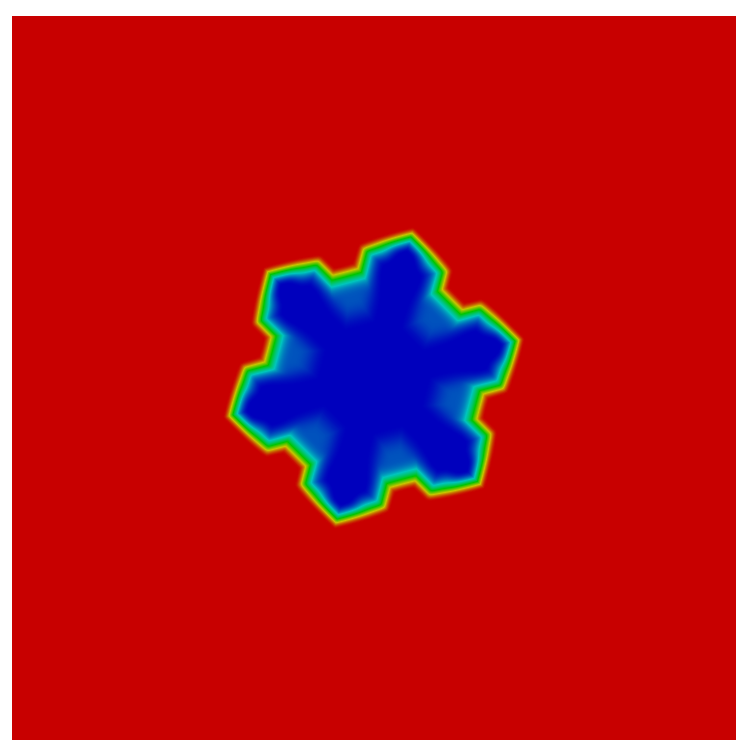}
\includegraphics[angle=-0,width=0.19\textwidth]{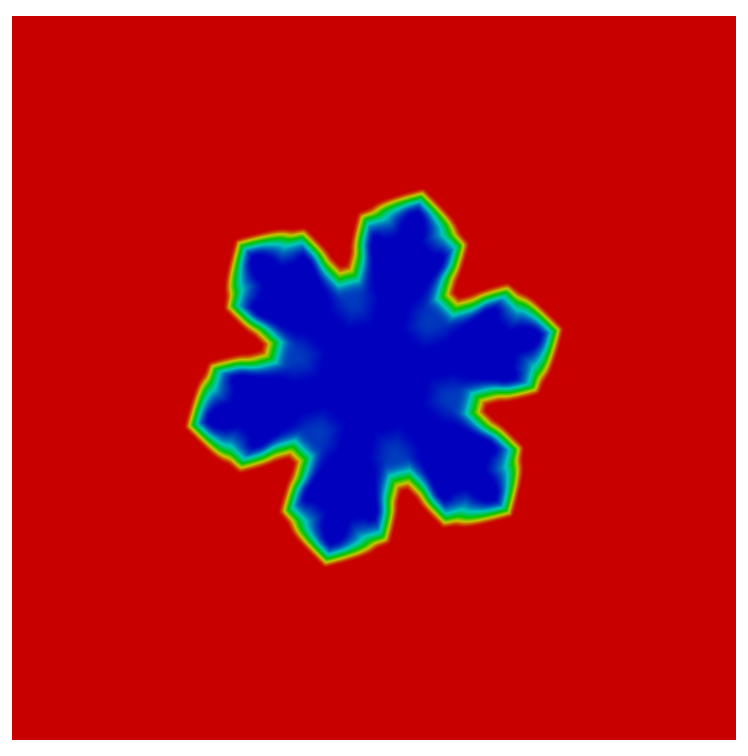}
\includegraphics[angle=-0,width=0.19\textwidth]{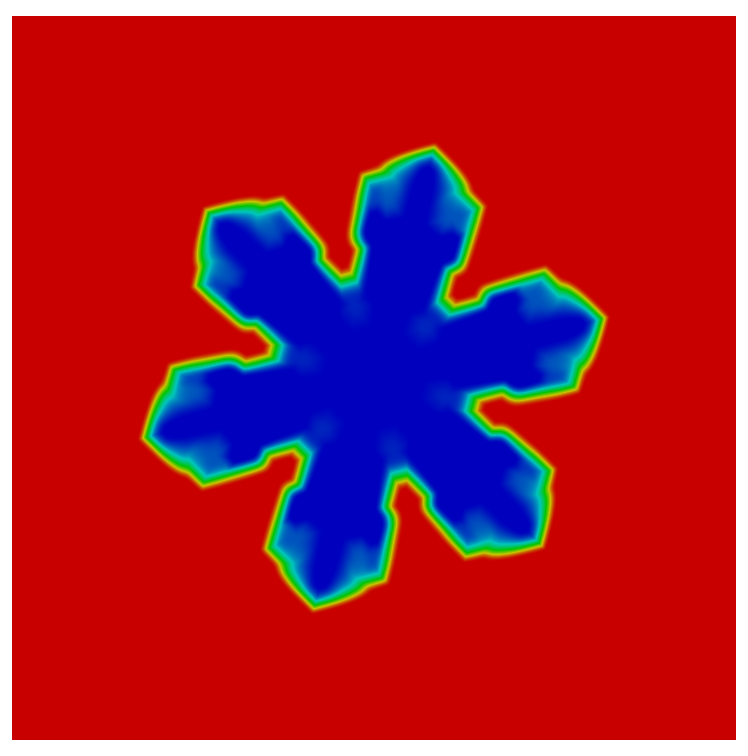}
\includegraphics[angle=-0,width=0.19\textwidth]{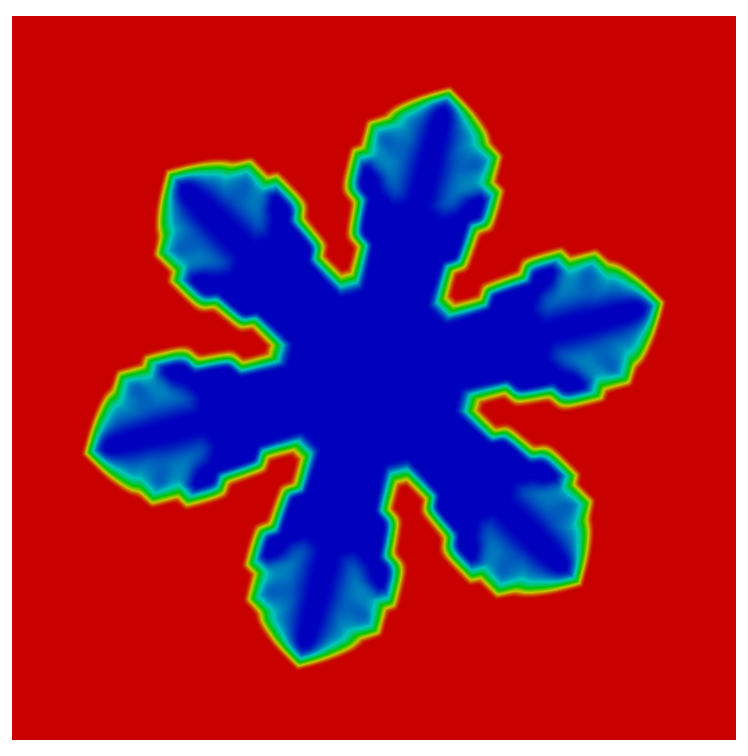}
\includegraphics[angle=-0,width=0.19\textwidth]{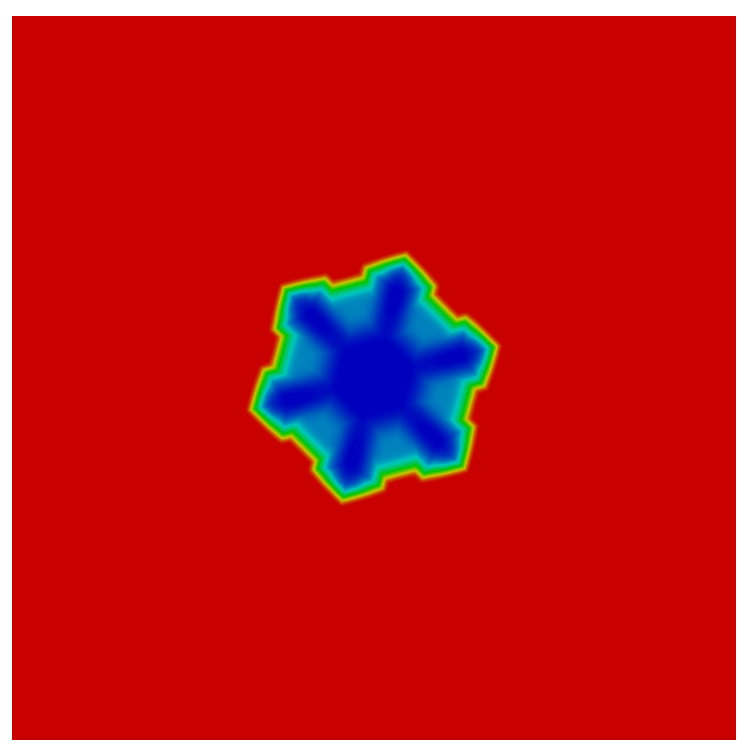}
\includegraphics[angle=-0,width=0.19\textwidth]{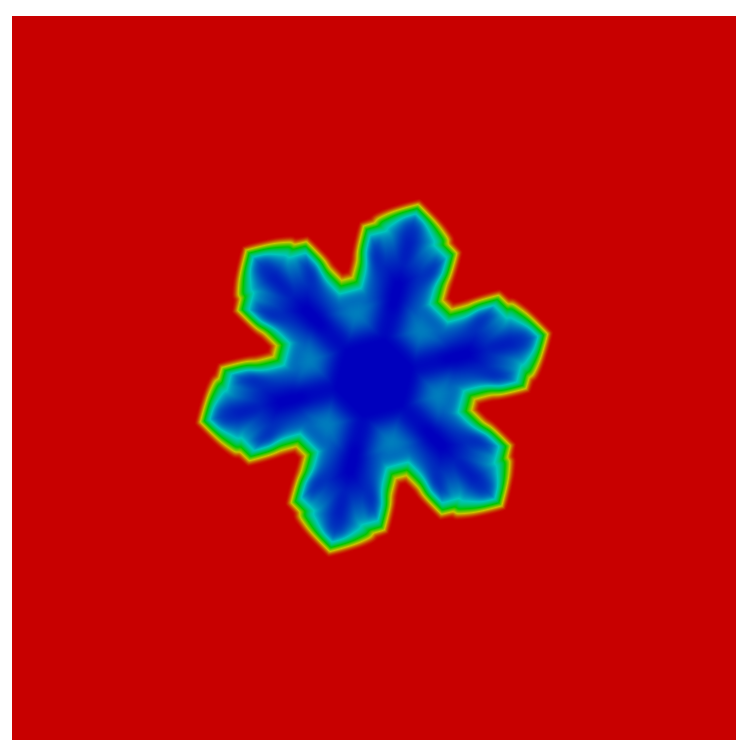}
\includegraphics[angle=-0,width=0.19\textwidth]{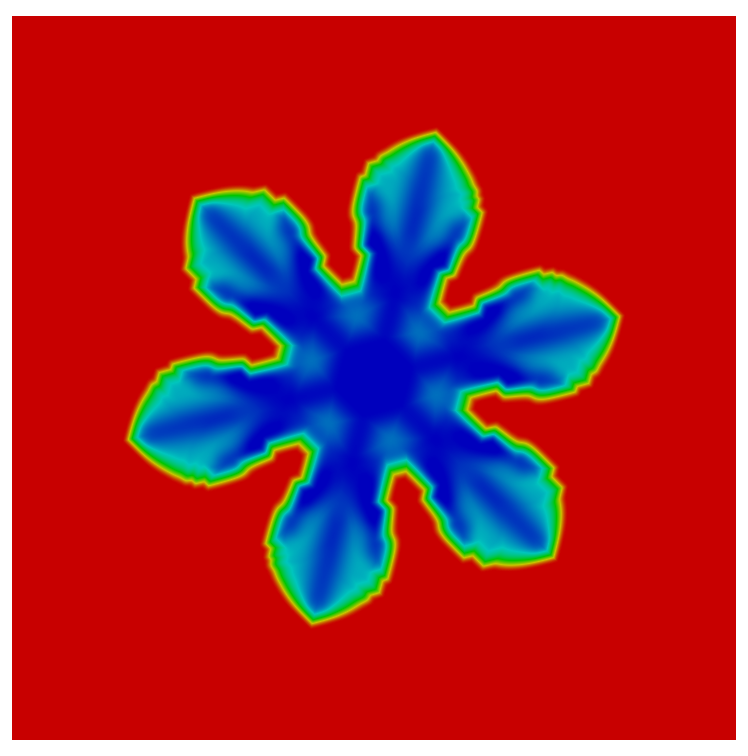}
\includegraphics[angle=-0,width=0.19\textwidth]{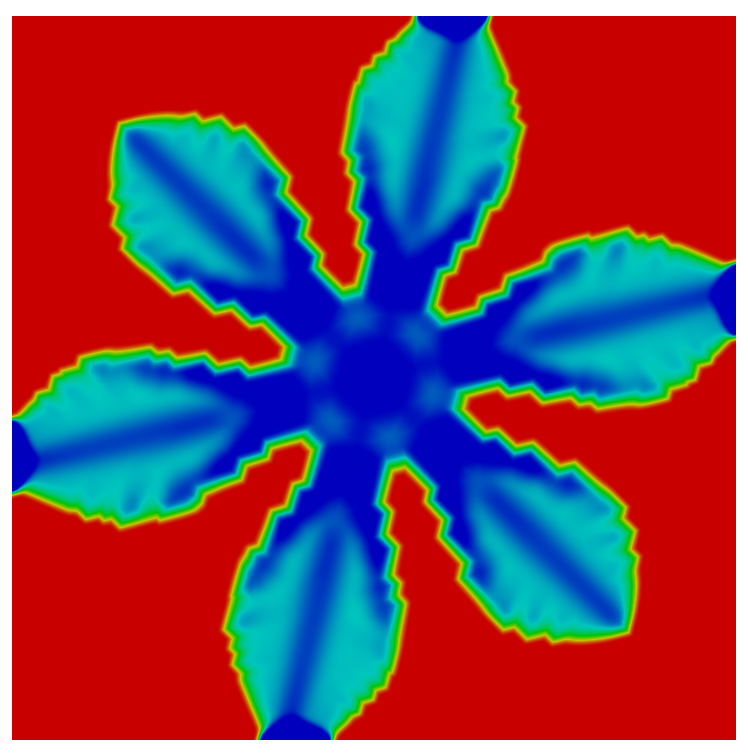}
\includegraphics[angle=-0,width=0.19\textwidth]{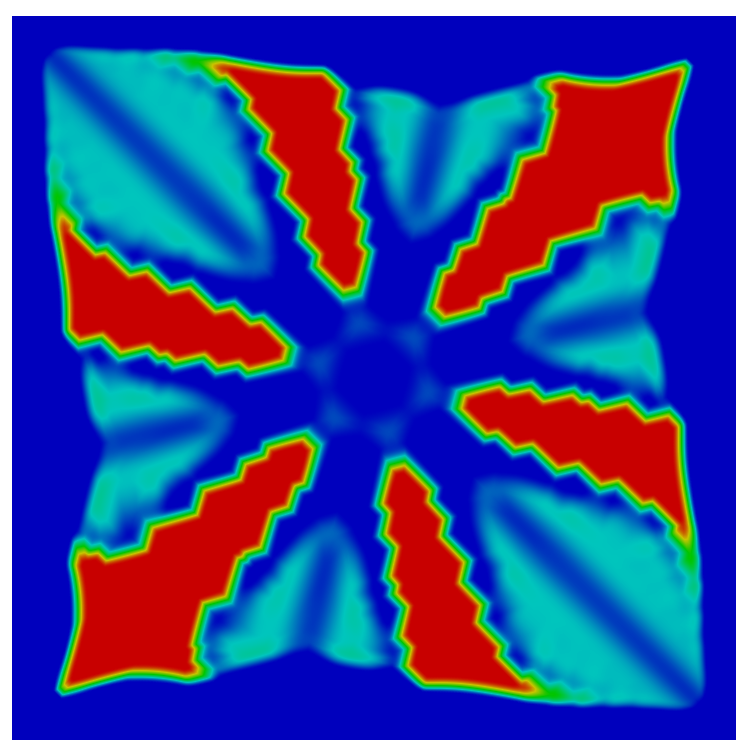}
\fi
\caption{(\PFii, $\epsilon^{-1} = 16\,\pi$, {\sc ani$_2$}, 
$\vartheta=1$, $\alpha=5\times10^{-4}$, $\rho = 0.01$, $\uD = -\frac12$, 
$\Omega=(-2,2)^2$)
Snapshots of the solution at times $t=0.3,\,0.4,\,0.5,\,0.6,\,0.7$. 
From top to bottom $\tau = 10^{-k}$, $k = 2 \to 4$.
[These computations took $29$ seconds, $26$ minutes and 
$381$ minutes, respectively.]
}
\label{fig:Stefanii_16pi}
\end{figure}

\begin{figure}
\center
\ifpdf
\includegraphics[angle=-0,width=0.19\textwidth]{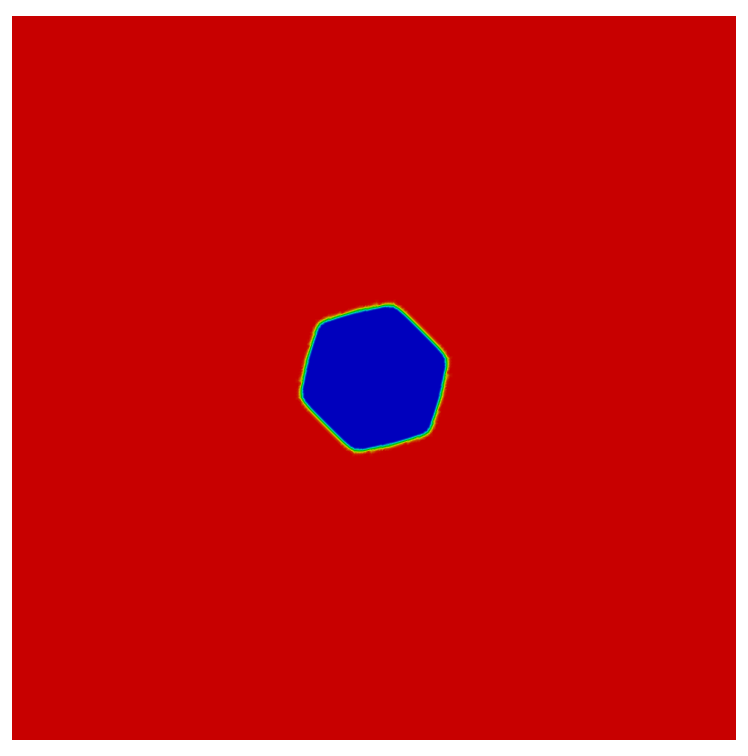}
\includegraphics[angle=-0,width=0.19\textwidth]{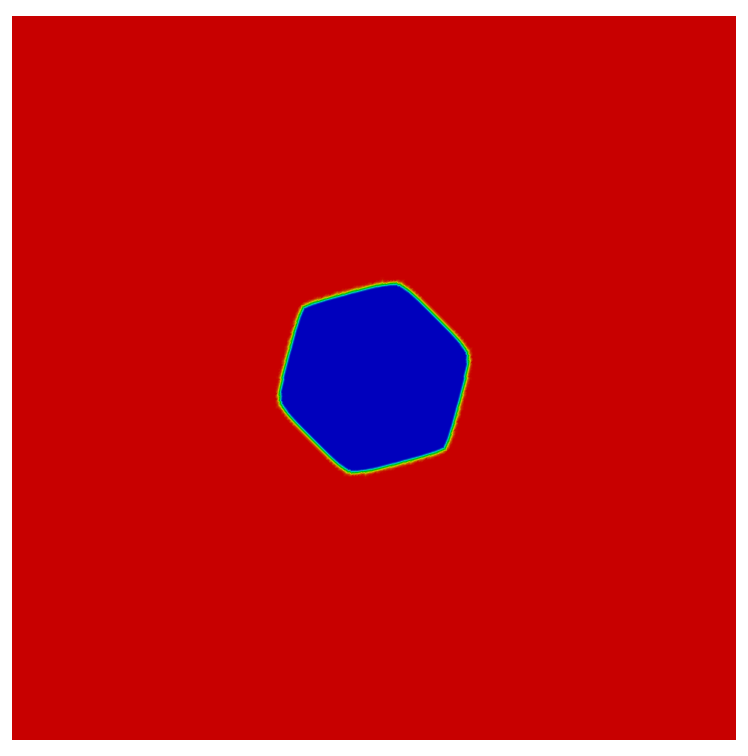}
\includegraphics[angle=-0,width=0.19\textwidth]{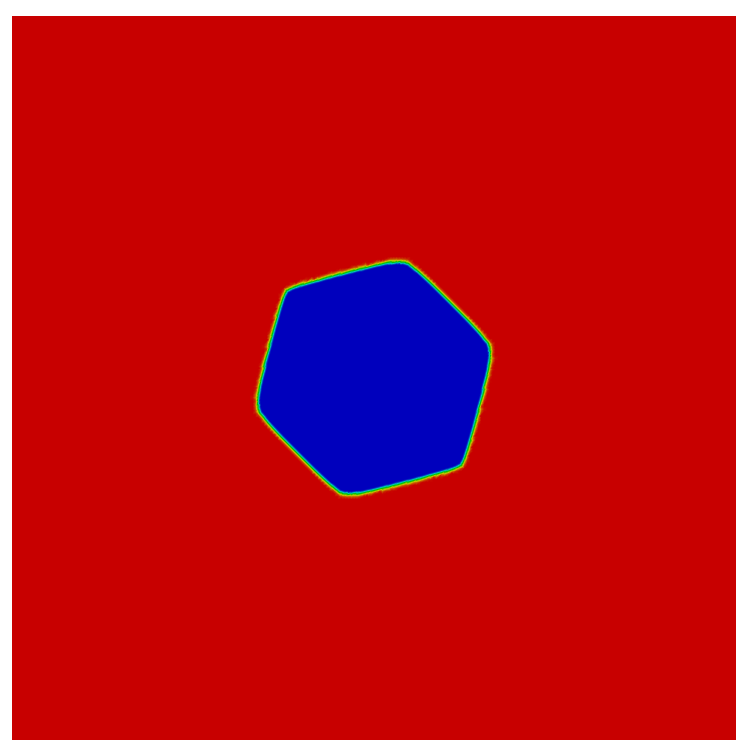}
\includegraphics[angle=-0,width=0.19\textwidth]{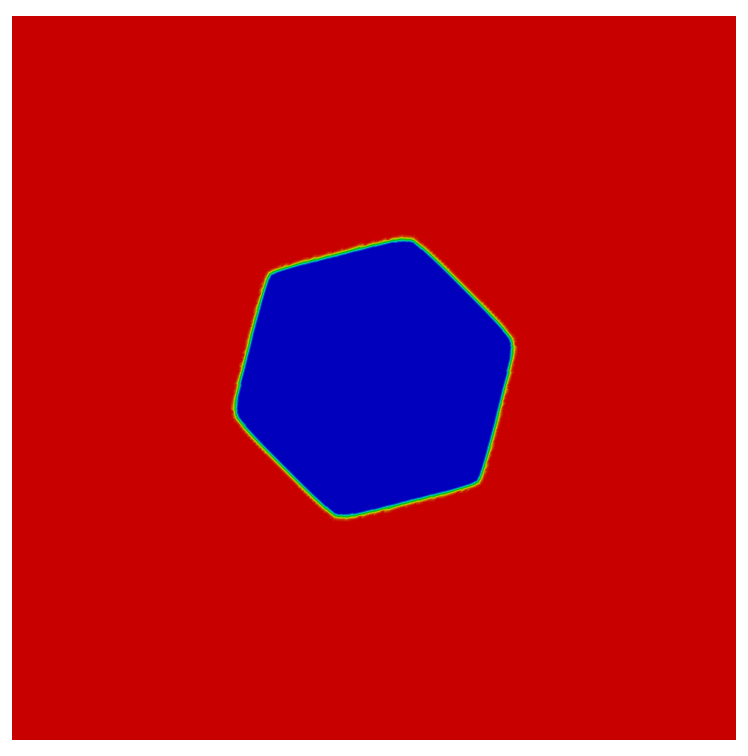}
\includegraphics[angle=-0,width=0.19\textwidth]{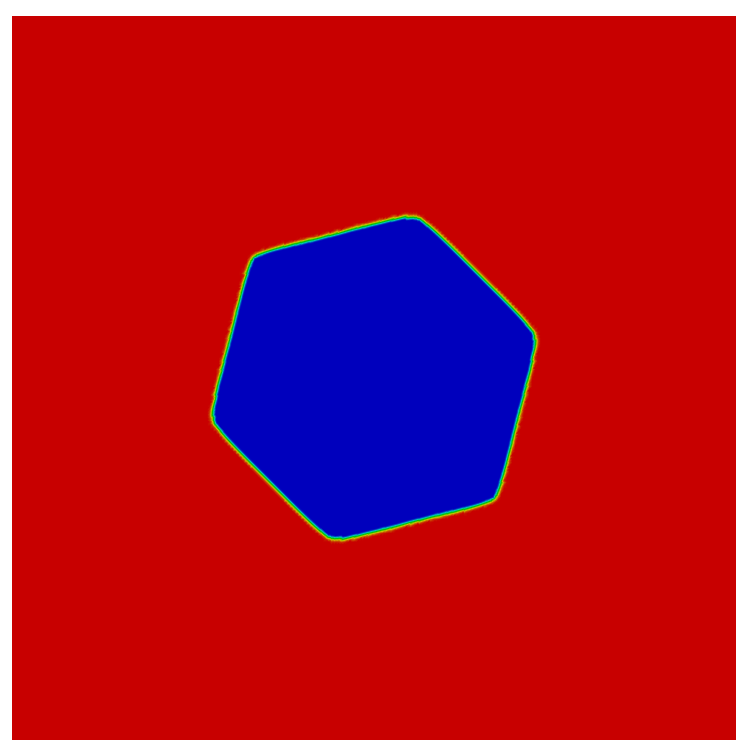}
\includegraphics[angle=-0,width=0.19\textwidth]{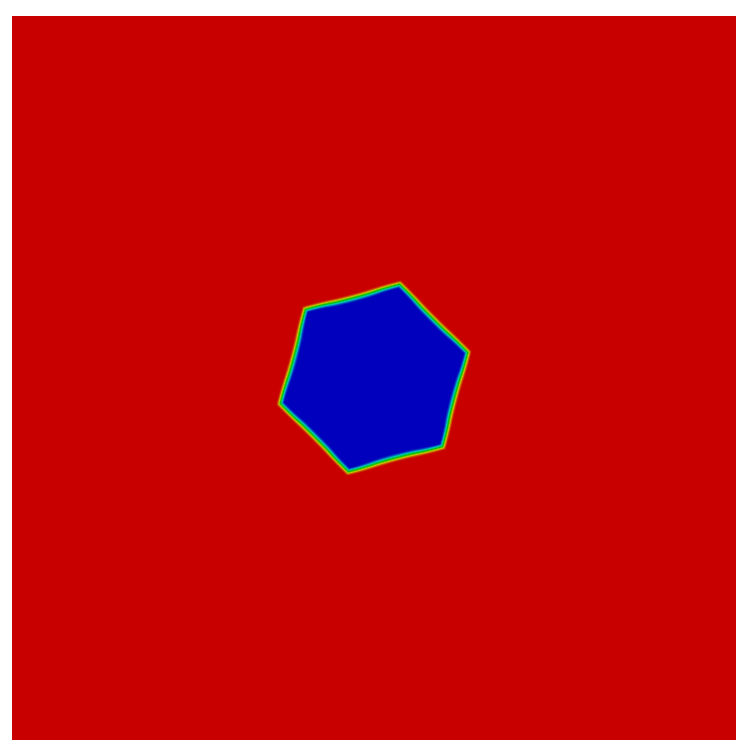}
\includegraphics[angle=-0,width=0.19\textwidth]{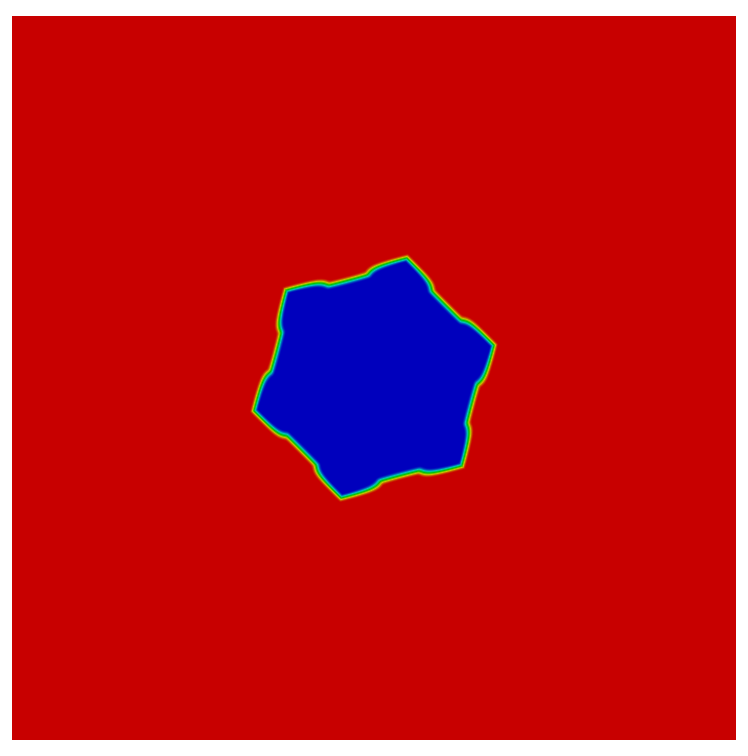}
\includegraphics[angle=-0,width=0.19\textwidth]{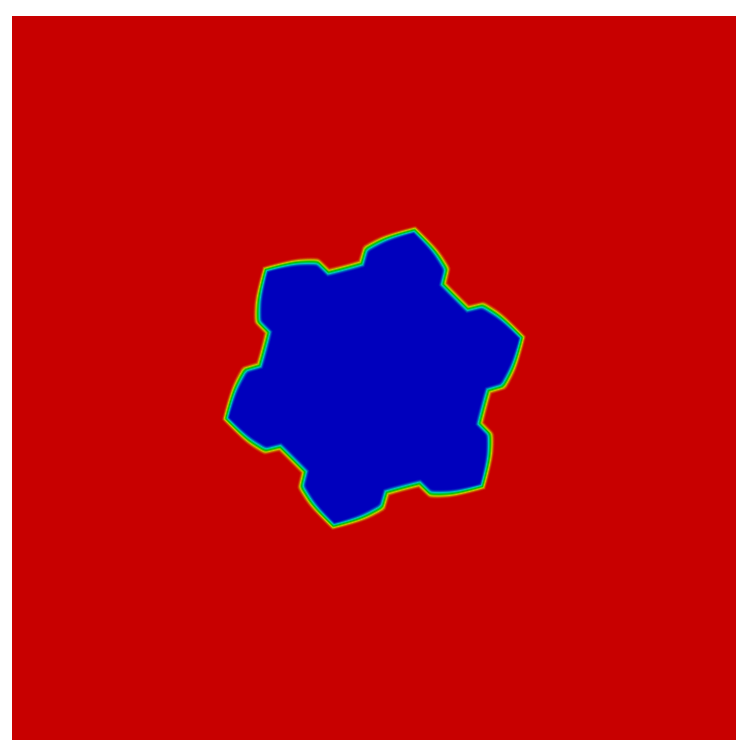}
\includegraphics[angle=-0,width=0.19\textwidth]{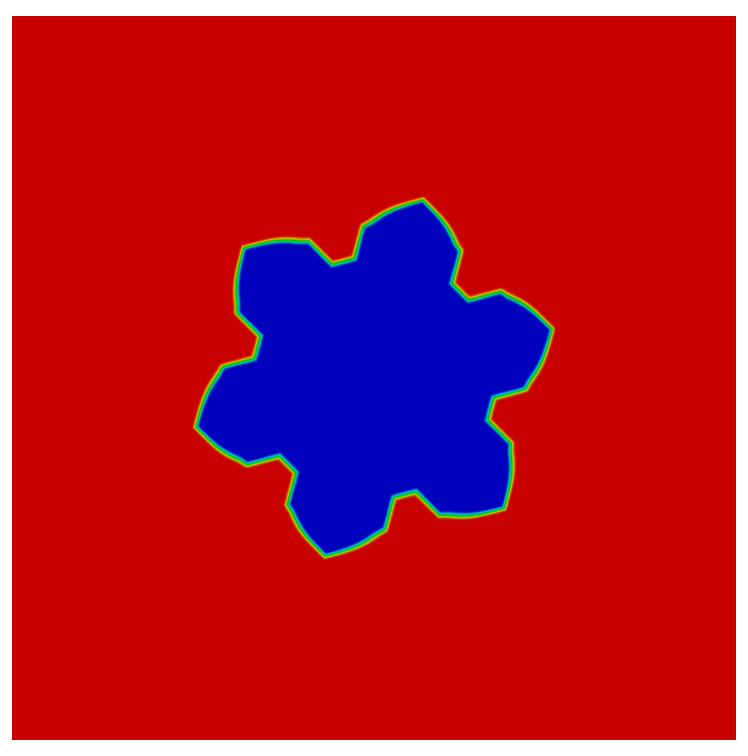}
\includegraphics[angle=-0,width=0.19\textwidth]{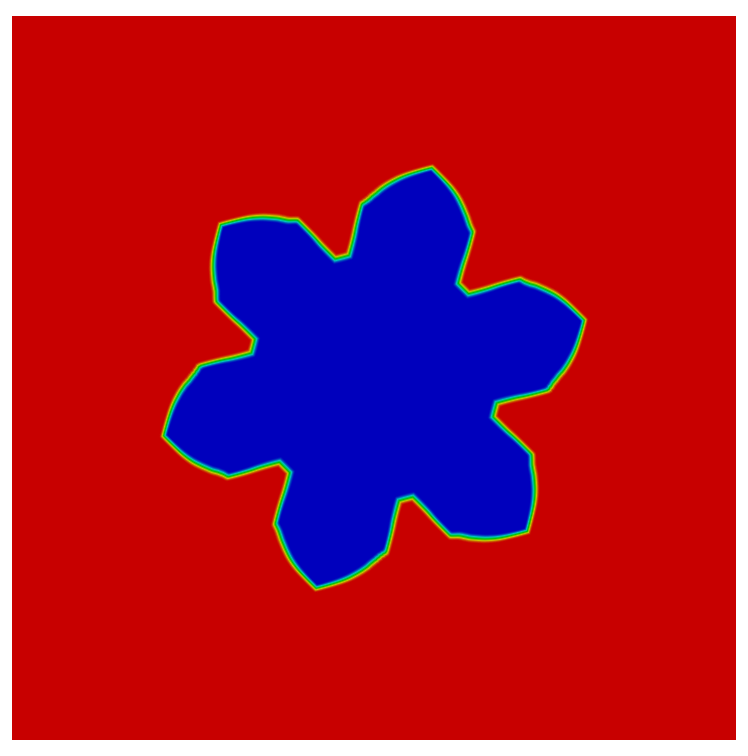}
\includegraphics[angle=-0,width=0.19\textwidth]{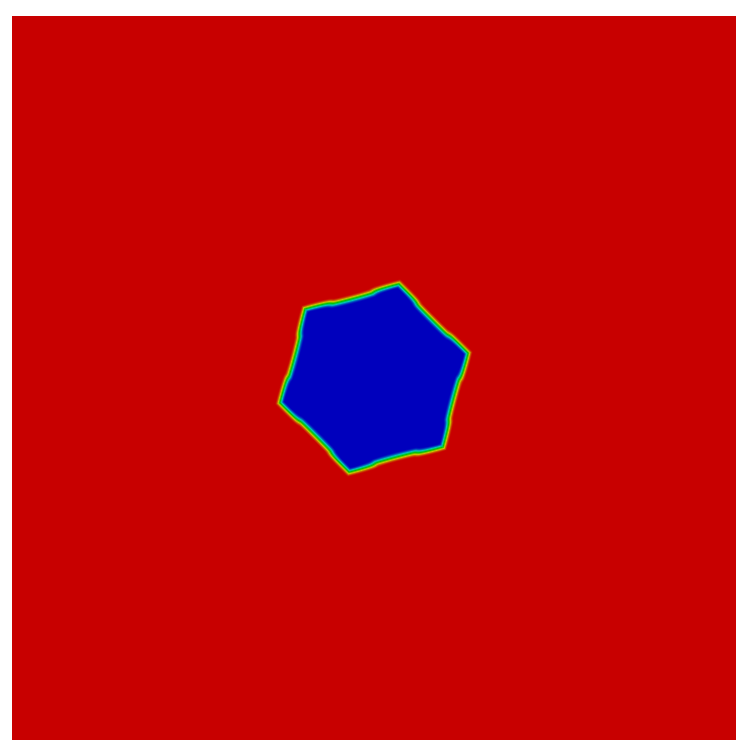}
\includegraphics[angle=-0,width=0.19\textwidth]{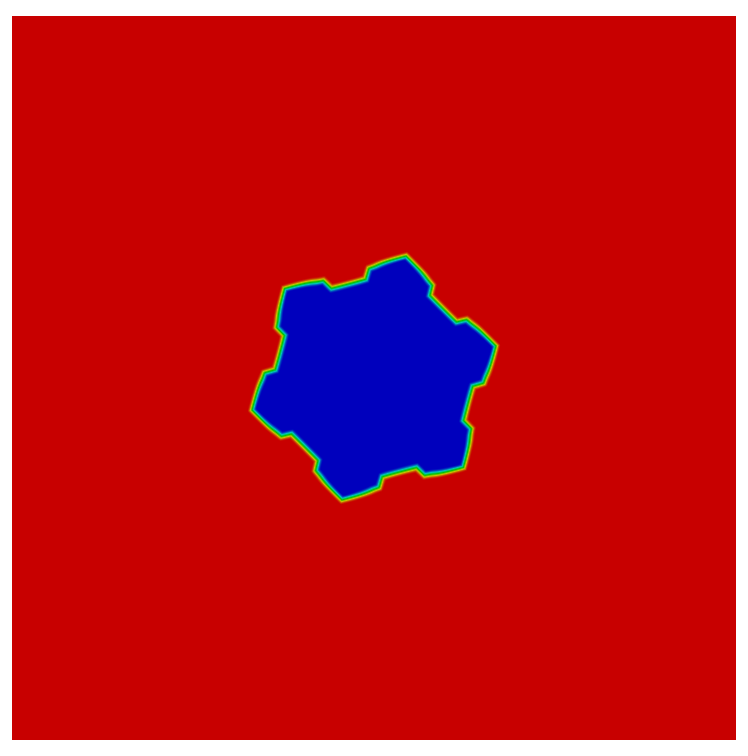}
\includegraphics[angle=-0,width=0.19\textwidth]{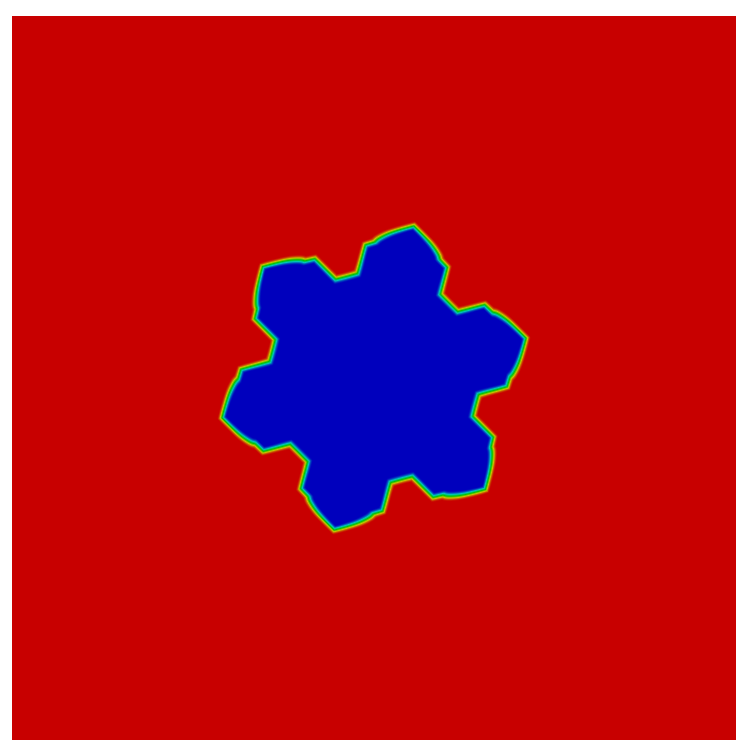}
\includegraphics[angle=-0,width=0.19\textwidth]{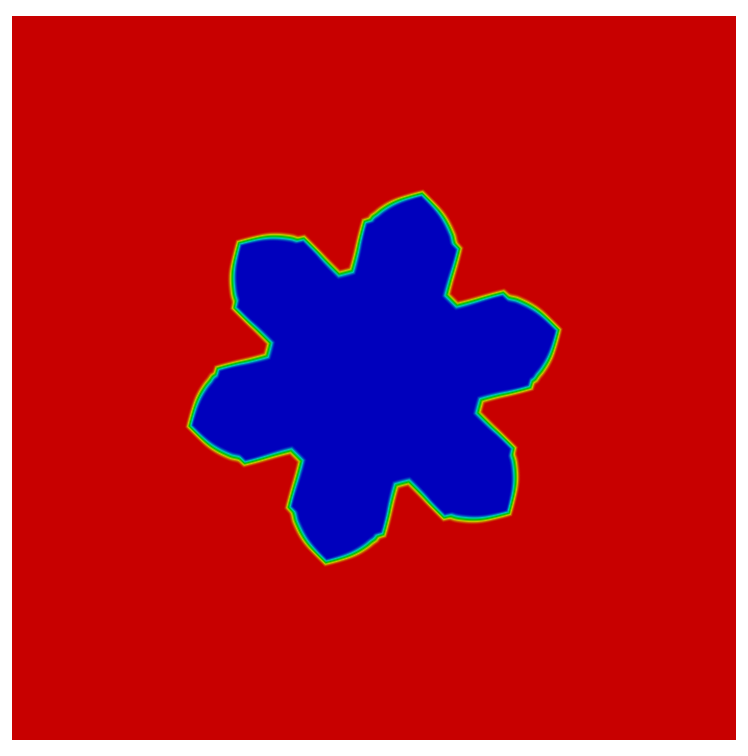}
\includegraphics[angle=-0,width=0.19\textwidth]{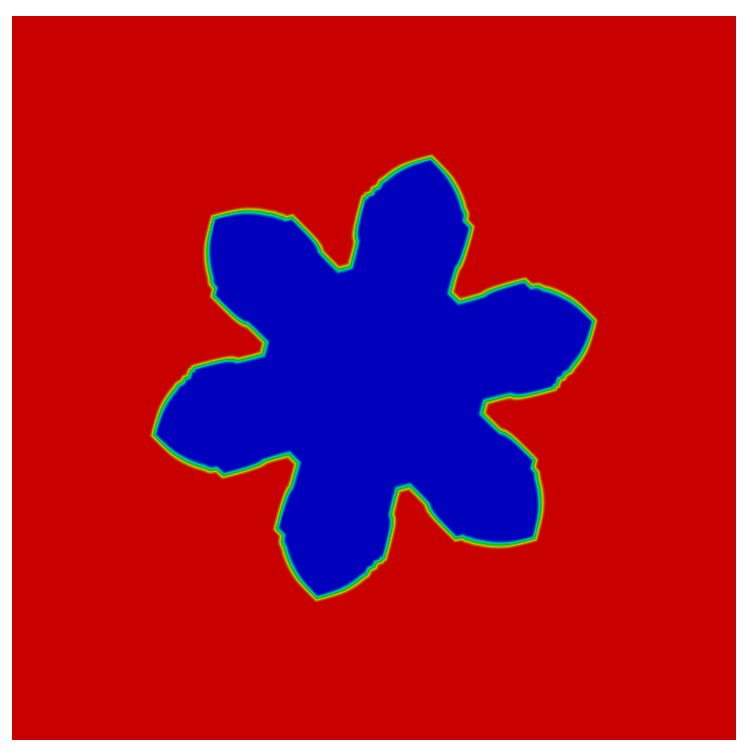}
\fi
\caption{(\PFii, $\epsilon^{-1} = 32\,\pi$, {\sc ani$_2$}, 
$\vartheta=1$, $\alpha=5\times10^{-4}$, $\rho = 0.01$, $\uD = -\frac12$, 
$\Omega=(-2,2)^2$)
Snapshots of the solution at times $t=0.3,\,0.4,\,0.5,\,0.6,\,0.7$. 
From top to bottom $\tau = 10^{-k}$, $k = 2 \to 4$.
[These computations took $38$ seconds, $8$ minutes and $69$ minutes,
respectively.]
}
\label{fig:Stefanii_32pi}
\end{figure}

Three numerical simulations for the scheme \PFii\ with the interfacial
parameter $\epsilon^{-1} = 16\,\pi$ can be seen in 
Figure~\ref{fig:Stefanii_16pi}. Observe that here we use a much smaller value
of $\epsilon$ than previously, because for larger values of $\epsilon$ large
mushy interfacial regions develop, which means that the phase field simulations
hold no value for the investigation of the underlying sharp interface problem.
The creation of small localized mushy interfacial regions can be observed in
Figure~\ref{fig:Stefanii_16pi} for the run with $\tau=10^{-3}$, while the run
with $\tau=10^{-4}$ shows larger such regions. In addition, in the latter run
the phase field approximation of the growing crystal's surface reaches the 
external boundary $\partial\Omega$, which results in the creation of 
artificial, nonphysical interfaces.
Repeating these simulations for a smaller interfacial parameter $\epsilon$
yields the results shown in Figure~\ref{fig:Stefanii_32pi}. Now for
sufficiently small values of the time step size $\tau$, the numerical results
appear to be converging.

\begin{figure}
\center
\ifpdf
\includegraphics[angle=-0,width=0.19\textwidth]{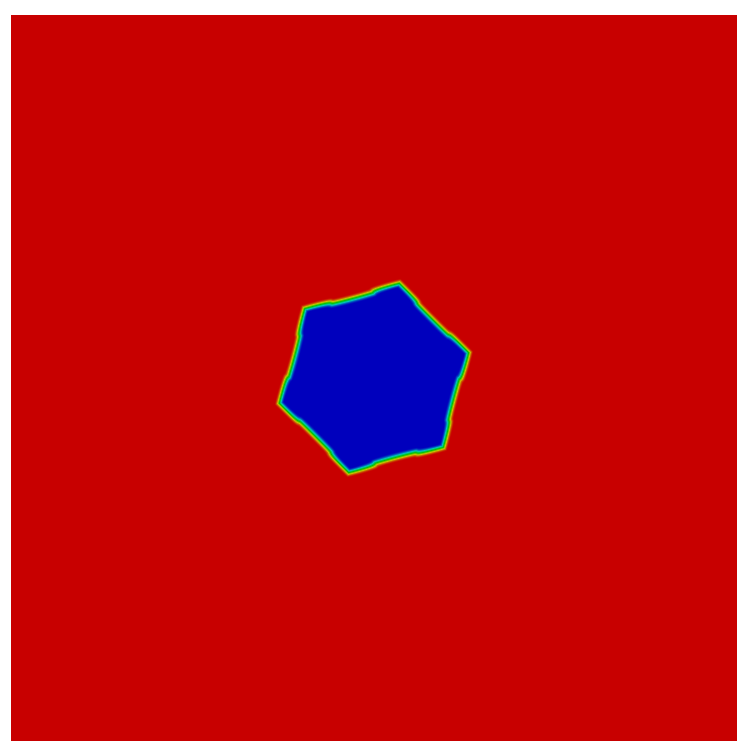}
\includegraphics[angle=-0,width=0.19\textwidth]{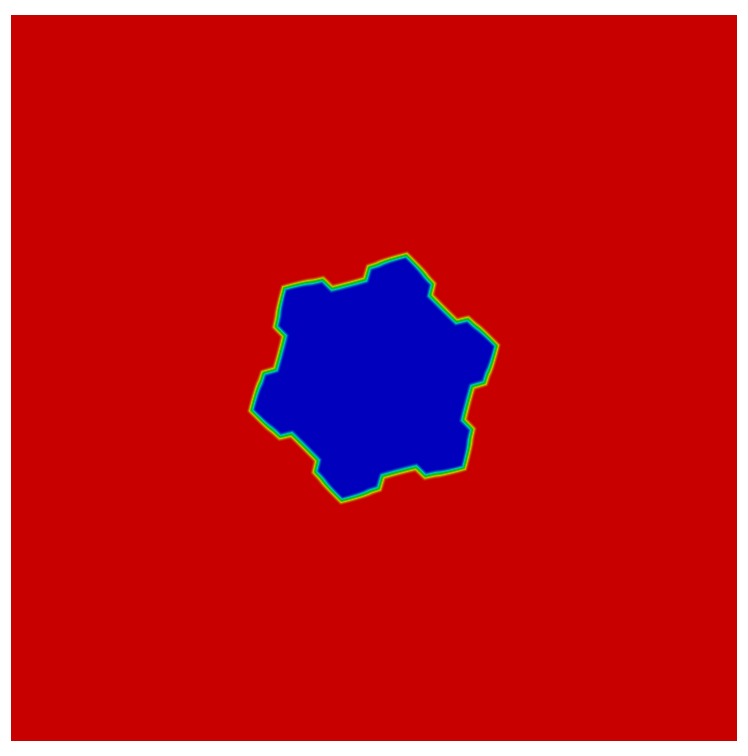}
\includegraphics[angle=-0,width=0.19\textwidth]{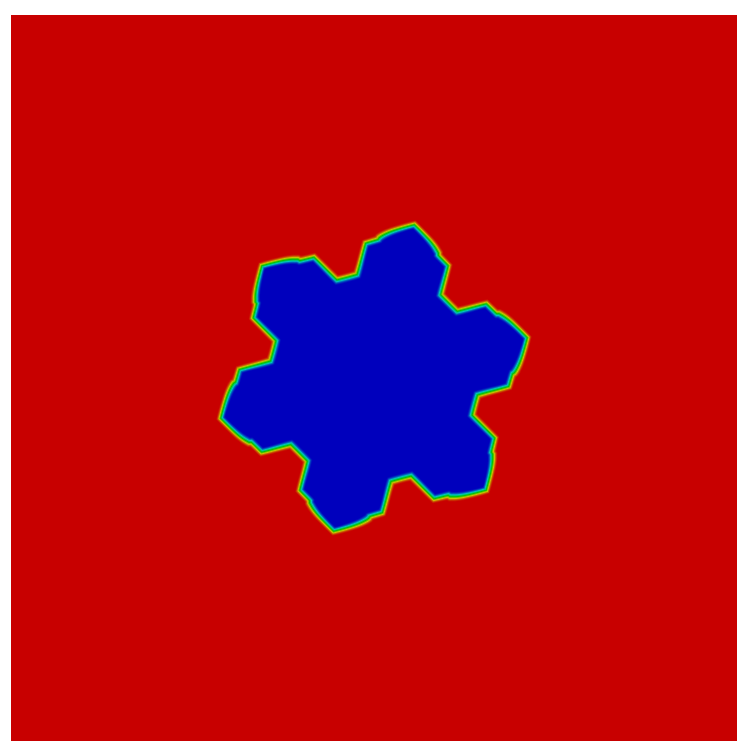}
\includegraphics[angle=-0,width=0.19\textwidth]{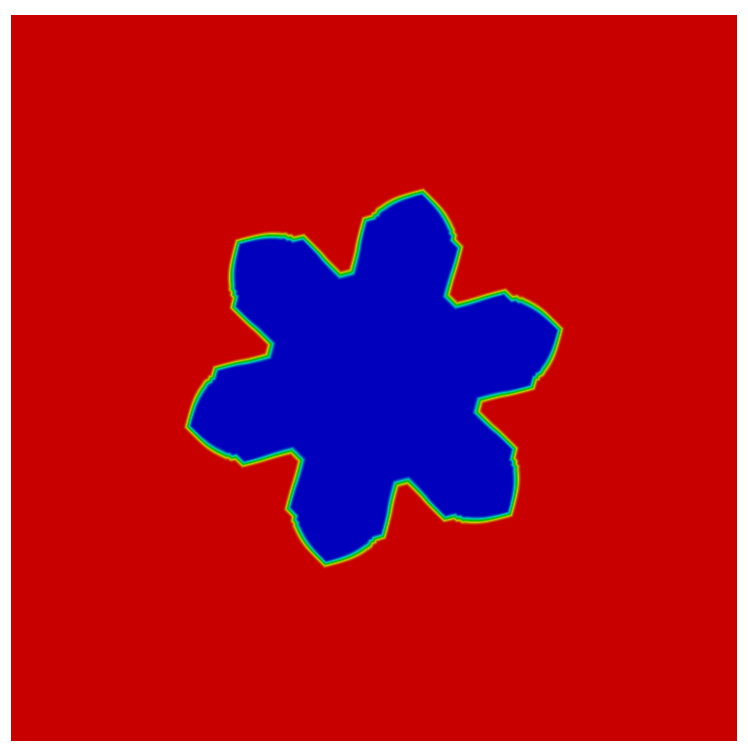}
\includegraphics[angle=-0,width=0.19\textwidth]{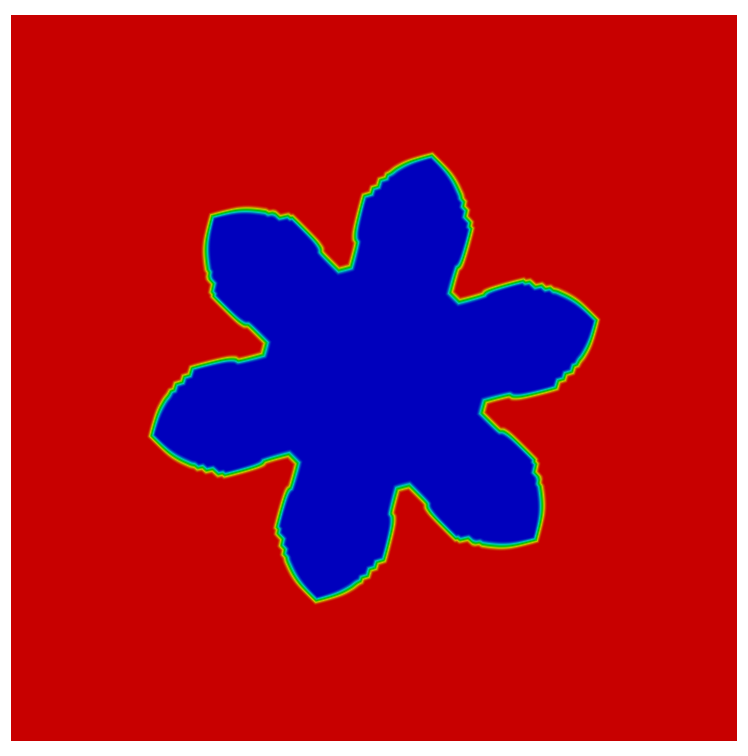}
\fi
\caption{(\PFii\ with the explicit time discretization from (\ref{eq:expl1},b), 
$\epsilon^{-1} = 32\,\pi$, {\sc ani$_2$}, 
$\vartheta=1$, $\alpha=5\times10^{-4}$, $\rho = 0.01$, $\uD = -\frac12$, 
$\Omega=(-2,2)^2$)
Snapshots of the solution at times $t=0.3,\,0.4,\,0.5,\,0.6,\,0.7$ for
$\tau = 10^{-6}$.
[This computation took $44$ hours.]
}
\label{fig:Stefanii_32piex}
\end{figure}

We also repeat the last computation for the scheme \PFii\ with the 
explicit time discretization from (\ref{eq:expl1},b). Here any computation with
a time step size $\tau \geq 10^{-5}$ was unstable, and so in 
Figure~\ref{fig:Stefanii_32piex} we only show a run for $\tau = 10^{-6}$. We
recall from Section~\ref{sec:42}, see Tables~\ref{tab:TSii1_expl} and 
\ref{tab:TSii1_expl_unif}, that in the interest of accuracy uniform meshes
should be employed for an explicit method. However, the large CPU times 
associated with a uniform grid mean that we are unable to complete the 
evolution within a reasonable amount of time. 
Hence in Figure~\ref{fig:Stefanii_32piex} we use the
same adaptive mesh strategy as in Figure~\ref{fig:Stefanii_32pi} for the
semi-implicit scheme \PFii. 
Note that while the finest run in Figure~\ref{fig:Stefanii_32pi} agrees
well with the results shown in Figure~\ref{fig:Stefanii_32piex}, the very small
time step size used for the latter means that the explicit scheme takes about
$40$ times as long as the semi-implicit scheme to compute the evolution. Hence,
without further code optimizations, the explicit scheme would need to be run 
in parallel on a cluster with at least $40$ nodes to become competitive.

\begin{figure}
\center
\ifpdf
\includegraphics[angle=-0,width=0.19\textwidth]{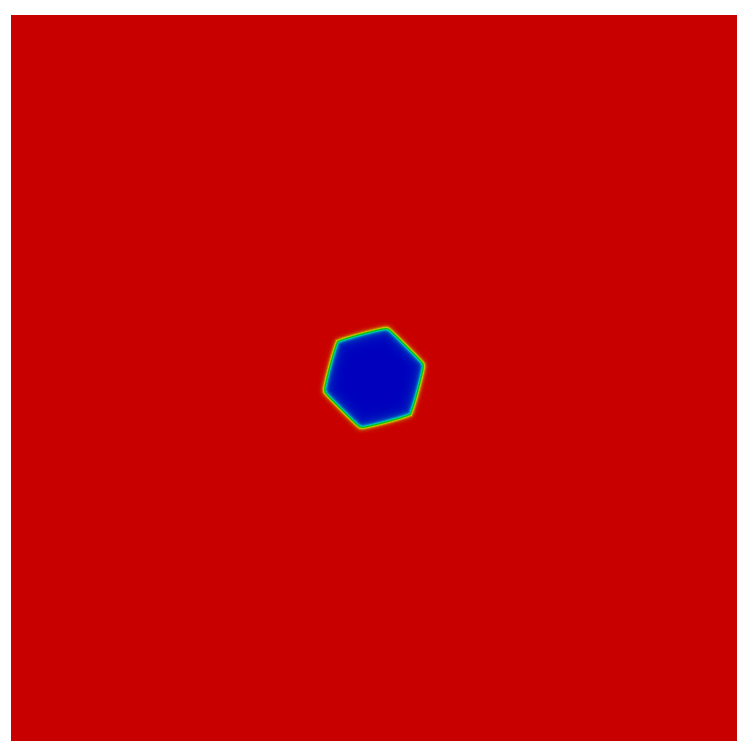}
\includegraphics[angle=-0,width=0.19\textwidth]{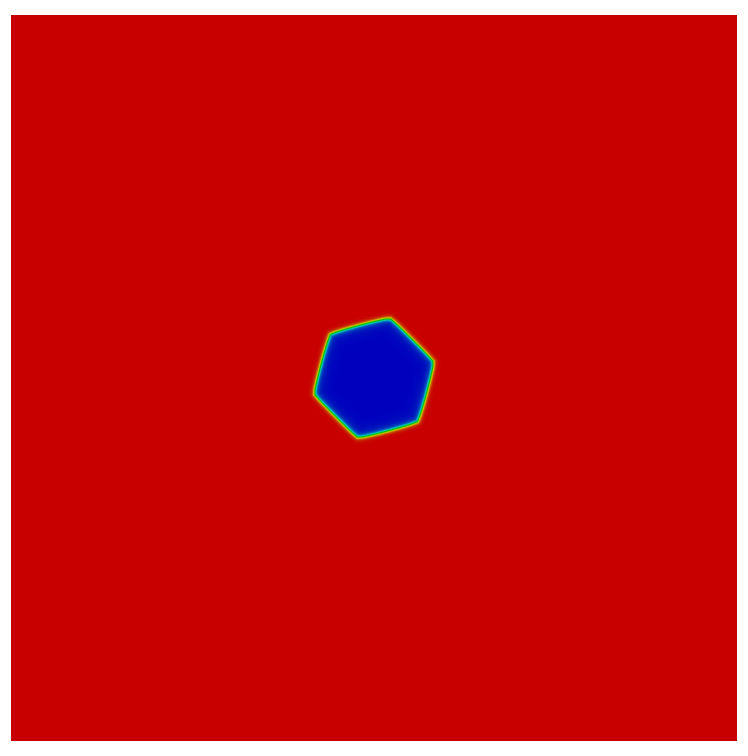}
\includegraphics[angle=-0,width=0.19\textwidth]{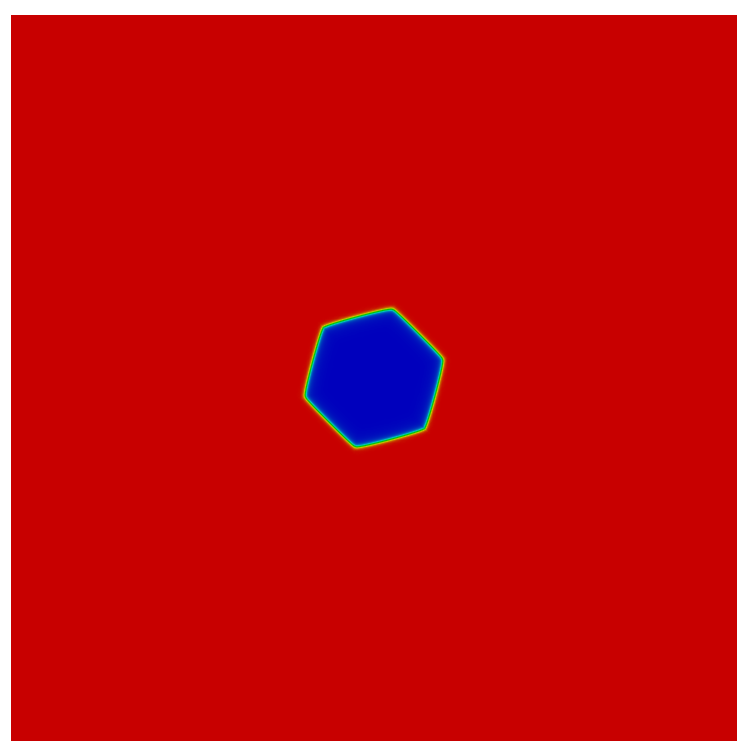}
\includegraphics[angle=-0,width=0.19\textwidth]{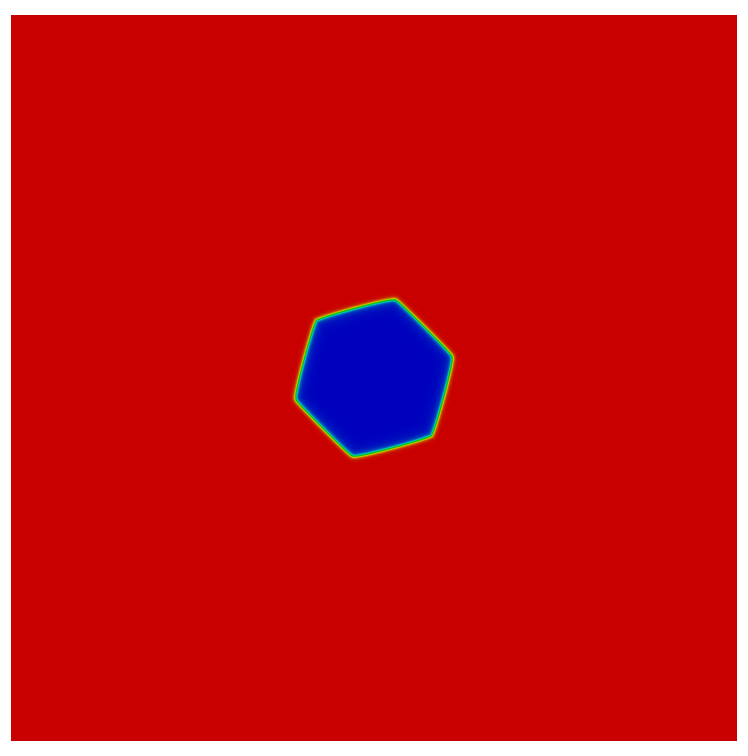}
\includegraphics[angle=-0,width=0.19\textwidth]{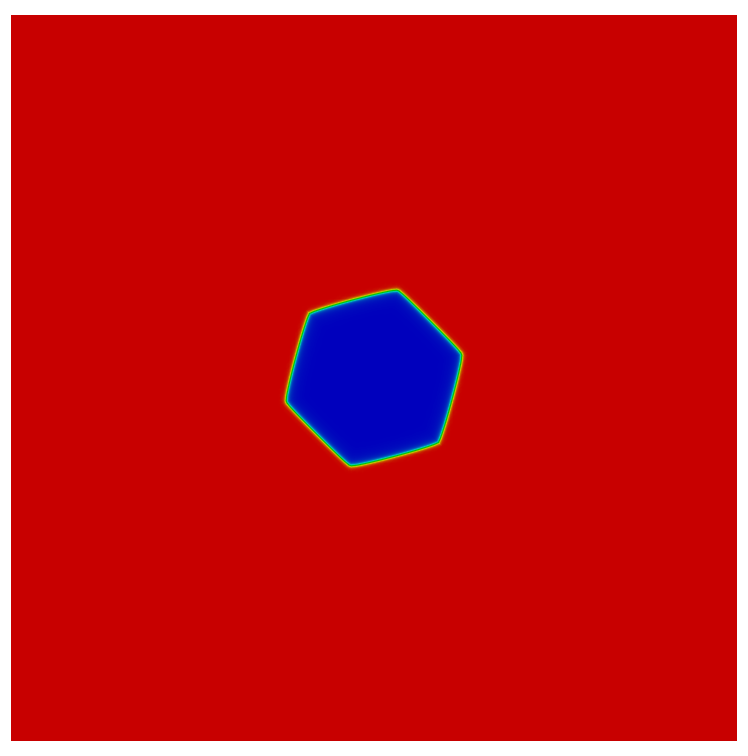}
\includegraphics[angle=-0,width=0.19\textwidth]{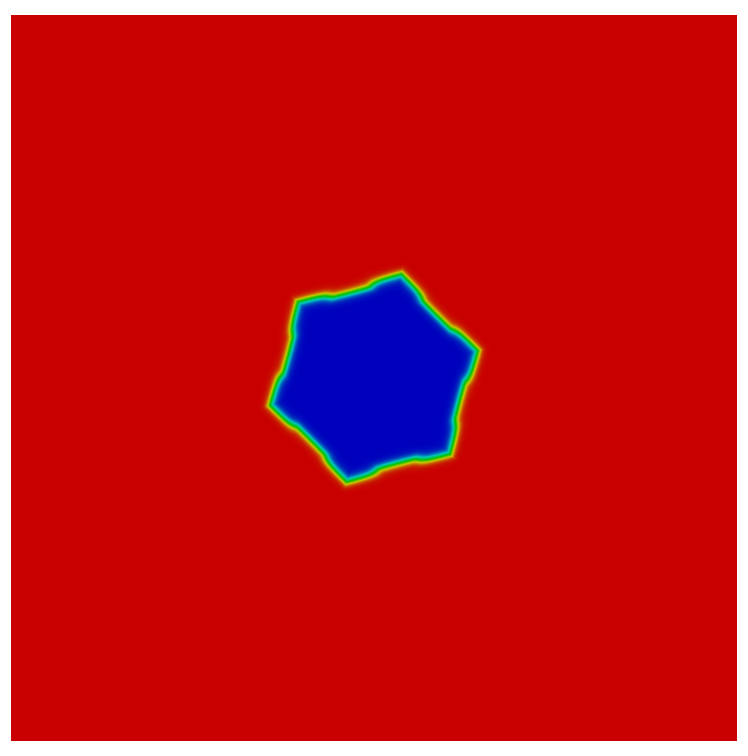}
\includegraphics[angle=-0,width=0.19\textwidth]{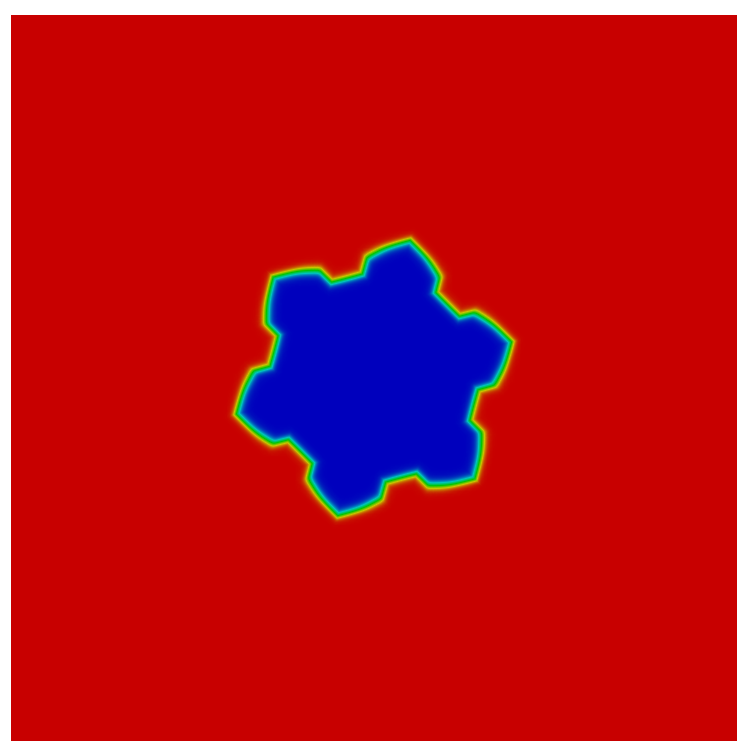}
\includegraphics[angle=-0,width=0.19\textwidth]{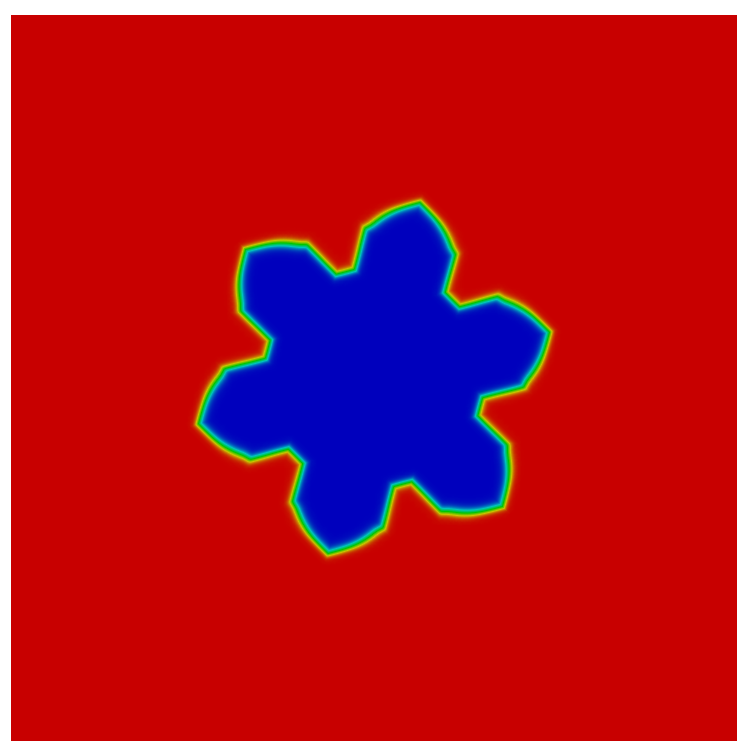}
\includegraphics[angle=-0,width=0.19\textwidth]{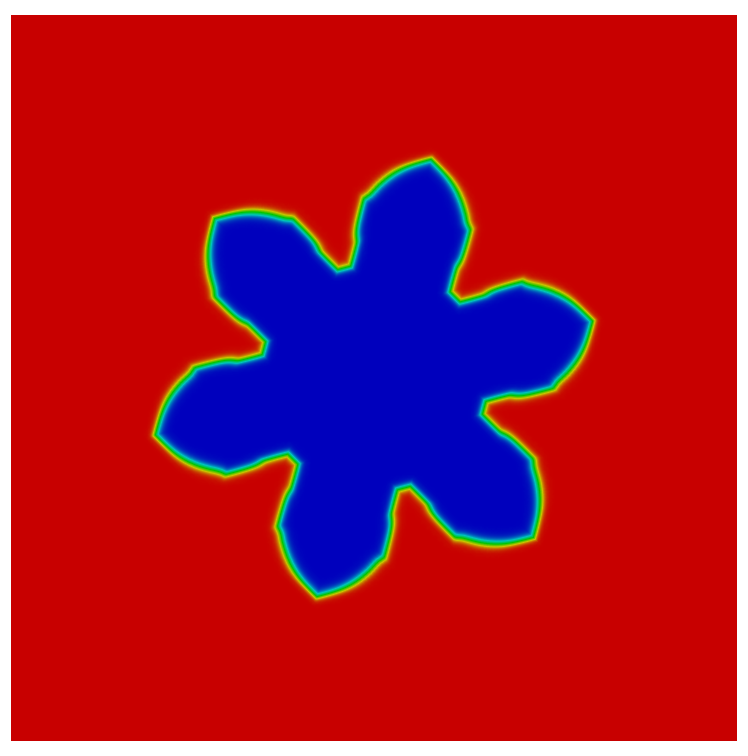}
\includegraphics[angle=-0,width=0.19\textwidth]{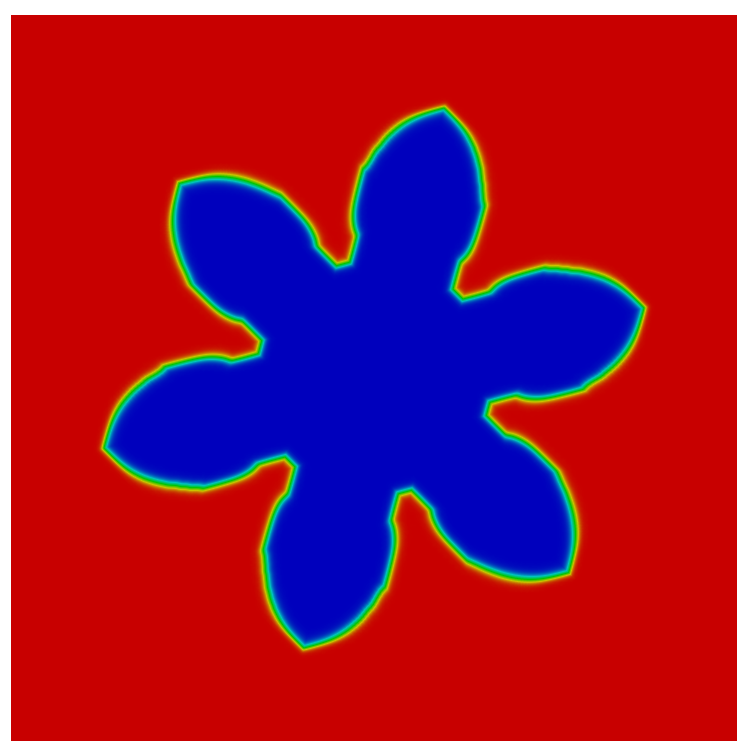}
\includegraphics[angle=-0,width=0.19\textwidth]{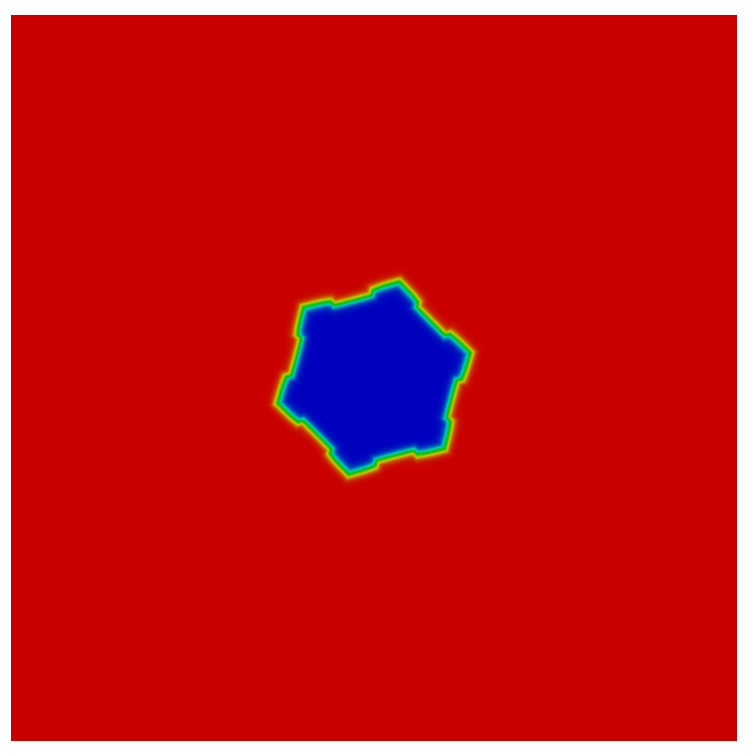}
\includegraphics[angle=-0,width=0.19\textwidth]{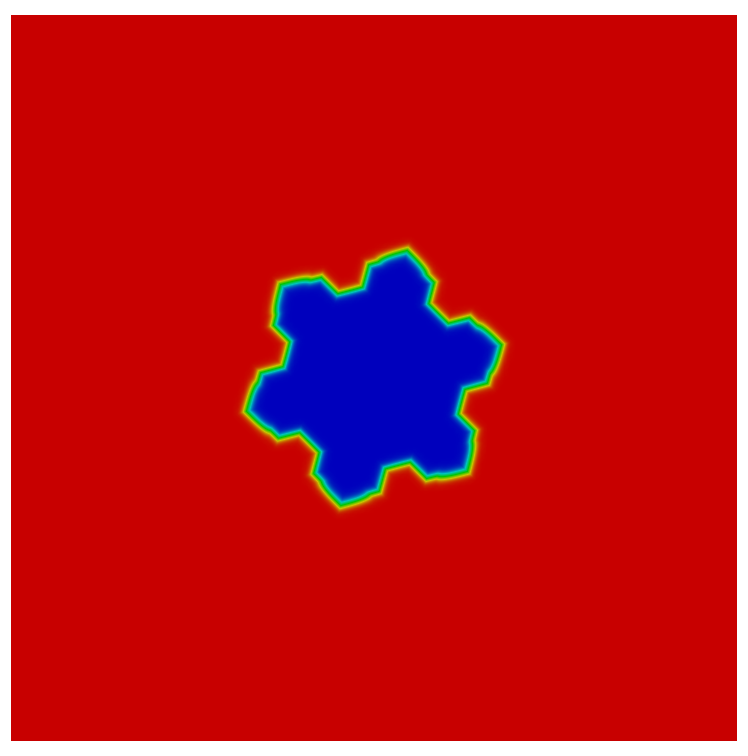}
\includegraphics[angle=-0,width=0.19\textwidth]{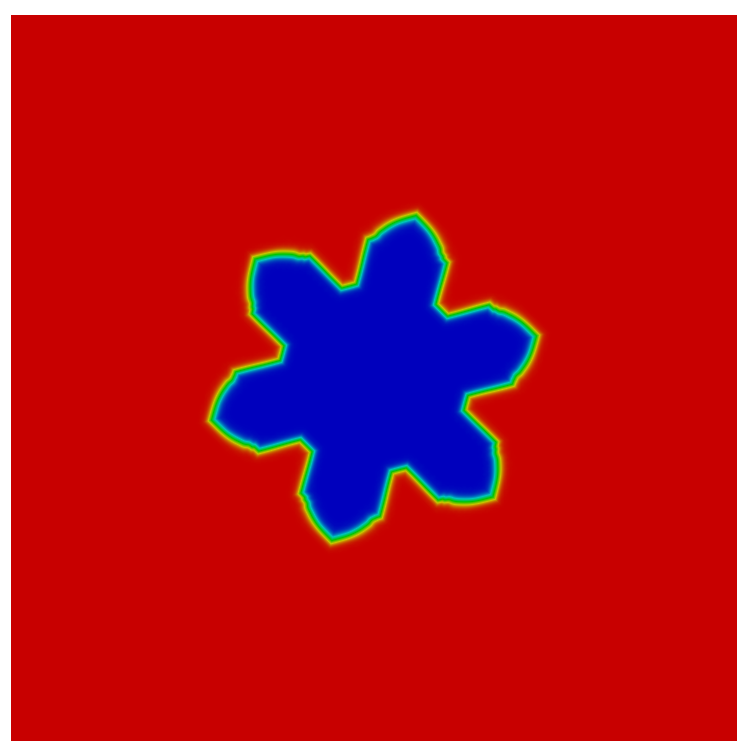}
\includegraphics[angle=-0,width=0.19\textwidth]{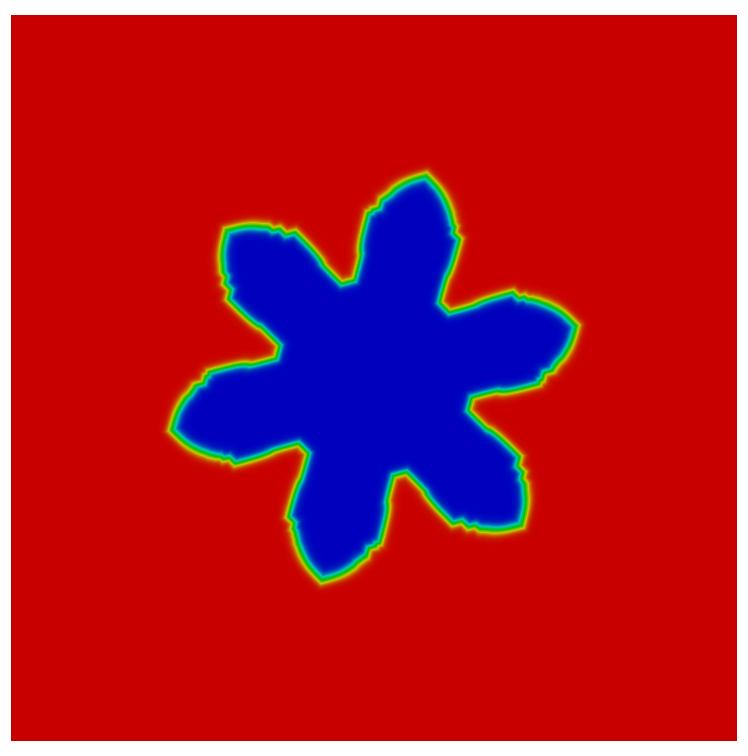}
\includegraphics[angle=-0,width=0.19\textwidth]{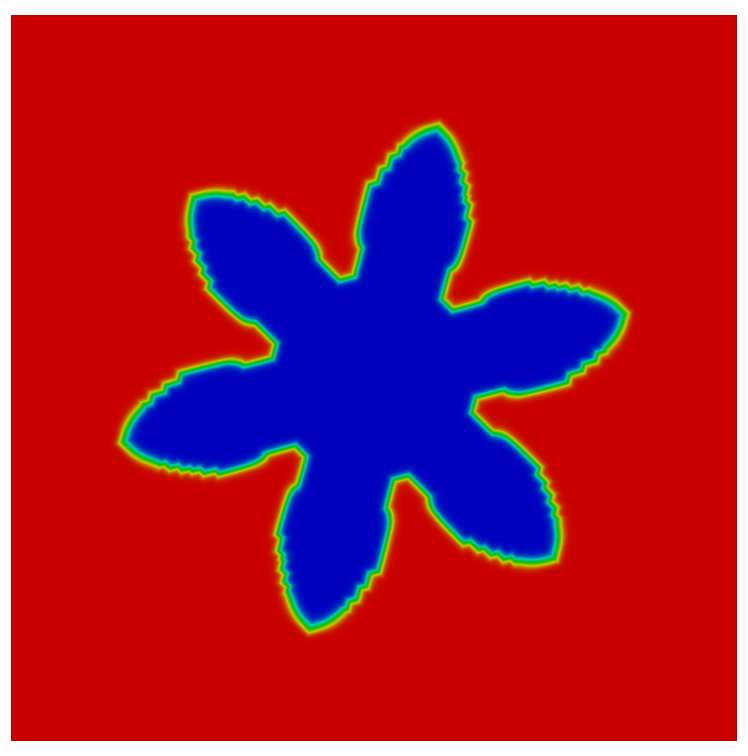}
\fi
\caption{(\PFqiii, $\epsilon^{-1} = 32\,\pi$, {\sc ani$_2$}, 
$\vartheta=1$, $\alpha=5\times10^{-4}$, $\rho = 0.01$, $\uD = -\frac12$, 
$\Omega=(-2,2)^2$)
Snapshots of the solution at times $t=0.3,\,0.4,\,0.5,\,0.6,\,0.7$. 
From top to bottom $\tau = 10^{-k}$, $k = 2 \to 4$.
[These computations took $34$ seconds, $16$ minutes and $101$ minutes, 
respectively.]
}
\label{fig:qvStefan_32pi}
\end{figure}

\begin{figure}
\center
\ifpdf
\includegraphics[angle=-0,width=0.19\textwidth]{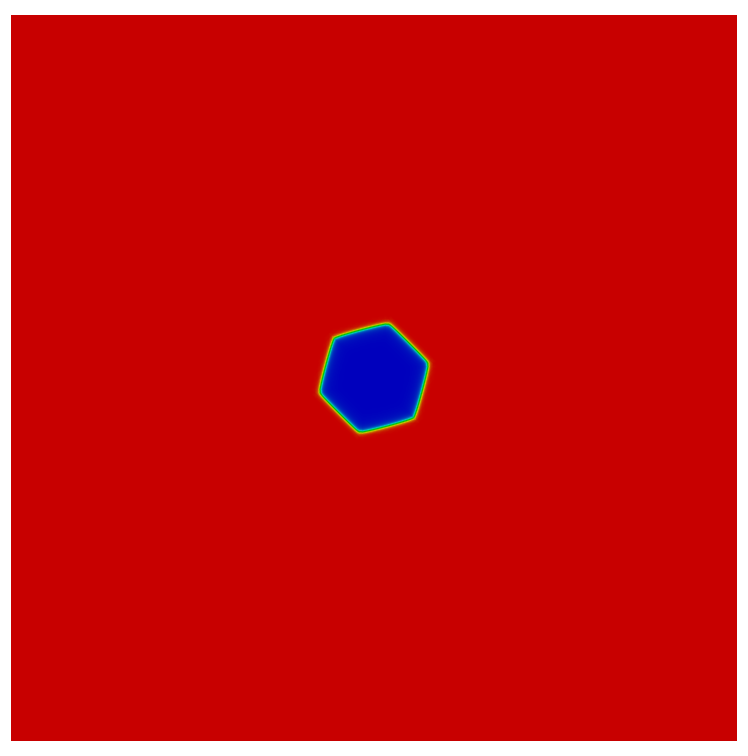}
\includegraphics[angle=-0,width=0.19\textwidth]{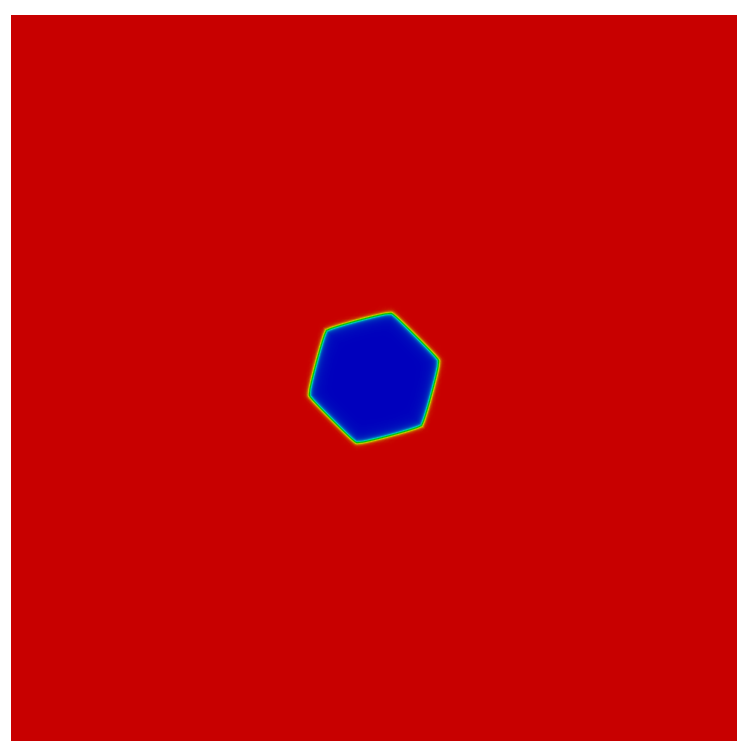}
\includegraphics[angle=-0,width=0.19\textwidth]{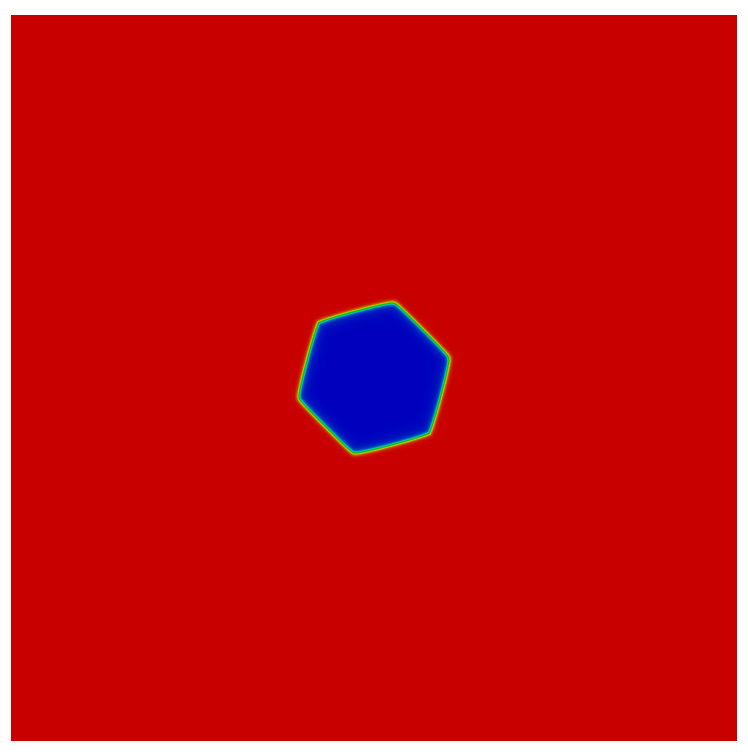}
\includegraphics[angle=-0,width=0.19\textwidth]{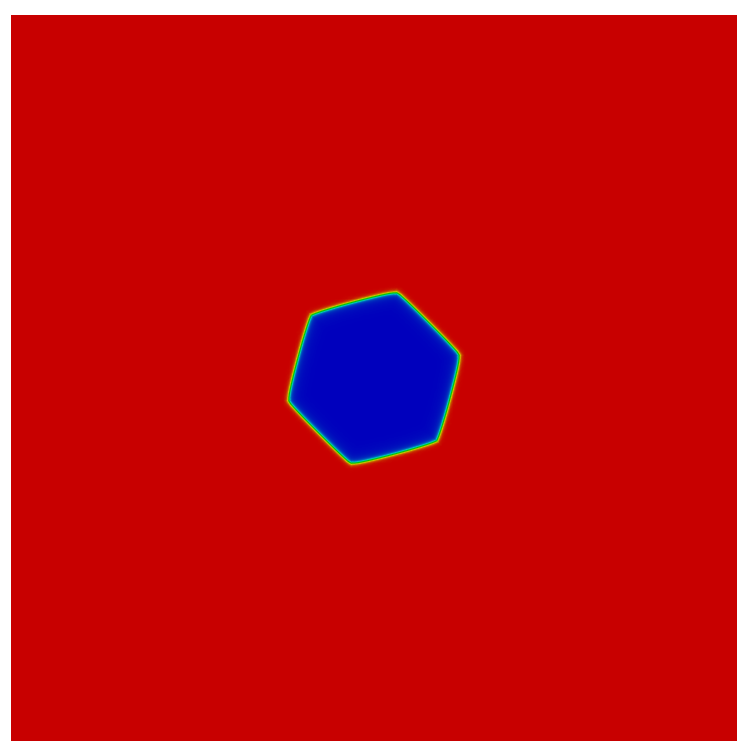}
\includegraphics[angle=-0,width=0.19\textwidth]{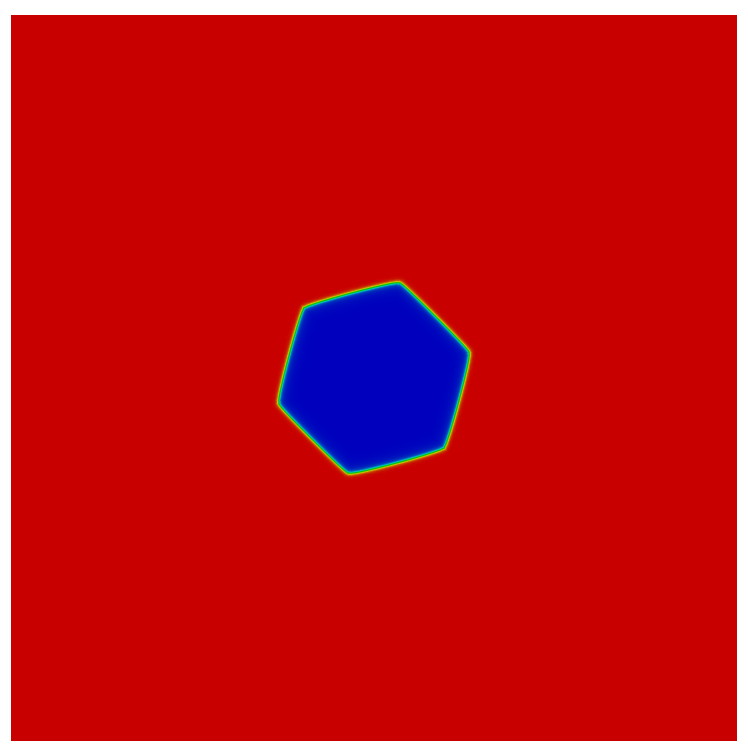}
\includegraphics[angle=-0,width=0.19\textwidth]{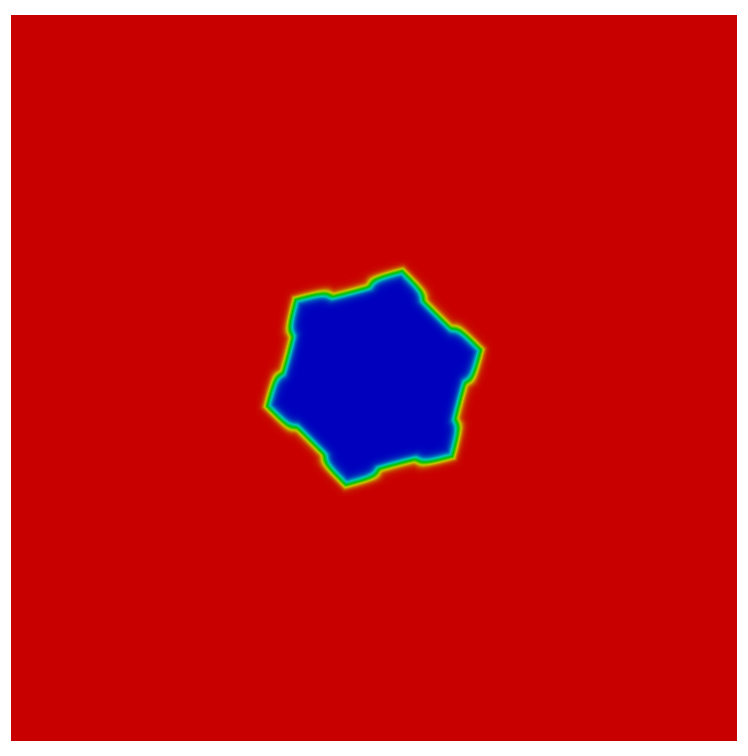}
\includegraphics[angle=-0,width=0.19\textwidth]{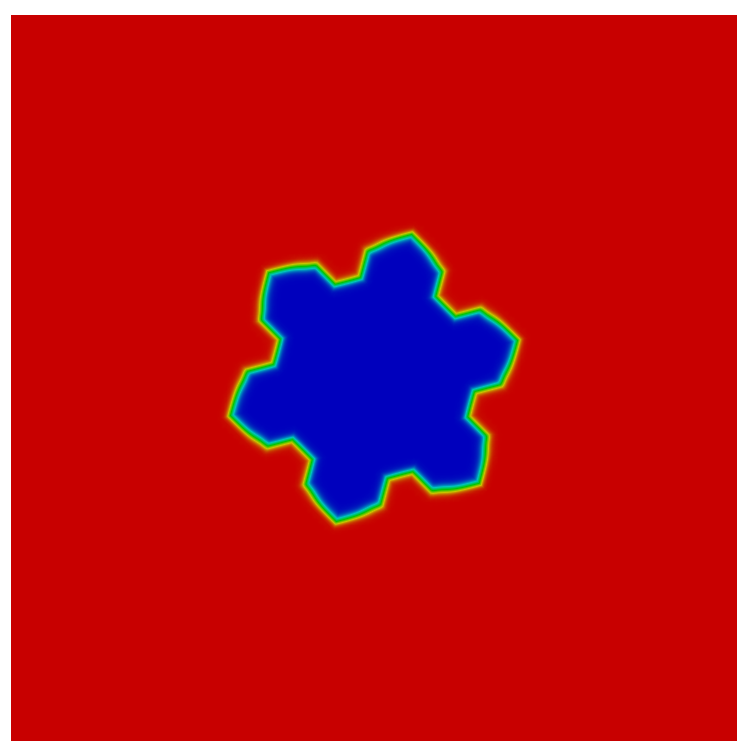}
\includegraphics[angle=-0,width=0.19\textwidth]{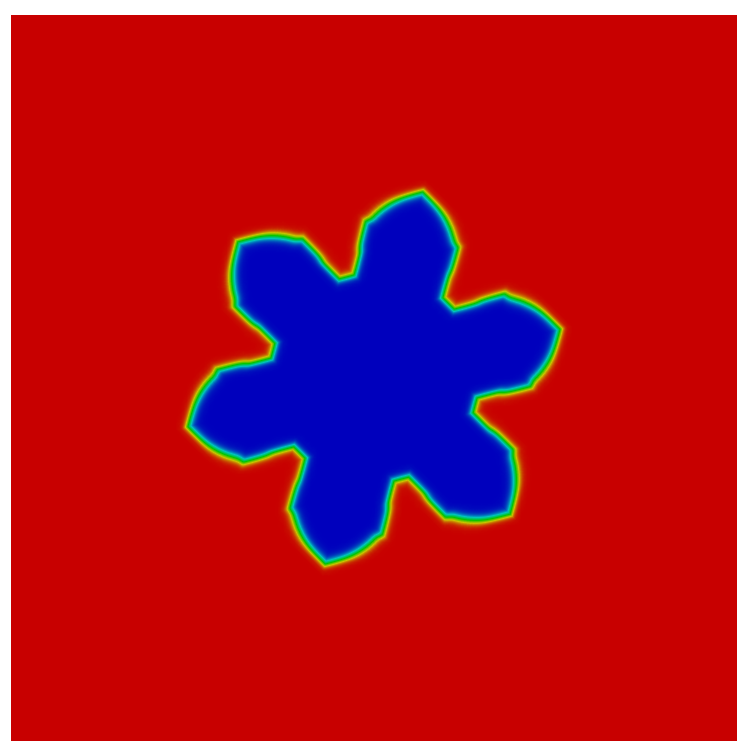}
\includegraphics[angle=-0,width=0.19\textwidth]{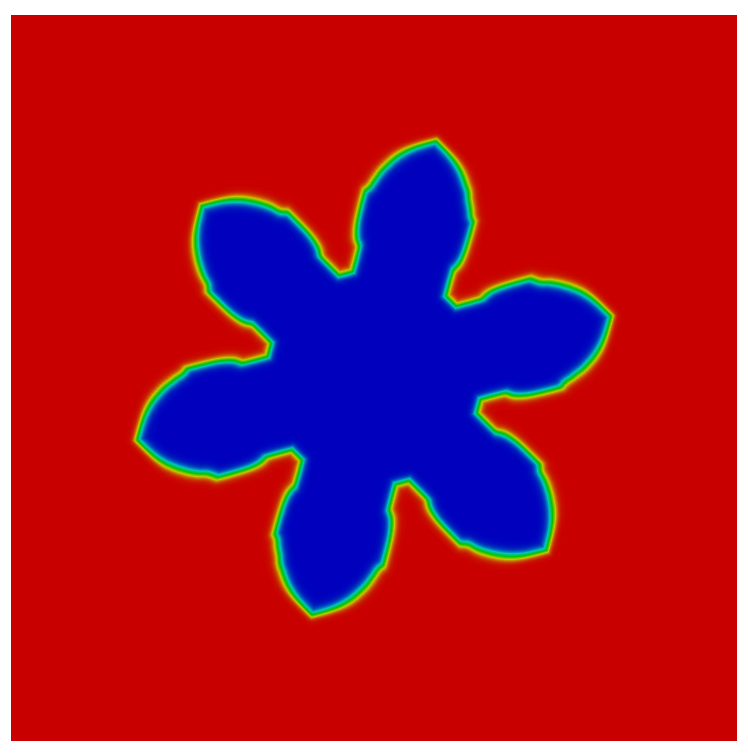}
\includegraphics[angle=-0,width=0.19\textwidth]{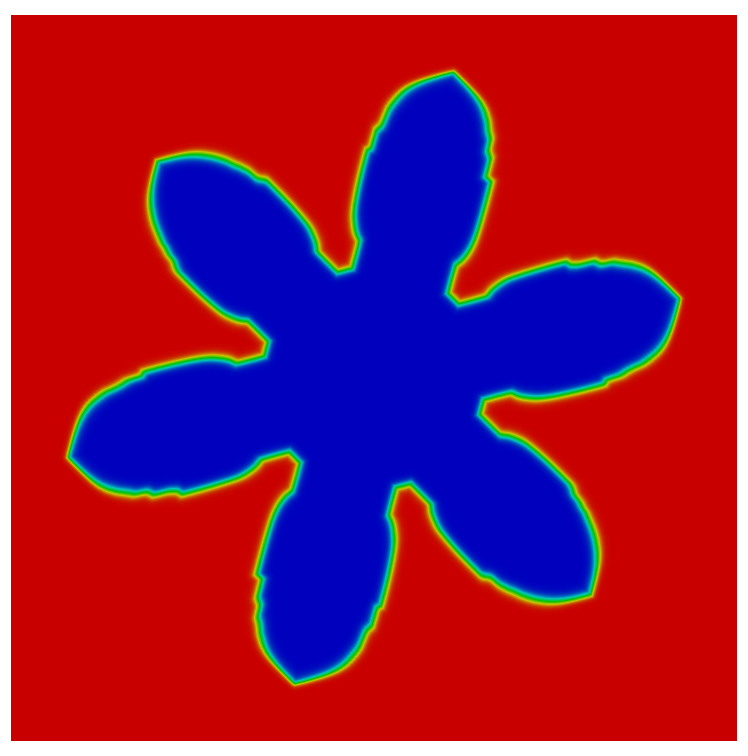}
\includegraphics[angle=-0,width=0.19\textwidth]{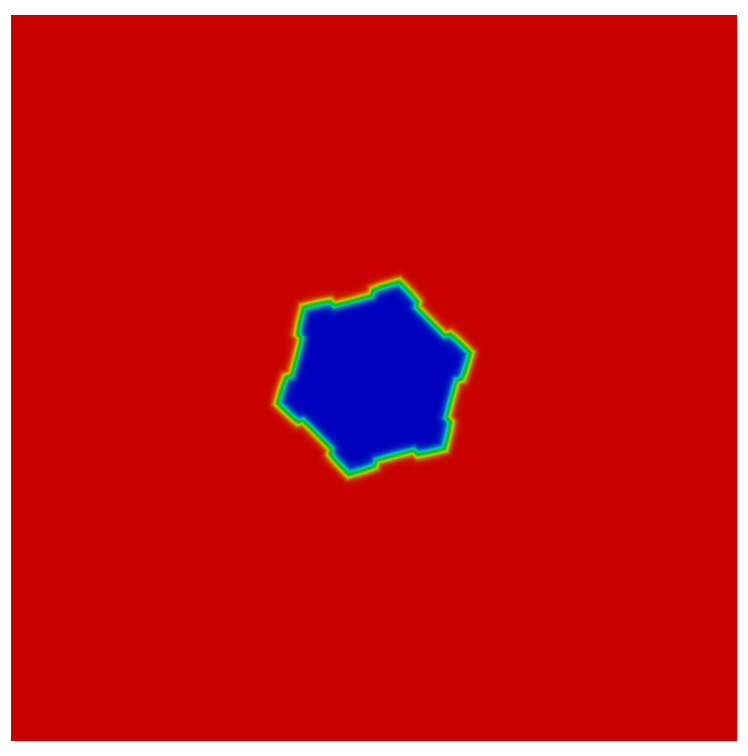}
\includegraphics[angle=-0,width=0.19\textwidth]{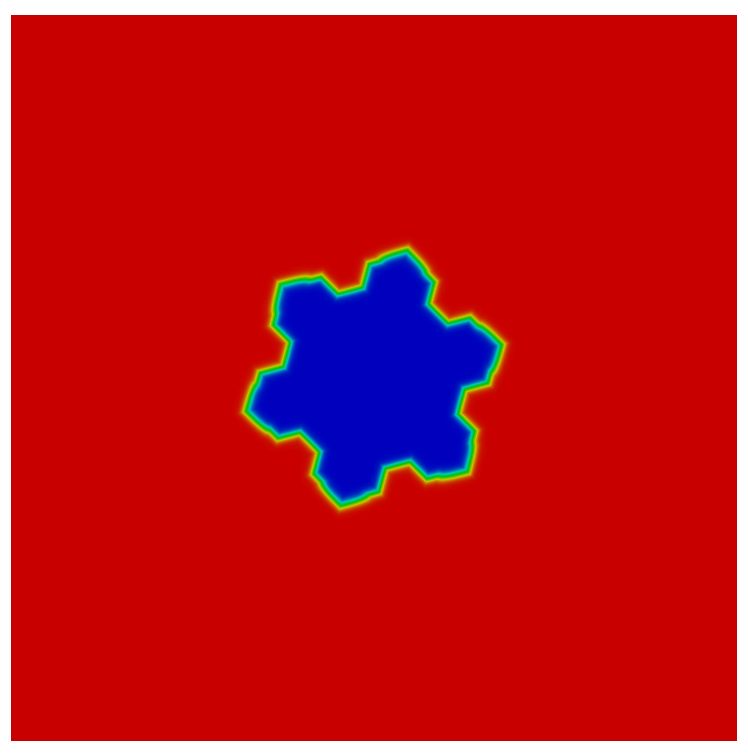}
\includegraphics[angle=-0,width=0.19\textwidth]{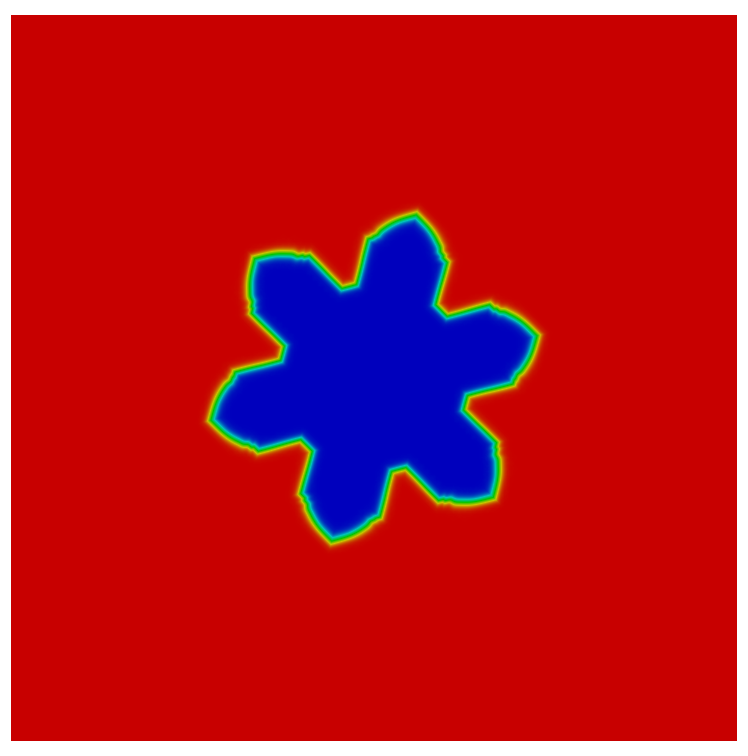}
\includegraphics[angle=-0,width=0.19\textwidth]{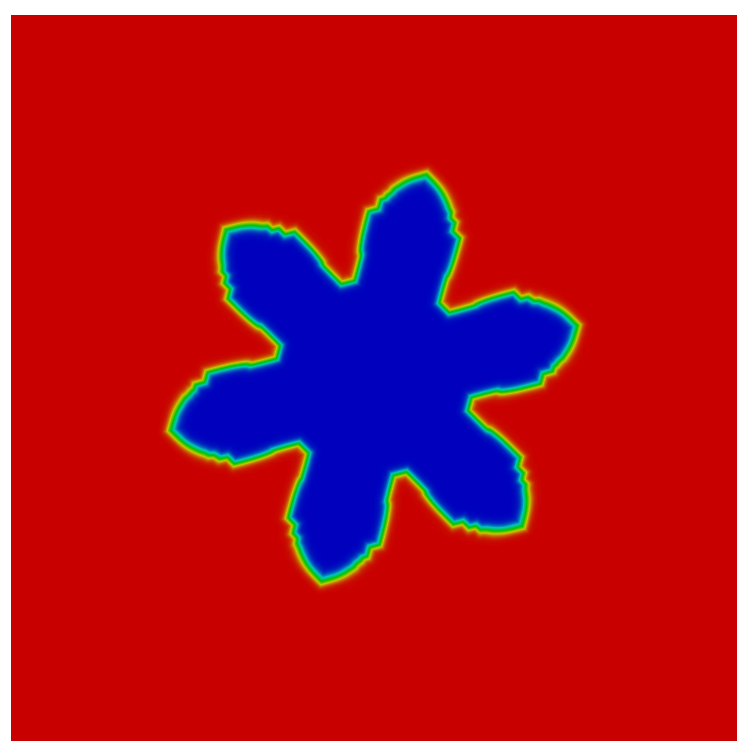}
\includegraphics[angle=-0,width=0.19\textwidth]{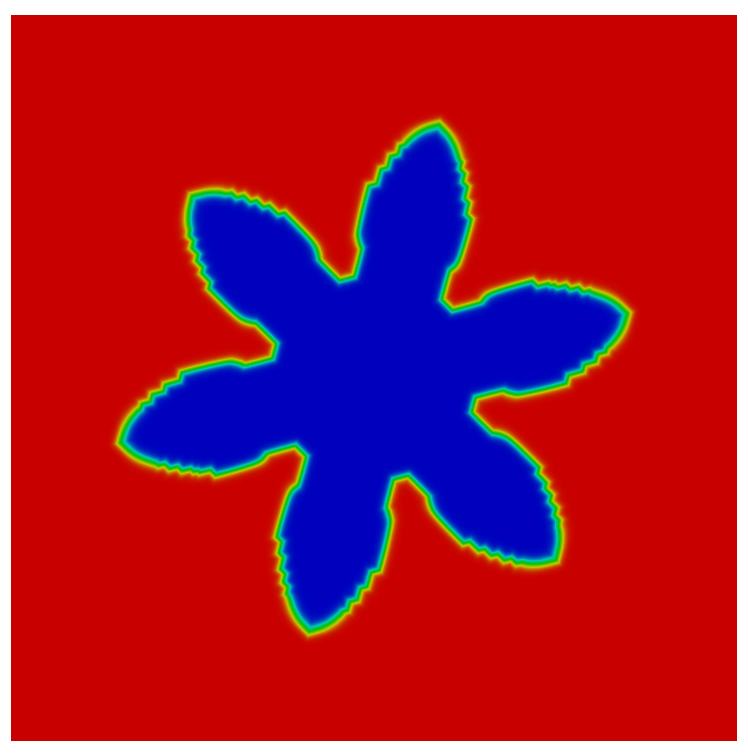}
\fi
\caption{(\PFqiii\ with the implicit time discretization from (\ref{eq:impl}), 
$\epsilon^{-1} = 32\,\pi$, {\sc ani$_2$}, 
$\vartheta=1$, $\alpha=5\times10^{-4}$, $\rho = 0.01$, $\uD = -\frac12$, 
$\Omega=(-2,2)^2$)
Snapshots of the solution at times $t=0.3,\,0.4,\,0.5,\,0.6,\,0.7$. 
From top to bottom $\tau = 10^{-k}$, $k = 2 \to 4$.
[These computations took $43$ seconds, $23$ minutes and $119$ minutes, 
respectively.]
}
\label{fig:qvStefan_32pii}
\end{figure}

When we repeat the simulations shown in Figures~\ref{fig:Stefanii_16pi} and
\ref{fig:Stefanii_32pi} for the scheme \PFqiii, then in the run for
$\epsilon^{-1} = 16\,\pi$ large mushy interfacial regions appear, which quickly
reach the boundary $\partial\Omega$. 
Hence we only present the simulations for
$\epsilon^{-1} = 32\,\pi$, see Figure~\ref{fig:qvStefan_32pi},
where we observe similar, but qualitatively quite different, 
results to the ones shown in Figure~\ref{fig:Stefanii_32pi} for the
scheme \PFii. In particular, the side arms in Figure~\ref{fig:qvStefan_32pi}
appear to be thinner than in Figure~\ref{fig:Stefanii_32pi}, and the
convergence as $\tau$ gets smaller appears to be slower.
We also repeat the computation from Figure~\ref{fig:qvStefan_32pi} for the
implicit time discretization from (\ref{eq:impl}), see 
Figure~\ref{fig:qvStefan_32pii}. 
We observe that, in contrast to the conclusions that could be drawn from the
isotropic experiments in Section~\ref{sec:4}, for the strongly anisotropic 
situation treated here there does not seem to be an advantage in using the 
implicit
time discretization from (\ref{eq:impl}) over the standard semi-implicit
discretization (\ref{eq:semi1},b) from \cite{vch}.

\begin{figure}
\center
\ifpdf
\includegraphics[angle=-0,width=0.19\textwidth]{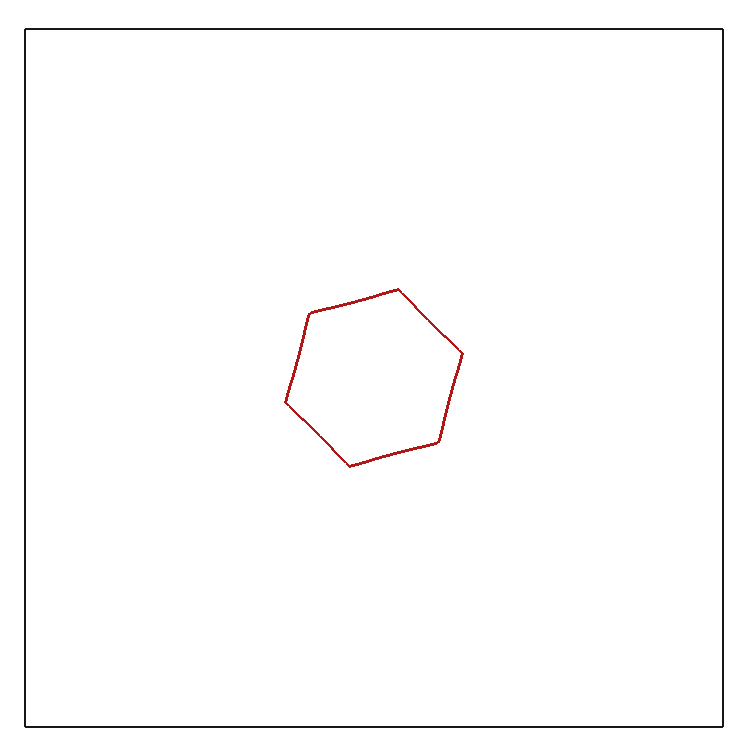}
\includegraphics[angle=-0,width=0.19\textwidth]{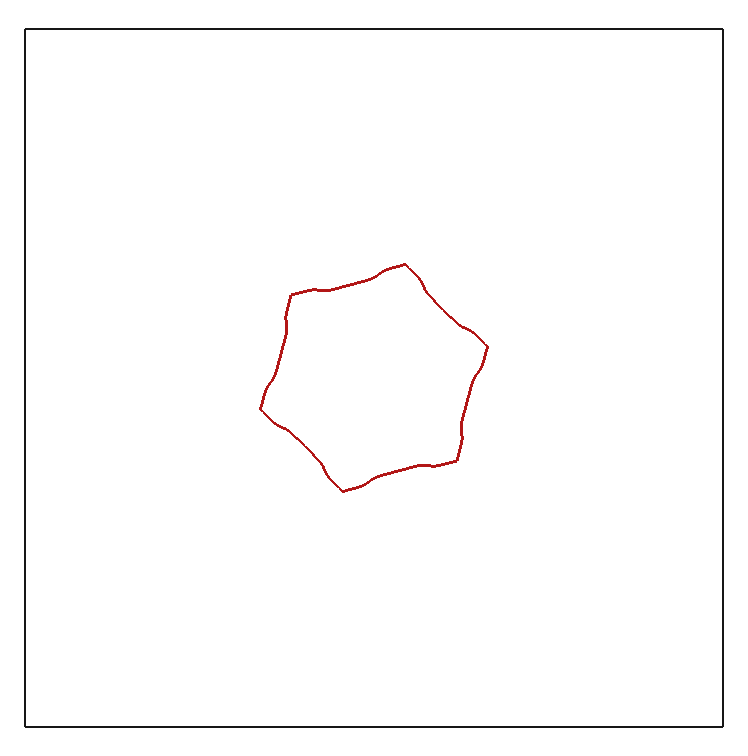}
\includegraphics[angle=-0,width=0.19\textwidth]{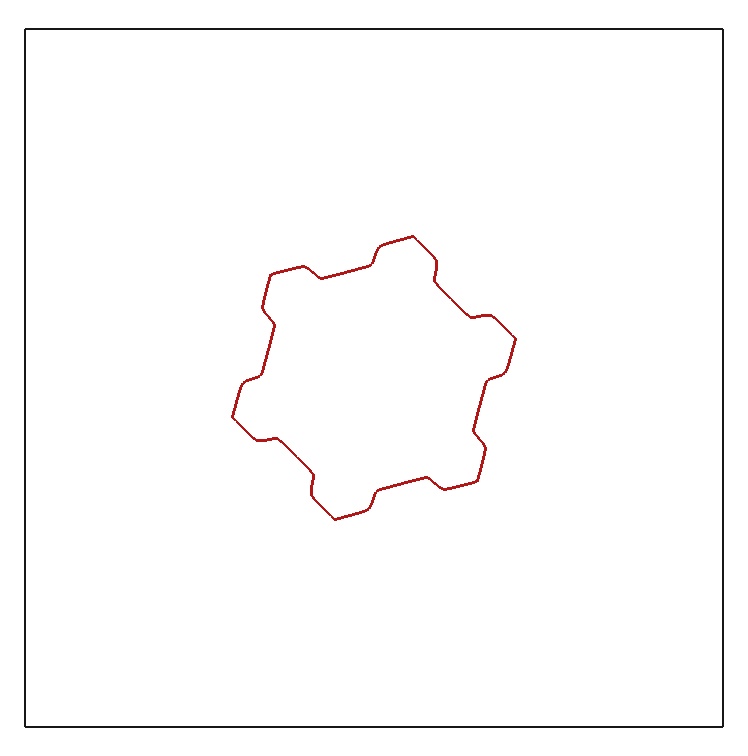}
\includegraphics[angle=-0,width=0.19\textwidth]{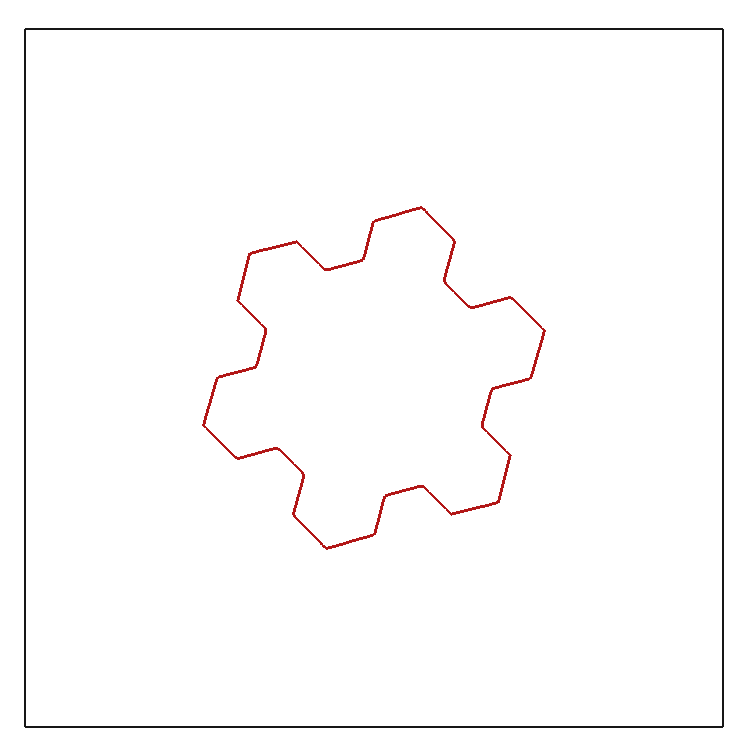}
\includegraphics[angle=-0,width=0.19\textwidth]{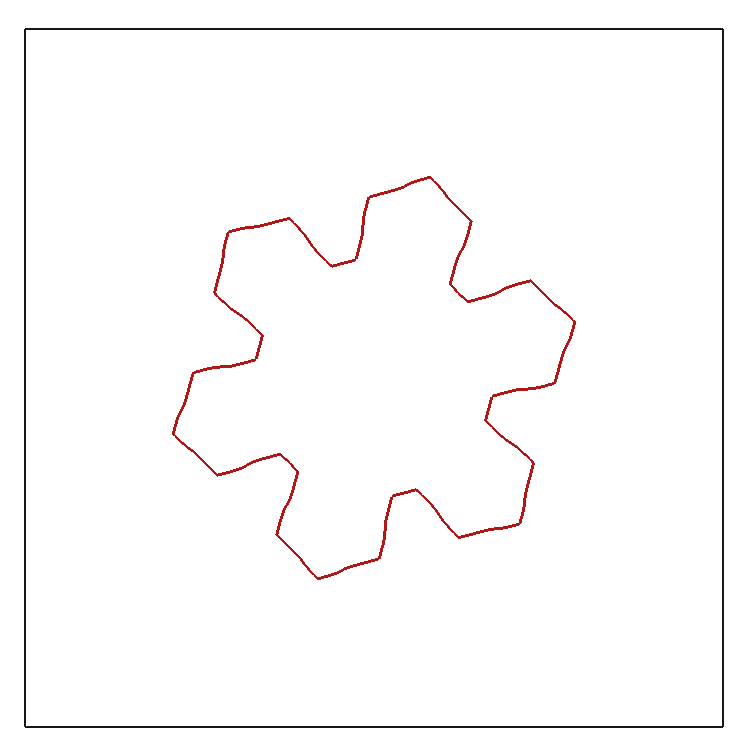}
\includegraphics[angle=-0,width=0.19\textwidth]{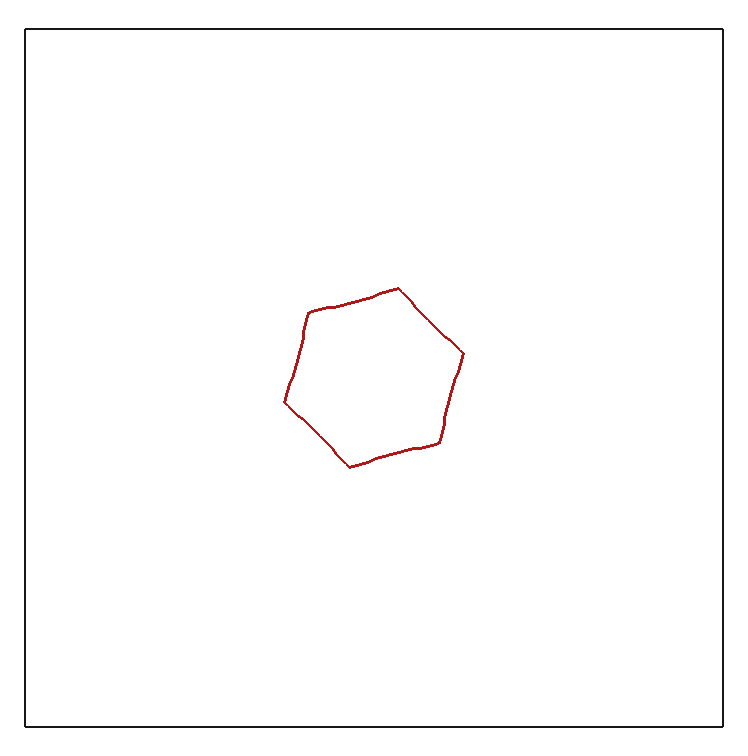}
\includegraphics[angle=-0,width=0.19\textwidth]{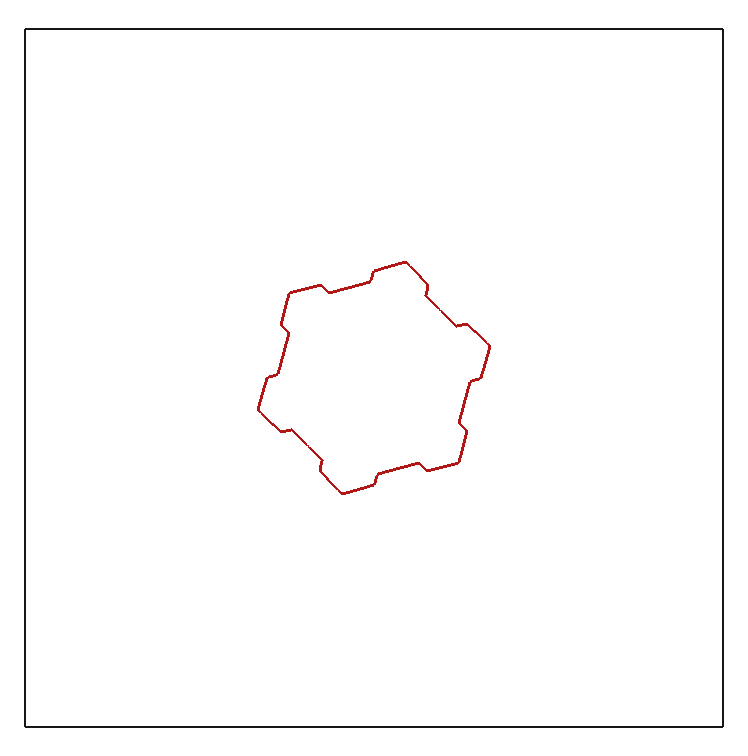}
\includegraphics[angle=-0,width=0.19\textwidth]{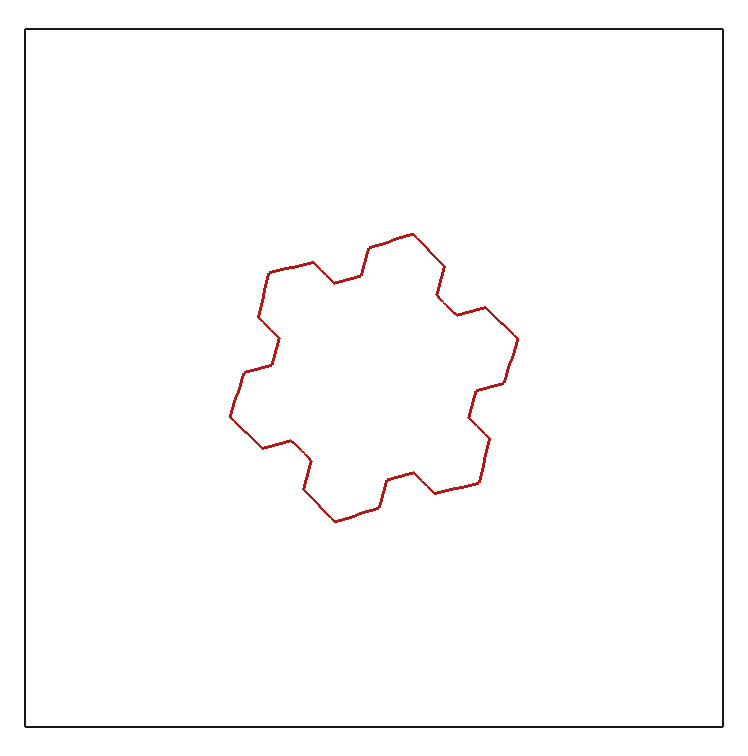}
\includegraphics[angle=-0,width=0.19\textwidth]{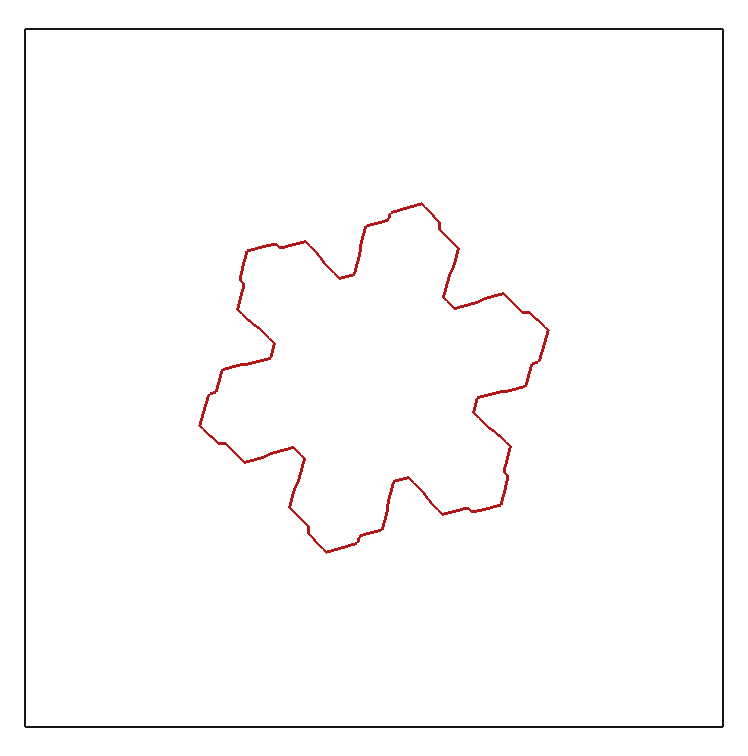}
\includegraphics[angle=-0,width=0.19\textwidth]{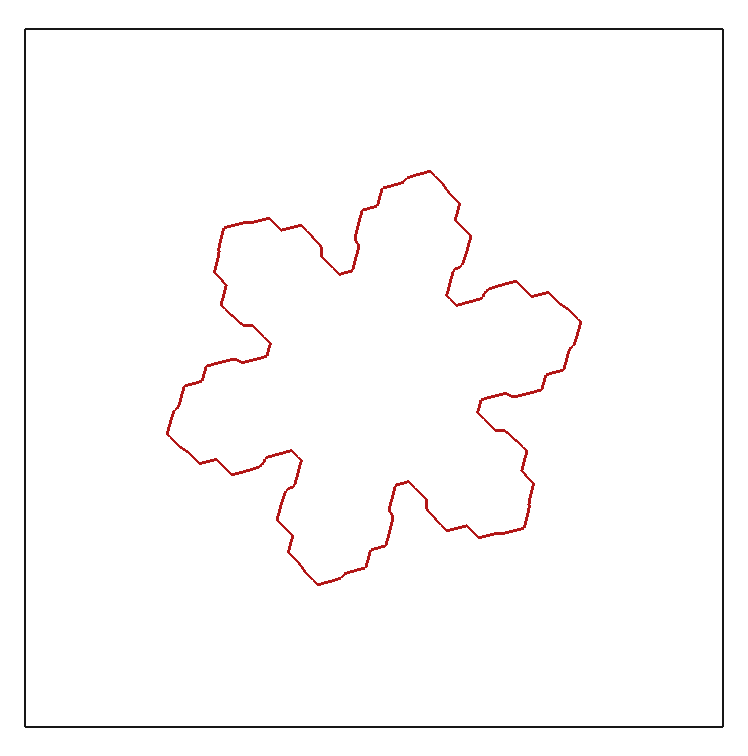}
\includegraphics[angle=-0,width=0.19\textwidth]{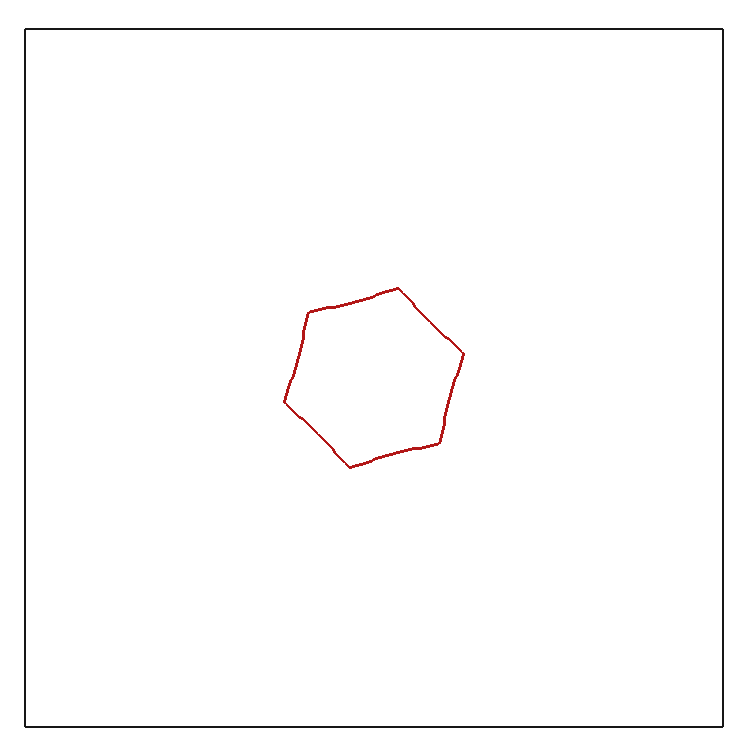}
\includegraphics[angle=-0,width=0.19\textwidth]{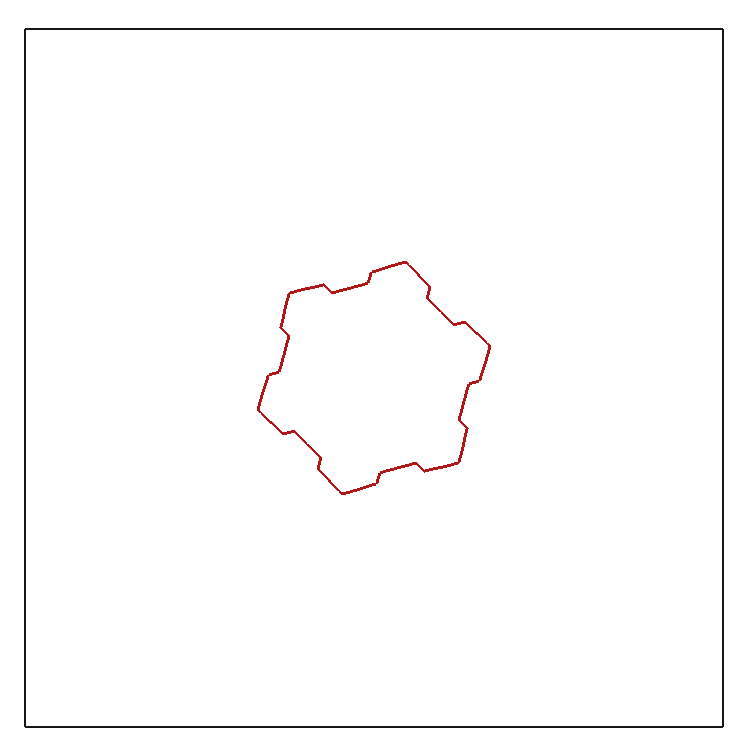}
\includegraphics[angle=-0,width=0.19\textwidth]{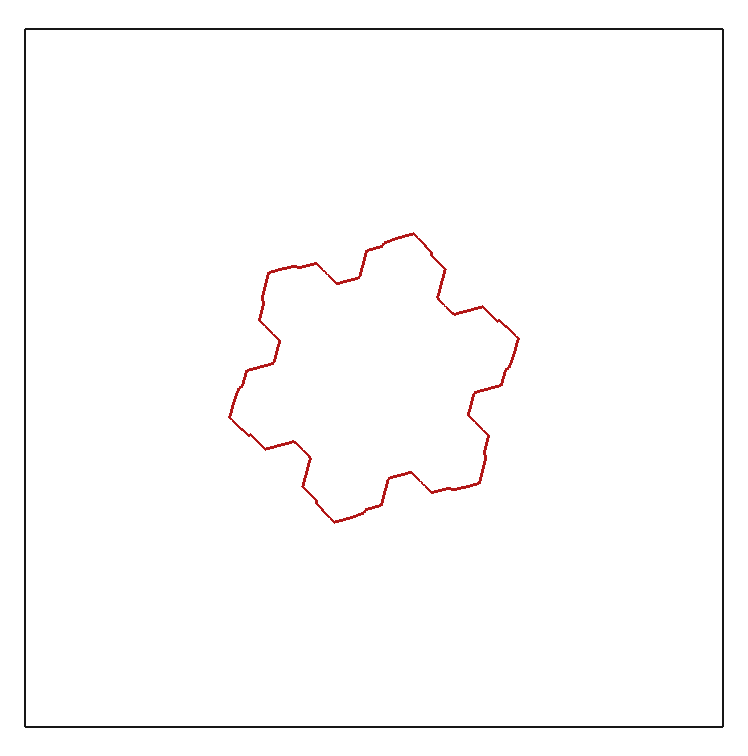}
\includegraphics[angle=-0,width=0.19\textwidth]{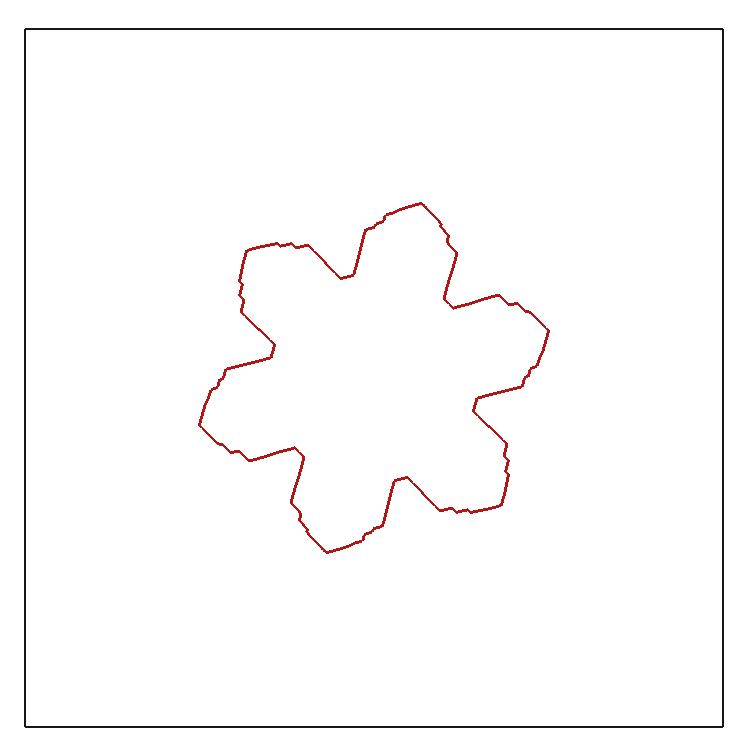}
\includegraphics[angle=-0,width=0.19\textwidth]{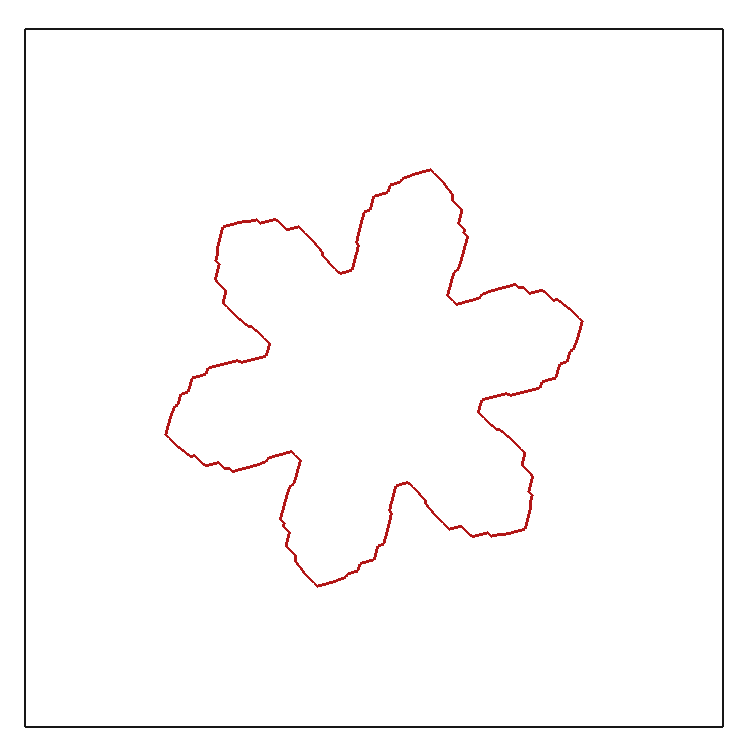}
\fi
\caption{(\PFEM, {\sc ani$_2$}, 
$\vartheta=1$, $\alpha=5\times10^{-4}$, $\rho = 0.01$, $\uD = -\frac12$, 
$\Omega=(-2,2)^2$)
Snapshots of the solution at times $t=0.3,\,0.4,\,0.5,\,0.6,\,0.7$. 
From top to bottom $\tau = 10^{-k}$, $k = 2 \to 4$.
[These computations took $1$, $9$ seconds and
$87$ seconds, respectively.]
}
\label{fig:Stefan}
\end{figure}

In addition, we present three simulations for the same physical problem for the
sharp interface approximation \PFEM\ in Figure~\ref{fig:Stefan}, where we fix
the spatial discretization parameters as
$h_\Gamma \approx h_f = \frac{\sqrt{2}}{64}$.
Here we observe once again that even for a very crude
time discretization, the evolution is captured remarkably well, and there is
very little variation in the numerical results from \PFEM\ when $\tau$ is
decreased. We also draw particular attention to the differing CPU times between
the sharp interface calculations in Figure~\ref{fig:Stefan} and the phase field
simulations in Figures~\ref{fig:Stefanii_32pi}--\ref{fig:qvStefan_32pii}.

\begin{figure}
\center
\ifpdf
\includegraphics[angle=-0,width=0.19\textwidth]{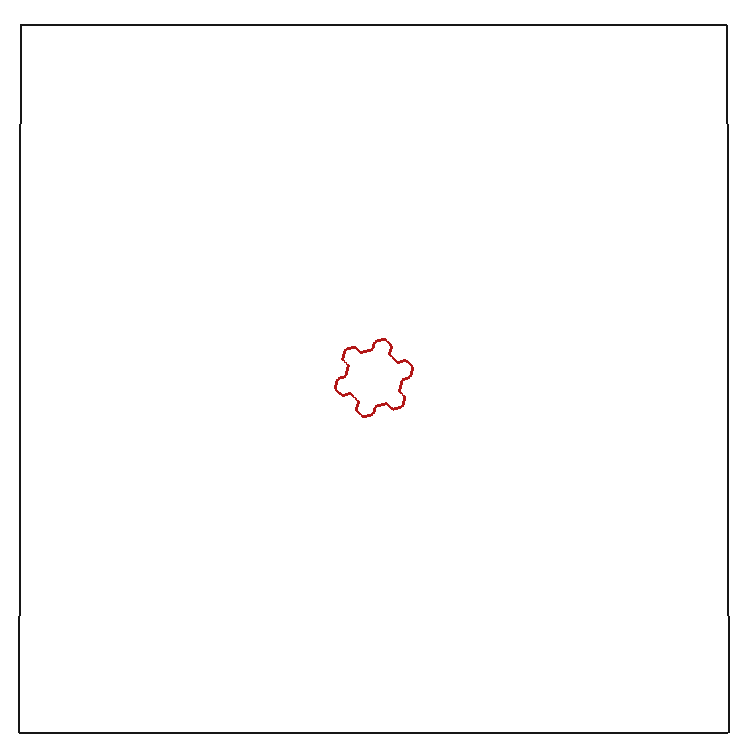}
\includegraphics[angle=-0,width=0.19\textwidth]{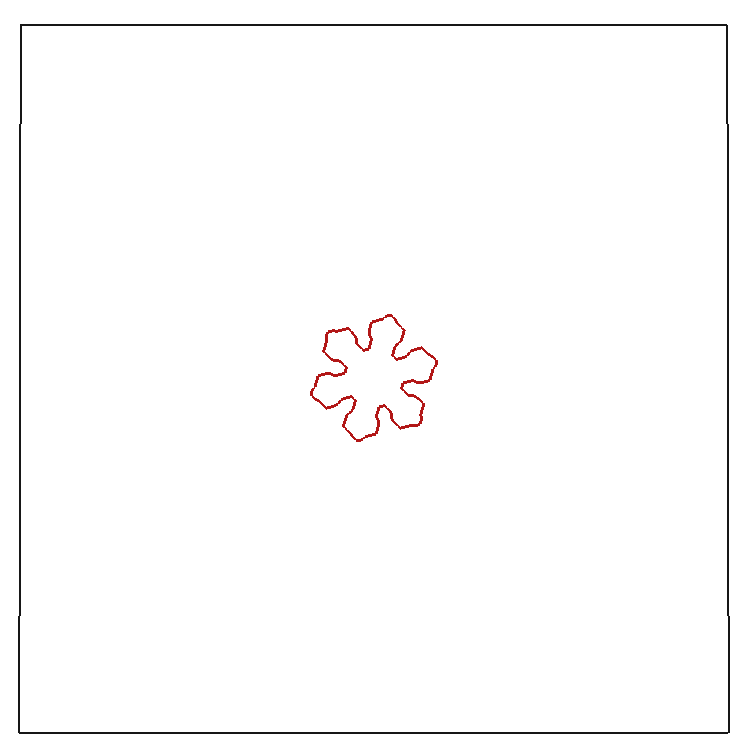}
\includegraphics[angle=-0,width=0.19\textwidth]{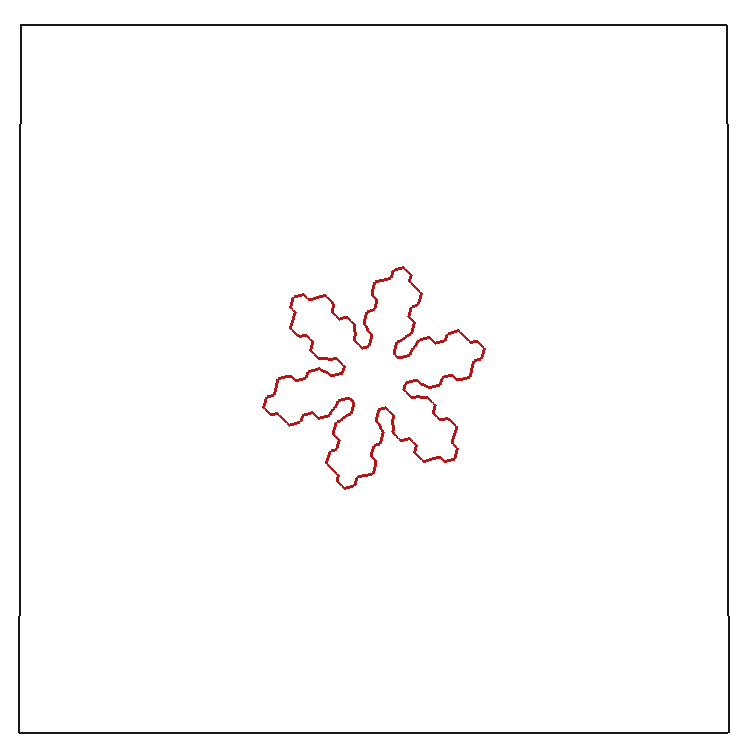}
\includegraphics[angle=-0,width=0.19\textwidth]{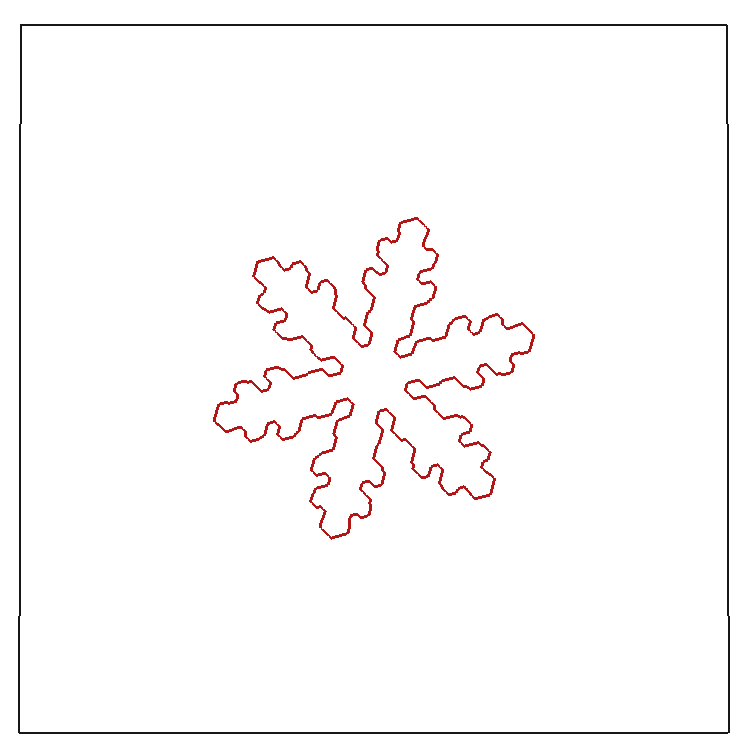}
\includegraphics[angle=-0,width=0.19\textwidth]{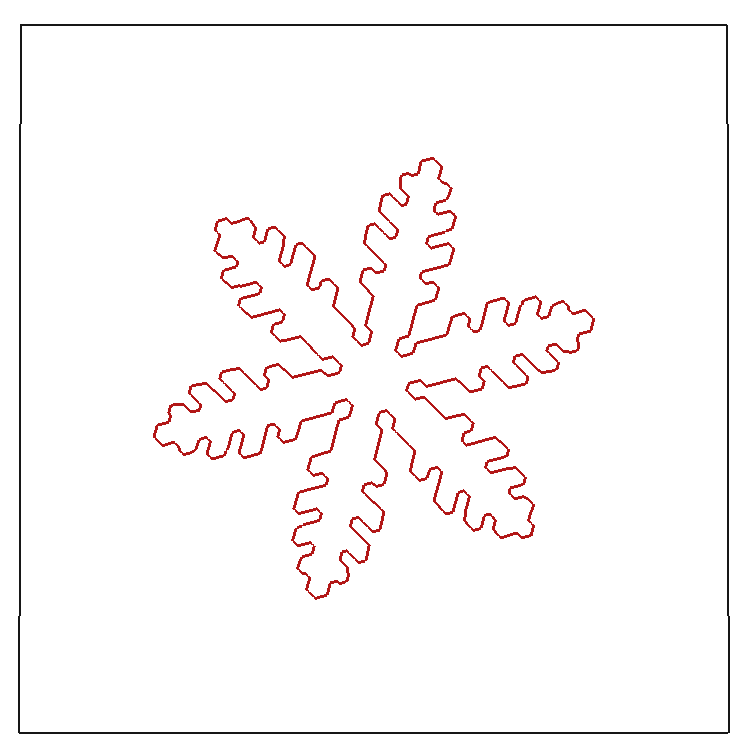}
\includegraphics[angle=-0,width=0.19\textwidth]{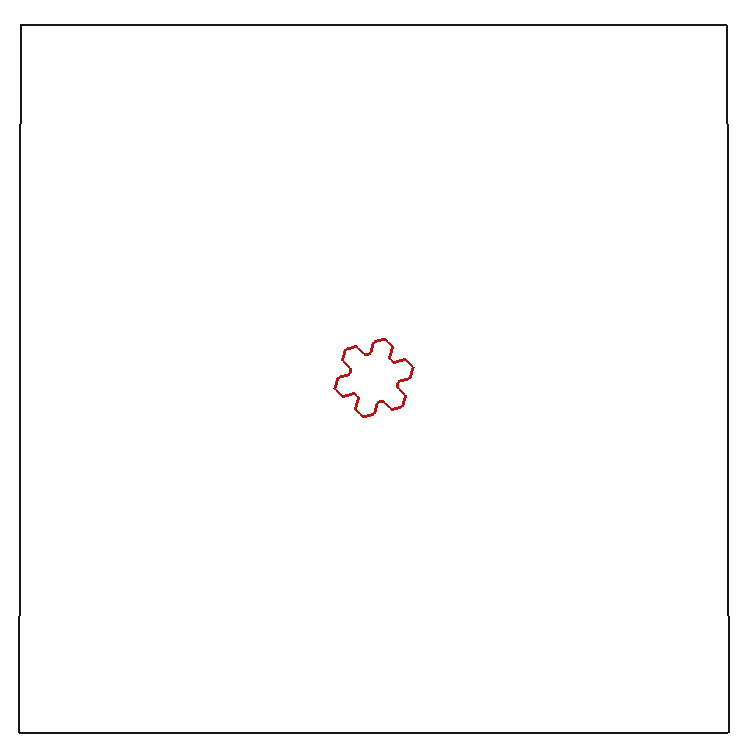}
\includegraphics[angle=-0,width=0.19\textwidth]{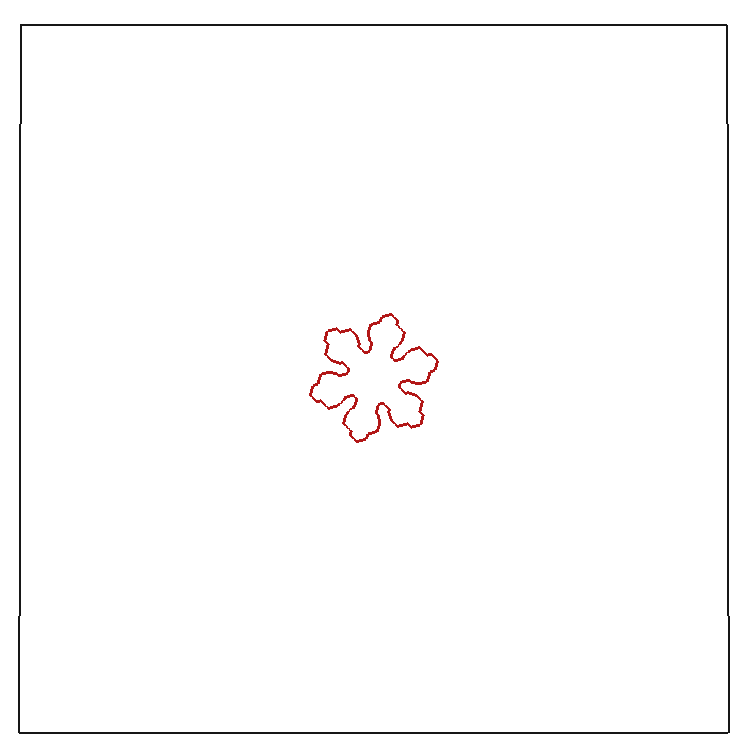}
\includegraphics[angle=-0,width=0.19\textwidth]{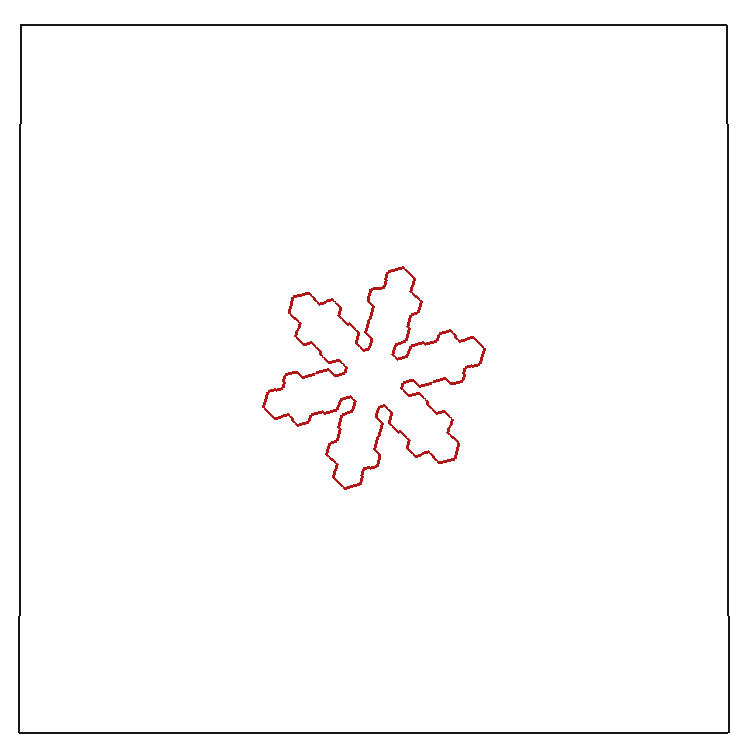}
\includegraphics[angle=-0,width=0.19\textwidth]{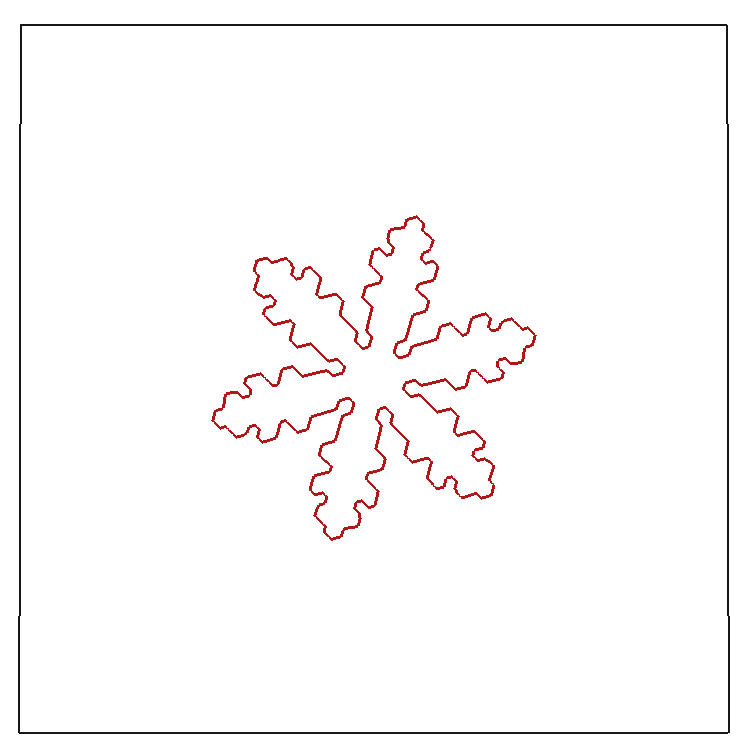}
\includegraphics[angle=-0,width=0.19\textwidth]{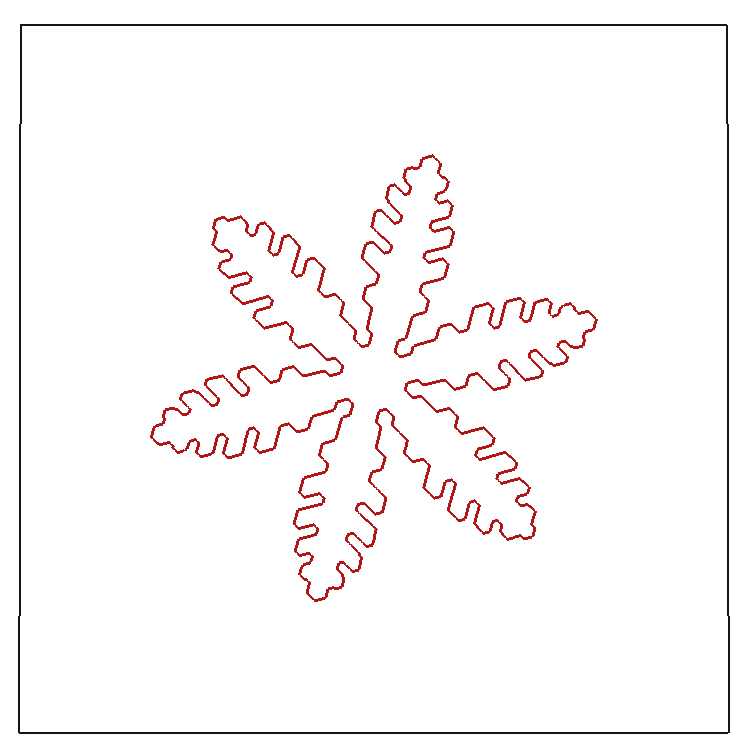}
\includegraphics[angle=-0,width=0.19\textwidth]{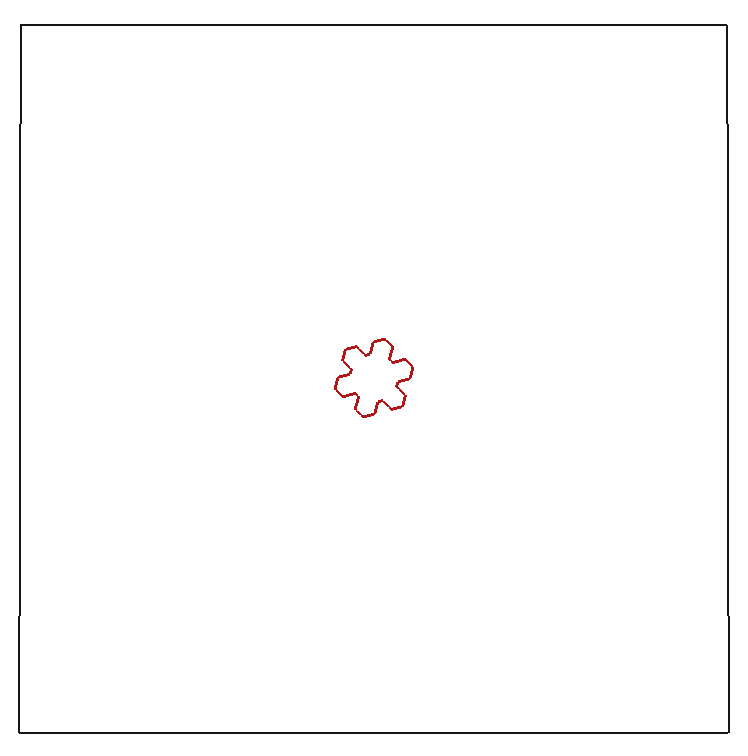}
\includegraphics[angle=-0,width=0.19\textwidth]{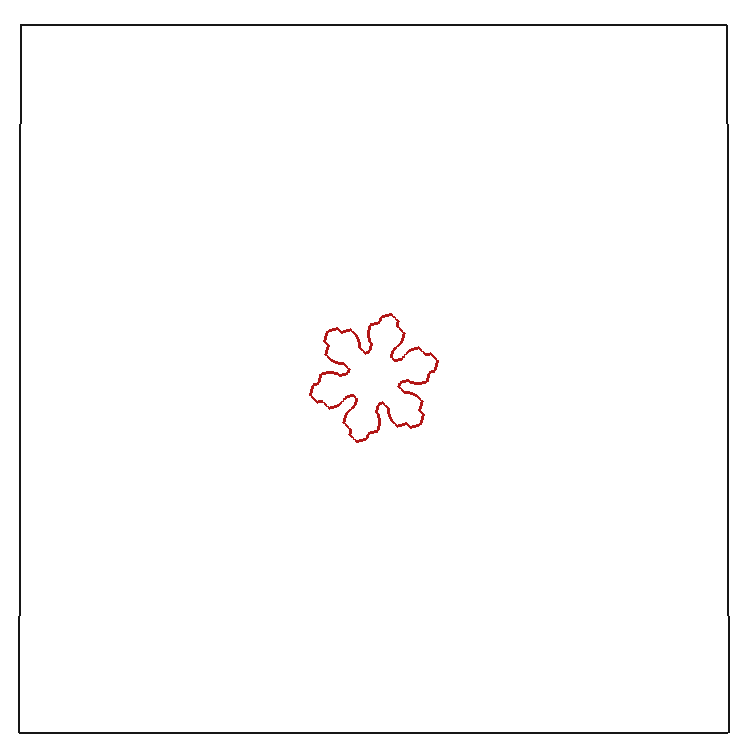}
\includegraphics[angle=-0,width=0.19\textwidth]{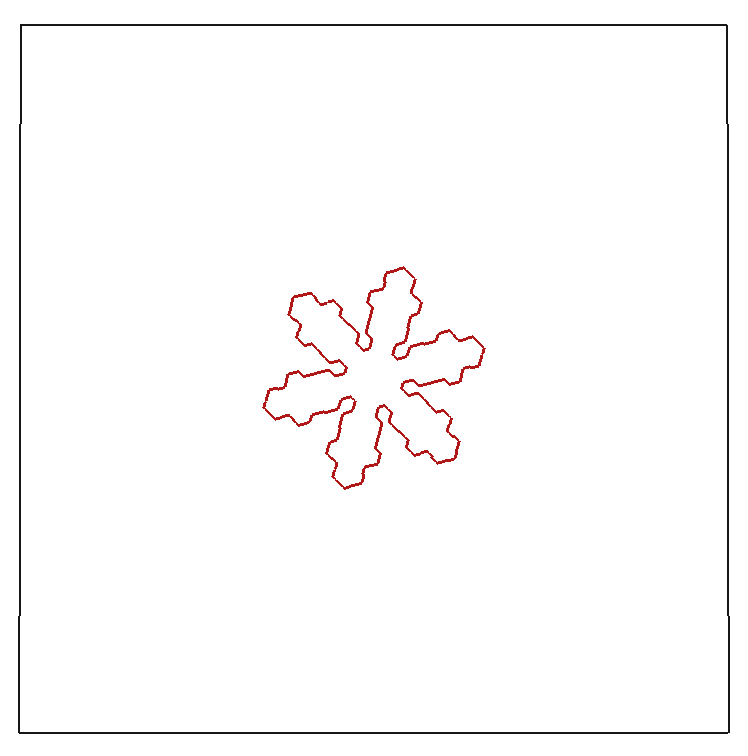}
\includegraphics[angle=-0,width=0.19\textwidth]{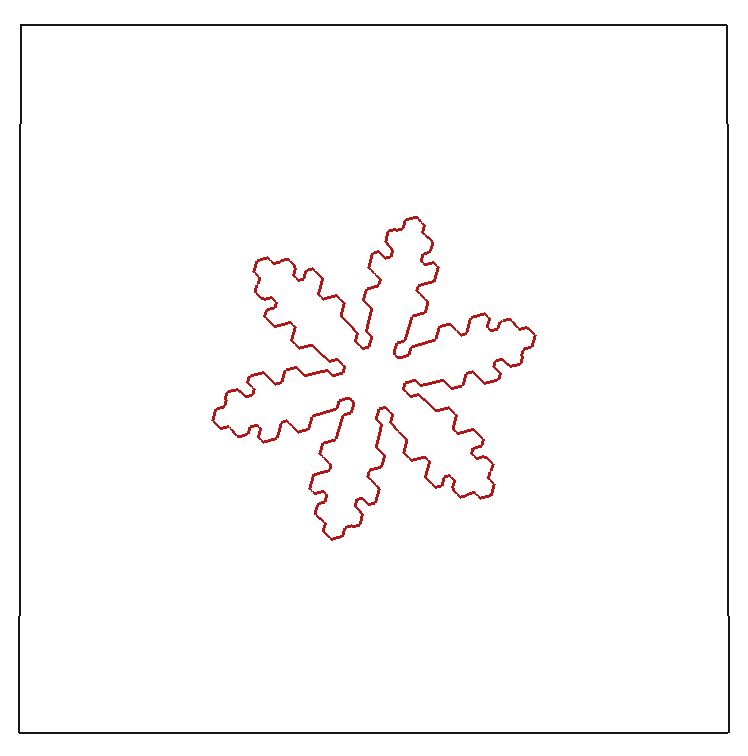}
\includegraphics[angle=-0,width=0.19\textwidth]{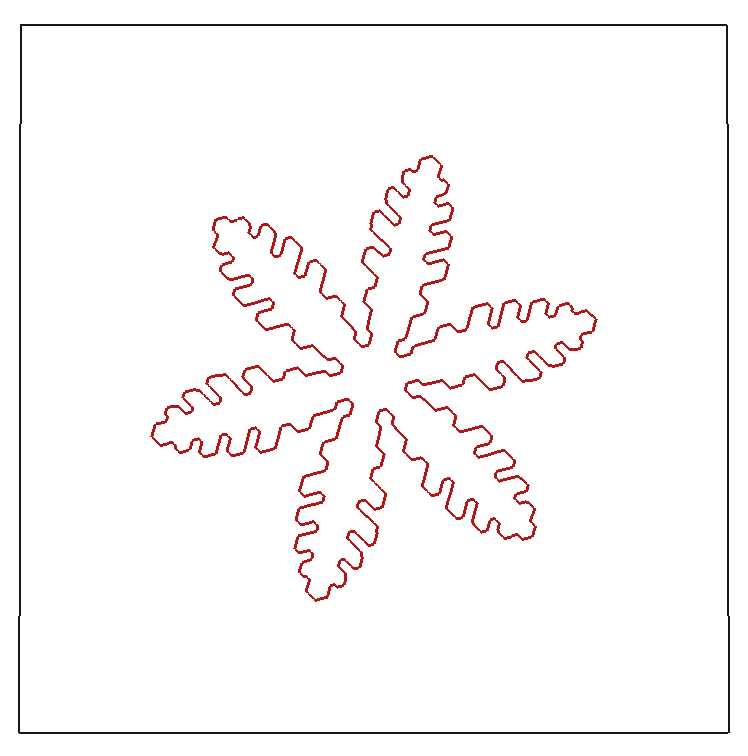}
\fi
\caption{(\PFEM, {\sc ani$_2$}, 
$\vartheta=0$, $\alpha=10^{-5}$, $\rho = 1.42\times10^{-3}$, $\uD = -0.04$, 
$\Omega=(-4,4)^2$)
Snapshots of the solution at times $t=5,\,10,\,20,\,30,\,40$. 
From top to bottom $\tau = 10^{-k}$, $k = 1 \to 3$.
[These computations took $6$, $124$ and $781$ minutes, respectively.] 
}
\label{fig:jcgtau}
\end{figure}

In the remainder of this section we consider
two simulations for the scheme \PFEM, which on present
computer hardware are virtually impossible to repeat to a desirable accuracy
with the phase field method. The first experiment is with the physical 
parameters from \cite[Fig.~7]{jcg},
and so is for the one-sided quasi-stationary problem (\ref{eq:1a}--e) with
$\vartheta = \conduct_- = 0$ and with $\gamma$ as in (\ref{eq:hexgamma2d}). 
The remaining parameters are chosen as
$\alpha = 10^{-5}$, $\rho= 1.42\times 10^{-3}$ and $\uD = -0.04$
on the boundary
$\partial_D\Omega=\partial\Omega$ of $\Omega=(-4,4)^2$. 
The radius of the initially circular seed is chosen as $0.05$.
See Figure~\ref{fig:jcgtau} for the results for different choices of the time 
step sizes $\tau$, and with
$h_\Gamma \approx h_f = \frac{\sqrt{2}}{128}$ fixed.
We see that, as before, there is hardly any variation in the
numerical results obtained from the three simulations with different values of
$\tau$ for \PFEM. Moreover, we note that due the choice of the physical
parameters much finer side branches appear in Figure~\ref{fig:jcgtau} compared
to the simulations in Figure~\ref{fig:Stefan}. To precisely capture these small
structures within a phase field computation would require very small values for
the interfacial parameter $\epsilon$, as well as correspondingly small
discretization parameters $h_f$ and $\tau$; recall (\ref{eq:hf}), 
and (\ref{eq:NV97a}). 
Taken together this means that we are currently unable to
present phase field computations for an evolution as shown in
Figure~\ref{fig:jcgtau}.

\begin{figure}
\center
\ifpdf
\includegraphics[angle=-0,width=0.24\textwidth]{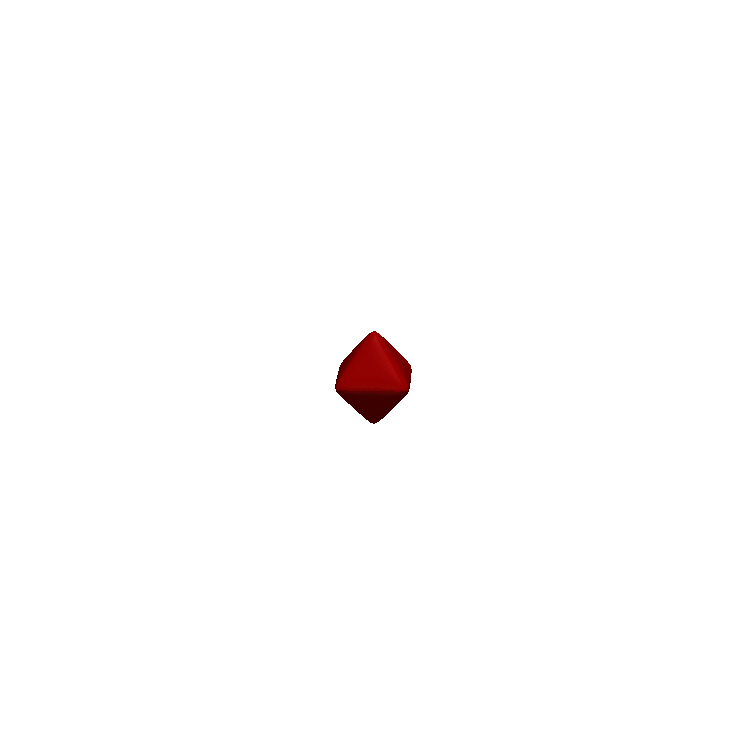}
\includegraphics[angle=-0,width=0.24\textwidth]{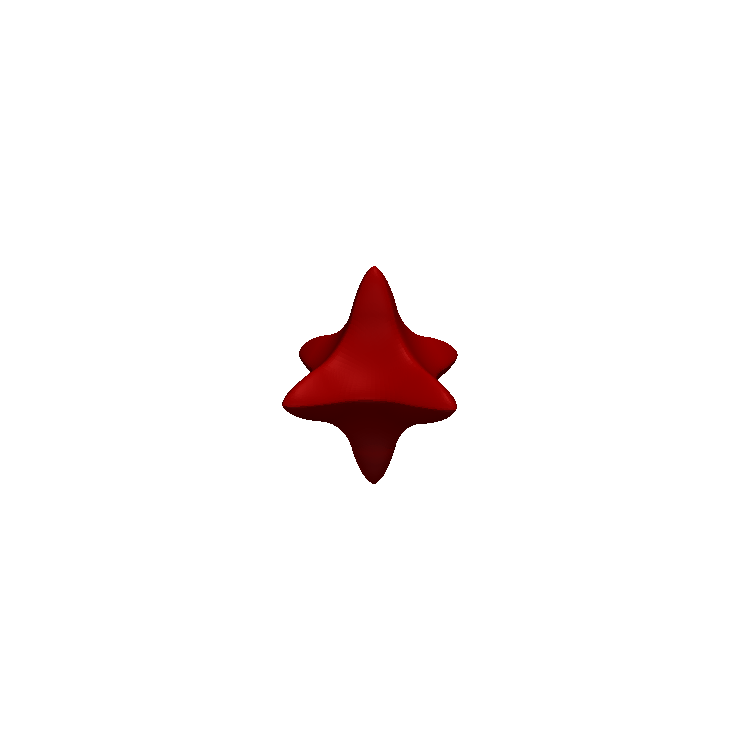}
\includegraphics[angle=-0,width=0.24\textwidth]{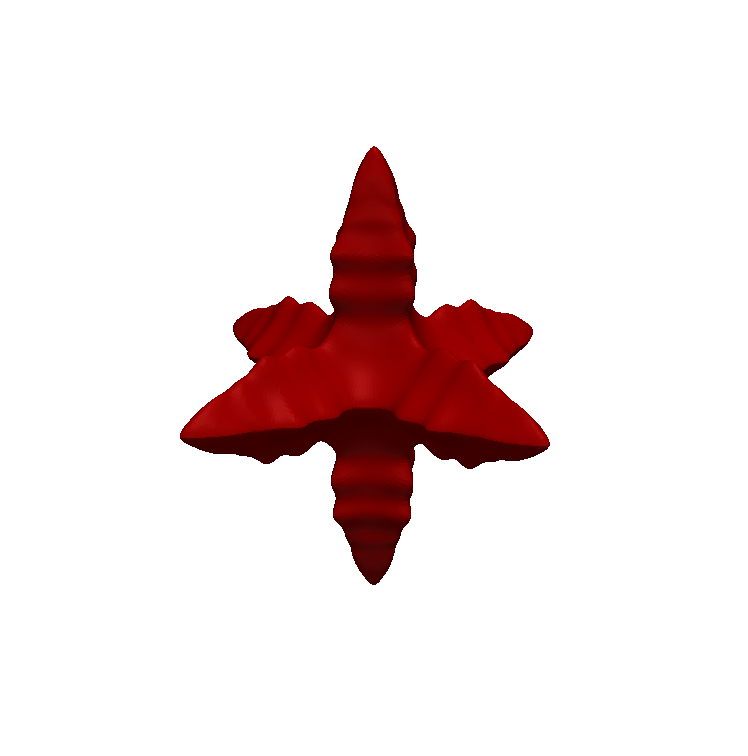}
\includegraphics[angle=-0,width=0.24\textwidth]{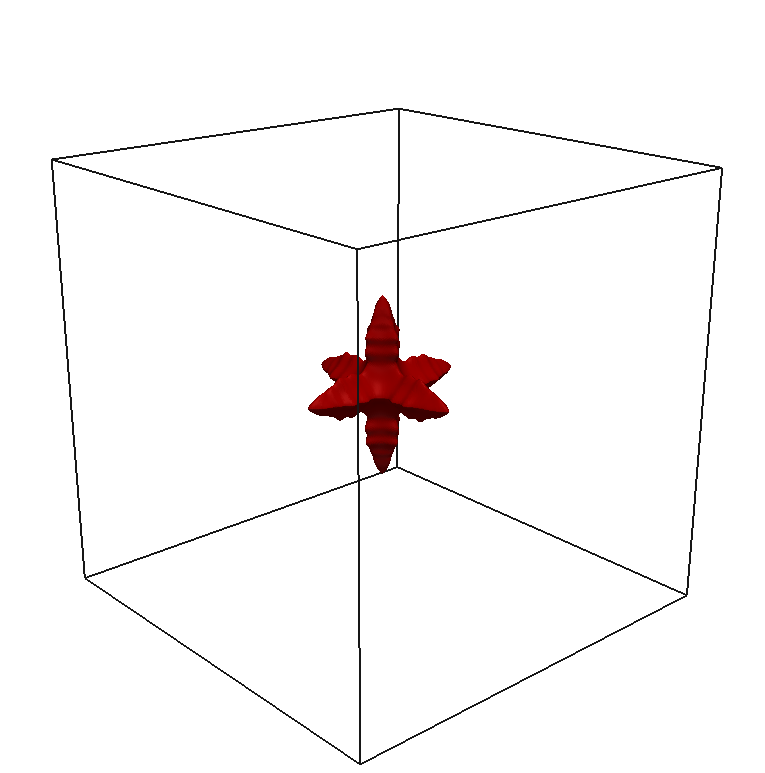}
\includegraphics[angle=-0,width=0.24\textwidth]{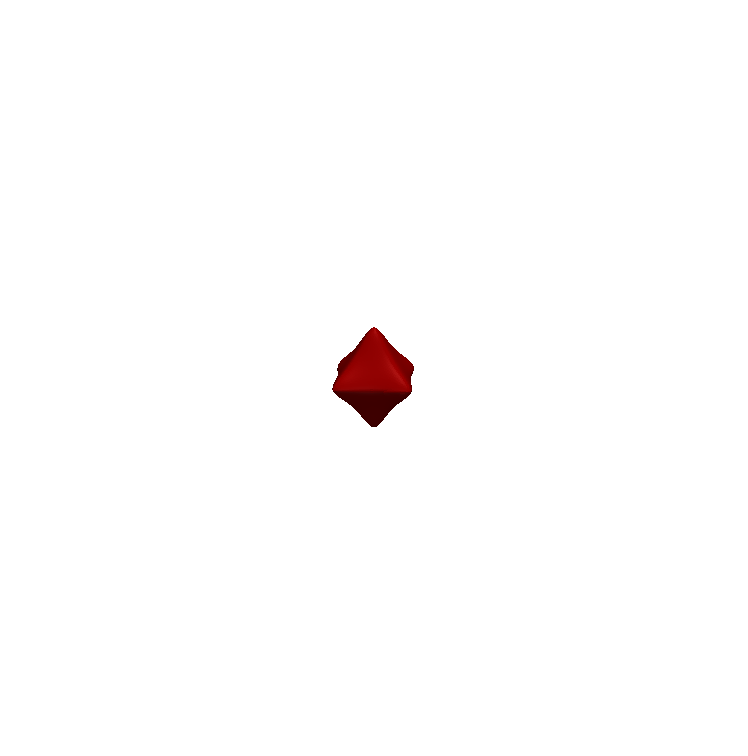}
\includegraphics[angle=-0,width=0.24\textwidth]{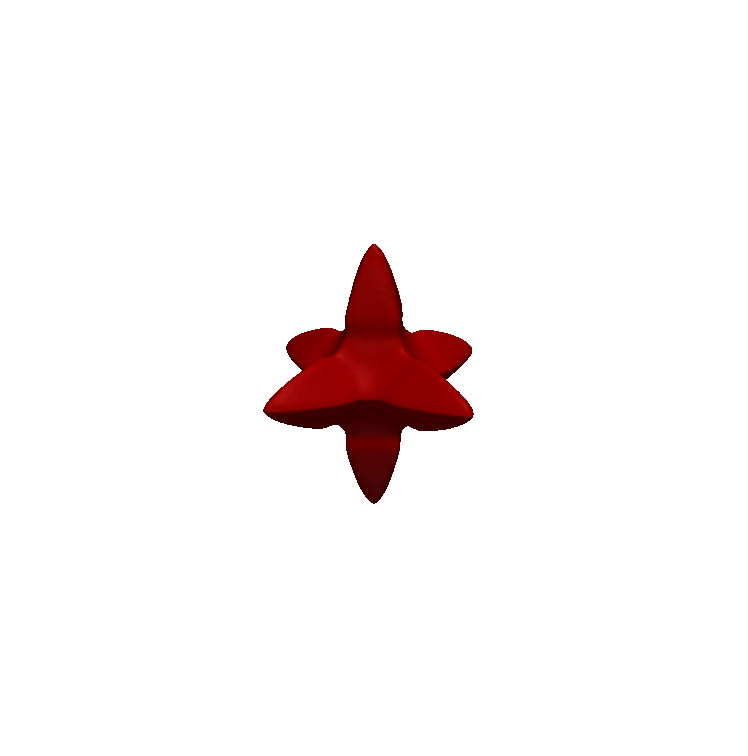}
\includegraphics[angle=-0,width=0.24\textwidth]{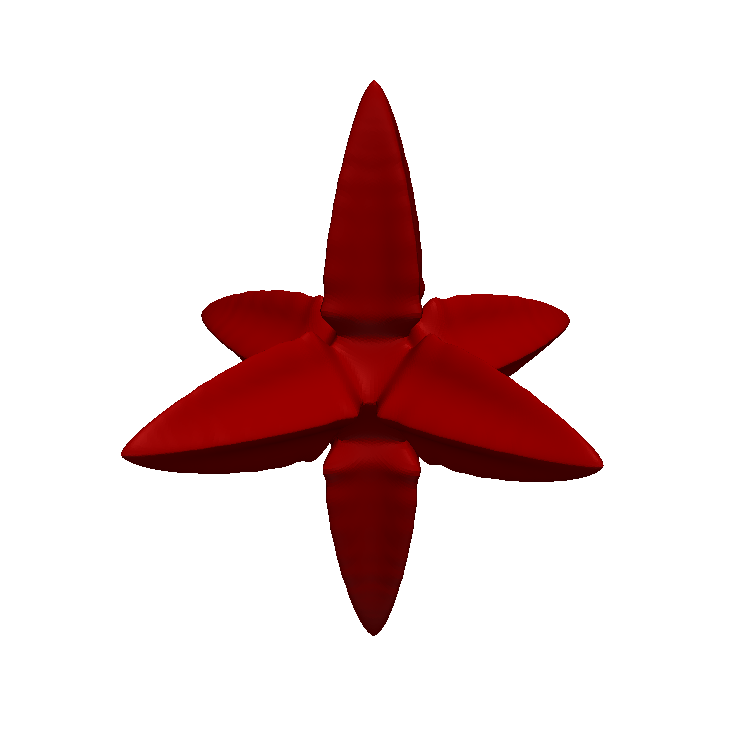}
\includegraphics[angle=-0,width=0.24\textwidth]{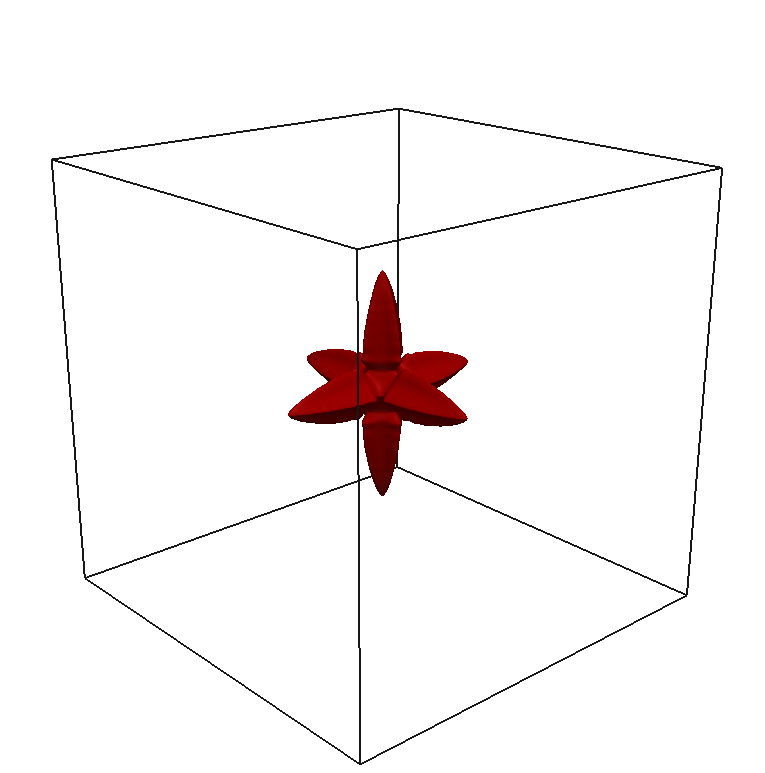}
\includegraphics[angle=-0,width=0.24\textwidth]{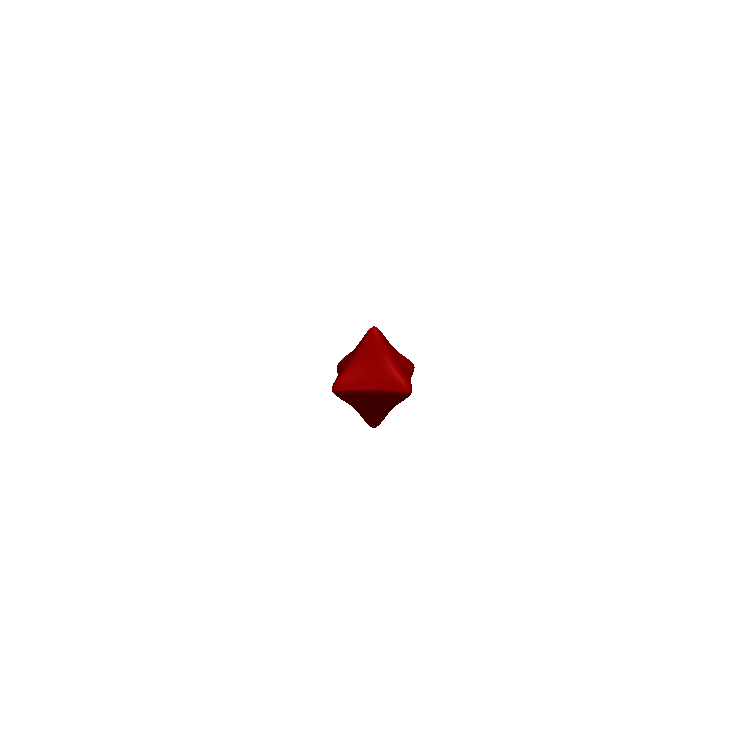}
\includegraphics[angle=-0,width=0.24\textwidth]{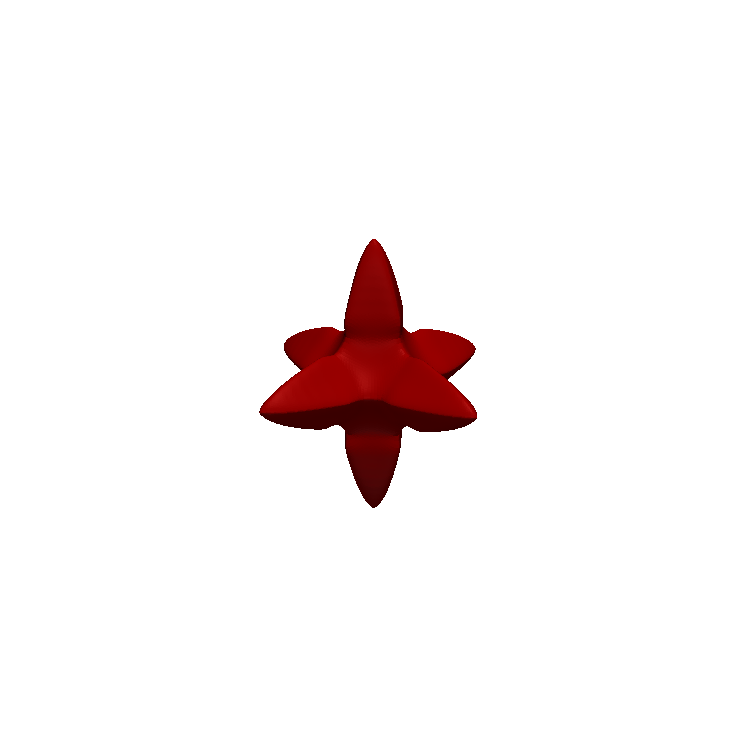}
\includegraphics[angle=-0,width=0.24\textwidth]{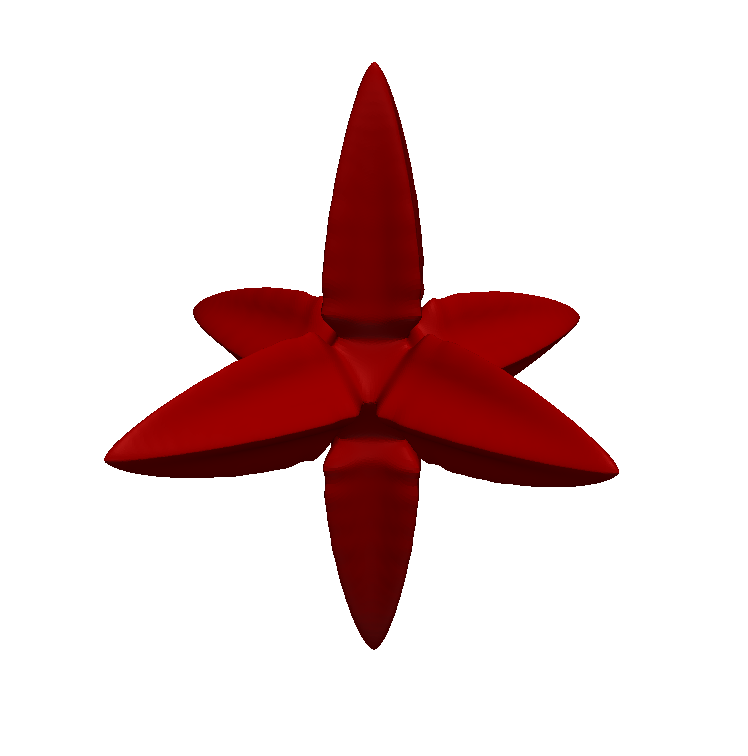}
\includegraphics[angle=-0,width=0.24\textwidth]{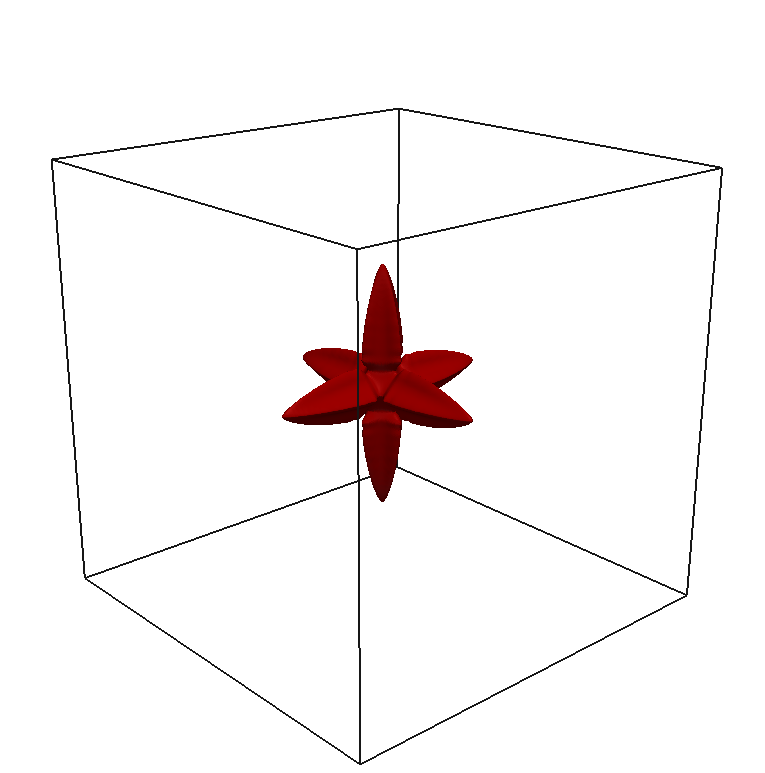}
\fi
\caption{(\PFEM, {\sc ani$_3$}, 
$\vartheta=1$, $\alpha=10^{-3}$, $\rho = 0.01$, $\uD = -\frac12$, 
$\Omega=(-4,4)^3$)
Snapshots of the solution at times $t=0.1,\,0.2,\,0.3$, as well as the
solution at time $T=0.3$ within $\overline\Omega$. From top
to bottom $\tau=10^{-2},\,10^{-3},\,2\times10^{-4}$.
[These computations took $2$ hours, $20$ hours and $50$ hours, respectively.]
}
\label{fig:dendritic}
\end{figure}

\begin{figure}
\center
\ifpdf
\includegraphics[angle=-0,width=0.35\textwidth]{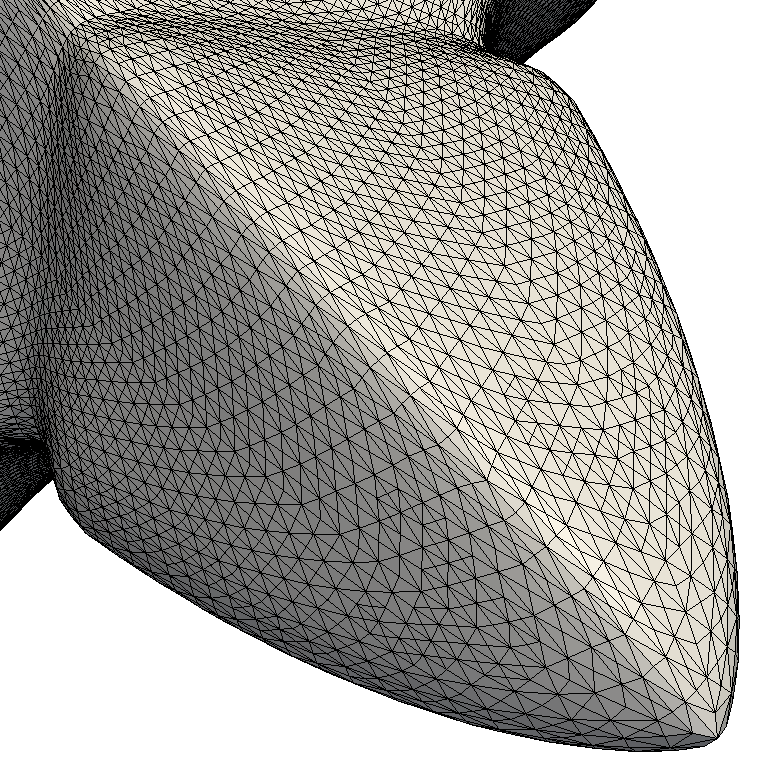}
\qquad
\includegraphics[angle=-0,width=0.35\textwidth]{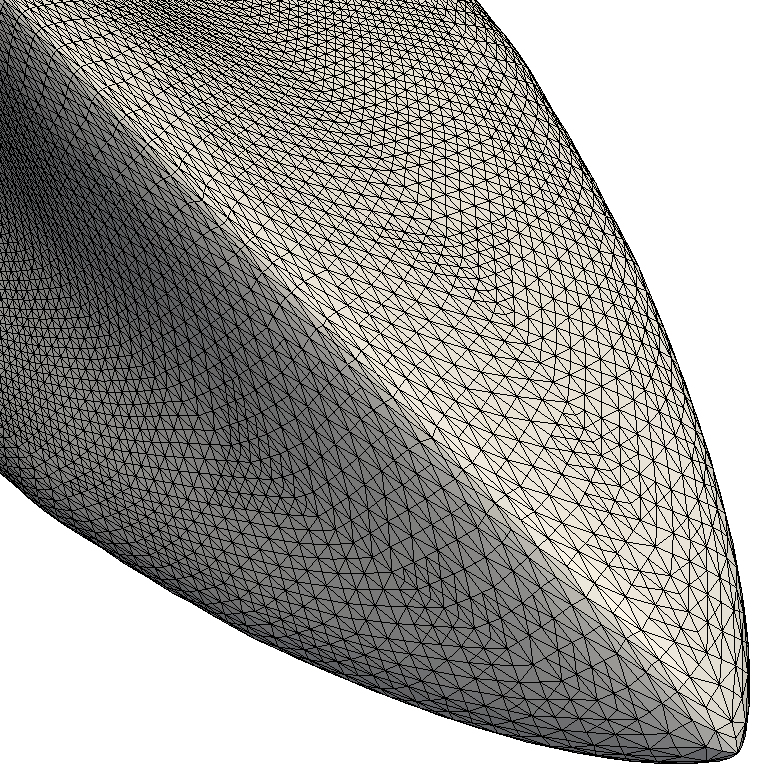}
\\[2mm]
\includegraphics[angle=-0,width=0.35\textwidth]{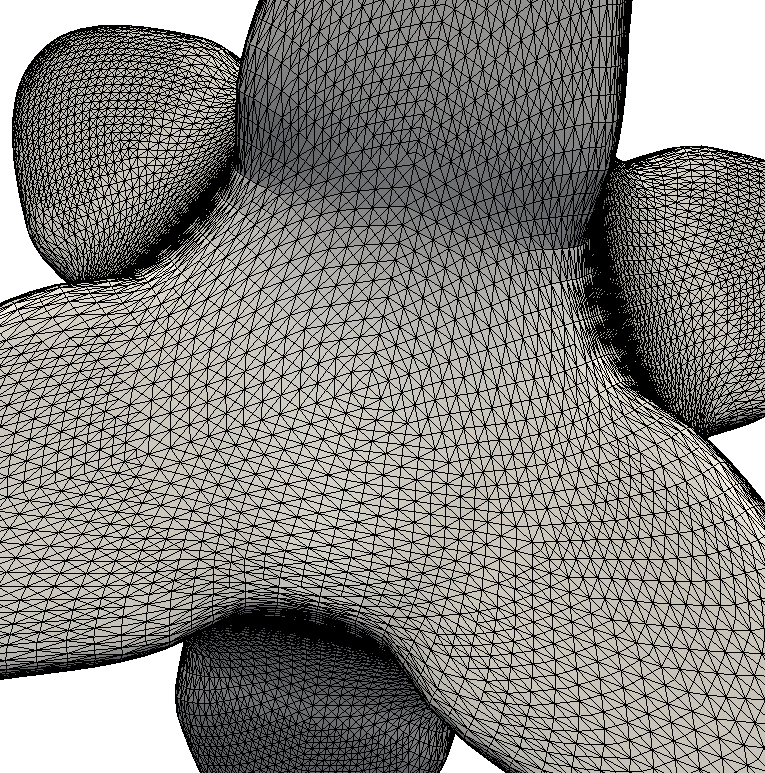}
\qquad
\includegraphics[angle=-0,width=0.35\textwidth]{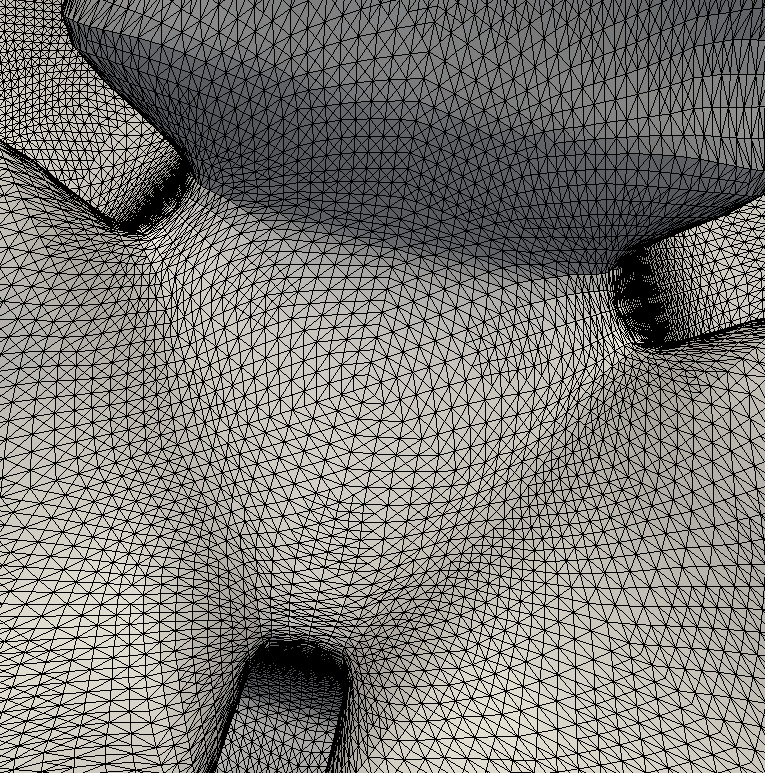} 
\fi
\caption{
Details of the triangulations of \PFEM\ in Figure~\ref{fig:dendritic} at times
$t=0.2$ (left) and $t=0.3$ (right).
}
\label{fig:mesh}
\end{figure}

The next computation is similar to the simulation shown in 
\cite[Fig.\ 14]{dendritic}, where here we take as anisotropy
\begin{equation} \label{eq:gocta}
 \text{\sc ani$_3$:} \quad
\gamma(\vec{p}) = \left( [g(\vec{p})]^9 +
[g(R_1\,\vec{p})]^9 +
[g(R_2\,\vec{p})]^9 \right)^\frac19\,,
\quad\text{where}~~
g(\vec{p}) := \left[ p_1^2 + \tfrac14\,(p_2^2 + p_3^2) \right]^{\frac12}\,,
\end{equation}
with the two rotation matrices defined as
$R_{1}:=\Bigl(\!\!\!\scriptsize
\begin{array}{rrr} 0 & 1 &0 \\
-1 & 0 & 0 \\ 0 & 0 & 1 \end{array}\!\! \Bigr)$ and 
$R_{2}:=\Bigl(\!\!\!\scriptsize
\begin{array}{rrr} 0 & 0 & 1 \\
0 & 1 & 0 \\ -1 & 0 & 0 \end{array}\!\! \Bigr)$. The Wulff shape of the
anisotropy (\ref{eq:gocta}) can be seen on the right of
Figure~3 in \cite{dendritic}.
Moreover, we use the parameters $\vartheta=1$, $\alpha=10^{-3}$,
$\rho = 0.01$ and $\uD = -\frac12$ on the boundary 
$\partial_D\Omega=\partial\Omega$ of $\Omega=(-4,4)^3$.
The radius of the initially spherical seed is chosen as $R_0=0.1$, and we 
let $u_0$ be defined by (\ref{eq:u0exp}) with $H=4$.
Three simulations for these parameters,
with the spatial discretization parameters fixed as
$h_\Gamma \approx 5\,h_f = \frac{\sqrt{2}}{64}$, 
are presented in Figure~\ref{fig:dendritic}.
We observe that the three simulations all show the same
general shape of the growing six-armed crystal, and for the smallest value of
$\tau$ the results appear to have converged. 
We also note that the small oscillations in the solution for 
the simulation with the largest time step size disappear as $\tau$ is 
decreased.
In order to demonstrate the good mesh properties of the parametric method
\PFEM, we show in Figure~\ref{fig:mesh} two details of the triangulated
approximation of $\Gamma(t)$ at times $t=0.2$ and $t=0.3$ for 
the finest time discretization in Figure~\ref{fig:dendritic}. 
We recall that the algorithm \PFEM\
does not employ any mesh-redistribution or mesh-smoothing methods. Rather it
relies solely on local mesh refinements, where individual elements of the
triangulation become too large, see \cite[\S5.2]{dendritic} for more details. 
The quality of the meshes shown in
Figure~\ref{fig:mesh} is excellent. 

\begin{figure}
\center
\ifpdf
\includegraphics[angle=-0,width=0.24\textwidth]{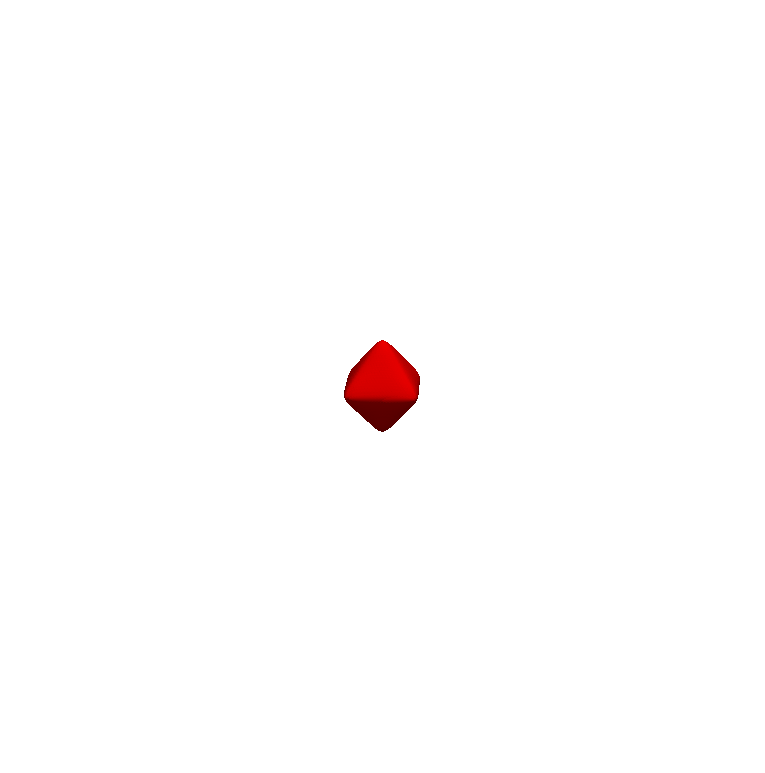}
\includegraphics[angle=-0,width=0.24\textwidth]{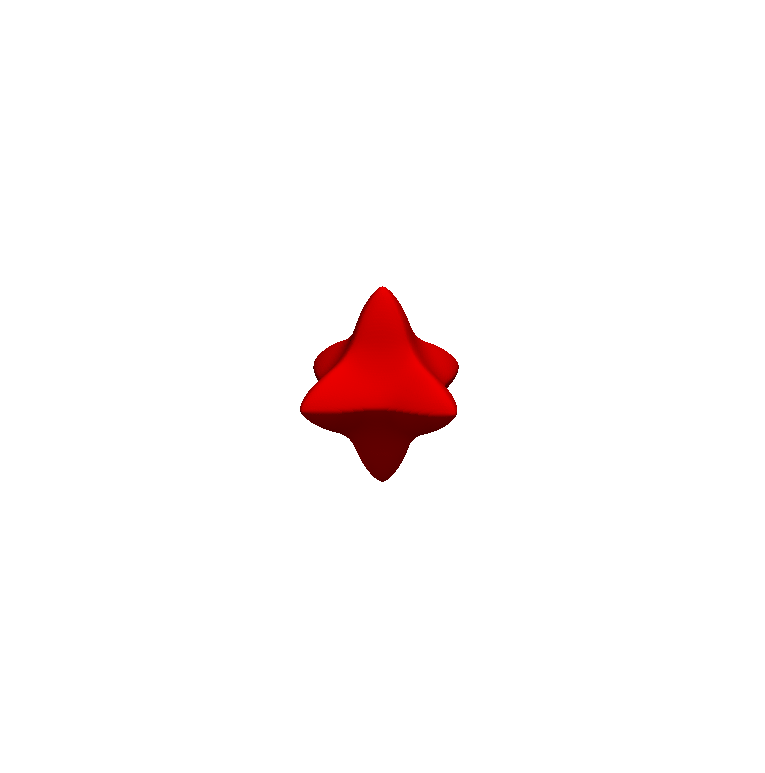}
\includegraphics[angle=-0,width=0.24\textwidth]{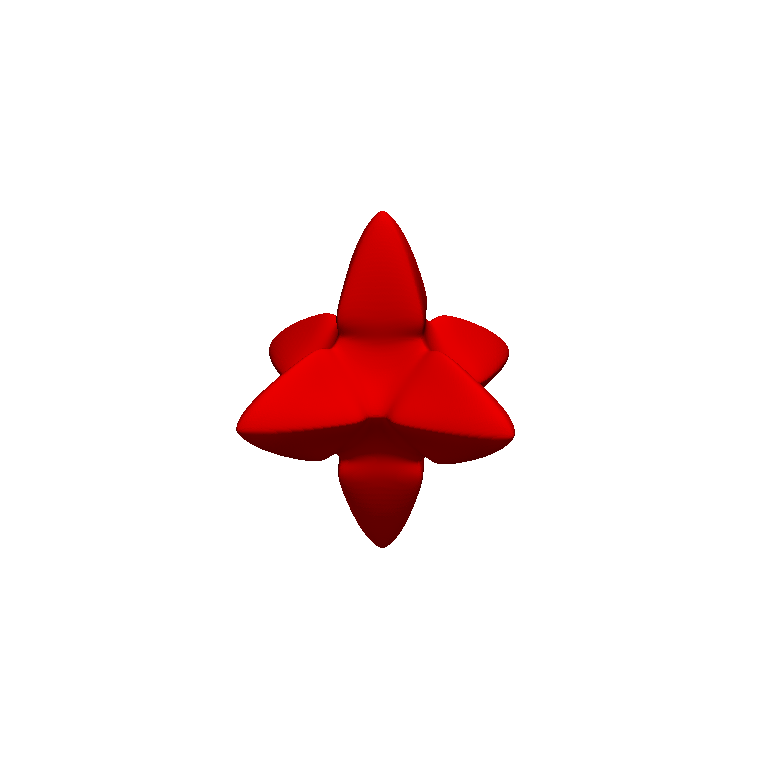}
\includegraphics[angle=-0,width=0.24\textwidth]{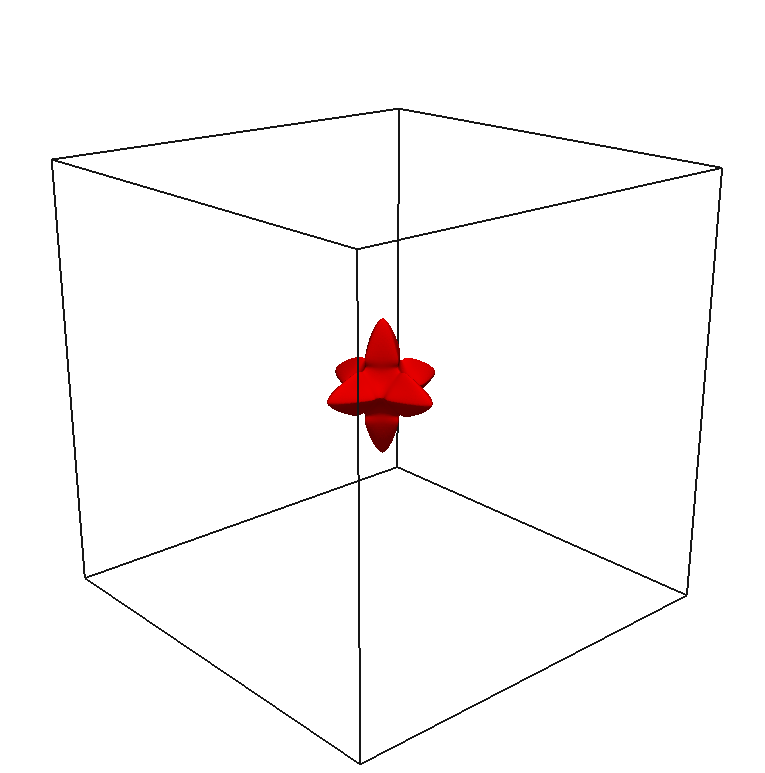}
\fi
\caption{(\PFii, $\epsilon^{-1} = 16\,\pi$, {\sc ani$_3$}, 
$\vartheta=1$, $\alpha=10^{-3}$, $\rho = 0.01$, $\uD = -\frac12$, 
$\Omega=(-4,4)^3$)
Snapshots of the solution at times $t=0.1,\,0.2,\,0.3$, as well as the
solution at time $T=0.3$ within $\overline\Omega$. This calculation uses
$\tau=10^{-4}$. 
[This computation took $8$ days.]
}
\label{fig:vch}
\end{figure}

We recall that a simulation for the phase field algorithm \PFii\ for the same
physical parameters has recently been performed in \cite[Fig.\ 23]{vch}. As a
comparison to the sharp interface calculations from Figure~\ref{fig:dendritic},
we present the results for the scheme \PFii\ in Figure~\ref{fig:vch}.
We note that the evolution shown for the phase field approximation in
Figure~\ref{fig:vch} is qualitatively very different from the sharp interface
simulations in Figure~\ref{fig:dendritic}. In all likelihood the physically
challenging parameters for the computation in Figure~\ref{fig:vch} mean that,
both in terms of the discretization parameters for the given
$\epsilon^{-1} = 16\,\pi$, e.g.\ the time step size $\tau$, 
as well as in terms of the interfacial parameter $\epsilon$ itself, the shown 
numerical results are still far from the true underlying solution to the sharp
interface problem (\ref{eq:1a}--e). Of course, a detailed numerical study into
this question is not yet possible due to the long time that such computations
would take. 

\setcounter{equation}{0}
\section*{Conclusions} \label{sec:concl}

While numerical simulations for phase field models in general show 
qualitatively correct behaviour, often such numerical results are far away 
from the true sharp interface evolution. In order to obtain accurate
simulations, the interface width $\epsilon$, as well as the spatial and 
temporal discretization parameters need to be chosen sufficiently small.
However, reducing these parameters to reach an acceptable accuracy
often requires very large computing times on even the most advanced of
today's desktop computers.

Direct sharp interface approximations, on the other hand, can provide a
computationally cheap method to compute interface evolutions in materials
science accurately. An example of such an algorithm is \PFEM\ from
\cite{dendritic,crystal}. In the computations presented in this paper we have
seen that even for very crude discretization parameters, the algorithm \PFEM\
provides surprisingly accurate approximations. 
The computational time needed to compute these sharp interface approximations 
is often negligible compared to the CPU times necessary for a corresponding 
phase field simulation.

The main problem of phase field methods is that the asymptotic error in
$\epsilon$, which in general is not known, plays a significant role in
determining the accuracy of phase field simulations. Small values of
$\epsilon$, in turn, require very small discretization parameters. 
Similarly, in second order convergent isotropic phase field models with a 
correction term, where the asymptotic error in $\epsilon$ may be assumed to 
be relatively smaller than in classical phase field models, very small
discretization parameters need to be employed in order to benefit from the
smaller asymptotic error in practice.
All of these issues do not arise in sharp interface approximations.

The main advantage of phase field methods over direct front tracking methods is
that they intrinsically allow for topological changes. However, for the problem
of solidification and dendritic growth as considered in this paper, topological
changes in general do not occur during the simulation of physically relevant
evolutions for single crystals. 

In the past, 
researchers and scientists may have been discouraged from
applying front tracking methods because of the difficulties in implementing
such methods and because of the deterioration of the mesh quality as the
approximated sharp interface evolves in time. 
However, assembling the system matrices in parametric finite element methods
for evolving manifolds is not much different from the assembly in standard
Cartesian problems, see e.g.\ \cite{DziukE07,DziukE07a}. Of course, the
coupling between a lower dimensional parametric mesh and a bulk mesh is
nontrivial, but successful implementations have been used in e.g.\
\cite{Schmidt93,Schmidt96,dendritic,crystal}.
Moreover, the good mesh properties of the scheme \PFEM\ from 
\cite{dendritic,crystal} mean that a good mesh quality is maintained throughout
the numerical simulations, and no remeshing is required in practice. In fact,
all the simulations presented in this paper were performed without any
remeshing, see \cite{dendritic} for more details.

We can summarize our main conclusions as follows:
\begin{itemize}
\item[(C1)] The parametric front tracking method \PFEM\ is more accurate and
computationally more efficient than phase field methods.
\item[(C2)] For isotropic problems, 
implicit time discretizations for phase field models
are often more accurate than semi-implicit time discretizations.
\item[(C3)] Explicit time discretizations for phase field models need very
small time steps in practice, and hence computations with explicit schemes
are only competitive if run in parallel on large clusters.
\item[(C4)] Second order accurate phase field models need finer discretization
parameters than classical phase field models in order to demonstrate their
superior approximation properties in practice.
\end{itemize}

Finally we note that while the focus of this paper has been the problem of
dendritic solidification, it is to be expected that similar conclusions can be
drawn when considering the respective merits of phase field models and sharp
interface methods for other free boundary problems in 
materials science, physics and
biology. As possible examples we refer to epitaxial growth, surface
diffusion, thermal grooving, sintering, vesicle dynamics and two phase
flow. 

It is our hope that the comparisons presented in this paper encourage a
discussion about the merits of phase field methods 
in general, and of the possible advantages of using sharp interface 
approximations instead. 

\end{document}